\documentclass{article}

\usepackage{PRIMEarxiv}

\usepackage[utf8]{inputenc} 
\usepackage[T1]{fontenc}    
\usepackage{hyperref}       
\usepackage{url}            
\usepackage{booktabs}       
\usepackage{amsfonts}       
\usepackage{nicefrac}       
\usepackage{microtype}      
\usepackage{lipsum}
\usepackage{fancyhdr}       
\usepackage{graphicx}       
\graphicspath{{media/}}     

\usepackage{amsmath}
\usepackage{graphicx,psfrag,epsf}
\usepackage{enumerate}
\usepackage{natbib}

\usepackage{siunitx}
\usepackage{multirow}
\usepackage{amssymb}
\usepackage{appendix}
\usepackage{xcolor}
\usepackage[linesnumbered,ruled,vlined]{algorithm2e}
\usepackage[switch,pagewise]{lineno}
\usepackage{makecell}

\newcommand{\KLD}[2]{\mathbb{D}_{\mathrm{KL}} \left( #1 \middle|\middle| #2 \right) }

\newcommand{\bez}{\mathbf{0}}

\newcommand{\diag}{\text{diag}}

\DeclareMathOperator*{\argmax}{argmax}

\title{Factor Analysis of Multivariate Stochastic Volatility Model}

\author{
  Taehee Lee \\
  Department of Earth, Environmental \& Planetary Sciences \\
  Brown University \\
  \texttt{taehee\_lee@brown.edu} \\
   \And
  Jun S. Liu \\
  Department of Statistics and Data Science \\
  Tsinghua University \\
  \texttt{junsliu@tsinghua.edu.cn} \\
}

\begin{document}
\maketitle

\begin{abstract}
Modeling the time-varying covariance structures of high-dimensional variables is critical across diverse scientific and industrial applications; however, existing approaches exhibit notable limitations in either modeling flexibility or inferential efficiency. For instance, change-point modeling fails to account for the continuous time-varying nature of covariance structures, while GARCH and stochastic volatility models suffer from over-parameterization and the risk of overfitting.
To address these challenges, we propose a Bayesian factor modeling framework designed to enable simultaneous inference of both the covariance structure of a high-dimensional time series and its time-varying dynamics. The associated Expectation-Maximization (EM) algorithm not only features an exact, closed-form update for the M-step but also is easily generalizable to more complex settings, such as spatiotemporal multivariate factor analysis. We validate our method through simulation studies and real-data experiments using climate and financial datasets.
\end{abstract}


\section{Introduction}\label{sec1}

Accurate modeling of covariance structures among large sets of measurements, where variables may exhibit time-varying correlations with each other, is critical in diverse scientific and industrial applications.
 In financial analysis, covariance estimation among asset returns is a key input for portfolio construction and risk management \citep{Markowitz1952, Fan2012}. In cognitive science, functional interaction of the human brain can be explained by estimating voxel-wise and region-wise correlations from fMRI time series \citep{Barnett2016}. In climatology, understanding the spatial and temporal correlation between multiple climate variables in various sites has been an important problem \citep{Banerjee2014}. In most of those applications, a key challenge is to understand how the covariance structure evolves as the time series unfolds by modeling and inferring the time-varying covariance function.

Estimating time-varying covariance matrices poses two key challenges.
First, the number of time series considered, $Q$, can be greater than the number of  observations, $N$, in each time series. This is known as the ``$Q>N$'' problem in high-dimensional statistics. Since the number of unknown parameters to estimate grows quadratically with $Q$,  the empirical estimate of the covariance matrix is singular without proper regularization in either the time domain or the structure of the covariance matrix itself. 
Second, the time-varying nature of the underlying covariance structure is often unknown {\it a priori}. An appropriate time-varying covariance model  should supply not only the richness in pattern matching, but also parsimony and interpretability of the inferred evolution over time.

In this work, we describe a Bayesian factor model for high-dimensional time series with  time-varying covariance structures, addressing both of the aforementioned challenges.
Our model assumes that changes in covariances are driven by those of the underlying low-dimensional factors. In contrast to  Principle Component Analysis (PCA)-type methods, our approach explicitly postulates a generative model driven by a few latent factors with a time-varying covariance structure. Our Bayesian analysis 
allows for simultaneous inference of the covariance structure in both the spatial and temporal domains by the expectation-maximization (EM) algorithm \citep{Dempster1977} with an \textit{exact} and \textit{closed-form} M-step update. Moreover, our simple factor model can be extended to more complicated settings, such as spatiotemporal factor analysis, by keeping both the interpretability and the computational convenience of a closed-form update through the expectation conditional maximization (ECM) algorithm \citep{Meng1993}.

\subsection{Review of Previous Studies}\label{sec1_1}

Statistical inference for time-varying high-dimensional covariance structures is a complex problem. Both the temporal and coordinate-wise aspects are to be simultaneously examined. We first review some existing approaches suggested by the literature to address either or both of these aspects as follows.

\paragraph{Stationary Factor Model.}

To address the high dimensionality of observations in the time-invariant covariance estimation, most current methodologies resort to factor models. A factor model assumes that individual observables are driven by a small number of common latent factors and a set of idiosyncratic errors \citep{Chen2020}. As a straightforward extension to time series analysis, factor models have been regarded as the primary method for covariance estimation of high-dimensional or matrix-variate time series in the statistical literature with adequate mathematical justifications \citep{Fan2013, Fan2021, Huang2021, Chen2023} and have been actively applied to economic and financial studies \citep{Chamberlain1983, Stock2002, Bai2002}.

\paragraph{Change-point Model.}
Change-point modeling might be the simplest method for considering the time-varying aspect of the covariance structure. A change-point model partitions the observed time frame into multiple windows and assumes that the time series is stationary within each one. Determining the number and exploring the locations of change points have been studied under general frameworks from both Bayesian and frequentist perspectives \citep{Fearnhead2006, Killick2012}, although only a few investigations have considered change-point models for the covariance structure of high-dimensional time series. For instance, \cite{Barigozzi2018} study high-dimensional time series factor models with multiple change-points in their second-order structure. \cite{Sun2017} and \cite{MA2018} consider a similar problem by exploiting multiple stage methodologies and combining PCA  analysis and change-point detection methods such as binary segmentation. \cite{Cabrieto2018} proposes a permutation-based significance test to detect abrupt correlation changes in multivariate time series. Despite its intuitiveness in modeling, such models are not designed to reflect continuity of the time-varying covariance structure and its proper inference is often computationally too expensive. Also, they are often incapable of detecting change points that only a small portion of variables show the covariance change.

\paragraph{GARCH and Stochastic Volatility Model.}
In the context of financial time series, the multivariate generalized autoregressive conditional heteroscedasticity (GARCH) and stochastic volatility (SV) model are popular alternatives to the change-point model for time-varying covariance modeling and has been much more thoroughly studied; see \cite{Asai2006} and \cite{Bauwens2006} for a review. Roughly speaking, the GARCH and SV models assume  an evolution process of the time-dependent covariance structure, thus the covariance at a given time point is a function of the previous time point and the realized observations. For example, the Dynamic Conditional Correlation model tackles the time-varying aspect of variances and correlations separately \citep{Engle2002, Tse2002}. 

However, when the dimensionality of multivariate time series exceeds a few hundred, most existing methods cannot provide satisfactory estimates of time-varying covariance structures, as highlighted in \cite{Pakel2021}. In particular, computational challenges arise when inverting large covariance matrices, which is required as part of the likelihood evaluation in many methods. Computationally cheaper variations often require additional assumptions. For example, \cite{Bollerslev1990} assumes time-varying covariance but constant conditional correlation, whereas \cite{Engle2007} requires that the selected dynamic conditional correlation model be correctly specified for all bivariate pairs. Another line of work adopts the factor model structure for multivariate time series and incorporates GARCH and SV modeling on the latent factor components. For example, \cite{Aguilar2000} and \cite{Chib2006} combine the features of the factor model with those of the univariate SV model. In practice, GARCH and SV models are generally preferred over change-point models for financial time series, due to the fact that volatility indeed varies continuously over time. However, over-parameterization and excessive assumptions of the GARCH and SV models are often inevitable to simultaneously sketch and estimate an ever-changing evolution of the covariance structure.

\subsection{Paper Organization}\label{sec1_2}

The remainder of this paper is organized as follows. Section \ref{sec2} introduces our Gaussian distribution-based Bayesian Factor (GBF) model for high-dimensional time series with time-varying volatility evolution. We discuss the prior assumptions on the model parameters, the precision process together with the weight functions, and the rule on deciding the number of factors. Section \ref{sec3} discusses the expectation-maximization (EM) algorithm \citep{Dempster1977} for simultaneously estimating the model parameters and inferring the precision process. Section \ref{sec4} deals with generalizations of Gaussian factor analysis, including Student's t-factor analysis and spatiotemporal factor analysis. Section \ref{sec5} presents simulation studies and Section \ref{sec6} showcases three applications of the proposed GBF model:  the daily temperature changes of the 50 US airports; a spatiotemporal comprehensive climate dataset, and the S\&P 100 stock daily returns (from 04/01/2019 to 10/01/2024). Section \ref{sec7} concludes with a short discussion.

\section{Bayesian Factor Model for Covariance Estimation}\label{sec2}
\subsection{A Bayesian factor model}\label{sec2_1}
Without loss of generality, we assume that the time series has mean $\bez$. For a given time $t\in\mathcal{T}=\left\{t_{1},\cdots,t_{N}\right\}$, the $Q$-dimensional time series $y_t\in\mathcal{Y}$ is driven by $K$ common factors, with $K \ll Q$. For the $Q\times K$ matrix of factor loadings $B$, we assume the following model structure for a $N$-long time series:
\begin{equation}
\begin{aligned}
y_t = Bf_t+\epsilon_t, \quad t \in \mathcal{T},
\end{aligned}
\label{equation_2_1_1}
\end{equation}
where $\epsilon_t$ is the vector of idiosyncratic errors, follows $\mathcal{N}\left(\bez,\Sigma\right)$ with a nonnegative \textit{diagonal} matrix $\Sigma=\text{diag}\left\{\Sigma_{1},\cdots,\Sigma_{Q}\right\}$, and the vector of latent factors $f_t\sim \mathcal{N}\left(\bez,\Lambda_t\right)$ is assumed to be distributed \textit{independently}
with a time-varying $K\times K$ covariance matrix $\Lambda_t$. Note that, by marginalizing over the latent factors $f_t$, we arrive at the following sampling model for the observable $y_t$ given $B$, $L$ and $\omega$:
\begin{equation}
\begin{aligned}
\left. y_t\mid B,L,\omega \right. \stackrel{i.i.d.}{\sim} \mathcal{N}\left(\bez,B\Lambda_t B^\top+\Sigma\right).
\end{aligned}
\label{equation_2_1_4}
\end{equation}

In the trivial case where we can observe $f_{t_1},\ldots, f_{t_N}$, we may estimate $\Lambda_t$ at any given time by local average. 
That is, we have a window $[t-h,t+h]$ for $h>0$, and use the data points in this window to estimate the covariance matrix $\Lambda_t$, assuming that it is approximately invariant in this region. An alternative  strategy that avoids explicitly selecting the window size is to use an exponential moving average approach \citep{Yin2010}. That is, the contribution of each data $y_{t_n}$ to the estimation of $\Lambda_t$ is weighted by $\exp\left(-\gamma |t-t_n|\right)$ for some $\gamma>0$. This latter approach can be viewed as a special case of the more general kernel-based method named the Nadaraya–Watson estimator \citep{Bierens1994,LANGRENE2019}, i.e.,
\begin{equation}
\begin{aligned}
\hat{\Lambda}_t = \frac{\sum_{n=1}^{N}{\exp\left(-\gamma |t-t_n|\right) f_{t_n} f_{t_n}^\top}}{\sum_{n=1}^{N}{\exp\left(-\gamma |t-t_n|\right)}} .
\end{aligned}
\label{equation_2_1_2}
\end{equation}

However, in our setting, this classic approach encounters two main issues: 
(a) the factors $f_{t_n}$ are not observable; and (b) this approach cannot be coherently incorporated into a likelihood-based estimation method.  To overcome these challenges in modeling the time-varying nature of $\Lambda_t$ flexibly, we assume that it can be represented as 
a weighted {\it harmonic average} of a set of ``basis covariance matrices.'' That is, we can write 
\begin{equation}
\begin{aligned}
\Lambda_t \triangleq \left(\sum_{d=1}^{D}{\omega_{d}} {\left(t\right) \lambda_{d}^{-1}}\right)^{-1},
\end{aligned}
\label{equation_2_1_3}
\end{equation}
where $\omega(t) = \left\{\omega_{d}(t)\right\}_{d=1}^{D}$ is a set of weight functions with $\omega_{d}(t)\geq 0$ 
such that $\sum_{d=1}^{D}\omega_{d}{\left(t\right)}$=1, $\forall t>0$, and $L=\left\{\lambda_d\right\}_{d=1}^{D}$ is a set of $K\times K$ basis covariance matrices, which will be  estimated from the data. 
Although this over-parameterized modeling approach offers a great flexibility, we need to 
control these $\lambda_d$'s in order for them to behave properly.  A natural strategy  is to assume a joint prior distribution for the $\lambda_d$'s as in \eqref{equation_2_2_1_2}.
 
Note that the model \eqref{equation_2_1_3} is not only defined for each $t \in \mathcal{T}$ but also for \textit{any} $t \in \mathbb{R}$ where we may not have an observation since the weight function $\omega_d(t)$ can extrapolate. Thus, although \eqref{equation_2_1_3}  takes the form of interpolation and does not describe how $\Lambda_t$ evolves to $\Lambda_{t+\Delta t}$ in an explicit and analytical way, it can still be used to predict future volatility. As shown in  Section~\ref{sec6_1}, our prediction for the covariance matrices underlying the  S\&P 100 stocks is superior to that based on the popular exponential moving average approach.

\subsection{Priors  for the  key parameters}\label{sec2_2} \label{sec2_2_4}\label{sec2_2_2}\label{sec2_2_3}

For now, let us assume that the dimension of the latent factor space $K$ is a known constant and the set of weight functions $\omega$  is given {\it a priori}. The choice of $K$ and the definition of the weight functions $\omega_d(t)$ will be addressed in Sections \ref{sec2_3}.

\paragraph{Basis covariances.}

We assume that the basis covariances $L=\left\{\lambda_d\right\}_{d=1}^{D}$ given the set of weight functions $\omega = \left\{\omega_d\right\}_{d=1}^{D}$ and time indices $\mathcal{T}$ follow the distribution \eqref{equation_2_2_1_2} {\it a priori}:
\begin{equation}
\begin{aligned}
p\left(L\mid \omega,\mathcal{T}\right)\propto \exp\left(\frac{1}{2}\sum_{n=1}^{N}\sum_{d=1}^{D}{\omega_{d}{\left(t_n\right)}\log{\left|\lambda_d^{-1}\right|}}-\frac{1}{2}\sum_{n=1}^{N}\log{\left|\sum_{d=1}^{D}{\omega_{d}{\left(t_n\right)}\lambda_d^{-1}}\right|}\right).
\end{aligned}
\label{equation_2_2_1_2}
\end{equation}
Although this  improper prior looks a bit arbitrary, it has the following properties:

\begin{enumerate}
  \item[(a)] If $D = 1$, then $\omega_1 \equiv 1$ by definition. Thus, $p\left(L\middle|\omega,\mathcal{T}\right) \propto 1$, so it is improper and the corresponding model is nothing but a homoscedastic factor model and $\Lambda \equiv \lambda_1$.
  \item[(b)] By Jensen's inequality, the following inequality always holds:
  \begin{equation}
 \sum_{d=1}^{D}{\omega_{d}{\left(t_n\right)}\log{\left|\lambda_d^{-1}\right|}} \leq 
\log{\left|\sum_{d=1}^{D}{\omega_{d}{\left(t_n\right)}\lambda_d^{-1}}\right|}, \ \ \forall n.
  \label{equation_2_2_1_3}
  \end{equation}
  Thus, \eqref{equation_2_2_1_2} has a mode when the equality holds in \eqref{equation_2_2_1_3}. If all $w_d(t)>0$, then the equality holds only when all the $\lambda_d$'s are identical (time-invariant). This implies that this prior encourages more tightly distributed $\lambda_d$'s.
  
  \item[(c)] The exponential term in \eqref{equation_2_2_1_2} is invariant under the transformation $L=\left\{\lambda_d\right\}_{d=1}^{D}\rightarrow C L C^\top\triangleq\left\{C \lambda_d C^\top\right\}_{d=1}^{D}$ for a $K\times K$ invertible matrix $C$. That is, our prior on $L$ does not affect the identifiability of $B \Lambda_t B^{\top}$ in the model \eqref{equation_2_1_4}. 

  \item[(d)] If all $\lambda_d$'s are strictly positive definite matrices,  so is every $\Lambda_t$, which then has the \textit{unique} Cholesky decomposition (up to the sign); therefore, $B \Lambda_t B^\top = \mathcal{B}_t \mathcal{B}_{t}^\top$ is identifiable without affecting the prior on $L$. In other words, our model can also be interpretable as a time-varying factor loading matrix model.

  \item[(e)] The form of prior \eqref{equation_2_2_1_2} ``replaces'' the log-of-sum with the sum-of-log in the objective function in estimation, thus contributing to a closed-form M-step update in the EM algorithm: see Section \ref{sec3} and Appendix \ref{secA} for more details.
\end{enumerate}

In Section~\ref{sec4_4}, we discuss more ways to restrict the $\lambda_d$'s. One possibility is to multiply independent inverse-Wishart priors, one for each $\lambda_d$, to \eqref{equation_2_2_1_2} so that the resulting form becomes a proper prior distribution. This modified prior can also facilitate a quick and explicit EM update.


\paragraph{Factor loading matrix $B$.}
Here we just give an improper prior $p\left(B\mid K\right)\propto 1$ for the factor loading matrix $B$. It is well-known that a strong non-identifiability issue exists in factor models: the factors can be arbitrarily rescaled and rotated (and the corresponding loading matrix counter-rotated and scaled) without affecting the likelihood of the observed data. Sparsity usually becomes a key to the identifiability of the model parameters. According to aforementioned property (c), together with our prior on $L$, our joint prior for $(L,B)$ is \textit{invariant} for any transformation of the form 
\begin{equation}
\left(L,B\right)\rightarrow\left(\hat{L},\hat{B}\right) \triangleq \left( C L C^{\top}, B C^{-1} \right)
\label{eq:transform}
\end{equation}
for any $K \times K$ invertible matrix $C$. 
Thus,  once all parameters are estimated using the EM algorithm (see Section~\ref{sec3}), we consider a matrix $C$  to do the transformation in \eqref{eq:transform} so that $\hat{B}$ becomes a sparse matrix (see Appendix \ref{secE} for details). 





\paragraph{Idiosyncratic covariance matrix $\Sigma$.}
As with the factor loading matrix $B$, we give an improper prior $p\left(\Sigma\right) \propto 1$ to the diagonal idiosyncratic covariance matrix $\Sigma$. We note that the model can be further extended to allow $\Sigma$ to be dependent on $t$, i.e., to be written as a time-varying covariance function $\Sigma_t$, similarly to how we treat the factor covariance matrix $\Lambda_t$  as time-varying: see Section \ref{sec4_2} for more details.

\subsection{Specifications for other parameters}\label{sec2_4}\label{sec2_3}

\paragraph{Weight functions.}
We define the weight functions in $\omega = \left\{\omega_{d}\right\}_{d=1}^{D}$ as follows:
\begin{equation}
\omega_{d}{\left(t\right)} \triangleq\frac{\mathcal{K}_{h_d}\left(t,s_d\right)}{\sum_{c=1}^{D}{\mathcal{K}_{h_c}\left(t,s_c\right)}}\propto\mathcal{K}_{h_d}\left(t,s_d\right),
\label{equation_2_3_1}
\end{equation}
for a density kernel $\mathcal{K}$ with a set of bandwidth parameters $\mathcal{H} = \left\{h_d\right\}_{d=1}^{D}$ with $h_d>0$ and $\mathcal{S} = \left\{s_d\right\}_{d=1}^{D}$ with $ s_d\in\mathbb{R}$. The smoothness of time-varying volatility is controlled (or tunable) by the choice of $\left(\mathcal{K},\mathcal{H},\mathcal{S}\right)$. We assume that the density kernel and $\mathcal{S}$ are given {\it a priori} and fixed, but the bandwidths are to be estimated (or tuned) by a version of leave-one-out cross-validation (see Appendix \ref{secD} for details). One practical setting is $\mathcal{K}_{h_d}{\left(t,s_d\right)} \triangleq \exp{-\left[\left(t-s_d\right)^{2}/h_d^2\right]} $, $\mathcal{S} \equiv \mathcal{T}$ and $h_0 \equiv h_d$ and just estimate a single $h_0$, which is our \textit{default setting} in Section \ref{sec5} and \ref{sec6}. Devising criteria for deciding the ``optimal'' $\mathcal{S}$ and $h$ is beyond the scope of this work.


\paragraph{Number of factors $K$.}
We have discussed the model under the assumption that the number of latent factors is known {\it a priori}, which is generally unrealistic. We now discuss how to determine the dimensionality $K$ of the latent factor space. For the \textit{time-invariant} factor model (i.e., $\Lambda_t \equiv \Lambda_0$), the number of free parameters is $Q\left(K+2\right) - K\left(K-1\right)/2$ \citep{Zhao2024}, which brings forth the following prior on the number of active factors based on the Bayesian information criterion (BIC) \citep{Schwarz1978}:
\begin{equation}
\begin{aligned}
p\left(K\right)\propto \exp\left(-\frac{1}{2}{\left(Q\left(K+2\right)-\frac{1}{2} K\left(K-1\right)\right)\log{N}}\right),
\end{aligned}
\label{equation_2_4_1}
\end{equation}
where $N$, $K$ and $Q$ are the length of the observed time series, the number of factors, and the dimensionality of time series, respectively. However, our model also assumes the \textit{time-varying} $\Lambda_t$ as in \eqref{equation_2_1_3}. Thus, we cannot simply apply \eqref{equation_2_4_1} in penalizing the number of factors $K$. Instead of justifying a ``plausible'' prior, we just employ a practical alternative, cross-validation, to determine the number of factors $K$. To be specific, first, we randomly split the dataset into two parts: one for training and the other for validation. Then, we compute the likelihood of the validation set based on the model and parameters estimated by the training dataset. After repeating this step multiple times (with a fixed portion, e.g., 20\%, of the whole data as the validation set) for each candidate of $K$, we pick the one that maximizes the average predictive likelihood. This is feasible because our method works for arbitrarily spaced time indices $\mathcal{T}$ and interpolates for unobserved $t$.

\section{The EM Approach to Bayesian Factor Analysis} \label{sec3}

For simplicity, we let $\mathcal{Y}_0$ be the set of observed time series in time indices $\mathcal{T}_0$, and split them into disjoint subsets: $\mathcal{Y}_0=\mathcal{Y}\cup\mathcal{Y}^{\prime}$ and $\mathcal{T}_0=\mathcal{T}\cup\mathcal{T}^{\prime}$, where $\mathcal{Y}=\left\{y_{t_n}\right\}_{n=1}^{N}$ and $\mathcal{T}=\left\{t_n\right\}_{n=1}^{N}$ are for training and $\mathcal{Y}^{\prime}$ and $\mathcal{T}^{\prime}$ are for validating.
Here, the split can be at random if we are not interested in preserving the time/spatial structure of the data, or can be blockwise if we want to retain structural information during estimation and validation. The ratio $\left|\mathcal{Y}^{\prime}\right| / \left|\mathcal{Y}_0\right|$ remains the same for different splits.

\begin{figure}[!htb]
\renewcommand{\baselinestretch}{1}
\centering
\includegraphics[width=0.3\textwidth]{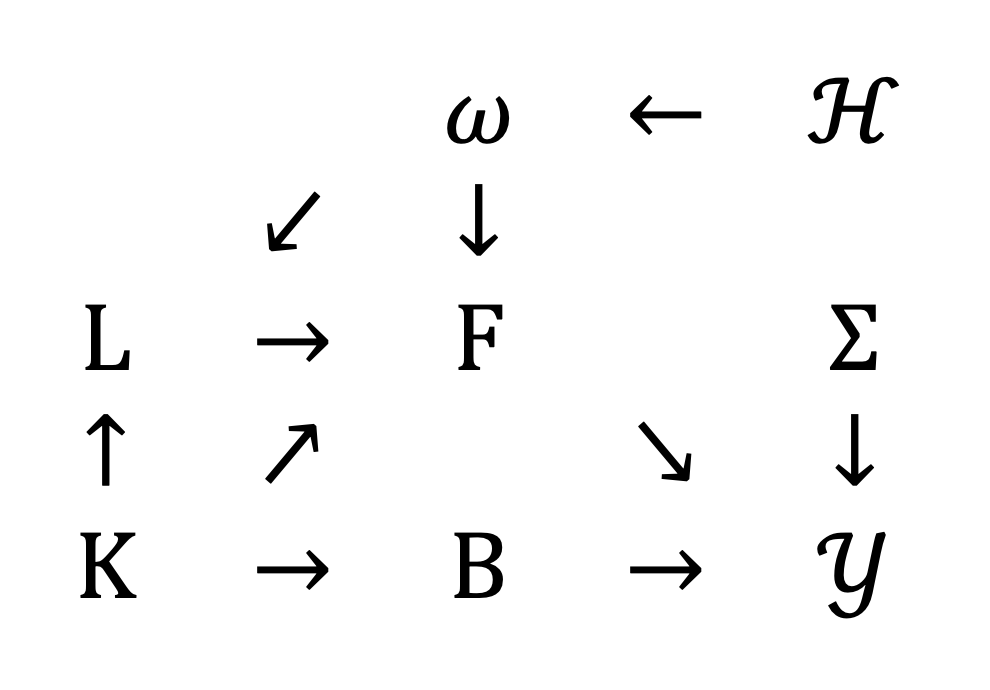}
\caption{A graphical depiction of the proposed Bayesian factor model.}
\label{fig_3_0_1}
\end{figure}

Let $F=\left\{f_n\right\}_{n=1}^{N}$ be the values of the latent factors associated with the set of time indices $\mathcal{T}$ and observations $\mathcal{Y}$, where each $f_n$ is of dimension $K$. For each $t \in \mathcal{T}$, the sampling distribution of $y_t$ is governed by parameters $L=\left\{\lambda_d\right\}_{d=1}^{D}$ and $B=\left[B_{qk}\right]_{Q\times K}$ as given in \eqref{equation_2_1_4}. We assume a finite candidate set $\mathbb{K}$ for the factor dimension $K$ (the number of factors). For each $K\in\mathbb{K}$, we 
treat the set of latent factors $F$ as missing variables and regard $L$, $B$ and $\Sigma$ as parameters of interest. The bandwidth $\mathcal{H}$ that defines the set of weight functions $\omega = \left\{\omega_{d}\right\}_{d=1}^{D}$ is assumed to be given, and we employ the priors introduced in Section~\ref{sec2_2}. The relationship of the model parameters is shown in Figure \ref{fig_3_0_1}.

Our model postulates the following specific relationships:  
\begin{equation}
\begin{aligned}
p\left(\mathcal{Y}\middle|F,B,\Sigma\right) & =\prod_{n=1}^{N}{\mathcal{N}\left(y_{t_n}\middle|Bf_n,\Sigma\right)} \\ p\left(F\middle|L,\mathcal{T},\omega,K\right) & =\prod_{n=1}^{N}{\mathcal{N}\left(f_n\middle|0,\Lambda_{t_n}\right)}=\prod_{n=1}^{N}{\mathcal{N}\left(f_n\middle|0,\left(\sum_{d=1}^{D}{\omega_{d}{\left(t_n\right)}\lambda_d^{-1}}\right)^{-1}\right)} \\ p\left(L\middle|\mathcal{T},\omega,K\right) & \propto \exp\left(\frac{1}{2}\sum_{n=1}^{N}\sum_{d=1}^{D}{\omega_{d}{\left(t_n\right)}\log{\left|\lambda_d^{-1}\right|}}-\frac{1}{2}\sum_{n=1}^{N}\log{\left|\sum_{d=1}^{D}{\omega_{d}{\left(t_n\right)}\lambda_d^{-1}}\right|}\right) \\ p\left(B\middle|K\right) & \propto 1, \quad p\left(\Sigma\right) \propto 1.
\end{aligned}
\label{equation_3_0_2}
\end{equation}
Let $\Omega \triangleq \left(L,B,\Sigma\right)$. The joint distribution of $(\mathcal{Y}, F, \Omega)$  conditional on $(\mathcal{T} , \omega,K)$ is
\begin{equation}
\begin{aligned}
p\left(\mathcal{Y},F,\Omega\middle|\mathcal{T},\omega,K\right) = p\left(\mathcal{Y}\middle|F,B,\Sigma\right)p\left(F\middle|L,\mathcal{T},\omega,K\right)p\left(L\middle|\mathcal{T},\omega,K\right) p\left(B\middle|K\right) p\left(\Sigma\right),
\end{aligned}
\label{equation_3_0_1}
\end{equation}
Given the previously updated set of parameters $\Omega_K^{\left(r\right)}$ at iteration $r$,
we have \textit{closed-form} M-step updates for each basis covariance matrices $\lambda_{d}^{\left(r+1\right)}\in L^{\left(r+1\right)}$, factor loading $B^{\left(r+1\right)}$, and each diagonal entry of $\Sigma^{\left(r+1\right)}$ as follows:

\begin{equation}
\begin{aligned}
\lambda_{d}^{\left(r+1\right)} & =\frac{\sum_{n=1}^{N}{\omega_{d}{\left(t_n\right)}\left({\hat{\eta}}_{n}{\hat{\eta}}_{n}^\top+{\hat{\Psi}}_{n}\right)}}{\sum_{n=1}^{N}\omega_{d}{\left(t_{n}\right)}} \\ B^{\left(r+1\right)} & =\left(\sum_{n=1}^{N}{y_{t_n}{\hat{\eta}}_{n}^\top}\right)\left(\sum_{n=1}^{N}{\left({\hat{\eta}}_{n}{\hat{\eta}}_{n}^\top+{\hat{\Psi}}_{n}\right)}\right)^{-1} \\ \Sigma_{q}^{\left(r+1\right)} & = \frac{1}{N} \sum_{n=1}^{N}{\left(\left(y_{{t_n}q}-B_{q,:}^{\left(r+1\right)}{\hat{\eta}}_{n}\right)^2+B_{q,:}^{\left(r+1\right)}{\hat{\Psi}}_{n}\left(B_{q,:}^{\left(r+1\right)}\right)^\top\right)},
\end{aligned}
\label{equation_3_2_1}
\end{equation}
where ${\hat{\eta}}_{n}\triangleq{\hat{\Psi}}_{n}\left(B^{\left(r\right)}\right)^\top\left(\Sigma^{\left(r\right)}\right)^{-1}y_{t_n}$ and ${\hat{\Psi}}_{n}\triangleq\left(\left(\Lambda_{t_n}^{\left(r\right)}\right)^{-1}+\left(B^{\left(r\right)}\right)^\top\left(\Sigma^{\left(r\right)}\right)^{-1}B^{\left(r\right)}\right)^{-1}$. Detailed derivation can be found in Appendix \ref{secA}.

Given $K\in\mathbb{K}$, once the set of parameters $\Omega={\hat{\Omega}}_K$ is estimated and the weight functions $\omega=\omega_K$ are chosen for each disjoint split $\mathcal{Y}_0=\mathcal{Y}\cup\mathcal{Y}^{\prime}$ and $\mathcal{T}_0=\mathcal{T}\cup\mathcal{T}^{\prime}$, we compute the following value:
\begin{equation}
\begin{aligned}
\mathcal{V}\left(\mathcal{Y}^{\prime},\mathcal{T}^{\prime},K\right) \triangleq \log{p\left(\mathcal{Y}^{\prime}\middle|{\hat{\Omega}}_K,\mathcal{T}^{\prime},\omega_K,K\right)} = 
\sum_{t\in\mathcal{T}^{\prime}}\log{\mathcal{N}\left(y_t\middle|0,{\hat{B}}_K{\hat{\Lambda}}_{K}{\left(t\right)}{\hat{B}}_K^\top+{\hat{\Sigma}}_K\right)}.
\end{aligned}
\label{equation_3_4_1}
\end{equation}

By repeating the EM algorithm for multiple random disjoint splits $\mathcal{Y}_0=\mathcal{Y}^{\left(s\right)}\cup{\mathcal{Y}^{\prime}}^{\left(s\right)}$ and $\mathcal{T}_0=\mathcal{T}^{\left(s\right)}\cup{\mathcal{T}^{\prime}}^{\left(s\right)}$, $s=1,\cdots,S$, we obtain the following empirical expectation $\mathcal{V}\left(K\right)$:
\begin{equation}
\begin{aligned}
\mathcal{V}\left(K\right) \triangleq \frac{1}{S}\sum_{s=1}^{S}\mathcal{V}\left({\mathcal{Y}^{\prime}}^{\left(s\right)},{\mathcal{T}^{\prime}}^{\left(s\right)},K\right)\approx \left<\mathcal{V}\left(\mathcal{Y}^{\prime},\mathcal{T}^{\prime},K\right)\right>_{\left(\mathcal{Y}^{\prime},\mathcal{T}^{\prime}\right)}.
\end{aligned}
\label{equation_3_4_2}
\end{equation}
We choose $K\in\mathbb{K}$ that maximizes $\mathcal{V}\left(K\right)$ and run the EM algorithm again for the full training dataset $\left(\mathcal{Y}_0,\mathcal{T}_0\right)$ to estimate the parameters and weight functions. Algorithm \ref{algorithm1} summarizes what has been described so far.

\begin{algorithm}
\SetAlgoLined
\DontPrintSemicolon
 \For{each $K\in\mathbb{K}$}{
  \For {each training-validating split $\mathcal{Y}_{0} = \mathcal{Y} \cup \mathcal{Y}^{\prime}$ and $\mathcal{T}_{0} = \mathcal{T} \cup \mathcal{T}^{\prime}$}{
  \textbf{initialize} the set of parameters $\Omega_K=\Omega_K^{\left(0\right)}$ and the bandwidth $\mathcal{H}$.\;
  \While{convergence}{
  \textbf{(Bandwidth) update} the bandwidth $\mathcal{H}$ that maximizes \eqref{equation_3_3_1}.\;
  \textbf{(E-step) compute} $\left\{{\hat{\eta}}_n,{\hat{\Psi}}_n\right\}_{n=1}^{N}$ given $\Omega_K^{\left(r\right)}$ by \eqref{equation_3_1_1}.\;
  \textbf{(M-step) update} $\Omega_K=\Omega_K^{\left(r+1\right)}$ given $\left\{{\hat{\eta}}_n,{\hat{\Psi}}_n\right\}_{n=1}^{N}$ by \eqref{equation_3_2_1}.\;
  }
  \textbf{compute} $\mathcal{V}\left( \mathcal{Y}^{\prime}, \mathcal{T}^{\prime} ,  K\right)$ given ${\hat{\Omega}}_K$ and $\omega_K$ (function of $\mathcal{H}$) by \eqref{equation_3_4_1}.\;
 }
 \textbf{compute} $\mathcal{V}\left(K\right)$ by \eqref{equation_3_4_2}.\;
 }
 \textbf{choose} $\hat{K}=\argmax_{K}{\mathcal{V}\left(K\right)}$. \;
 \textbf{run} the above algorithm for the fixed $K=\hat{K}$ and the whole $\mathcal{Y}_{0}$ and $\mathcal{T}_{0}$ as the training dataset to estimate $\Omega_{\hat{K}} ={\hat{\Omega}}_{\hat{K}}$. \;
 \textbf{identify} $\left(L,B\right)\in{\hat{\Omega}}_{\hat{K}}$ by following steps in Appendix \ref{secE}.\;
 \caption{Gaussian Factor Analysis Model}
\label{algorithm1}\end{algorithm}

\section{Model Extensions} \label{sec4}

Our Gaussian factor model in Section \ref{sec2} is the simplest setting designed for conveying the core idea. We here discuss several generalizations of this base model.

\subsection{Robust Factor Model} \label{sec4_0}

In Section \ref{sec2}, we assume that $y_t \stackrel{i.i.d.}{\sim} \mathcal{N}\left(0,B\Lambda_tB^\top+\Sigma\right)$. This may not be proper in practice, especially if a portion of the observed $y_t$’s are outliers or when the error distribution has a heavier tail than Gaussian. A natural idea for making the Gaussian model robust is to replace it with a (multivariate) Student’s t-distribution, i.e., $y_t \stackrel{i.i.d.}{\sim} \mathcal{T}_\upsilon\left(0,B\Lambda_t B^\top+\Sigma\right)$, where $\upsilon$ is its degrees of freedom.

It is well-known that a Student’s t-distribution can be represented as an average of scaled Gaussian distributions. That is, $Z \sim  \mathcal{T}_\upsilon\left(\mu,\Sigma\right)$ can be perceived as having a latent scale $a$, i.e., $Z\left|a\right. \sim \mathcal{N}\left(\mu,a\cdot\Sigma\right)$  and $a \sim \mathcal{IG}\left(\frac{\upsilon}{2},\frac{\upsilon}{2}\right)$ \citep{Peel2000}. Thus, we have the following extended model for $\mathcal{Y}=\left\{y_t\right\}_{t \in \mathcal{T}}$: for a new set of positive random variables $A=\left\{\alpha_t\right\}_{t \in \mathcal{T}}$,
\begin{equation}
\begin{aligned}
\begin{matrix}y_t \mid f_t, \alpha_t \sim \mathcal{N}\left(Bf_t,\alpha_t\cdot\Sigma\right)\\
f_t \mid \alpha_t \sim \mathcal{N}\left(0,\alpha_t\cdot\Lambda_t\right)\\
\alpha_t \sim \mathcal{IG}\left(\frac{\upsilon}{2},\frac{\upsilon}{2}\right)\\\end{matrix}, \quad t \in \mathcal{T}.
\end{aligned}
\label{equation_4_0_1}
\end{equation}
By treating both $A$ and $F$ as missing variables and $\upsilon$ as the given value, we are still able to derive the \textit{closed}-form solution for the M-step: details can be found in Appendix \ref{secC_1}.

\subsection{Spatiotemporal Multivariate Factor Model} \label{sec4_1}

So far, we have discussed the case that each $y_t \in \mathcal{Y} \subset \mathbb{R}^{Q}$ consists of $Q$ time-varying correlated values. Now suppose that each $y_t$ is a $Q \times P$ \textit{matrix} for $P>1$ and
\begin{equation}
{\rm{vec}}{\left(y_t\right)} \ \sim \ \mathcal{N}{\left( 0, \left(C \otimes B \right) \Lambda_{t} {\left(C \otimes B \right)}^\top + \Phi \otimes \Sigma \right)},
\label{equation_4_1_0}
\end{equation}
where $P \times K_P$ and $Q \times K_Q$ matrix parameters $C$ and $B$ are factor loadings, $\Phi$ and $\Sigma$ are $P \times P$ and $Q \times Q$ diagonal matrix parameters with positive diagonal entries, and $\Lambda_t$ is a time-varying $K_P K_Q \times K_P K_Q$ covariance matrix function as \eqref{equation_2_1_3}. In other words, $\mathcal{Y}$ forms a \textit{spatiotemporal} time series with $Q$ outputs over $P$ spatial and $N$ temporal specifications.

Furthermore, one can instead consider the following formulation:
\begin{equation}
{\rm{vec}}{\left(y_t\right)} \ \sim \ \mathcal{N}{\left( 0, C \Gamma_{t} C^\top \otimes B \Lambda_{t} B^\top + \Phi \otimes \Sigma \right)},
\label{equation_4_1_1}
\end{equation}
where $\Gamma_t$ and $\Lambda_t$ are time-varying $K_P \times K_P$ and $K_Q \times K_Q$ covariance matrix functions defined as follows:
\begin{equation}
\begin{aligned}
\Gamma_t \triangleq \left(\sum_{d^{\prime}=1}^{D_{\Gamma}}{\rho_{d^{\prime}}} {\left(t\right) \cdot \gamma_{d^{\prime}}^{-1}}\right)^{-1}, \quad \Lambda_t \triangleq \left(\sum_{d=1}^{D_{\Lambda}}{\omega_{d}} {\left(t\right) \cdot \lambda_{d}^{-1}}\right)^{-1},
\end{aligned}
\label{equation_4_1_2}
\end{equation}
where $\rho = \left\{\rho_{d^{\prime}}\right\}_{d^{\prime}=1}^{D_{\Gamma}}$ and $\omega = \left\{\omega_{d}\right\}_{d=1}^{D_{\Lambda}}$ are sets of weight functions $\rho_{d^{\prime}}, \omega_{d}:\mathbb{R}\rightarrow\mathbb{R}_{\geq0}$ such that $\sum_{d^{\prime}=1}^{D_{\Gamma}}\rho_{d^{\prime}}{\left(t\right)} = \sum_{d=1}^{D_{\Lambda}}\omega_{d}{\left(t\right)} \equiv 1$ for every $t$, and $G=\left\{\gamma_{d^{\prime}}\right\}_{d^{\prime}=1}^{D_{\Gamma}}$ and $L=\left\{\lambda_d\right\}_{d=1}^{D_{\Lambda}}$ are sets of $K_P \times K_P$ and $K_Q \times K_Q$ basis covariance matrices to estimate, respectively. This formulation provides a more flexible and interpretable time-varying spatiotemporal covariance function than \eqref{equation_4_1_0}. Although the EM algorithm does not have closed-form M-step updates for either of the above, closed-form \textit{conditional  Maximization} updates  \citep{Meng1993} are available for both cases and the algorithm's monotone convergence is also guaranteed (Appendix \ref{secC_2}).

\subsection{Time-varying Idiosyncratic Error Covariance Function} \label{sec4_2}

In the model \eqref{equation_2_1_4}, we assumed that the covariance matrix $\Sigma$ for idiosyncratic errors does not vary over time. By defining each of its diagonal entry $\Sigma_{q}$ as a function like \eqref{equation_2_1_3}, we obtain a time-varying extension $\Sigma_t = \text{diag}\left\{\Sigma_{t1},\cdots,\Sigma_{tQ}\right\}$ of $\Sigma$, i.e., for each $q$, for a set of basis positive scalars $U_q = \left\{\nu_{qd}\right\}_{d=1}^{D_q}$ and weight functions $\tilde{\omega}_q = \left\{\tilde{\omega}_{qd}\right\}_{d=1}^{D_q}$, we have:
  \begin{equation}
  \begin{aligned}
  \Sigma_{tq} \triangleq \left(\sum_{d=1}^{D_q}{\tilde{\omega}_{qd}} {\left(t\right) \cdot \nu_{qd}^{-1}}\right)^{-1}, \quad \tilde{\omega}_{qd}{\left(t\right)} \triangleq \frac{\tilde{\mathcal{K}}_{\tilde{h}_q}^{(q)}\left(t,t_d\right)}{\sum_{c=1}^{D_q}{\tilde{\mathcal{K}}_{\tilde{h}_q}^{(q)}\left(t,t_c\right)}}\propto \tilde{\mathcal{K}}_{\tilde{h}_q}^{(q)} \left(t,t_d\right),
  \end{aligned}
  \label{equation_2_5_3}
  \end{equation}
  with a prior on $U_q = \left\{\nu_{qd}\right\}_{d=1}^{D_q}$ as follows:
  \begin{equation}
  \begin{aligned}
p\left(U_q\middle|\tilde{\omega}_q,\mathcal{T}\right)\propto \exp\left(\frac{1}{2}\sum_{n=1}^{N}\sum_{d=1}^{D_q}{\tilde{\omega}_{qd}{\left(t_n\right)}\log{\left|\nu_{qd}^{-1}\right|}}-\frac{1}{2}\sum_{n=1}^{N}\log{\left|\sum_{d=1}^{D_q}{\tilde{\omega}_{qd}{\left(t_n\right)}\nu_{qd}^{-1}}\right|}\right).
  \end{aligned}
  \label{equation_2_5_4}
  \end{equation}

In this setting, we are still able to derive the conditional closed-form M-step updates: details can be found in Appendix \ref{secC_3}. Also, one can easily derive the extensions of the robust factor model in Section \ref{sec4_0} and spatiotemporal factor model in Section \ref{sec4_1} with the above $\Sigma_t$ (and $\Phi_t$ similarly) and prior, and their \textit{closed}-form conditional M-step updates.

\subsection{Regularized Modeling} \label{sec4_4}

To reduce the model complexity further, one can restrict each $\lambda_d$ to be diagonal matrices. Then, we still have the closed-form M-step updates for
$\lambda_d$'s:
\begin{equation}
\begin{aligned}
\lambda_{d}^{\left(r+1\right)} & = \frac{\sum_{n=1}^{N}{\omega_{d}{\left(t_n\right)} \cdot \diag{\left({\hat{\eta}}_{n}{\hat{\eta}}_{n}^\top+{\hat{\Psi}}_{n}\right)}}}{\sum_{n=1}^{N}\omega_{d}{\left(t_{n}\right)}},
\end{aligned}
\label{equation_4_4_1}
\end{equation}
which replaces $\lambda_{d}^{\left(r+1\right)}$ in Equation \eqref{equation_3_2_1}. 

Moreover, we can further impose an inverse-Wishart prior on each $\lambda_d$ to guarantee that they are strictly positive definite. That is, for a degree of freedom $\zeta > K - 1$ and a non-degenerate $K \times K$ covariance matrix $\Theta$, if we assume $\lambda_d \stackrel{i.i.d.}{\sim} \mathcal{W}^{-1}\left(\Theta,\zeta \right)$, then the M-step update for each $\lambda_d$ is given as the maximum {\it a posteriori} estimate:
\begin{equation}
\begin{aligned}
\lambda_{d}^{\left(r+1\right)} & = \frac{\Theta + \sum_{n=1}^{N}{\omega_{d}{\left(t_n\right)} \left({\hat{\eta}}_{n}{\hat{\eta}}_{n}^\top+{\hat{\Psi}}_{n}\right)}}{\zeta + K + 1 + \sum_{n=1}^{N}\omega_{d}{\left(t_{n}\right)}}.
\end{aligned}
\label{equation_4_4_2}
\end{equation} 
We may also want to prescribe a diffused (improper) prior, such as letting
 $\zeta + K + 1 = 10^{-8}$ and $\Theta = 10^{-8} \cdot \mathbb{I} $, so that one may only guarantee that  each $\lambda_d$ is non-degenerate. Clearly, as long as $\zeta + K + 1 > 0$, each nonnegative weight function $\omega_d$ does not need to be always strictly positive.

So far, we have considered $L=\left\{\lambda_d\right\}_{d=1}^{D}$ to be a set of free covariance matrices. In practice, each $\lambda_d$ can be parameterized as a covariance matrix function for reducing the model complexity. Also, the factor loading $B$ has been assumed to be a free matrix parameter. Because each row of $B$ can be regarded as an embedding of the corresponding coordinate of the output $y$, one can parametrize $B$ according to the prior knowledge of the coordinates, which allows us to extend the factor loading $B$ by ``predicting'' rows corresponding to the coordinates that lack of observations. In either of the cases, the exact closed-form M step update might not be available anymore, but one can instead numerically estimate $\lambda_d$'s and/or the factor loading $B$ in each M step using the chain rule with partial derivatives \eqref{equation_A_2_1} in Appendix \ref{secA} and those of $\lambda_d$'s and/or $B$ with respect to their parameters.

\subsection{Non-factor Simplification} \label{sec4_5}

So far, we have considered the stochastic \textit{factor} model $y_t \stackrel{i.i.d.}{\sim} \mathcal{N}\left(0,B\Lambda_t B^\top+\Sigma\right)$ for a $Q \times K$ factor loading matrix $B$, a $K \times K$ covariance matrix function $\Lambda_t$, and a $Q \times Q$ diagonal matrix $\Sigma$, where $K \ll Q$ for a large $Q$. Here, we simplify the model to $y_t \stackrel{i.i.d.}{\sim} \mathcal{N}\left(0,\Lambda_t\right)$ for a $Q \times Q$ covariance matrix function $\Lambda_t$ defined by \eqref{equation_2_1_3} with the same prior \eqref{equation_2_2_1_2} for $L = \left\{\lambda_d\right\}_{d=1}^{D}$ for a small $Q$. Then, we have instead the following {\it maximum a posteriori} (MAP) estimate for each $\lambda_d$:
\begin{equation}
\begin{aligned}
\hat{\lambda}_{d} & = \frac{\sum_{n=1}^{N}{\omega_{d}{\left(t_n\right)} \cdot y_{t_n} y_{t_n}^\top}}{\sum_{n=1}^{N}\omega_{d}{\left(t_{n}\right)}}.
\end{aligned}
\label{equation_4_5_1}
\end{equation}

In other words, given a set of weight functions $\omega_d$'s, we first calculate each $\lambda_d$ in $L$ by the weighted average of $\left\{y_{t_n} y_{t_n}^\top\right\}_{n=1}^{N}$ with the weights proportional to $\left\{\omega_{d}{\left(t_n\right)}\right\}_{n=1}^{N}$, and then define each $\Lambda_t$ by a weighted average of those $\lambda_d$'s with weights $\left\{\omega_{d}{\left(t\right)}\right\}_{d=1}^{D}$, i.e., ours is, roughly speaking, an ``average of average'' modeling, compared to the ``average'' modeling of \cite{Yin2010}, who directly estimate each $\Lambda_t$ as a weighted average of $\left\{y_{t_n} y_{t_n}^\top\right\}_{n=1}^{N}$.

\section{Simulation Study} \label{sec5}

In this section, we simulate a dataset with a sparse factor loading and study the effectiveness of our proposed factor model in Section \ref{sec2} in detecting the volatility changes and recovering the number of factors, compared with the existing method, the exponentially-weighted moving average (EWMA) covariance model \citep{Engle2009, Tsay2010}, described in Appendix \ref{secF}.

Firstly, we generated a simulated dataset with $N=300$ observations, $Q=130$ as the number of variables and $K=5$ as the true number of factors. The true factor loading matrix $B$ is a matrix which has a block-diagonal pattern as shown in Figure \ref{fig_5_1_1}, Appendix \ref{secG}, as the factor loadings are either $0$, or follow distribution $\mathcal{N}\left(1,{0.1}^2\right)$. We set $\mathcal{T} = \left\{t_1,t_2,\cdots,t_N\right\} = \left\{1,2,\cdots,300\right\}$ and defined the covariance matrix function $\Lambda : \mathbb{R} \rightarrow \mathbb{R}^{K \times K} $ for $\gamma \in \left\{ 3, 4, 5 \right\}$ as follows:

\begin{enumerate}
  \item Let $\mathbb{K}$ be the squared-exponential kernel $\mathbb{K}\left(t_1, t_2\right) \triangleq \exp\left(- \frac{1}{2} \cdot 10^{-\gamma} \cdot \left(t_1 - t_2\right)^2\right)$.
  \item We defined a $K \times K$ matrix function $R: \mathbb{R} \rightarrow \mathbb{R}^{K \times K}$ where each coordinate $R_{k_1, k_2}$ is a random function independently drawn from a Gaussian process  $\mathcal{GP} \left(0, \mathbb{K} \right)$ \citep{Rasmussen2006}.
  \item Then, we have a $K \times K$ matrix function $\Lambda\left(t\right) \triangleq corr\left( R\left(t\right) R\left(t\right)^\top \right)$ for each $t \in \mathbb{R}$.
\end{enumerate}
Examples of random functions drawn from the above Gaussian process with different $\gamma$ can be found in Figure \ref{fig_5_1_0} of Appendix \ref{secG}. The covariance matrix for the idiosyncratic errors is $\Sigma = \text{diag}\left\{\Sigma_{1},\cdots,\Sigma_{Q}\right\}$, of which the diagonal entries are i.i.d. from $ \text{Unif}\left(0.5s^2,1.5s^2\right)$ for $s^2\in\left\{0.25,0.5,1,2\right\}$. We ran 12 simulations by varying $\gamma$ and $s^2$. Clearly, a pair of smaller $\gamma$ and $s^2$ makes the simulated data set more heteroscedastic over time.

We let the candidate set for the numbers of factors be $\left\{1,\cdots,12\right\}$. We call our proposed Algorithm \ref{algorithm1} for the factor analysis modeling in Section \ref{sec2} as ``GHeFM'', which stands for ``Gaussian heteroscedastic factor model''.  One can also consider a time-invariant alternative, i.e., $\Lambda_t\equiv\Lambda_0$ for some matrix parameter $\Lambda_0$: we call this simplified model as ``GHoFM'', which stands for ``Gaussian homoscedastic factor model''. For validation, we have randomly split the given dataset into training and validation sets 12 times. To check the stability, we have run two sets of those simulations: the first ones are with the 9:1 ratio of split training and validation sets and the second with 8:2.

The purposes of our simulations are to check whether 1) our algorithm can detect the true number of factors, five; and 2) our algorithm can predict covariances at missing observations well. For the second purpose, we compute the average KL divergences from the simulation distribution to the predictions obtained by GHoFM, GHeFM and EWMA covariance models with decay rate $\alpha \in \left\{0.98,0.96,0.94,0.92,0.90,0.88,0.86,0.84,0.82,0.80\right\}$.

In both ratios of split training and validation sets, all $12$ cases over $\left(\gamma,s^2\right)$ detected the true number of factors, five, by GHeFM: Figure \ref{fig_5_1_2} in Appendix \ref{secG} shows the standardized values of $\mathcal{V}\left(K\right)$ from \eqref{equation_3_4_1} over $\mathbb{K}$ for two different ratios of validation sets. An example of the estimated factor loading matrix $B$ for each case $\left(\gamma,s^2\right)$ can be found in Figure \ref{fig_5_1_3}, Appendix \ref{secG}. In any case, the estimated $B$ is consistent with the true factor loadings up to the column permutations. Finally, we compared the average KL divergences from the simulation distribution to GHeFM, GHoFM and EWMA covariance models, and it turns out that the time-varying GHeFM model is mostly superior to the GHoFM and EWMA covariance models, as shown in Table \ref{table_5_1_1} and Table \ref{table_5_1_2}, Appendix \ref{secG}, while all models detected the true number of factors.

For the same simulated $\left(\Lambda,B,\Sigma\right)$ above, we also generated $y_t \sim \mathcal{T}_\nu\left(0,B\Lambda_t B^\top+\Sigma\right)$ for $\nu=6$ degree of freedom and run both the GHeFM which assumes each $y_t$ to follow a multivariate Gaussian distribution, and the robust (heteroscedastic) factor model (RHeFM) in Section \ref{sec4_0}, on those simulated dataset to compare their performances, given the true degree of freedom. While the RHeFM detects the true number of factors for all cases over $\left(\gamma,s^2\right)$, the GHeFM sometimes overestimates it. Figure \ref{fig_5_2_1} in Appendix \ref{secG} visualizes an example of the estimated factor loading matrices by the GHeFM that could not even find the true factors and it regards some noises as signals for some cases. RHeFM, however, restored the factor loadings consistently, as shown in Figure \ref{fig_5_2_2}, Appendix \ref{secG}.

\section{Three Real-Data Examples} \label{sec6}

\subsection{Daily average temperature changes of 50 US airports}\label{sec6_2}

This example studies the daily average temperature changes of 50 US airports  from January 1, 2021, to January 1, 2023. We compute the daily log-changes of the NOAA daily average temperatures (TAVG) of each airport and then normalize them by their total mean and standard deviation over all airports.
We applied GHeFM and set the candidate set for the number of factors as $\mathbb{K}=\left\{1,\cdots,30\right\}$. The estimated number of factors is $\hat{K}=19$ and some of the resulting time-varying nature of the correlations are visualized in Figures \ref{fig_6_2_1} and \ref{fig_6_2_2} of Appendix \ref{secG}.

\begin{figure}[!htb]
\renewcommand{\baselinestretch}{1}
\centering
\includegraphics[width=0.44\textwidth]{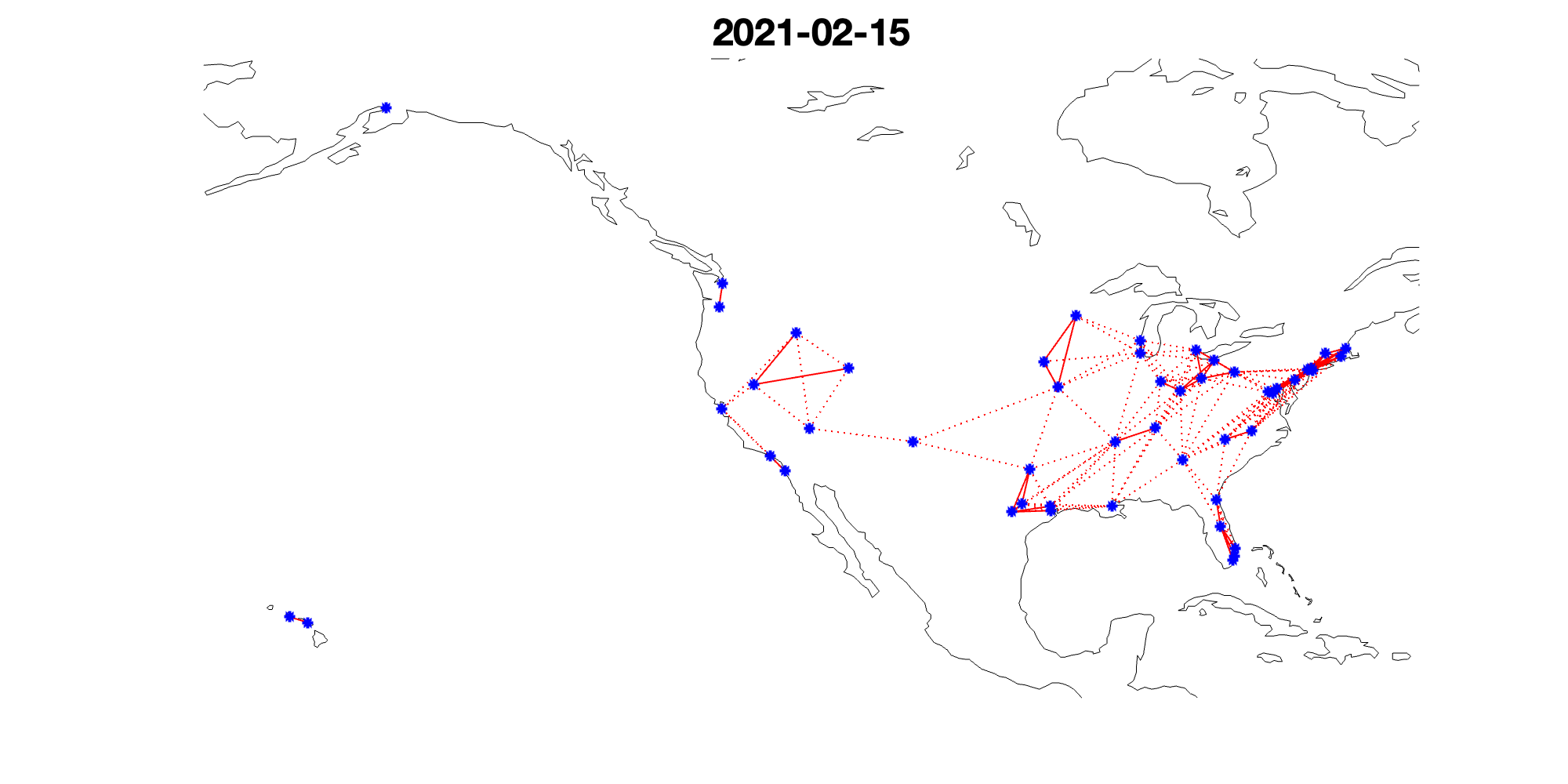}
\includegraphics[width=0.44\textwidth]{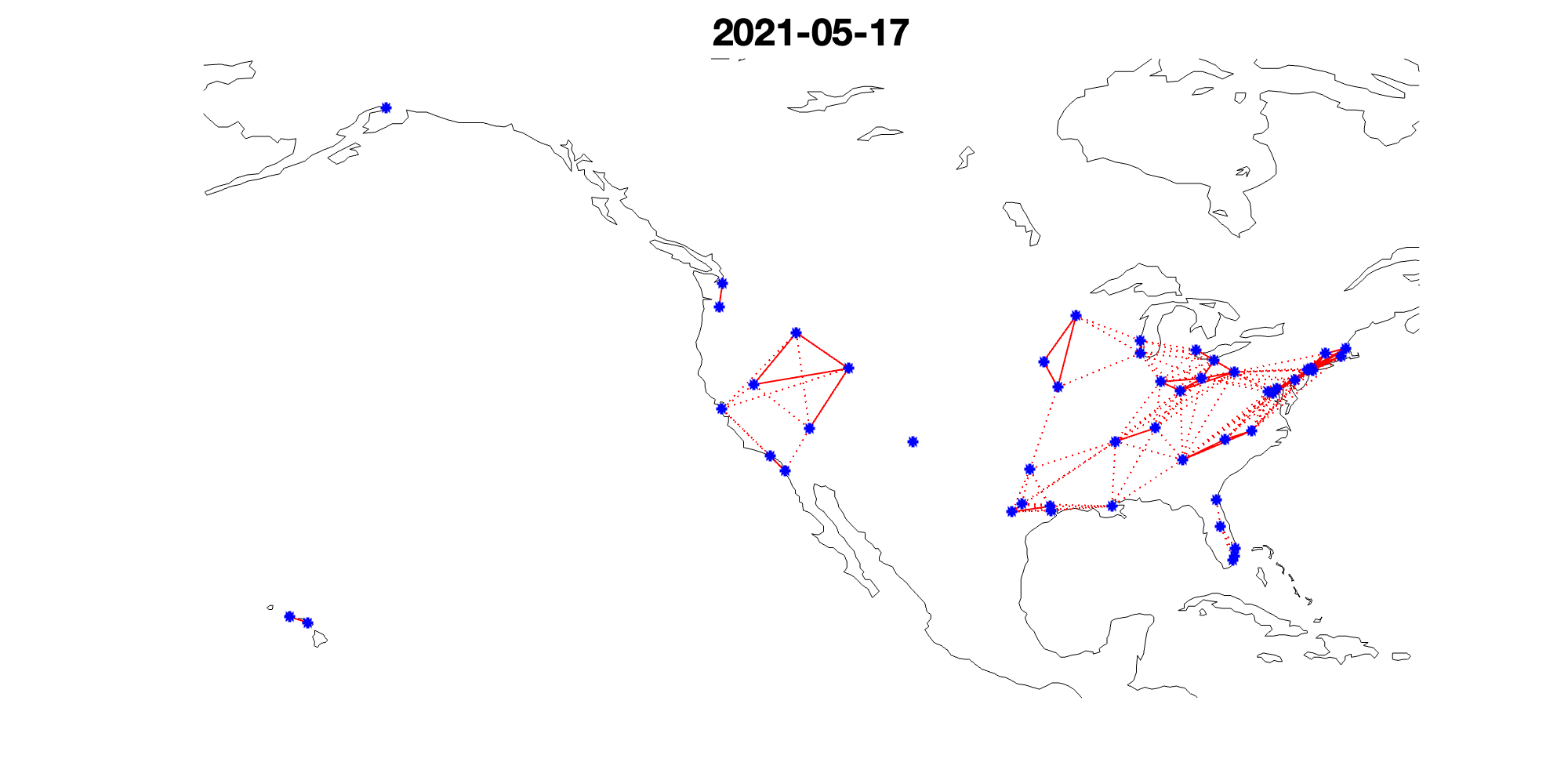}
\includegraphics[width=0.44\textwidth]{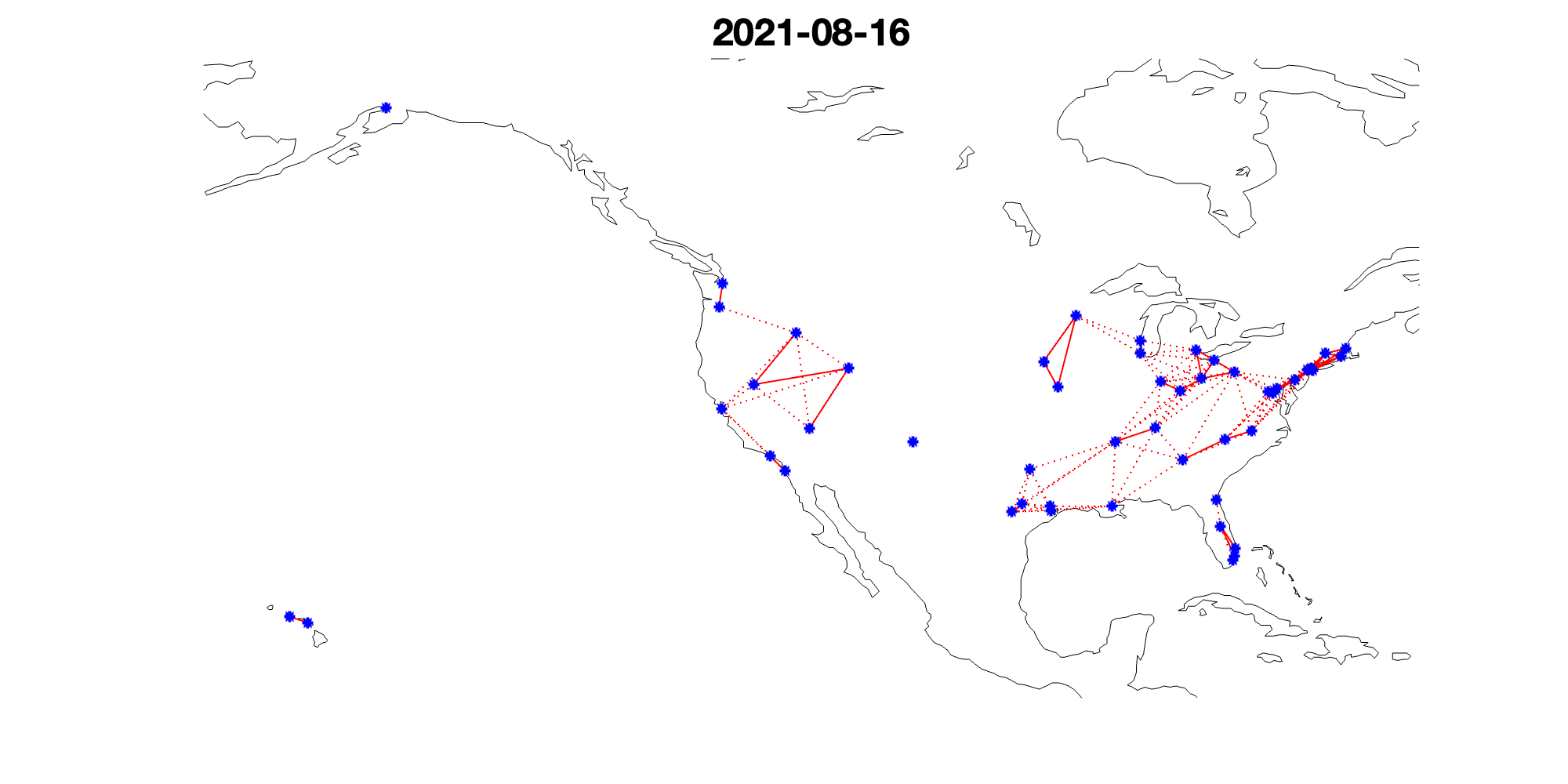}
\includegraphics[width=0.44\textwidth]{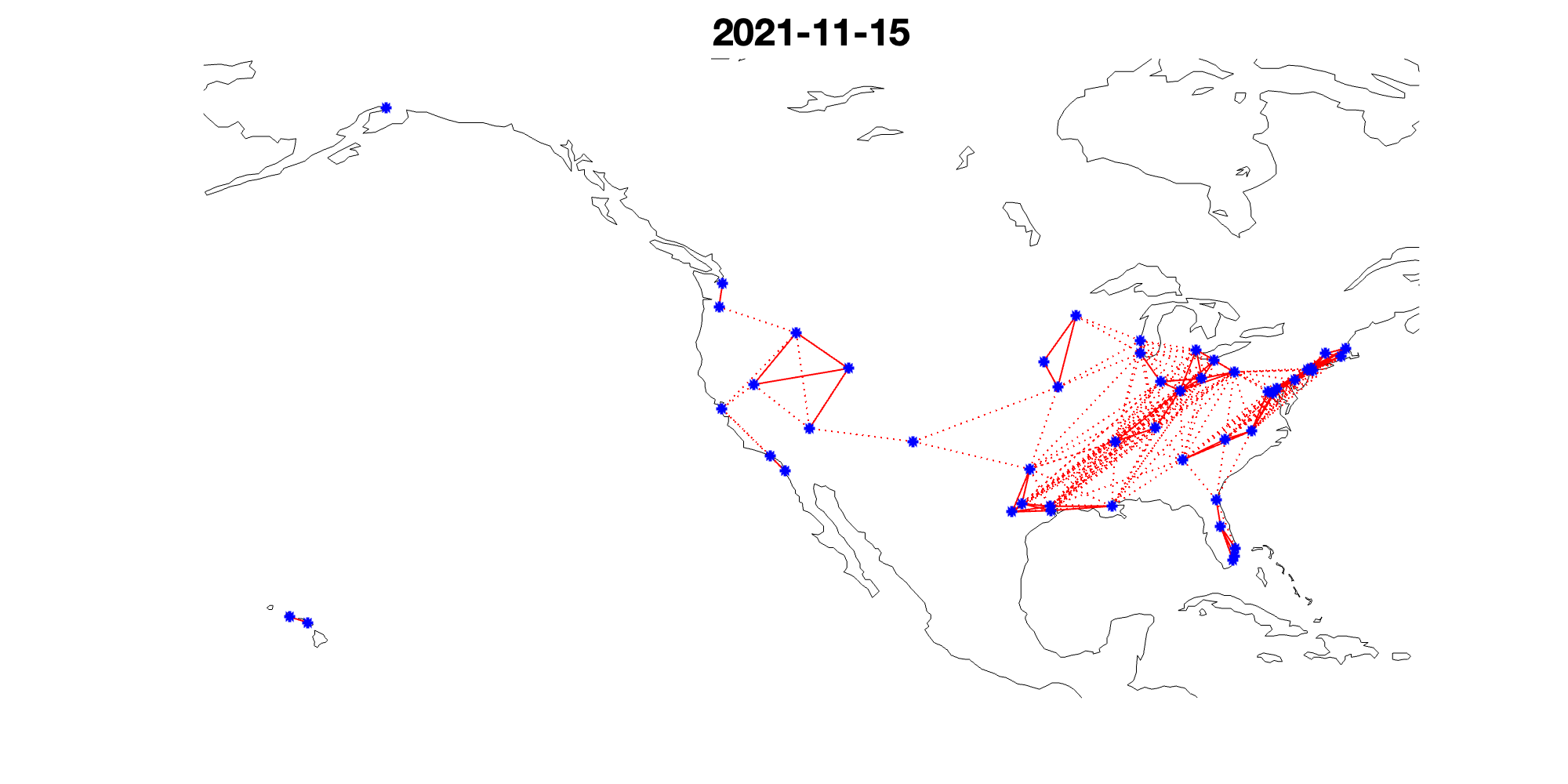}
\includegraphics[width=0.44\textwidth]{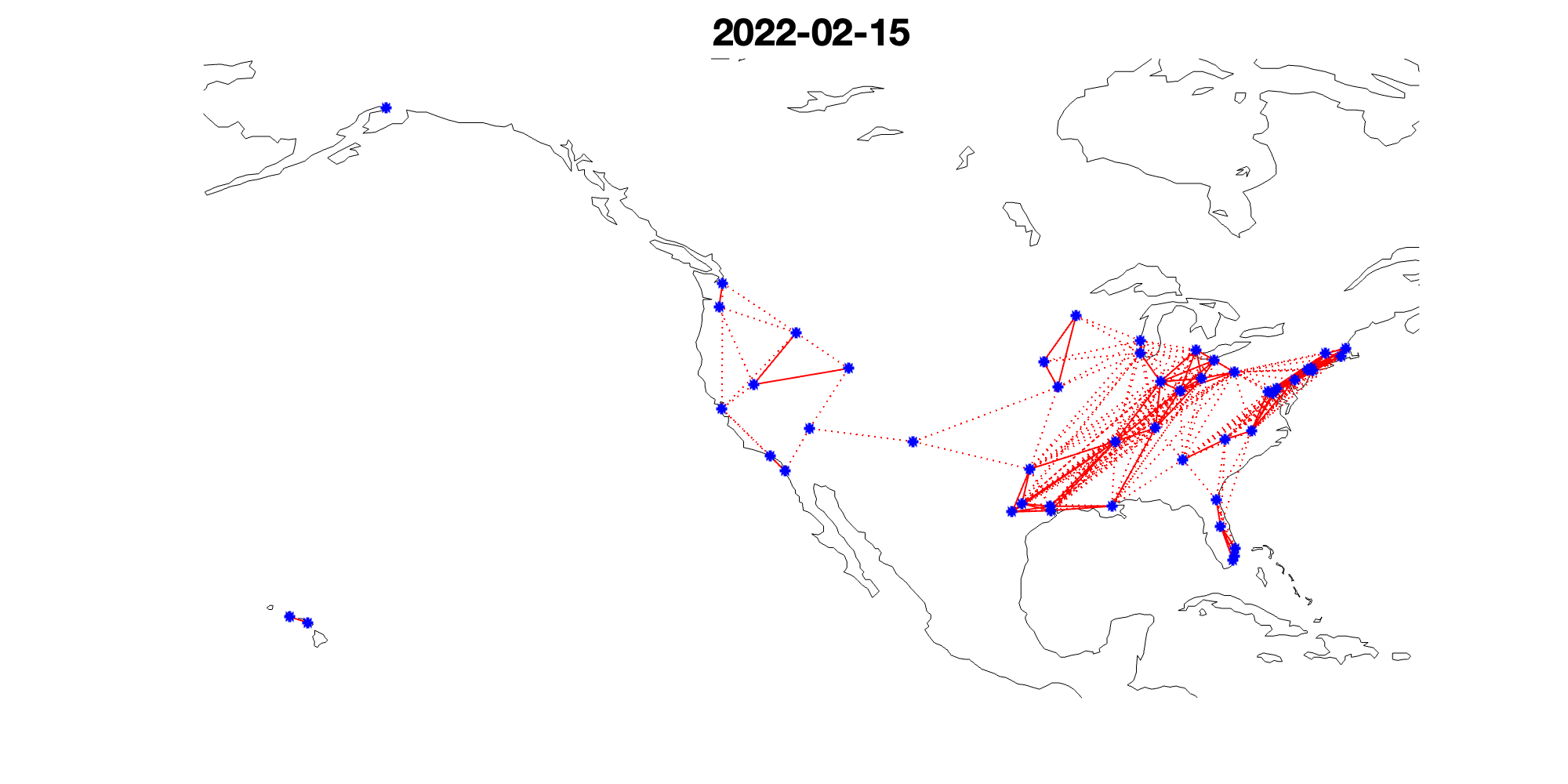}
\includegraphics[width=0.44\textwidth]{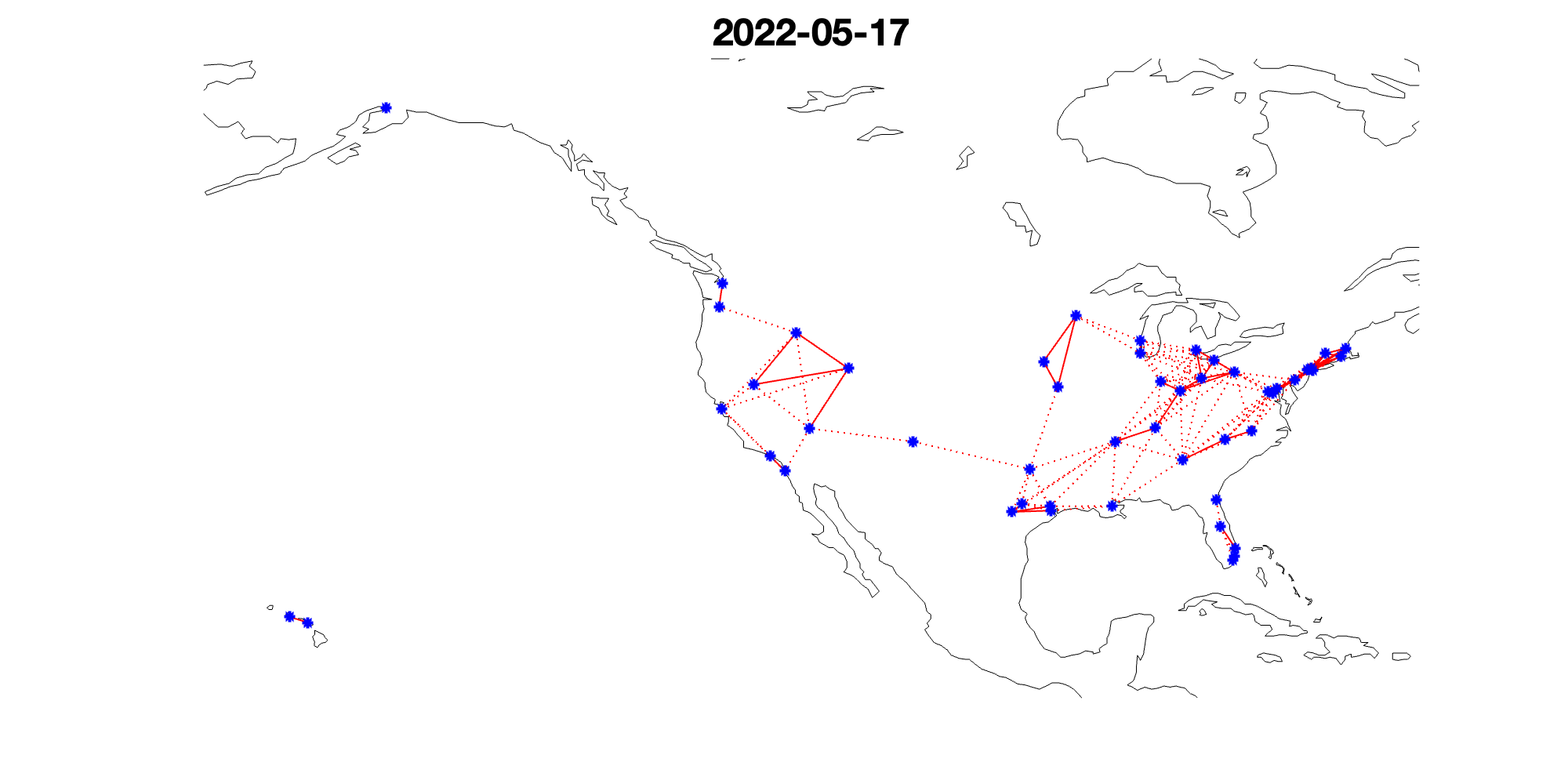}
\includegraphics[width=0.44\textwidth]{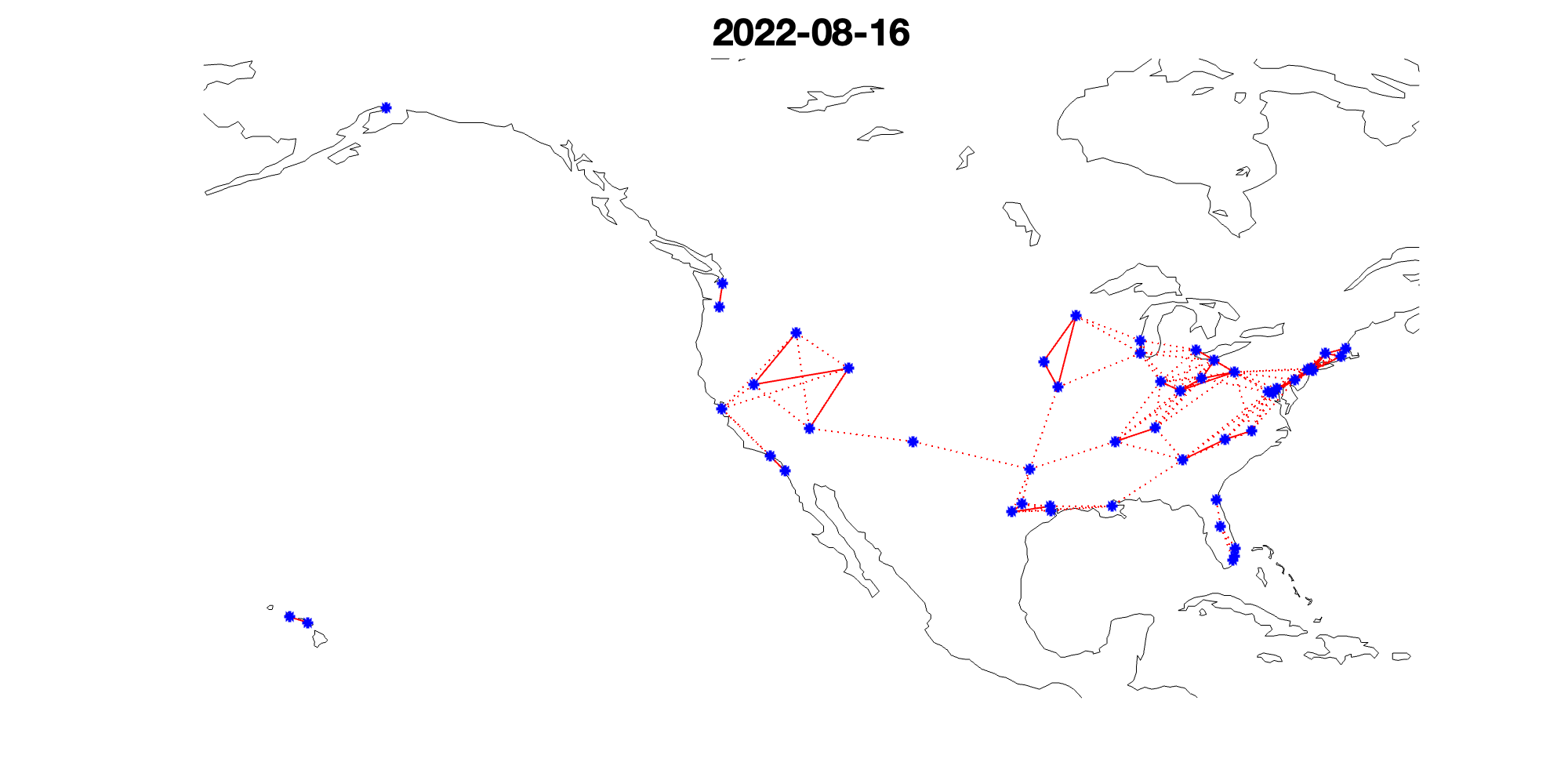}
\includegraphics[width=0.44\textwidth]{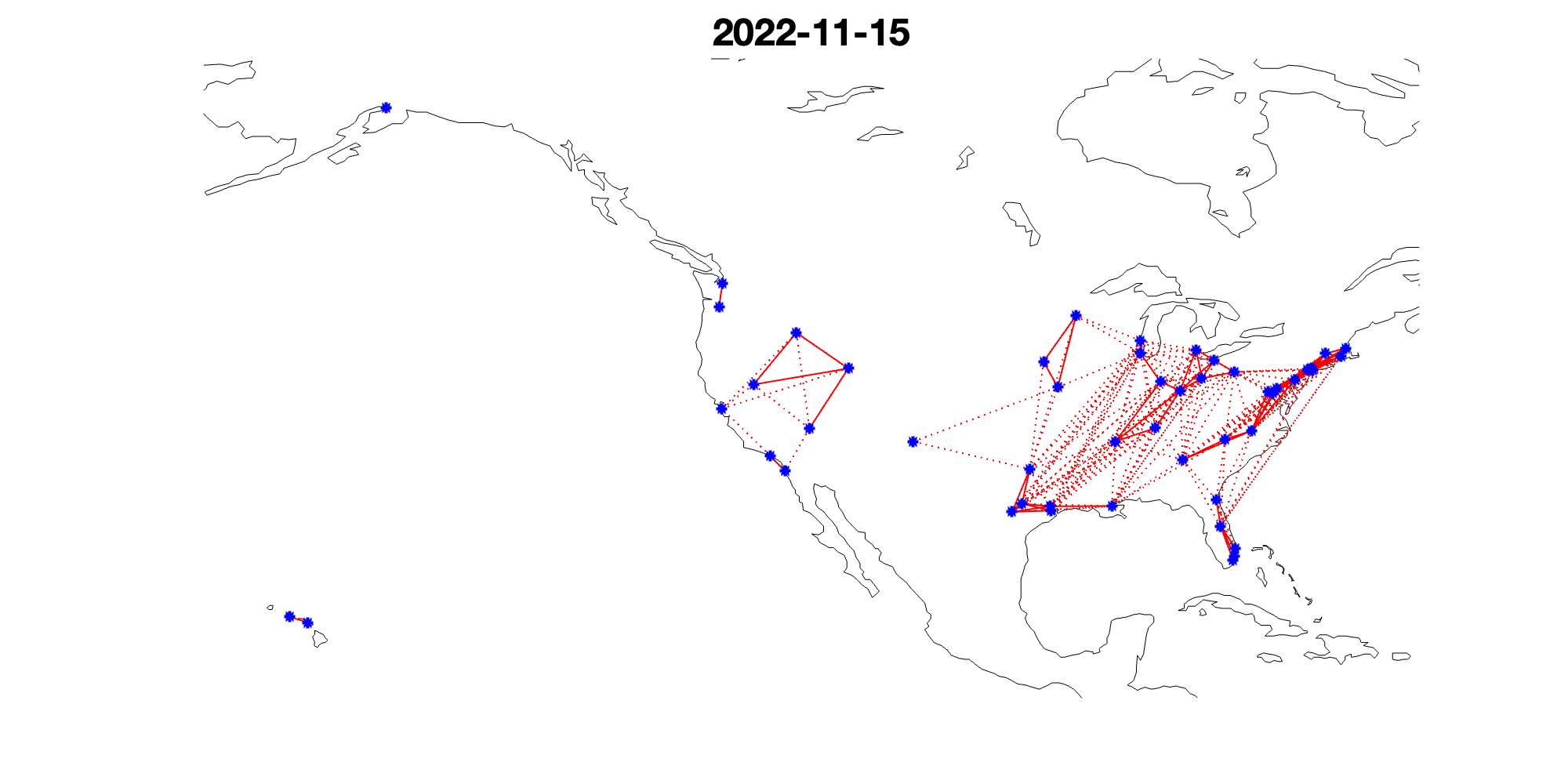}
\caption{Locations of the 50 US airports (blue dots) over 8 time windows. A pair of airports is connected by a solid red line if their cosine measure exceeds 5/6 and by a dashed red line if it exceeds 1/2 but less than 5/6.}
\label{fig_6_2_0}
\end{figure}

We are particularly interested in clustering airports with respect to the estimated factor loading matrix $\hat{B}$. That is, we interpret the rows of $\hat{B}$ as the embeddings of individual airports. To utilize these embeddings for similarity-based clustering, we rely on the cosine of the angle between two row vectors ${\hat{B}}_{p,:}^\top$ and ${\hat{B}}_{q,:}^\top$, defined as 
\begin{equation}
\begin{aligned}
\mathfrak{W}\left(p,q\right)\triangleq\frac{{\hat{B}}_{p,:}{\hat{B}}_{q,:}^\top}{\left|{\hat{B}}_{p,:}^\top\right|\cdot\left|{\hat{B}}_{q,:}^\top\right|}.
\end{aligned}
\label{equation_6_1_2}
\end{equation}
Once we estimated the factor loading matrix $B$ and time-varying matrix function $\Lambda_t$, we define a time-varying factor loading matrix function $\mathcal{B}_t$, which satisfies $B \Lambda_t B^\top = \mathcal{B}_t \mathcal{B}_t^\top$ for each $t$, according to (d) in Section \ref{sec2_2}. Then, we consider rows of $\mathcal{B}_t$ as the time-varying embeddings of the airports. That is, we cluster the airports with respect to the \textit{time-varying} estimated factor loading matrix thus the clustering is also time-varying, not time-invariant.

Figure \ref{fig_6_2_0} shows eight clustering results based on the time-varying factor loading matrix function that we estimated from the data: the clusters are consistent with their geographical characteristics, although no spatial information was assumed during factor analysis. In addition, they show a stronger correlation in winter than in summer over two years.

\subsection{Comprehensive Climate Dataset}\label{sec6_3}

\begin{figure}
\renewcommand{\baselinestretch}{1}
\centering
\includegraphics[width=0.24\textwidth]{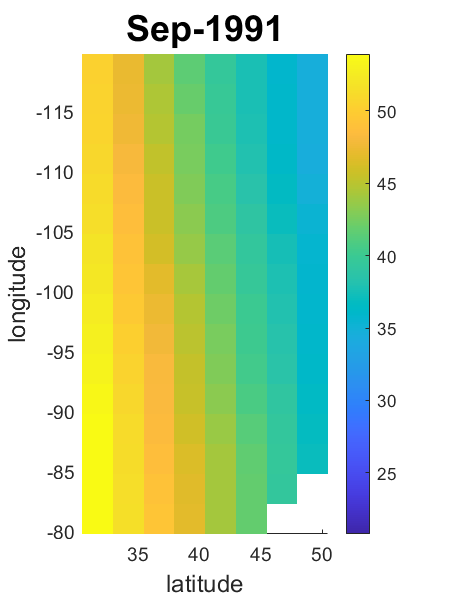}
\includegraphics[width=0.24\textwidth]{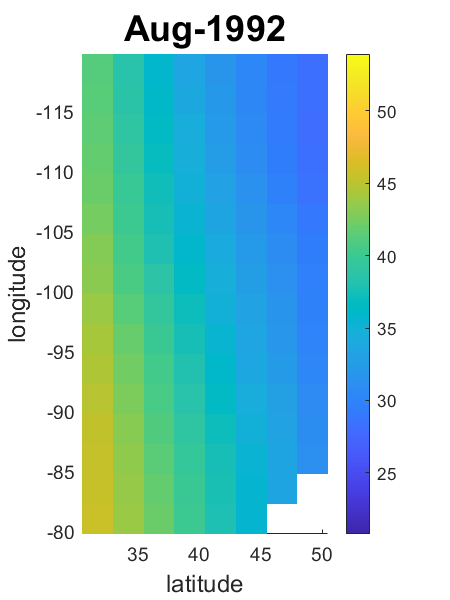}
\includegraphics[width=0.24\textwidth]{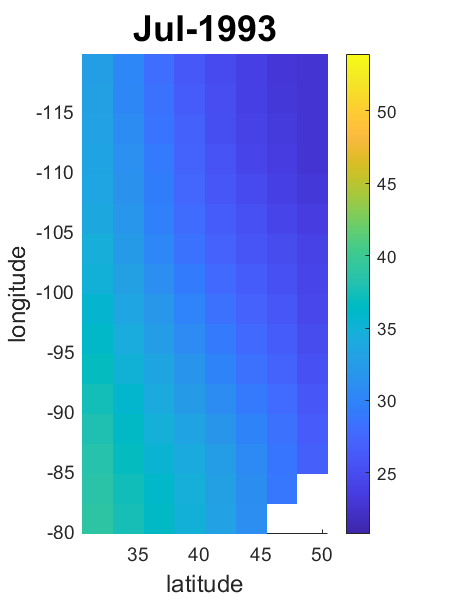}
\includegraphics[width=0.24\textwidth]{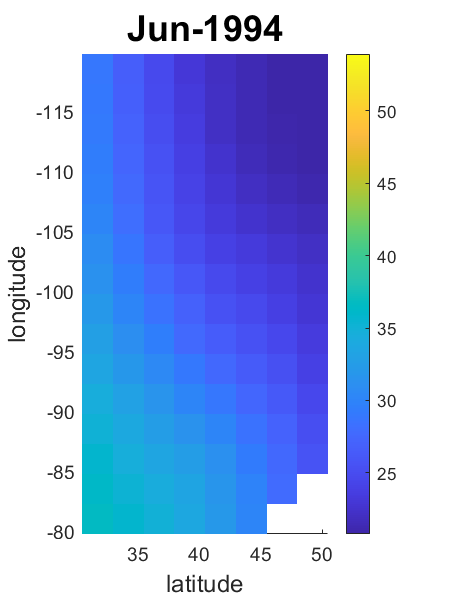}
\includegraphics[width=0.24\textwidth]{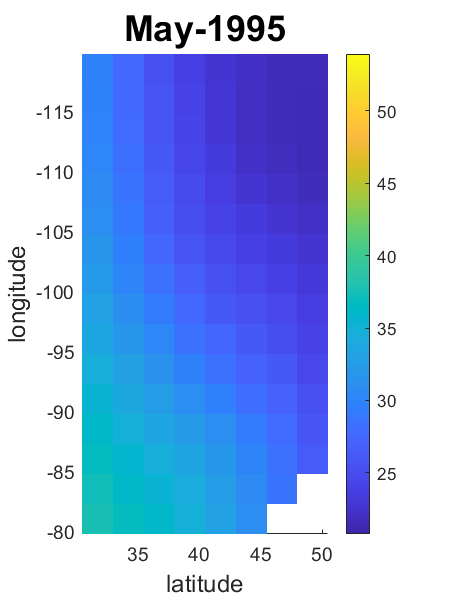}
\includegraphics[width=0.24\textwidth]{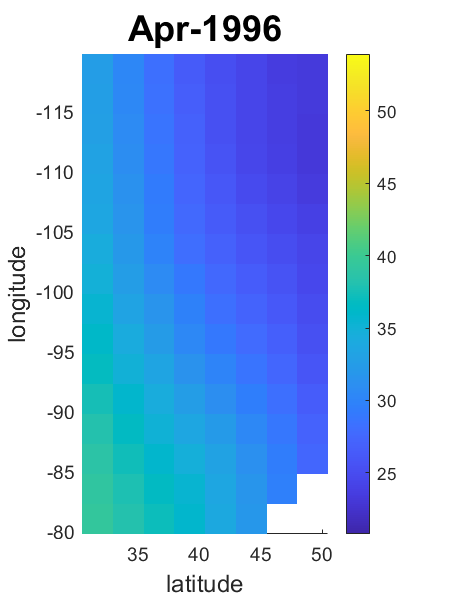}
\includegraphics[width=0.24\textwidth]{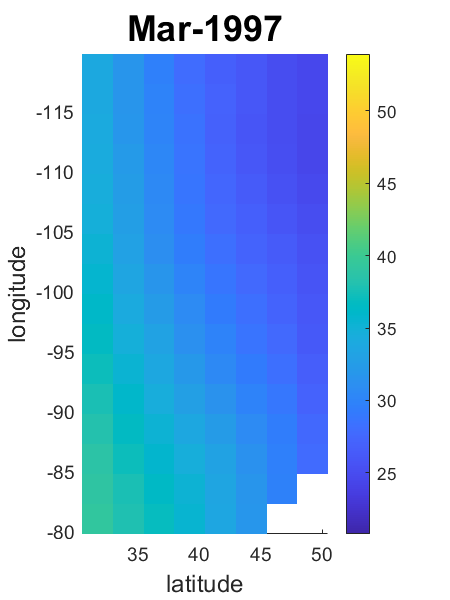}
\includegraphics[width=0.24\textwidth]{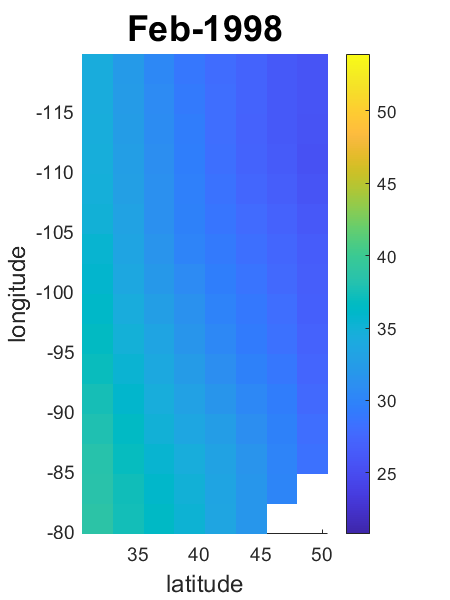}
\includegraphics[width=0.24\textwidth]{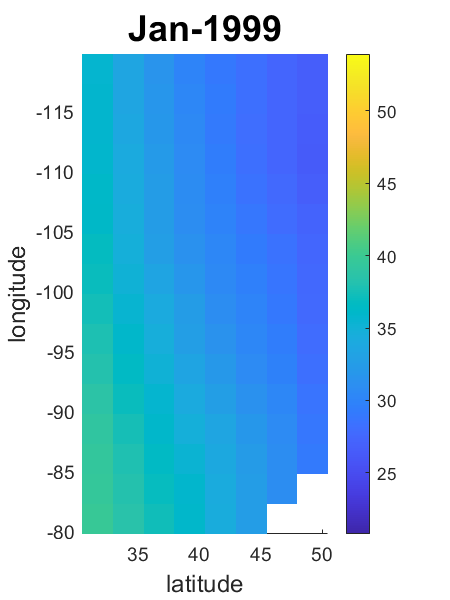}
\includegraphics[width=0.24\textwidth]{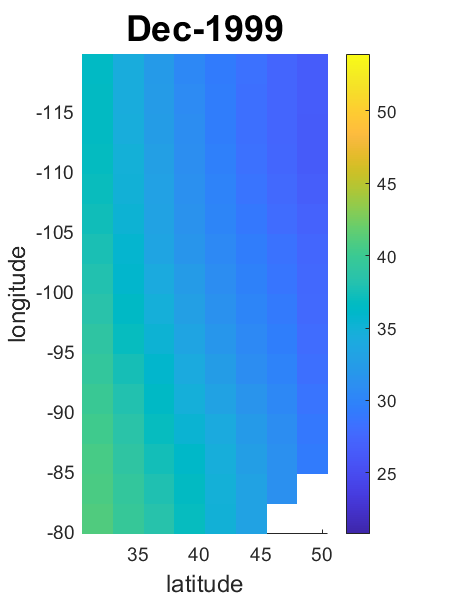}
\includegraphics[width=0.24\textwidth]{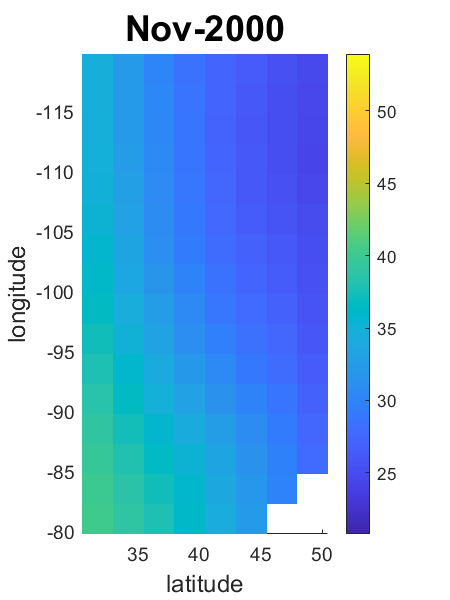}
\includegraphics[width=0.24\textwidth]{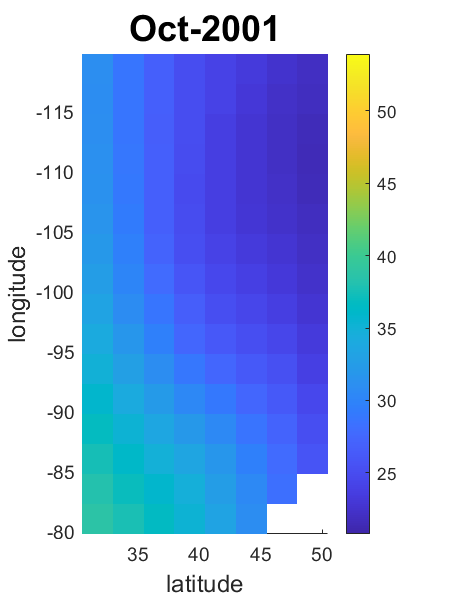}
\caption{The first spatial factor loadings over locations and time.}
\label{fig_6_3_1_1}
\end{figure}

The Comprehensive Climate Dataset (CCDS) is a collection of climate records of North America \citep{Lozano2009} and was used by \cite{Chen2020} for the verification of their multivariate spatial-temporal modeling. The spatiotemporal dataset consists of monthly observations of 16 climate variables from January 1991 to June 2002 (i.e., 138 time indices) over 125 regularly-spaced locations. \cite{Chen2020} first removed the seasonality and normalized the raw dataset;  then, they chose 8  latent variables and 4 spatial dimensions, respectively (i.e., $Q = 16$, $P = 125$, $N = 138$, $K_1 = 8$ and $K_2 = 4$ as in Section \ref{sec4_1}), under the assumption that the covariance is time-invariant. However, the KCP permutation test \citep{Cabrieto2018} rejects the null hypothesis that there is no change point in correlation over time with the window sizes bigger than 14 (\cite{Cabrieto2018} suggest 25 as the size of window, and the the p-value of the variance test is 0 and that of the variance drop test is 0.0320 for that window size), and significance level of 0.05. Also, the running correlation (with the window size of 15) between ETR and ETRN clearly demonstrates its time-varying nature as shown in Figure \ref{fig_6_2_3} in Appendix \ref{secG}. We applied the second spatiotemporal factor model in Section \ref{sec4_1}, which assumes that the time-varying covariance is the Kronecker product of two smaller ones, as Equation \eqref{equation_4_1_1}, with the same $K_1 = 8$ and $K_2 = 4$ as in \cite{Chen2020}.

Figure \ref{fig_6_3_1_1} (and Figures \ref{fig_6_3_1_2}, \ref{fig_6_3_1_3}, \ref{fig_6_3_1_4} in Appendix \ref{secG}) shows the estimated time-varying spatial factor loadings $\mathcal{C}_t$ defined as $\mathcal{B}_t$ in Section \ref{sec6_2} at time $t$ for every 11 months: each estimated spatial factor loading is spatially correlated. Except for the first factor loading at the early moment of time, the estimated spatial factor loadings do not change much over time. However, correlation matrices of the estimated $B \Lambda_t B^\top$ in \eqref{equation_4_1_1} and \eqref{equation_4_1_2}, undergo significant changes over time, as shown in Figure \ref{fig_6_3_4} of Appendix \ref{secG}. 
The estimated correlations are consistent with common sense, e.g., CH4 is highly correlated with CO with \textit{time-varying} coefficients. In contrast, \cite{Chen2020} only provided  \textit{time-invariant} estimates.

\subsection{S\&P 100 Stock Return} \label{sec6_1}

This example studies the daily returns of the S\&P 100 stocks in the time period from April 1, 2019 to October 1, 2024. Although the companies included in S\&P 100 change over time, the time window of our analysis is short enough so that the S\&P 100 company list used in our study is stable. The stock symbols of the companies are listed in Table \ref{table_6} of the Appendix \ref{secG}. Because symbols GOOG and GOOGL refer to the same company and their stock movements are identical, we dropped GOOGL from the list.

Here we consider the logarithmic daily returns for all stocks in S\&P 100. The log-return $r_t$ of a stock, whose price is denoted as $s_t$ at day $t$, is defined as $r_t=\log(s_t/s_{t-1})$.
Employing RHeFM (which can deal with extreme log-returns) with degree of freedom $\nu=10$., we first estimated the number of factors and all model parameters using the training data of S\&P 100 from April 1, 2019 to March 28, 2024, a total of 1258 trading days.  Then, with the estimated number of factors, weight functions, and factor loading matrix parameter fixed, we performed the following one-step forward prediction and testing recursively from April 1 to October 1, 2024. In other words, we regarded the log-daily returns from April 1st to October 1st, 2024, as the test data to validate our factor modeling strategy.

\begin{itemize}
\item (Predict) We compute $\Lambda_{t_{next}}$ in Equation \eqref{equation_2_1_3} at time $t_{next}$ as the very next date.
\item (Test) To measure the performance, we compute the likelihood of the newly observed $y_{t_{next}}$ of the predictive model $\mathcal{T}_{\nu}\left(0,B\Lambda_{t_{next}} B^\top+\Sigma\right)$ in Section \ref{sec4_0}.
\item (Update) To proceed to the next date, we first renew the set of covariance basis matrices $L$ by adding a randomly initialized basis at $t_{next}$ and removing one at the very first date, and then run the Robust Factor Model to update those in the renewed $L$ only.
\end{itemize}

\begin{figure}
\renewcommand{\baselinestretch}{1}
\centering
\includegraphics[width=0.32\textwidth]{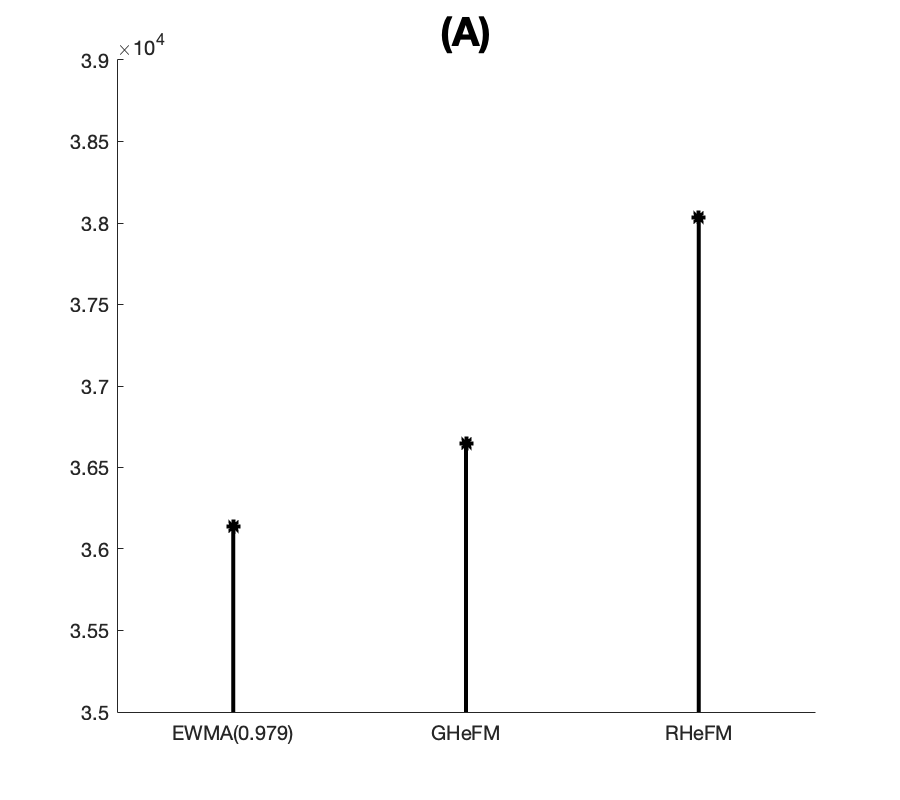}
\includegraphics[width=0.32\textwidth]{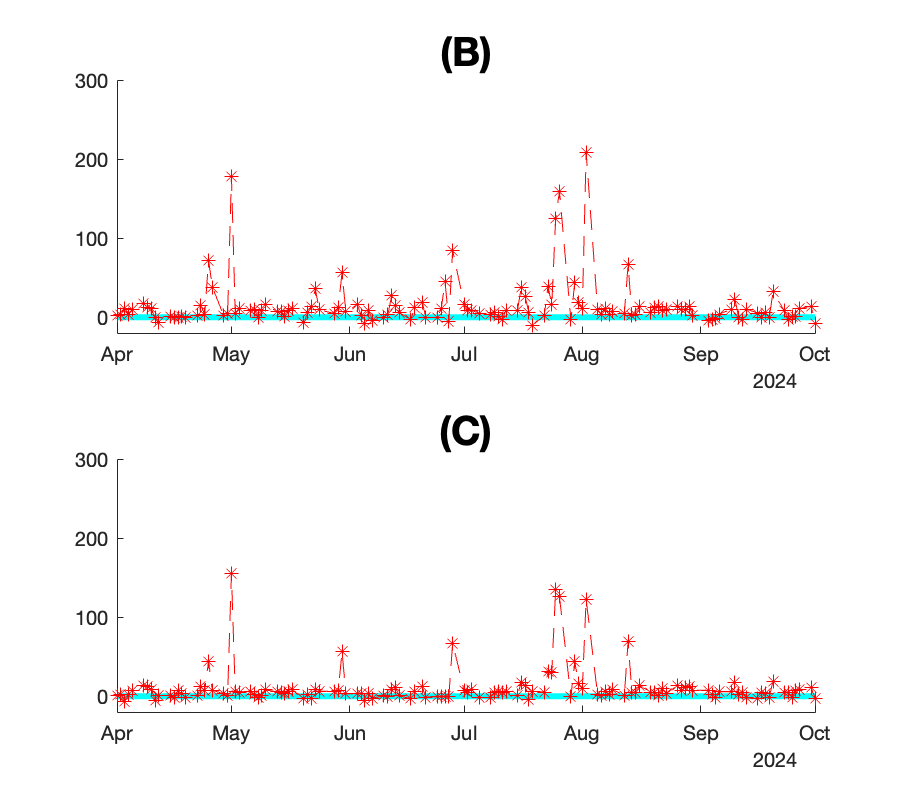}
\includegraphics[width=0.32\textwidth]{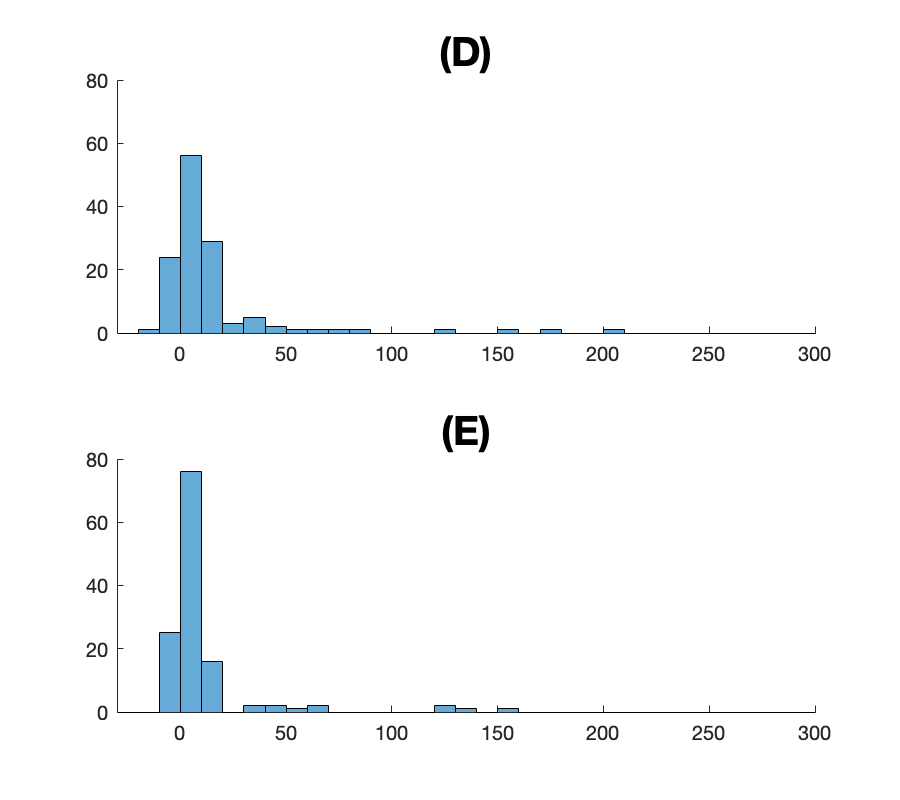}
\caption{(A) Sums of the predictive log-likelihoods (i.e., that evaluated by the true log-daily returns) of various models. (B): log-likelihoods of the true log-daily returns of the RHeFM minus those of the EWMA covariance model with $\alpha = 0.979$ over dates. (D): log-likelihoods of the true log-daily returns of the RHeFM minus those of the GHeFM over dates. (C) and (E): Histograms of the respective log-likelihood differences in the middle panels. Among 128 test business days, the RHeFM beats the GHeFM and EWMA with $\alpha=0.979$ for 103 days each.}
\label{fig_6_4_1}
\end{figure}

The estimated number of factors is $\hat{K}=32$. Figure \ref{fig_6_4_1} shows the comparison between the predictive performance of  RHeFM and those of GHeFM and EWMA covariance estimations (the detailed formulation of EWMA can be found in Appendix \ref{secF}). For EWMA, we estimated $\hat{\alpha} = 0.979 \in \left\{1.000, 0.999, 0.998, \cdots , 0.950 \right\}$ and $\hat{K} = 23$ by leave-one-out cross-validation. For GHeFM, the estimated number of factors was $\hat{K}=23$.

Panel (A) of Figure \ref{fig_6_4_1} supports that the performance of our RHeFM is significantly higher than that of the common EWMA approach in terms of their predictive likelihoods. RHeFM outperforms the EWMA model consistently and significantly, as shown in panels (B) and (D). Also, due to the nature of financial data, i.e., occasionally exposed to the extreme changes, RHeFM performs better than GHeFM, as shown in (C) and (E).

\begin{table}
\small
\begin{center}
\begin{tabular}{ |c||c|c||c|c||c|c| } 
  \hline
   & EWMA (1.000) & EWMA (0.979) & GHoFM & GHeFM & RHoFM & RHeFM \\ \hline 
  $\hat{K}$ & 21 & 23 & 24 & 23 & 34 & 32 \\
  \hline Scores & 36148 & 36136 & 36643 & 36647 & 37895 & \textbf{38035} \\
  \hline
  
\end{tabular}
\caption{Performance comparison between homoscedastic and heteroscedastic factor analysis models. Scores are the sums of the predictive log-likelihoods (the higher is better).}
\label{table_6_1_1}
\end{center}
\end{table}

We also compare the performance between each model and its time-invariant (homoscedastic) alternative, as shown in Table \ref{table_6_1_1}: EWMA (1.000) is the model with $\alpha=1.000$ and GHoFM and RHoFM are of the time-invariant multivariate normal and Student's t-distributions, respectively. While EWMA (0.979) and GHeFM show similar performance than their time-invariant versions, RHeFM performs better than RHoFM.

Some examples of the estimated time-varying correlation between log-returns of two individual companies can be found in Figure \ref{fig_6_4_3} and a company and all the others in Figure \ref{fig_6_4_4}. One common characteristic is that the time-varying correlations have two peaks in  early 2020 and early 2022,  corresponding to the outbreaks of COVID-19 and the Russo-Ukrainian war, respectively. Whereas the current section addresses raw log-returns, Appendix \ref{secH} applies the factor analysis models to the trimmed-down log-returns for dealing with outliers, and it shows consistent relative performances as well.


\section{Discussion and Conclusion} \label{sec7}

We proposed and formulated a factor analysis model on a multivariate time series covariance structure. The time-varying covariance matrix function is formulated as a harmonic weighted average of the basis covariance matrices over the time-varying weight function.

Our method has the following advantages: (1) the time-varying covariance matrix function can be defined not only on the discrete time indices but also on the continuous time space; (2) the basis covariance matrices, factor loading matrix and idiosyncratic covariance matrix can be estimated by an EM algorithm with closed M-step updates; (3) the original Gaussian model is generalized to Student's t-distribution, spatiotemporal factor modeling, and time-varying idiosyncratic error covariance modeling with closed M-step updates of the conditional maximization algorithm. The method shows superior performance on both the simulated and real-data examples over the EWMA covariance modeling.

So far, we have assumed that the mean of the time series is given as $\bez$. In practice, however, both time-varying mean and covariance functions could be unknown a priori. While the number of factors and bandwidths of the weight functions are estimated by the cross-validation, the number of basis covariance matrices is supposed to be given {\it a priori} and we have not discussed any criterion on its determination. Moreover, any theory regarding the convergence is not treated in the work, e.g., what is the relationship between the number of basis covariance matrices and the convergence to the real one under what conditions? All these questions will be addressed in future work.

\section*{Acknowledgments}
This work was supported in part by the NSF  Grants DMS-2015411;  OCE-2410906; and  OCE-2508422. Most of the work was done while JSL was on the faculty of Department of Statistics, Harvard University. No potential competing interest was reported by the authors.

\section*{Data Availability Statement}
The data and codes that support the findings of this study are openly available in GitHub at \\  https://github.com/eilion/FAMSVM.

\bibliographystyle{unsrt}  
\bibliography{references}

\newpage

\counterwithin{figure}{section}
\counterwithin{table}{section}

\appendixpage
\appendix
\section{The Detailed EM Algorithm in Section \ref{sec3}}\label{secA}

Starting with an initialization $\Omega_K^{\left(0\right)}$, our EM algorithm in Section \ref{sec3} searches for ${\hat{\Omega}}_K$ by iteratively applying the following two steps:
\begin{itemize}
  \item Expectation step (E-step): to calculate the expected logarithm of the augmented posterior with respect to the conditional distribution of unobserved latent data $F$ given $\mathcal{Y}$, $\mathcal{T}$, $\omega$, $K$ and $\Omega_K^{\left(r\right)}$ at the $r$th iteration.
  \begin{equation}
\begin{aligned}
\mathcal{Q}\left(\Omega\middle|\Omega_K^{\left(r\right)}\right)=\mathbb{E}_{p\left(F\middle|\mathcal{Y},\mathcal{T},\omega,K,\Omega_K^{\left(r\right)}\right)}\left[\log{p\left(\mathcal{Y},F,\Omega\mid\mathcal{T},\omega,K\right)}\right].
\end{aligned}
\label{equation_3_0_3}
\end{equation}
  \item Maximization step (M-step): to update parameters $\Omega_K^{\left(r+1\right)}=\argmax_{\Omega}{\mathcal{Q}\left(\Omega\middle|\Omega_K^{\left(r\right)}\right)}$.
\end{itemize}

In the calculation of $\mathcal{Q}\left(\Omega\middle|\Omega_K^{\left(r\right)}\right)$, $\log{p\left(\mathcal{Y},F,\Omega\middle|\mathcal{T},\omega,K\right)}$ is decomposed into:
\begin{equation}
\begin{aligned}
\log{p\left(\mathcal{Y},F,\Omega\middle|\mathcal{T},\omega,K\right)} = & \left(\log{p\left(F\middle|L,\mathcal{T},\omega,K\right)}+\log{p\left(L\middle|\mathcal{T},\omega,K\right)}\right) \\ & +\left(\log{p\left(\mathcal{Y}\middle|F,B,\Sigma\right)}+\log{p\left(B\middle|K\right)}+\log{p\left(\Sigma\right)}\right),
\end{aligned}
\label{equation_3_0_4}
\end{equation}
i.e., the parameters are separated into two disconnected parts $L$ and $\left(B,\Sigma\right)$:
\begin{equation}
\begin{aligned}
\mathcal{Q}\left(\Omega\middle|\Omega_K^{\left(r\right)}\right)=\mathcal{C}^{\left(r\right)}+\mathcal{Q}_1^{\left(r\right)}\left(L\right)+\mathcal{Q}_2^{\left(r\right)}\left(B,\Sigma\right),
\end{aligned}
\label{equation_3_0_5}
\end{equation}
where $\mathcal{Q}_1^{\left(r\right)}$ and $\mathcal{Q}_2^{\left(r\right)}$ are conditional expectations $\mathbb{E}_{p\left(F\middle|\mathcal{Y},\mathcal{T},\omega,K,\Omega_K^{\left(r\right)}\right)}\left[\log{p\left(F,L\middle|\mathcal{T},\omega,K\right)}\right]$ and $\mathbb{E}_{p\left(F\middle|\mathcal{Y},\mathcal{T},\omega,K,\Omega_K^{\left(r\right)}\right)}\left[\log{p\left(\mathcal{Y},B,\Sigma\middle|F,K\right)}\right]$ respectively, and $\mathcal{C}^{\left(r\right)}$ is a constant not involving $\Omega$.

We show the details about the Q-functions in the E-step. First of all, the posterior distribution $p\left(F\middle|\mathcal{Y},\mathcal{T},\omega,K,\Omega^{\left(r\right)}\right)$ can be written in a closed form. To be specific, we have:
\begin{equation}
\begin{aligned}
p\left(F\middle|\mathcal{Y},\mathcal{T},\omega,K,\Omega^{\left(r\right)}\right) & \propto p\left(\mathcal{Y}\middle|F,B^{\left(r\right)},\Sigma,\mathcal{T}\right)p\left(F\middle|L^{\left(r\right)},\mathcal{T},\omega,K\right) \\ & =\prod_{n=1}^{N}\left(\mathcal{N}\left(y_{t_n}\middle|B^{\left(r\right)}f_{n},\Sigma^{\left(r\right)}\right) \cdot \mathcal{N}\left(f_n\middle|0,\Lambda_{t_n}^{\left(r\right)}\right)\right),
\end{aligned}
\label{equation_A_1_1}
\end{equation}
and $\mathcal{N}\left(f_n\middle|{\hat{\eta}}_n,{\hat{\Psi}}_n\right)\propto\mathcal{N}\left(y_{t_n}\middle|B^{\left(r\right)}f_n,\Sigma^{\left(r\right)}\right)\cdot\mathcal{N}\left(f_{n}\middle|0,\Lambda_{t_n}^{\left(r\right)}\right)$ for $\left({\hat{\eta}}_n,{\hat{\Psi}}_n\right)$ defined as below:

\begin{equation}
\begin{aligned}
{\hat{\eta}}_{n}\triangleq{\hat{\Psi}}_{n}\left(B^{\left(r\right)}\right)^\top\left(\Sigma^{\left(r\right)}\right)^{-1}y_{t_n}, \quad {\hat{\Psi}}_{n}\triangleq\left(\left(\Lambda_{t_n}^{\left(r\right)}\right)^{-1}+\left(B^{\left(r\right)}\right)^\top\left(\Sigma^{\left(r\right)}\right)^{-1}B^{\left(r\right)}\right)^{-1}.
\end{aligned}
\label{equation_3_1_1}
\end{equation}

Therefore, we have:
\begin{equation}
\begin{aligned}
p\left(F\middle|\mathcal{Y},\mathcal{T},\omega,K,\Omega^{\left(r\right)}\right)=\prod_{n=1}^{N}{\mathcal{N}\left(f_n\middle|{\hat{\eta}}_n,{\hat{\Psi}}_n\right)}.
\end{aligned}
\label{equation_A_1_2}
\end{equation}

Because $\log{p\left(F,L\middle|\mathcal{T},\omega,K\right)}=\log{p\left(F\middle|L,\mathcal{T},\omega,K\right)}+\log{p\left(L\middle|\mathcal{T},\omega,K\right)}$, we can rewrite it as follows:
\begin{equation}
\begin{aligned}
\log{p\left(F,L\middle|\mathcal{T},\omega,K\right)} & ={\widetilde{\mathcal{C}}}_1^{\left(r\right)}+\sum_{n=1}^{N}\log{\mathcal{N}\left(f_n\middle|0,\left(\sum_{d=1}^{D}{\omega_{d}{\left(t_n\right)}\lambda_d^{-1}}\right)^{-1}\right)} \\ & +\frac{1}{2}\sum_{n=1}^{N}\sum_{d=1}^{D}{\omega_{d}{\left(t_n\right)}\log{\left|\lambda_d^{-1}\right|}}-\frac{1}{2}\sum_{n=1}^{N}\log{\left|\sum_{d=1}^{D}{\omega_{d}{\left(t_n\right)}\lambda_d^{-1}}\right|} \\ & =\mathcal{C}_1^{\left(r\right)}-\frac{1}{2}\sum_{n=1}^{N}\sum_{d=1}^{D}{\omega_{d}{\left(t_n\right)}f_n^\top\lambda_d^{-1}f_n}+\frac{1}{2}\sum_{n=1}^{N}\sum_{d=1}^{D}{\omega_{d}{\left(t_n\right)}\log{\left|\lambda_d^{-1}\right|}}.
\end{aligned}
\label{equation_A_1_3}
\end{equation}
By $\mathbb{E}_{\mathcal{N}\left(f_n\middle|{\hat{\eta}}_n,{\hat{\Psi}}_n\right)}\left[f_n^\top\lambda_d^{-1}f_n\right]={\hat{\eta}}_n^\top\lambda_d^{-1}{\hat{\eta}}_n+tr\left(\lambda_d^{-1}{\hat{\Psi}}_n\right)=tr\left(\left({\hat{\eta}}_n{\hat{\eta}}_n^\top+{\hat{\Psi}}_n\right)\lambda_d^{-1}\right)$, we derived $\mathcal{Q}_1^{\left(r\right)}\left(L\right)$ as below:

\begin{equation}
\begin{aligned}
\mathcal{Q}_1^{\left(r\right)}\left(L\right)=\mathcal{C}_1^{\left(r\right)}-\frac{1}{2}\sum_{n=1}^{N}\sum_{d=1}^{D}{\omega_{d}{\left(t_n\right)} tr \left(\left({\hat{\eta}}_{n}{\hat{\eta}}_{n}^\top+{\hat{\Psi}}_{n}\right)\lambda_{d}^{-1}\right)}-\frac{1}{2}\sum_{n=1}^{N}\sum_{d=1}^{D}{\omega_{d}{\left(t_n\right)}\log{\left|\lambda_d\right|}},
\end{aligned}
\label{equation_3_1_2}
\end{equation}

Also, for $\log{p\left(\mathcal{Y},B,\Sigma\middle|F,\mathcal{T}\right)}=\log{p\left(\mathcal{Y}\middle|F,B,\Sigma,\mathcal{T}\right)}+\log{p\left(B\middle|K\right)}+\log{p\left(\Sigma\right)}$, we have:
\begin{equation}
\begin{aligned}
\log{p\left(\mathcal{Y},B,\Sigma\middle|F,\mathcal{T}\right)} & =\sum_{n=1}^{N}\sum_{q=1}^{Q}\log{\mathcal{N}\left(y_{{t_n}q}\middle|B_{q,:}f_n,\Sigma_q\right)}  \\ & =\mathcal{C}_2^{\left(r\right)}-\frac{1}{2}\sum_{q=1}^{Q}{\Sigma_{q}^{-1}\sum_{n=1}^{N}{\left(y_{{t_n}q}-B_{q,:}f_n\right)^2}} - \frac{N}{2}{\sum_{q=1}^{Q}{\log{\Sigma_q}}}.
\end{aligned}
\label{equation_A_1_4}
\end{equation}
By $\mathbb{E}_{\mathcal{N}\left(f_n\middle|{\hat{\eta}}_n,{\hat{\Psi}}_n\right)}\left[\left(y_{{t_n}q}-B_{q,:}f_n\right)^2\right]=\left(y_{{t_n}q}-B_{q,:}{\hat{\eta}}_n\right)^2+B_{q,:}{\hat{\Psi}}_n B_{q,:}^\top$, we derived $\mathcal{Q}_2^{\left(r\right)}\left(B,\Sigma\right)$ as below:

\begin{equation}
\begin{aligned}
\mathcal{Q}_2^{\left(r\right)}\left(B,\Sigma\right)=\mathcal{C}_2^{\left(r\right)} & -\frac{1}{2}\sum_{q=1}^{Q}{\Sigma_q^{-1}\sum_{n=1}^{N}\left(\left(y_{{t_n}q}-B_{q,:}{\hat{\eta}}_{n}\right)^2+B_{q,:}{\hat{\Psi}}_{n}B_{q,:}^\top\right)}-\frac{N}{2}\sum_{q=1}^{Q}\log{\Sigma_q}.
\end{aligned}
\label{equation_3_1_3}
\end{equation}

In each E-step, we update the terms $\hat{\eta}=\left\{{\hat{\eta}}_n\right\}_{n=1}^{N}$ and $\hat{\Psi}=\left\{{\hat{\Psi}}_{n}\right\}_{n=1}^{N}$ in \eqref{equation_3_1_1} given $\Omega_K^{\left(r\right)}$, and then compute $\mathcal{Q}_1^{\left(r\right)}\left(L\right)$ and $\mathcal{Q}_2^{\left(r\right)}\left(B,\Sigma\right)$ given $\left(\hat{\eta},\hat{\Psi}\right)$ as \eqref{equation_3_1_2} and \eqref{equation_3_1_3}, respectively.

Once the Q-functions $\mathcal{Q}_1^{\left(r\right)}\left(L\right)$ and $\mathcal{Q}_2^{\left(r\right)}\left(B,\Sigma\right)$ have been updated, the M-step consists of maximizing \eqref{equation_3_0_5} with respect to the unknown parameters. The matrix partial derivatives of the Q-functions are computed as follows:
\begin{equation}
\begin{aligned}
\frac{\partial\mathcal{Q}_1^{\left(r\right)}\left(L\right)}{\partial\lambda_d} & =\frac{1}{2}\lambda_d^{-1}\sum_{n=1}^{N}{\omega_{d}{\left(t_n\right)}\left({\hat{\eta}}_n{\hat{\eta}}_n^\top+{\hat{\Psi}}_n\right)}\lambda_d^{-1}-\frac{1}{2}\sum_{n=1}^{N}{\omega_{d}{\left(t_n\right)}\lambda_d^{-1}} \\ \frac{\partial\mathcal{Q}_2^{\left(r\right)}\left(B,\Sigma\right)}{\partial B_{q,:}} & = \Sigma_q^{-1} \sum_{n=1}^{N}\left({y_{{t_n}q}{\hat{\eta}}_n^\top}-B_{q,:}{\left({\hat{\eta}}_n{\hat{\eta}}_n^\top+{\hat{\Psi}}_n \right)}\right) \\ \frac{\partial\mathcal{Q}_2^{\left(r\right)}\left(B,\Sigma\right)}{\partial\Sigma_{q}} & =\frac{1}{2}{\Sigma_q^{-2} \sum_{n=1}^{N}{\left(\left(y_{{t_n}q}-B_{q,:}{\hat{\eta}}_n\right)^2+B_{q,:}{\hat{\Psi}}_n B_{q,:}^\top\right)}} -\frac{N}{2}{\Sigma_{q}^{-1}},
\end{aligned}
\label{equation_A_2_1}
\end{equation}
and the maximizers of the Q-functions are the M-step updates in \eqref{equation_3_2_1} in Section \ref{sec3}.

\section{The Detailed Robust Factor Model in Section \ref{sec4_0}}\label{secC_1}

The full joint distribution of $\mathcal{Y}$, $A$, $F$, and $\Omega$ given $\mathcal{T}$, $\omega$ and $K$ from \eqref{equation_4_0_1} is:
\begin{equation}
\begin{aligned}
p\left(\mathcal{Y},A,F,\Omega\middle|\mathcal{T},\omega,K\right) & =p\left(\mathcal{Y}\middle|A,F,B,\Sigma,\mathcal{T}\right)p\left(F\middle|A,L,\mathcal{T},\omega,K\right)p\left(L\middle|\mathcal{T},\omega,K\right) \\ & \times p\left(B\middle|K\right) p\left(\Sigma\right)  p\left(A\right),
\end{aligned}
\label{equation_A_3_1}
\end{equation}
where:
\begin{equation}
\begin{aligned}
p\left(\mathcal{Y}\middle|A,F,B,\Sigma,\mathcal{T}\right) & =\prod_{n=1}^{N}{\mathcal{N}\left(y_{t_n}\middle|Bf_n,\alpha_n\cdot\Sigma\right)} \\ p\left(F\middle|A,L,\mathcal{T},\omega,K\right) & =\prod_{n=1}^{N}{\mathcal{N}\left(f_n\middle|0,\alpha_n\cdot\Lambda_{t_n}\right)} \\ p\left(L\middle|\mathcal{T},\omega,K\right) & \propto\exp\left(\frac{1}{2}\sum_{n=1}^{N}\sum_{m=1}^{N}{\omega_{m}{\left(t_n\right)}\log{\left|\lambda_m^{-1}\right|}}-\frac{1}{2}\sum_{n=1}^{N}\log{\left|\sum_{m=1}^{N}{\omega_{m}{\left(t_n\right)}\lambda_m^{-1}}\right|}\right) \\ p\left(B\middle|K\right) & \propto 1, \quad p\left(\Sigma\right) \propto 1, \quad  p\left(A\right)=\prod_{n=1}^{N}{\mathcal{IG}\left(\alpha_n\middle|\frac{\nu}{2},\frac{\nu}{2}\right)}.
\end{aligned}
\label{equation_A_3_2}
\end{equation}

In the EM algorithm, we treat both $A$ and $F$ as missing values and the degree of freedom $\upsilon$ is assumed to be given a priori and fixed: note that the posterior distribution $p\left(A,F\middle|\mathcal{Y},\mathcal{T},\omega,K,\Omega^{\left(r\right)}\right)$ can be written in a closed form. To be specific, we have:
\begin{equation}
\begin{aligned}
p & \left(A,F\middle|\mathcal{Y},\mathcal{T},\omega,K,\Omega^{\left(r\right)}\right) \propto p\left(\mathcal{Y}\middle|A,F,B^{\left(r\right)},\Sigma^{\left(r\right)},\mathcal{T}\right)p\left(F\middle|A,L^{\left(r\right)},\mathcal{T},\omega,K\right)p\left(A\right) \\ & =\prod_{n=1}^{N}\left(\mathcal{N}\left(y_{t_n}\middle|B^{\left(r\right)}f_n,\alpha_n\cdot\Sigma^{\left(r\right)}\right)\cdot\mathcal{N}\left(f_n\middle|0,\alpha_n\cdot\Lambda_{t_n}^{\left(r\right)}\right)\cdot\mathcal{IG}\left(\alpha_n\middle|\frac{\nu}{2},\frac{\nu}{2}\right)\right),
\end{aligned}
\label{equation_A_3_3}
\end{equation}
and:
\begin{equation}
\begin{aligned}
\mathcal{IG} & \left(\alpha_n\middle|\frac{\nu+Q}{2},\frac{\nu+Q}{2}{\hat{\xi}}_n^{-2}\right)\cdot\mathcal{N}\left(f_n\middle|{\hat{\eta}}_n,\alpha_n\cdot{\hat{\Psi}}_n\right) \\ & \propto\mathcal{N}\left(y_{t_n}\middle|B^{\left(r\right)}f_n,\alpha_n\cdot\Sigma^{\left(r\right)}\right)\cdot\mathcal{N}\left(f_n\middle|0,\alpha_n\cdot\Lambda^{\left(r\right)}\right)\cdot\mathcal{IG}\left(\alpha_n\middle|\frac{\nu}{2},\frac{\nu}{2}\right),
\end{aligned}
\label{equation_A_3_3_1}
\end{equation}
for $\left({\hat{\eta}}_n,{\hat{\Psi}}_n\right)$ in \eqref{equation_3_1_1} and ${\hat{\xi}}_n^2$ defined as follows:
\begin{equation}
\begin{aligned}
{\hat{\xi}}_{n}^2\triangleq\frac{\upsilon+Q}{\upsilon+y_{t_n}^\top\left(B^{\left(r\right)}\Lambda_{t_n}^{\left(r\right)}\left(B^{\left(r\right)}\right)^\top+\Sigma^{\left(r\right)}\right)^{-1}y_{t_n}}.
\end{aligned}
\label{equation_4_0_2}
\end{equation}

Therefore, we have:
\begin{equation}
\begin{aligned}
p\left(A,F\middle|\mathcal{Y},\mathcal{T},\omega,K,\Omega^{\left(r\right)}\right)=\prod_{n=1}^{N}\left(\mathcal{IG}\left(\alpha_n\middle|\frac{\nu+Q}{2},\frac{\nu+Q}{2}{\hat{\xi}}_n^{-2}\right)\cdot\mathcal{N}\left(f_n\middle|{\hat{\eta}}_n,\alpha_n\cdot{\hat{\Psi}}_n\right)\right).
\end{aligned}
\label{equation_A_3_4}
\end{equation}

Since $\log{p\left(F,L\middle|A,\mathcal{T},\omega,K\right)}=\log{p\left(F\middle|A,L,\mathcal{T},\omega,K\right)}+\log{p\left(L\middle|\mathcal{T},\omega,K\right)}$, we can rewrite it as follows:
\begin{equation}
\begin{aligned}
\log{p\left(F,L\middle|A,\mathcal{T},\omega,K\right)} & ={\widetilde{\mathcal{C}}}_1^{\left(r\right)}+\sum_{n=1}^{N}\log{\mathcal{N}\left(f_n\middle|0,\alpha_n\cdot\left(\sum_{m=1}^{N}{\omega_{m}{\left(t_n\right)}\lambda_m^{-1}}\right)^{-1}\right)} \\ & +\frac{1}{2}\sum_{n=1}^{N}\sum_{m=1}^{N}{\omega_{m}{\left(t_n\right)}\log{\left|\lambda_m^{-1}\right|}}-\frac{1}{2}\sum_{n=1}^{N}\log{\left|\sum_{m=1}^{N}{\omega_{m}{\left(t_n\right)}\lambda_m^{-1}}\right|} \\ & =\mathcal{C}_1^{\left(r\right)}-\frac{1}{2}\sum_{n=1}^{N}\sum_{m=1}^{N}{\omega_{m}{\left(t_n\right)}\alpha_n^{-1}f_n^\top\lambda_m^{-1}f_n} \\ & +\frac{1}{2}\sum_{n=1}^{N}\sum_{m=1}^{N}{\omega_{m}{\left(t_n\right)}\log{\left|\lambda_m^{-1}\right|}} -\frac{K}{2}\sum_{n=1}^{N}\log{\alpha_n}.
\end{aligned}
\label{equation_A_3_5}
\end{equation}

By the following equalities: 
\begin{equation}
\begin{aligned}
\mathbb{E}_{\mathcal{N}\left(f_n\middle|{\hat{\eta}}_n,\alpha_n\cdot{\hat{\Psi}}_n\right)}\left[\alpha_n^{-1}f_n^\top\lambda_m^{-1}f_n\right] & =\alpha_n^{-1}{\hat{\eta}}_n^\top\lambda_m^{-1}{\hat{\eta}}_n+tr\left(\lambda_m^{-1}{\hat{\Psi}}_n\right) \\ \mathbb{E}_{\mathcal{IG}\left(\alpha_n\middle|\frac{\nu+Q}{2},\frac{\nu+Q}{2}{\hat{\xi}}_n^{-2}\right)}\left[\alpha_n^{-1}{\hat{\eta}}_n^\top\lambda_m^{-1}{\hat{\eta}}_n\right] & ={\hat{\xi}}_n^2{\hat{\eta}}_n^\top\lambda_m^{-1}{\hat{\eta}}_n \\ \mathbb{E}_{\mathcal{IG}\left(\alpha_n\middle|\frac{\nu+Q}{2},\frac{\nu+Q}{2}{\hat{\xi}}_n^{-2}\right)}\left[\log{\alpha_n}\right] & =\log{\left(\frac{\nu+Q}{2}{\hat{\xi}}_n^{-2}\right)}-\psi\left(\frac{\nu+Q}{2}\right),
\end{aligned}
\label{equation_A_3_5_1}
\end{equation}
we have derived $\mathcal{Q}_1^{\left(r\right)}\left(L\right) \triangleq \mathbb{E}_{p\left(A,F\middle|\mathcal{Y},\mathcal{T},\omega,K,\Omega^{\left(r\right)}\right)}\left[\log{p\left(F,L\middle|A,\mathcal{T},\omega,K\right)}\right]$ as follows:
\begin{equation}
\begin{aligned}
\mathcal{Q}_1^{\left(r\right)}\left(L\right)=\mathcal{C}_1^{\left(r\right)} -\frac{1}{2}\sum_{n=1}^{N}\sum_{m=1}^{N}{\omega_{m}{\left(t_n\right)} tr{ \left(\left({\hat{\xi}}_{n}^2{\hat{\eta}}_{n}{\hat{\eta}}_{n}^\top+{\hat{\Psi}}_{n}\right)\lambda_{m}^{-1}\right)}} +\frac{1}{2}\sum_{n=1}^{N}\sum_{m=1}^{N}{\omega_{m}{\left(t_n\right)}\log{\left|\lambda_{m}^{-1}\right|}}.
\end{aligned}
\label{equation_4_0_3}
\end{equation}

For $\log{p\left(\mathcal{Y},B,\Sigma\middle|A,F,\mathcal{T}\right)}=\log{p\left(\mathcal{Y}\middle|A,F,B,U,\mathcal{T}\right)}+\log{p\left(B\middle|K\right)}+\log{p\left(\Sigma\right)}$, we have:
\begin{equation}
\begin{aligned}
& \log{p\left(\mathcal{Y},B,\Sigma\middle|A,F,\mathcal{T}\right)}  =\sum_{n=1}^{N}\sum_{q=1}^{Q}\log{\mathcal{N}\left(y_{{t_n}q}\middle|B_{q,:}f_n,\alpha_n\cdot\Sigma\right)} \\ & =\mathcal{C}_2^{\left(r\right)}-\frac{1}{2}\sum_{q=1}^{Q}{\Sigma_{q}^{-1} \sum_{n=1}^{N}{\alpha_n^{-1}\left(y_{{t_n}q}-B_{q,:}f_n\right)^2}} - \frac{N}{2}\sum_{q=1}^{Q}{\log{\Sigma_{q}}} - \frac{Q}{2}\sum_{n=1}^{N}\log{\alpha_n}.
\end{aligned}
\label{equation_A_3_6}
\end{equation}

By the following equalities:
\begin{equation}
\begin{aligned}
\mathbb{E}_{\mathcal{N}\left(f_n\middle|{\hat{\eta}}_n,\alpha_n\cdot {\hat{\Psi}}_n\right)}\left[\alpha_n^{-1}\left(y_{{t_n}q}-B_{q,:}f_n\right)^2\right] & =\alpha_n^{-1}\left(y_{{t_n}q}-B_{q,:}{\hat{\eta}}_n\right)^2+B_{q,:}{\hat{\Psi}}_n B_{q,:}^\top \\ \mathbb{E}_{\mathcal{IG}\left(\alpha_n\middle|\frac{\nu+Q}{2},\frac{\nu+Q}{2}{\hat{\xi}}_n^{-2}\right)}\left[\alpha_n^{-1}\left(y_{{t_n}q}-B_{q,:}{\hat{\eta}}_n\right)^2\right] & ={\hat{\xi}}_n^2\left(y_{{t_n}q}-B_{q,:}{\hat{\eta}}_n\right)^2 \\ \mathbb{E}_{\mathcal{IG}\left(\alpha_n\middle|\frac{\nu+Q}{2},\frac{\nu+Q}{2}{\hat{\xi}}_n^{-2}\right)}\left[\log{\alpha_n}\right] & =\log{\left(\frac{\nu+Q}{2}{\hat{\xi}}_n^{-2}\right)}-\psi\left(\frac{\nu+Q}{2}\right),
\end{aligned}
\label{equation_A_3_6_1}
\end{equation}
we have derived $\mathcal{Q}_2^{\left(r\right)}\left(B,\Sigma\right) \triangleq \mathbb{E}_{p\left(A,F\middle|\mathcal{Y},\mathcal{T},\omega,K,\Omega^{\left(r\right)}\right)}\left[\log{p\left(\mathcal{Y},B,\Sigma\middle|A,F,\mathcal{T}\right)}\right]$ as follows:
\begin{equation}
\begin{aligned}
\mathcal{Q}_2^{\left(r\right)}\left(B,\Sigma\right)=\mathcal{C}_2^{\left(r\right)} & -\frac{1}{2}\sum_{q=1}^{Q}{\sigma_q^{-2}\sum_{n=1}^{N}{\left({\hat{\xi}}_{n}^{2}\left(y_{{t_n}q}-B_{q,:}{\hat{\eta}}_{n}\right)^2+B_{q,:}{\hat{\Psi}}_{n}B_{q,:}^\top\right)}}-N\sum_{q=1}^{Q}\log{\sigma_q},
\end{aligned}
\label{equation_4_0_4}
\end{equation}

Finally, from \eqref{equation_4_0_3} and \eqref{equation_4_0_4}, we have M-step updates for $\left(L^{\left(r+1\right)},B^{\left(r+1\right)},\Sigma^{\left(r+1\right)}\right)$, similar to those in \eqref{equation_3_2_1}, as follows:
\begin{equation}
\begin{aligned}
\lambda_{m}^{\left(r+1\right)} & =\frac{\sum_{n=1}^{N}{\omega_{m}{\left(t_n\right)}\left({\hat{\xi}}_n^2{\hat{\eta}}_n{\hat{\eta}}_n^\top+{\hat{\Psi}}_n\right)}}{\sum_{n=1}\omega_{m}{\left(t_n\right)}} \\ B_{q,:}^{\left(r+1\right)} & =\left(\sum_{n=1}^{N}{\hat{\xi}_{n}^{2}y_{{t_n}q}{\hat{\eta}}_{n}^\top}\right)\left(\sum_{n=1}^{N}{\left(\hat{\xi}_{n}^{2}{\hat{\eta}}_{n}{\hat{\eta}}_{n}^\top+{\hat{\Psi}}_{n}\right)}\right)^{-1} \\ \Sigma_{q}^{\left(r+1\right)} & = \frac{1}{N} \sum_{n=1}^{N}{\left(\hat{\xi}_{n}^{2}\left(y_{{t_n}q}-B_{q,:}^{\left(r+1\right)}{\hat{\eta}}_{n}\right)^2+B_{q,:}^{\left(r+1\right)}{\hat{\Psi}}_{n}\left(B_{q,:}^{\left(r+1\right)}\right)^\top\right)}.
\end{aligned}
\label{equation_4_0_5}
\end{equation}

Note that the above forms are restored to those in Section \ref{sec3} if ${\hat{\xi}}_{n}^2\equiv 1$, which is consistent with an intuition that the Gaussian distribution is a special case of a Student’s t-distribution with an $\upsilon=\infty$ degree of freedom. The estimated ${\hat{\xi}}_{n}^2$’s can be understood as a quantification of the magnitude of not being an outlier of the corresponding $y_{t_n}$. Deciding the optimal $K$ is similar to Section \ref{sec3}, with the Student's t-distribution instead of Gaussian in \eqref{equation_3_4_1}.

\section{The Detailed Spatiotemporal Factor Model in Section \ref{sec4_1}}\label{secC_2}

\subsection{Model \eqref{equation_4_1_0}}\label{secC_2_1}

Let $\Omega=\left(L,B,C,\Sigma,\Phi\right)$, and the full joint distribution of $\mathcal{Y}$, a set of $K_1 K_2 \times 1$ latent factor matrices $F$, and $\Omega$ - given $\mathcal{T}$, $\omega$, $K_1$ and $K_2$ - is defined as follows:
\begin{equation}
\begin{aligned}
p\left(\mathcal{Y},F,\Omega\middle|\mathcal{T},\omega,K_1,K_2\right) = & p\left(\mathcal{Y}\middle|F,B,C,\Sigma,\Phi\right) p\left(F\middle|L,\mathcal{T},\omega,K\right) p\left(L\middle|\mathcal{T},\omega,K_1 K_2\right)   \\ & \times  p\left(B\middle|K_1\right) p\left(C\middle|K_2\right) p\left(\Sigma\right) p\left(\Phi\right),
\end{aligned}
\label{equation_C_2_1_1}
\end{equation}
where:
\begin{equation}
\begin{aligned}
p\left(\mathcal{Y}\middle|F,B,C,\Sigma,\Phi\right) & = \prod_{n=1}^{N}{\mathcal{N}\left({\rm{vec}}{\left(y_{t_n}\right)}\middle|\left(C\otimes B\right)f_n,\Phi\otimes\Sigma\right)} \\ p\left(F\middle|L,\mathcal{T},\omega,K\right) & =\prod_{n=1}^{N}{\mathcal{N}\left(f_n\middle|0,\Lambda_{t_n}\right)} = \prod_{n=1}^{N}{\mathcal{N}\left(f_n\middle|0, \left(\sum_{m=1}^{N}{\omega_{m}{\left(t_n\right)}\lambda_m^{-1}}\right)^{-1}\right)} \\ p\left(L\middle|\mathcal{T},\omega,K_1 K_2\right) & \propto \exp\left(\frac{P}{2}\sum_{m=1}^{N}\sum_{n=1}^{N}{\omega_{m}{\left(t_n\right)}\log{\left|\lambda_m^{-1}\right|}}-\frac{P}{2}\sum_{m=1}^{N}\log{\left|\sum_{n=1}^{N}{\omega_{m}{\left(t_n\right)}\lambda_m^{-1}}\right|}\right) \\ p\left(B\middle|K_1\right) & \propto 1, \quad p\left(C\middle|K_2\right) \propto 1, \quad  p\left(\Sigma\right) \propto 1, \quad  p\left(\Phi\right) \propto 1.
\end{aligned}
\label{equation_C_2_1_2}
\end{equation}

In the EM algorithm, we treat $F$ as missing values. For each EM iteration $r$, we first define:
\begin{equation}
{\hat{\Psi}}_{n} \triangleq\left(\left(\Lambda_{t_n}^{\left(r\right)}\right)^{-1} + \left(C^{\left(r\right)} \otimes B^{\left(r\right)}\right)^\top\left(\Phi^{\left(r\right)} \otimes \Sigma^{\left(r\right)}\right)^{-1} \left(C^{\left(r\right)} \otimes B^{\left(r\right)}\right)\right)^{-1},
\label{equation_C_2_1_3}
\end{equation}
and let $\left\{{\hat{\Psi}}_n\right\}_{ij}$ be a $K_1 \times K_1$ block of the $K_1 K_2 \times K_1 K_2$ matrix ${\hat{\Psi}}_n$, from the ($\left(i-1\right)K_1+1$)-th to $i K_1$-th row, from the ($\left(j-1\right)K_1+1$)-th to $j K_1$-th column, for each $i,j = 1,2,\cdots,K_2$. Then, we define each column of the $K_1 \times K_2$ matrix ${\hat{\eta}}_{n}$ as follows:
\begin{equation}
\left[{\hat{\eta}}_n\right]_{:,k} \triangleq \sum_{l=1}^{K_2}{\left\{{\hat{\Psi}}_n\right\}_{kl}\left[{B^{\left(r\right)}}^\top\left(\Sigma^{\left(r\right)}\right)^{-1}y_{t_n}\left(\Phi^{\left(r\right)}\right)^{-1}C^{\left(r\right)}\right]_{:,l}}.
\label{equation_C_2_1_4}
\end{equation}

Note that ${\rm{vec}}{\left(\hat{\eta}_n\right)}$ and $\hat{\Psi}_n$ are the posterior mean and covariance of $f_n$, i.e.,
\begin{equation}
p\left(F\middle|\mathcal{Y},\mathcal{T},\omega,K_1,K_2,\Omega^{\left(r\right)}\right)=\prod_{n=1}^{N}{\mathcal{N}\left(f_n\middle|{\rm{vec}}{\left({\hat{\eta}}_n\right)},{\hat{\Psi}}_n\right)},
\label{equation_C_2_1_5}
\end{equation}
and it is straightforward to derive the following Q-functions just as Appendix \ref{secA}:
\begin{equation}
\begin{aligned}
& \mathcal{Q}_1^{\left(r\right)}\left(L\right) \triangleq  \mathbb{E}_{p\left(F\middle|\mathcal{Y},\mathcal{T},\omega,K_1,K_2,\Omega^{\left(r\right)}\right)}\left[\log{p\left(F,L\middle|\mathcal{T},\omega,K_1,K_2\right)}\right] \\ = & -\frac{1}{2}\sum_{n=1}^{N}tr\left(\left({\rm{vec}}\left({\hat{\eta}}_n\right){{\rm{vec}}\left({\hat{\eta}}_n\right)}^\top+{\hat{\Psi}}_n\right)\left(\sum_{m=1}^{N}{\omega_m\left(t_n\right)\lambda_m^{-1}}\right)\right) - \frac{1}{2}\sum_{m=1}^{N}\sum_{n=1}^{N}{\omega_m\left(t_n\right)\log{\left|\lambda_m\right|}} + \mathcal{C}_1^{\left(r\right)} \\ & \mathcal{Q}_2^{\left(r\right)}\left(C,B,\Phi,\Sigma\right) \triangleq \mathbb{E}_{p\left(F\middle|\mathcal{Y},\mathcal{T},\omega,K_1,K_2,\Omega^{\left(r\right)}\right)}\left[\log{p\left(Y,C,B,\Phi,\Sigma\middle|F\right)}\right] \\ = & -\frac{1}{2}\sum_{n=1}^{N}tr\left(\left({\rm{vec}}\left(y_{t_n}-B{\hat{\eta}}_nC^\top\right){{\rm{vec}}\left(y_{t_n}-B{\hat{\eta}}_nC^\top\right)}^\top+\left(C\otimes B\right){\hat{\Psi}}_n\left(C\otimes B\right)^\top\right)\left(\Phi\otimes\Sigma\right)^{-1}\right) \\ & - \frac{QN}{2}\sum_{p=1}^{P}{\log{\Phi_p}} - \frac{PN}{2}\sum_{q=1}^{Q}{\log{\Sigma_q}} + \mathcal{C}_2^{\left(r\right)},
\end{aligned}
\label{equation_C_2_1_6}
\end{equation}
and the following conditional M-step updates are the maximizer of those Q-functions:
\begin{equation}
\begin{aligned}
\lambda_{m}^{\left(r+1\right)} = & \frac{\sum_{n=1}^{N}{\omega_{m}{\left(t_n\right)}\left({\rm{vec}}\left({\hat{\eta}}_n\right){{\rm{vec}}\left({\hat{\eta}}_n\right)}^\top+{\hat{\Psi}}_n\right)}}{\sum_{n=1}^{N}\omega_{m}{\left(t_{n}\right)}}. \\ 
\Phi_{pp}^{\left(r+1\right)} = & \frac{1}{QN}\sum_{n=1}^{N}\left[\left(y_{t_n}-B^{\left(r\right)}{\hat{\eta}}_n{C^{\left(r\right)}}^\top\right)^\top\left(\Sigma^{\left(r\right)}\right)^{-1}\left(y_{t_n}-B^{\left(r\right)}{\hat{\eta}}_n{C^{\left(r\right)}}^\top\right)\right]_{pp} \\ & + \frac{1}{QN}\sum_{i=1}^{K_2}\sum_{j=1}^{K_2}{\left[C^{\left(r\right)}\right]_{pi}\left[C^{\left(r\right)}\right]_{pj}tr\left({B^{\left(r\right)}}^\top\left(\Sigma^{\left(r\right)}\right)^{-1}B^{\left(r\right)}\sum_{n=1}^{N}\left\{{\hat{\Psi}}_n\right\}_{ij}\right)} \\ \Sigma_{qq}^{\left(r+1\right)} = & \frac{1}{PN}\sum_{n=1}^{N}\left[\left(y_{t_n}-B^{\left(r\right)}{\hat{\eta}}_n{C^{\left(r\right)}}^\top\right)\left(\Phi^{\left(r+1\right)}\right)^{-1}\left(y_{t_n}-B^{\left(r\right)}{\hat{\eta}}_n{C^{\left(r\right)}}^\top\right)^\top\right]_{qq} \\ & + \frac{1}{PN}\sum_{i=1}^{K_2}\sum_{j=1}^{K_2}{\left[{C^{\left(r\right)}}^\top\left(\Phi^{\left(r+1\right)}\right)^{-1}C^{\left(r\right)}\right]_{ij}\left[B^{\left(r\right)}\sum_{n=1}^{N}\left\{{\hat{\Psi}}_n\right\}_{ij}{B^{\left(r\right)}}^\top\right]_{qq}} \\ C^{\left(r+1\right)} = & \left(\sum_{n=1}^{N}{y_{t_n}^\top\left(\Sigma^{\left(r+1\right)}\right)^{-1}B^{\left(r\right)}{\hat{\eta}}_n}\right) \\ & \cdot \left(\sum_{n=1}^{N}{{\hat{\eta}}_n^\top{B^{\left(r\right)}}^\top\left(\Sigma^{\left(r+1\right)}\right)^{-1}B^{\left(r\right)}{\hat{\eta}}_n}+\sum_{i=1}^{K_2}\sum_{j=1}^{K_2}{tr\left({B^{\left(r\right)}}^\top\left(\Sigma^{\left(r+1\right)}\right)^{-1}B^{\left(r\right)}\sum_{n=1}^{N}\left\{{\hat{\Psi}}_n\right\}_{ij}\right)\mathcal{J}_{ij}^{K_2K_2}}\right)^{-1} \\ B^{\left(r+1\right)} = & \left(\sum_{n=1}^{N}{y_{t_n}\left(\Phi^{\left(r+1\right)}\right)^{-1}C^{\left(r+1\right)}{\hat{\eta}}_n^\top}\right)  \\ & \cdot \left(\sum_{n=1}^{N}{{\hat{\eta}}_n{C^{\left(r+1\right)}}^\top\left(\Phi^{\left(r+1\right)}\right)^{-1}C^{\left(r+1\right)}{\hat{\eta}}_n^\top}+\sum_{i=1}^{K_2}\sum_{j=1}^{K_2}{\left[{C^{\left(r+1\right)}}^\top\left(\Phi^{\left(r+1\right)}\right)^{-1}C^{\left(r+1\right)}\right]_{ij}\sum_{n=1}^{N}\left\{{\hat{\Psi}}_n\right\}_{ij}}\right)^{-1}.
\end{aligned}
\label{equation_C_2_1_7}
\end{equation}

\subsection{Model \eqref{equation_4_1_1}}\label{secC_2_2}

Let $\Omega=\left(L,B,G,C,\Sigma,\Phi\right)$, and the full joint distribution of $\mathcal{Y}$, a set of $K_1 \times K_2$ latent factor matrices $F$, and $\Omega$ - given $\mathcal{T}$, $\omega$, $\rho$, $K_1$ and $K_2$ - is defined as follows:
\begin{equation}
\begin{aligned}
p\left(\mathcal{Y},F,\Omega\middle|\mathcal{T},\rho,\omega,K_1,K_2\right) = & p\left(\mathcal{Y}\middle|F,B,C,\Sigma,\Phi\right) p\left(F\middle|L,G,\mathcal{T},\omega,\rho,K\right) p\left(L\middle|\mathcal{T},\omega,K_1,K_2\right)   \\ & \times p\left(G\middle|\mathcal{T},\rho,K_1,K_2\right) p\left(B\middle|K_1\right) p\left(C\middle|K_2\right) p\left(\Sigma\right) p\left(\Phi\right),
\end{aligned}
\label{equation_4_1_3}
\end{equation}
where:
\begin{equation}
\begin{aligned}
p\left(\mathcal{Y}\middle|F,B,C,\Sigma,\Phi\right) & = \prod_{n=1}^{N}{\mathcal{N}\left({\rm{vec}}{\left(y_{t_n}\right)}\middle|\left(C\otimes B\right){\rm{vec}}{\left(f_n\right)},\Phi\otimes\Sigma\right)} \\ p\left(F\middle|L,G,\mathcal{T},\omega,\rho,K\right) & =\prod_{n=1}^{N}{\mathcal{N}\left({\rm{vec}}{\left(f_n\right)}\middle|0,\Gamma_{t_n}\otimes\Lambda_{t_n}\right)} \\ & =\prod_{n=1}^{N}{\mathcal{N}\left({\rm{vec}}{\left(f_n\right)}\middle|0,\left(\sum_{m=1}^{N}{\rho_{m}{\left(t_n\right)}\gamma_m^{-1}}\right)^{-1} \otimes \left(\sum_{m=1}^{N}{\omega_{m}{\left(t_n\right)}\lambda_m^{-1}}\right)^{-1}\right)} \\ p\left(L\middle|\mathcal{T},\omega,K_1,K_2\right) & \propto \exp\left(\frac{K_2}{2}\sum_{m=1}^{N}\sum_{n=1}^{N}{\omega_{m}{\left(t_n\right)}\log{\left|\lambda_m^{-1}\right|}}-\frac{K_2}{2}\sum_{m=1}^{N}\log{\left|\sum_{n=1}^{N}{\omega_{m}{\left(t_n\right)}\lambda_m^{-1}}\right|}\right) \\  p\left(G\middle|\mathcal{T},\rho,K_1,K_2\right) & \propto \exp\left(\frac{K_1}{2}\sum_{m=1}^{N}\sum_{n=1}^{N}{\rho_{m}{\left(t_n\right)}\log{\left|\gamma_m^{-1}\right|}}-\frac{K_1}{2}\sum_{m=1}^{N}\log{\left|\sum_{n=1}^{N}{\rho_{m}{\left(t_n\right)}\gamma_m^{-1}}\right|}\right) \\ p\left(B\middle|K_1\right) & \propto 1, \quad p\left(C\middle|K_2\right) \propto 1, \quad  p\left(\Sigma\right) \propto 1, \quad  p\left(\Phi\right) \propto 1.
\end{aligned}
\label{equation_4_1_4}
\end{equation}

In the EM algorithm, we treat $F$ as missing values. For each EM iteration $r$, we first define:
\begin{equation}
{\hat{\Psi}}_{n} \triangleq\left(\left(\Gamma_{t_n}^{\left(r\right)} \otimes \Lambda_{t_n}^{\left(r\right)}\right)^{-1} + \left(C^{\left(r\right)} \otimes B^{\left(r\right)}\right)^\top\left(\Phi^{\left(r\right)} \otimes \Sigma^{\left(r\right)}\right)^{-1} \left(C^{\left(r\right)} \otimes B^{\left(r\right)}\right)\right)^{-1},
\label{equation_4_1_5}
\end{equation}
and let $\left\{{\hat{\Psi}}_n\right\}_{ij}$ be a $K_1 \times K_1$ block of the $K_1 K_2 \times K_1 K_2$ matrix ${\hat{\Psi}}_n$, from the ($\left(i-1\right)K_1+1$)-th to $i K_1$-th row, from the ($\left(j-1\right)K_1+1$)-th to $j K_1$-th column, for each $i,j = 1,2,\cdots,K_2$. Then, we define each column of the $K_1 \times K_2$ matrix ${\hat{\eta}}_{n}$ as follows:
\begin{equation}
\left[{\hat{\eta}}_n\right]_{:,k} \triangleq \sum_{l=1}^{K_2}{\left\{{\hat{\Psi}}_n\right\}_{kl}\left[{B^{\left(r\right)}}^\top\left(\Sigma^{\left(r\right)}\right)^{-1}y_{t_n}\left(\Phi^{\left(r\right)}\right)^{-1}C^{\left(r\right)}\right]_{:,l}}.
\label{equation_4_1_6}
\end{equation}

Note that ${\rm{vec}}{\left(\hat{\eta}_n\right)}$ and $\hat{\Psi}_n$ are the posterior mean and covariance of ${\rm{vec}}{\left(f_n\right)}$, i.e.,
\begin{equation}
p\left(F\middle|\mathcal{Y},\mathcal{T},\rho,\omega,K_1,K_2,\Omega^{\left(r\right)}\right)=\prod_{n=1}^{N}{\mathcal{N}\left({\rm{vec}}{\left(f_n\right)}\middle|{\rm{vec}}{\left({\hat{\eta}}_n\right)},{\hat{\Psi}}_n\right)},
\label{equation_4_1_7}
\end{equation}
and it is straightforward to derive the following Q-functions just as Appendix \ref{secA}:
\begin{equation}
\begin{aligned}
& \mathcal{Q}_1^{\left(r\right)}\left(L,G\right) \triangleq  \mathbb{E}_{p\left(F\middle|\mathcal{Y},\mathcal{T},\rho,\omega,K_1,K_2,\Omega^{\left(r\right)}\right)}\left[\log{p\left(F,L,G\middle|\mathcal{T},\rho,\omega,K_1,K_2\right)}\right] \\ = & -\frac{1}{2}\sum_{n=1}^{N}tr\left(\left({\rm{vec}}\left({\hat{\eta}}_n\right){{\rm{vec}}\left({\hat{\eta}}_n\right)}^\top+{\hat{\Psi}}_n\right)\left(\left(\sum_{m=1}^{N}{\rho_m\left(t_n\right)\gamma_m^{-1}}\right)\otimes\left(\sum_{m=1}^{N}{\omega_m\left(t_n\right)\lambda_m^{-1}}\right)\right)\right) \\ & - \frac{K_1}{2}\sum_{m=1}^{N}\sum_{n=1}^{N}{\rho_m\left(t_n\right)\log{\left|\gamma_m\right|}}-\frac{K_2}{2}\sum_{m=1}^{N}\sum_{n=1}^{N}{\omega_m\left(t_n\right)\log{\left|\lambda_m\right|}} + \mathcal{C}_1^{\left(r\right)} \\ & \mathcal{Q}_2^{\left(r\right)}\left(C,B,\Phi,\Sigma\right) \triangleq \mathbb{E}_{p\left(F\middle|\mathcal{Y},\mathcal{T},\rho,\omega,K_1,K_2,\Omega^{\left(r\right)}\right)}\left[\log{p\left(Y,C,B,\Phi,\Sigma\middle|F\right)}\right] \\ = & -\frac{1}{2}\sum_{n=1}^{N}tr\left(\left({\rm{vec}}\left(y_{t_n}-B{\hat{\eta}}_nC^\top\right){{\rm{vec}}\left(y_{t_n}-B{\hat{\eta}}_nC^\top\right)}^\top+\left(C\otimes B\right){\hat{\Psi}}_n\left(C\otimes B\right)^\top\right)\left(\Phi\otimes\Sigma\right)^{-1}\right) \\ & - \frac{QN}{2}\sum_{p=1}^{P}{\log{\Phi_p}} - \frac{PN}{2}\sum_{q=1}^{Q}{\log{\Sigma_q}} + \mathcal{C}_2^{\left(r\right)},
\end{aligned}
\label{equation_4_1_8}
\end{equation}
and the following conditional M-step updates are the maximizer of those Q-functions:
\begin{equation}
\begin{aligned}
\gamma_m^{\left(r+1\right)} = & \frac{\sum_{n=1}^{N}{\rho_m\left(t_n\right)\left({\hat{\eta}}_n^\top\left(\Lambda_{t_n}^{\left(r\right)}\right)^{-1}{\hat{\eta}}_n+\sum_{i=1}^{K_2}\sum_{j=1}^{K_2}{tr\left(\left\{{\hat{\Psi}}_n\right\}_{ij}\left(\Lambda_{t_n}^{\left(r\right)}\right)^{-1}\right)\mathcal{J}_{ij}^{K_2 K_2}}\right)}}{K_1 \sum_{n=1}^{N}{\rho_m\left(t_n\right)}} \\ \lambda_m^{\left(r+1\right)} = & \frac{\sum_{n=1}^{N}{\omega_m\left(t_n\right)\left({\hat{\eta}}_n\left(\Gamma_{t_n}^{\left(r+1\right)}\right)^{-1}{\hat{\eta}}_n^\top+\sum_{i=1}^{K_2}\sum_{j=1}^{K_2}{\left[\left(\Gamma_{t_n}^{\left(r+1\right)}\right)^{-1}\right]_{ij}\left\{{\hat{\Psi}}_n\right\}_{ij}}\right)}}{K_2 \sum_{n=1}^{N}{\omega_m\left(t_n\right)}} \\ 
\Phi_{pp}^{\left(r+1\right)} = & \frac{1}{QN}\sum_{n=1}^{N}\left[\left(y_{t_n}-B^{\left(r\right)}{\hat{\eta}}_n{C^{\left(r\right)}}^\top\right)^\top\left(\Sigma^{\left(r\right)}\right)^{-1}\left(y_{t_n}-B^{\left(r\right)}{\hat{\eta}}_n{C^{\left(r\right)}}^\top\right)\right]_{pp} \\ & + \frac{1}{QN}\sum_{i=1}^{K_2}\sum_{j=1}^{K_2}{\left[C^{\left(r\right)}\right]_{pi}\left[C^{\left(r\right)}\right]_{pj}tr\left({B^{\left(r\right)}}^\top\left(\Sigma^{\left(r\right)}\right)^{-1}B^{\left(r\right)}\sum_{n=1}^{N}\left\{{\hat{\Psi}}_n\right\}_{ij}\right)} \\ \Sigma_{qq}^{\left(r+1\right)} = & \frac{1}{PN}\sum_{n=1}^{N}\left[\left(y_{t_n}-B^{\left(r\right)}{\hat{\eta}}_n{C^{\left(r\right)}}^\top\right)\left(\Phi^{\left(r+1\right)}\right)^{-1}\left(y_{t_n}-B^{\left(r\right)}{\hat{\eta}}_n{C^{\left(r\right)}}^\top\right)^\top\right]_{qq} \\ & + \frac{1}{PN}\sum_{i=1}^{K_2}\sum_{j=1}^{K_2}{\left[{C^{\left(r\right)}}^\top\left(\Phi^{\left(r+1\right)}\right)^{-1}C^{\left(r\right)}\right]_{ij}\left[B^{\left(r\right)}\sum_{n=1}^{N}\left\{{\hat{\Psi}}_n\right\}_{ij}{B^{\left(r\right)}}^\top\right]_{qq}} \\ C^{\left(r+1\right)} = & \left(\sum_{n=1}^{N}{y_{t_n}^\top\left(\Sigma^{\left(r+1\right)}\right)^{-1}B^{\left(r\right)}{\hat{\eta}}_n}\right) \\ & \cdot \left(\sum_{n=1}^{N}{{\hat{\eta}}_n^\top{B^{\left(r\right)}}^\top\left(\Sigma^{\left(r+1\right)}\right)^{-1}B^{\left(r\right)}{\hat{\eta}}_n}+\sum_{i=1}^{K_2}\sum_{j=1}^{K_2}{tr\left({B^{\left(r\right)}}^\top\left(\Sigma^{\left(r+1\right)}\right)^{-1}B^{\left(r\right)}\sum_{n=1}^{N}\left\{{\hat{\Psi}}_n\right\}_{ij}\right)\mathcal{J}_{ij}^{K_2K_2}}\right)^{-1} \\ B^{\left(r+1\right)} = & \left(\sum_{n=1}^{N}{y_{t_n}\left(\Phi^{\left(r+1\right)}\right)^{-1}C^{\left(r+1\right)}{\hat{\eta}}_n^\top}\right)  \\ & \cdot \left(\sum_{n=1}^{N}{{\hat{\eta}}_n{C^{\left(r+1\right)}}^\top\left(\Phi^{\left(r+1\right)}\right)^{-1}C^{\left(r+1\right)}{\hat{\eta}}_n^\top}+\sum_{i=1}^{K_2}\sum_{j=1}^{K_2}{\left[{C^{\left(r+1\right)}}^\top\left(\Phi^{\left(r+1\right)}\right)^{-1}C^{\left(r+1\right)}\right]_{ij}\sum_{n=1}^{N}\left\{{\hat{\Psi}}_n\right\}_{ij}}\right)^{-1}.
\end{aligned}
\label{equation_4_1_11}
\end{equation}

Note that the M-step updates of $\lambda_m^{\left(r+1\right)}$, $B_{q,:}^{\left(r+1\right)}$ and $\Sigma_{q}^{\left(r+1\right)}$ in \eqref{equation_3_2_1} are the special cases of the corresponding in \eqref{equation_4_1_11} where $\gamma = C = \Phi \equiv 1$.

\section{The Detailed Model with Time-varying Idiosyncratic Error Covariance Function in Section \ref{sec4_2}}\label{secC_3}

With \eqref{equation_2_5_3} and \eqref{equation_2_5_4}, we can extend the full joint distribution \eqref{equation_3_0_1} to the following:
\begin{equation}
\begin{aligned}
p\left(\mathcal{Y},F,\Omega\middle|\mathcal{T},\tilde{\omega},\omega,K\right) = p\left(\mathcal{Y}\middle|F,B,U,\mathcal{T},\tilde{\omega}\right)p\left(F\middle|L,\mathcal{T},\omega,K\right) p\left(L\middle|\mathcal{T},\omega,K\right) p\left(U\middle|\mathcal{T},\tilde{\omega}\right) p\left(B\middle|K\right),
\end{aligned}
\label{equation_C_3_1}
\end{equation}
where:
\begin{equation}
\begin{aligned}
p\left(\mathcal{Y}\middle|F,B,U,\mathcal{T},\tilde{\omega}\right) & = \prod_{n=1}^{N}{\prod_{q=1}^{Q}{\mathcal{N}\left(y_{{t_n}q}\middle|B_{q,:}f_n,\Sigma_{{t_n} q}\right)}} \\ & = \prod_{n=1}^{N}{\prod_{q=1}^{Q}{\mathcal{N}\left(y_{{t_n}q}\middle|B_{q,:}f_n,\left(\sum_{m=1}^{N}{\tilde{\omega}_{qm}} {\left(t_n\right) \cdot \nu_{qm}^{-1}}\right)^{-1}\right)}} \\ p\left(F\middle|L,\mathcal{T},\omega,K\right) & =\prod_{n=1}^{N}{\mathcal{N}\left(f_n\middle|0,\Lambda_{t_n}\right)}=\prod_{n=1}^{N}{\mathcal{N}\left(f_n\middle|0,\left(\sum_{m=1}^{N}{\omega_{m}{\left(t_n\right)}\lambda_m^{-1}}\right)^{-1}\right)} \\ p\left(L\middle|\mathcal{T},\omega,K\right) & \propto \exp\left(\frac{1}{2}\sum_{m=1}^{N}\sum_{n=1}^{N}{\omega_{m}{\left(t_n\right)}\log{\left|\lambda_m^{-1}\right|}}-\frac{1}{2}\sum_{m=1}^{N}\log{\left|\sum_{n=1}^{N}{\omega_{m}{\left(t_n\right)}\lambda_m^{-1}}\right|}\right) \\ p\left(U\middle|\tilde{\omega},\mathcal{T}\right) & = \prod_{q=1}^{Q}{p\left(U_q\middle|\tilde{\omega}_q,\mathcal{T}\right)}  \\ & \propto \exp\left(\frac{1}{2}\sum_{q=1}^{Q}\sum_{n=1}^{N}\sum_{m=1}^{N}{\tilde{\omega}_{qm}{\left(t_n\right)}\log{\left|\nu_{qm}^{-1}\right|}}-\frac{1}{2}\sum_{q=1}^{Q}\sum_{n=1}^{N}\log{\left|\sum_{m=1}^{N}{\tilde{\omega}_{qm}{\left(t_n\right)}\nu_{qm}^{-1}}\right|}\right). \\ p\left(B\middle|K\right) & \propto 1.
\end{aligned}
\label{equation_C_3_2}
\end{equation}

With the above full joint distribution and considering $F$ as the missing values, derivation of the Q-functions in the E-step and conditional M-step updates is analogous to Appendix \ref{secA}. Therefore, here we just state the results.

Firstly, the Q-functions are given as follows:
\begin{equation}
\begin{aligned}
\mathcal{Q}_1^{\left(r\right)}\left(L\right) = \mathcal{C}_1^{\left(r\right)} & -\frac{1}{2}\sum_{n=1}^{N}\sum_{m=1}^{N}{\omega_{m}{\left(t_n\right)} tr \left(\left({\hat{\eta}}_{n}{\hat{\eta}}_{n}^\top+{\hat{\Psi}}_{n}\right)\lambda_{m}^{-1}\right)}-\frac{1}{2}\sum_{n=1}^{N}\sum_{m=1}^{N}{\omega_{m}{\left(t_n\right)}\log{\left|\lambda_m\right|}} \\ \mathcal{Q}_2^{\left(r\right)}\left(B,U\right) = \mathcal{C}_2^{\left(r\right)} & - \frac{1}{2}\sum_{q=1}^{Q}\sum_{n=1}^{N}\sum_{m=1}^{N}{{\tilde{\omega}}_{qm}{\left(t_n\right)} \left(\left(y_{{t_n}q}-B_{q,:}{\hat{\eta}}_n\right)^2+B_{q,:}{\hat{\Psi}}_n B_{q,:}^\top\right)\nu_{qm}^{-1} } \\ & - \frac{1}{2}\sum_{n=1}^{N}\sum_{m=1}^{N}\sum_{q=1}^{Q}{{\tilde{\omega}}_{qm}{\left(t_n\right)}\log{\nu_{qm}}}.
\end{aligned}
\label{equation_C_3_3}
\end{equation}

Secondly, the partial derivatives of the above Q-functions are computed as follows:
\begin{equation}
\begin{aligned}
\frac{\partial\mathcal{Q}_1^{\left(r\right)}\left(L\right)}{\partial\lambda_m} & =\frac{1}{2}\lambda_m^{-1}\sum_{n=1}^{N}{\omega_{m}{\left(t_n\right)}\left({\hat{\eta}}_n{\hat{\eta}}_n^\top+{\hat{\Psi}}_n\right)}\lambda_m^{-1}-\frac{1}{2}\sum_{n=1}^{N}\omega_{m}{\left(t_n\right)}\lambda_m^{-1} \\ \frac{\partial\mathcal{Q}_2^{\left(r\right)}\left(B,U^{\left(r\right)}\right)}{\partial B_{q,:}} & =\sum_{n=1}^{N}\left(\frac{y_{{t_n}q}{\hat{\eta}}_n^\top}{\Sigma_{{t_n}q}^{\left(r\right)}}-B_{q,:}\frac{{\hat{\eta}}_n{\hat{\eta}}_n^\top+{\hat{\Psi}}_n}{\Sigma_{{t_n}q}^{\left(r\right)}}\right) \\ \frac{\partial\mathcal{Q}_2^{\left(r\right)}\left(B^{\left(r+1\right)},U\right)}{\partial\nu_{qm}} & =\frac{1}{2}\sum_{n=1}^{N}{{\tilde{\omega}}_{qm}{\left(t_n\right)}\left(\left(y_{{t_n}q}-B_{q,:}^{\left(r+1\right)}{\hat{\eta}}_n\right)^2+B_{q,:}^{\left(r+1\right)}{\hat{\Psi}}_n\left(B_{q,:}^{\left(r+1\right)}\right)^\top\right)\nu_{qm}^{-2}} \\ & -\frac{1}{2}\sum_{n=1}^{N}{{\tilde{\omega}}_{qm}{\left(t_n\right)}\nu_{qm}^{-1}}.
\end{aligned}
\label{equation_C_3_4}
\end{equation}

Thirdly, the conditional M-step updates are given as follows:
\begin{equation}
\begin{aligned}
\lambda_{m}^{\left(r+1\right)} & =\frac{\sum_{n=1}^{N}{\omega_{m}{\left(t_n\right)}\left({\hat{\eta}}_{n}{\hat{\eta}}_{n}^\top+{\hat{\Psi}}_{n}\right)}}{\sum_{n=1}^{N}\omega_{m}{\left(t_{n}\right)}} \\ 
B_{q,:}^{\left(r+1\right)} & =\left(\sum_{n=1}^{N}\frac{y_{{t_n}q}{\hat{\eta}}_n^\top}{\Sigma_{{t_n}q}^{\left(r\right)}}\right)\left(\sum_{n=1}^{N}\frac{{\hat{\eta}}_n{\hat{\eta}}_n^\top+{\hat{\Psi}}_n}{\Sigma_{{t_n}q}^{\left(r\right)}}\right)^{-1} \\ \nu_{qm}^{\left(r+1\right)} & =\frac{\sum_{n=1}^{N}{{\tilde{\omega}}_{qm}{\left(t_n\right)}\left(\left(y_{{t_n}q}-B_{q,:}^{\left(r+1\right)}{\hat{\eta}}_n\right)^2+B_{q,:}^{\left(r+1\right)}{\hat{\Psi}}_n\left(B_{q,:}^{\left(r+1\right)}\right)^\top\right)}}{\sum_{n=1}^{N}{\tilde{\omega}}_{qm}{\left(t_n\right)}}.
\end{aligned}
\label{equation_C_3_5}
\end{equation}

\section{Determination of $\mathcal{H}$} \label{secD}

The EM algorithm in Section \ref{sec3} assumes that the weight function $\omega$ is given a priori, and in Section \ref{sec2_3} we discussed that it is defined by the choice of the set of bandwidth parameters $\mathcal{H} = \left\{h_d\right\}_{d=1}^{D}$. The reason why we do not include the determination of $\mathcal{H}$ in Section \ref{sec3} is that they cannot be estimated based on the likelihood of time series $\left(\mathcal{T},\mathcal{Y}\right)$: smaller elements in $\mathcal{H}$ are always preferred, just as the nonparametric kernel density estimation problems.

Firstly, for each $n = 1,2,\cdots,N$, we define the alternative set of basis covariance matrices $\tilde{\lambda}_{d,n}$'s as follows:
\begin{equation}
\tilde{\lambda}_{d,n} \triangleq \frac{\sum_{m=1}^{N}{\omega_{d}{\left(t_m\right)}\left({\hat{\eta}}_{m}{\hat{\eta}}_{m}^\top+{\hat{\Psi}}_{m}\right)} - \omega_{d}{\left(t_n\right)}\left({\hat{\eta}}_{n}{\hat{\eta}}_{n}^\top+{\hat{\Psi}}_{n}\right) }{\sum_{m=1}^{N}{\omega_{d}{\left(t_{m}\right)}} - \omega_{d}{\left(t_{n}\right)}},
\label{equation_3_3_0}
\end{equation}
and the alternative $\tilde{\Lambda}{\left(t_n\right)} \triangleq \left(\sum_{d=1}^{D}{\omega_{d}} {\left(t_n\right) \cdot \tilde{\lambda}_{d,n}^{-1}}\right)^{-1}$, i.e., we completely remove the influence of ${\hat{\eta}}_{n}{\hat{\eta}}_{n}^\top+{\hat{\Psi}}_{n}$ in defining $\tilde{\Lambda}{\left(t_n\right)}$ for each $n$ (leave-one-out-cross-validation).

However, computing the inverse $\tilde{\lambda}_{d,n}^{-1}$ for every $d$ and $n$ is involved with time complexity of $\mathcal{O}\left(K^3 ND\right)$, which is intractable in practice. To circumvent it, we consider the following approximation:
\begin{equation}
\begin{aligned}
\tilde{\lambda}_{d,n} & \approx \frac{\sum_{m=1}^{N}{\omega_{d}{\left(t_m\right)}\cdot{\tilde{b}}_{m}{\tilde{b}}_{m}^\top} - \omega_{d}{\left(t_n\right)}\cdot{\tilde{b}}_{n}{\tilde{b}}_{n}^\top }{\sum_{m=1}^{N}{\omega_{d}{\left(t_{m}\right)}} - \omega_{d}{\left(t_{n}\right)}} \\ & = \frac{\sum_{m=1}^{N}{\omega_d\left(t_m\right)}}{\sum_{m=1}^{N}{\omega_d\left(t_m\right)}-\omega_d\left(t_n\right)} \cdot {\tilde{\lambda}}_d - \frac{\omega_d\left(t_n\right)}{\sum_{m=1}^{N}{\omega_d\left(t_m\right)}-\omega_d\left(t_n\right)} \cdot {\tilde{b}}_{n}{\tilde{b}}_{n}^\top,
\label{equation_3_3_2}
\end{aligned}
\end{equation}
where each $\tilde{b}_n \sim \mathcal{N}{\left(\hat{\eta}_n , \hat{\Psi}_n \right)}$ and:
\begin{equation}
\begin{aligned}
\tilde{\lambda}_{d} \triangleq \frac{\sum_{n=1}^{N}{\omega_{d}{\left(t_n\right)}\cdot{\tilde{b}}_{n}{\tilde{b}}_{n}^\top}}{\sum_{n=1}^{N}\omega_{d}{\left(t_{n}\right)}}.
\end{aligned}
\label{equation_3_2_3}
\end{equation}
By the Sherman-Morrison formula \citep{Hager1989}, we only need to invert $\tilde{\lambda}_d$'s for computing $\tilde{\lambda}_{d,n}^{-1}$'s, instead of directly inverting all of them.

\begin{figure}[!htb]
\renewcommand{\baselinestretch}{1}
\centering
\includegraphics[width=0.48\textwidth]{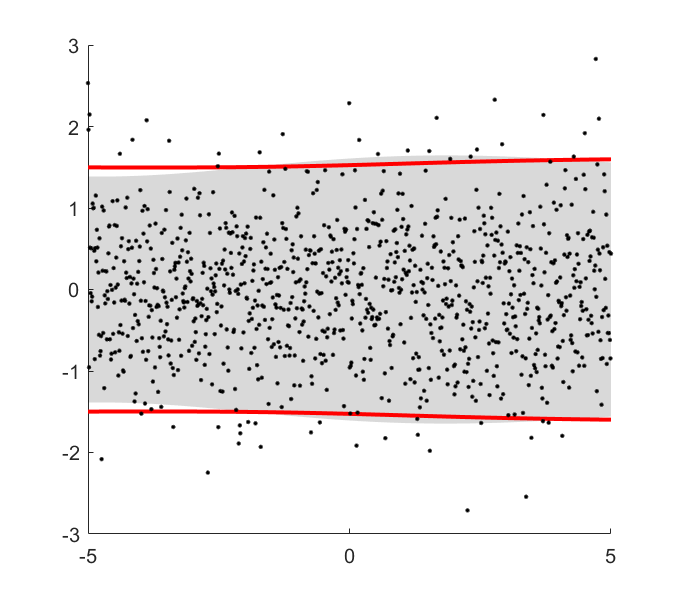}
\includegraphics[width=0.48\textwidth]{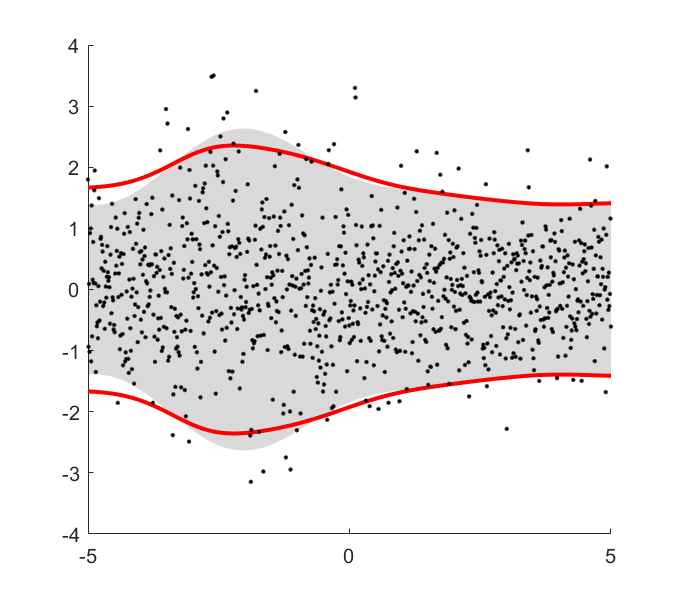}
\includegraphics[width=0.48\textwidth]{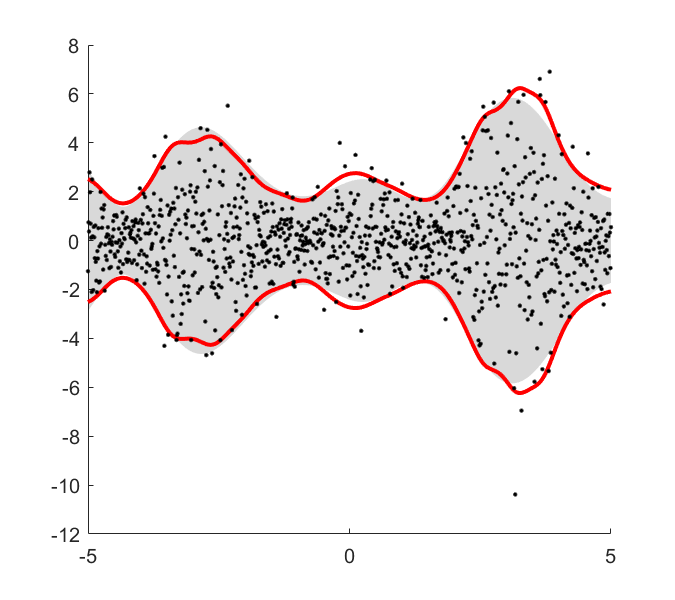}
\includegraphics[width=0.48\textwidth]{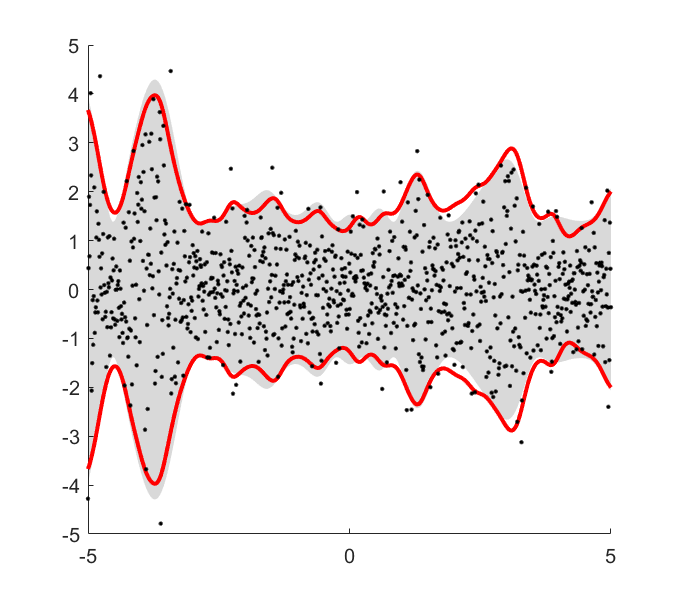}
\caption{Some simulated datasets with $Q=1$ under the time-varying volatility. Black dots and gray regions are the simulated datasets and true 95\% confidence bands, respectively. Red curves are the estimated 95\% confidence bands by the simplified model in Section \ref{sec4_5} and the criterion to determine the bandwidth in Appendix \ref{secD}.}
\label{fig_E}
\end{figure}

As discussed in Section \ref{sec2_3}, we set $h_0 \equiv h_d$ for a single bandwidth parameter $h_0$ and let $\mathbb{H}_0$ be the set of its candidates. For each candidate $h_0 \in \mathbb{H}_0$, we define $\omega_d = \omega_{d,h_0}$ and compute the following objective function with that weight function:
\begin{equation}
\begin{aligned}
\mathcal{H}\left(\mathcal{Y},\mathcal{T},K,h_0\right) \triangleq  \sum_{n=1}^{N}\log{\mathcal{N}\left(y_{t_n}\middle|0,{\hat{B}}_K{\tilde{\Lambda}}_{K,h_0}{\left(t_n\right)}{\hat{B}}_K^\top+{\hat{\Sigma}}_K\right)},
\end{aligned}
\label{equation_3_3_1}
\end{equation}
to pick the maximizer $\hat{h}_0 \triangleq \argmax_{h_0 \in \mathcal{H}_0}{\mathcal{H}\left(\mathcal{Y},\mathcal{T},K,h_0\right)}$ as the chosen bandwidth.

Then, the time complexity is linear to the number of candidates of $K$ times that of $h_0$. To be more efficient in time, one can instead compute \eqref{equation_3_3_1} for each EM iteration and choose $h_0$ \textit{dynamically} for each $K$ to let the time complexity just linear to the number of candidates of $K$: Algorithm \ref{algorithm1} follows this time-efficient way. Similar criteria are available for the models in Section \ref{sec4}. We applied the simplified modeling in Section \ref{sec4_5} and this criterion to some simulated $Q=1$ datasets with different volatility smoothness: see Figure \ref{fig_E} for the visualized results.

\section{Factor Loading Identification} \label{secE}

Because the covariance in Equation \eqref{equation_2_1_4} is $B\Lambda_t B^\top+\Sigma$, we have an identifiability issue in $B\Lambda_t B^\top$: as discussed in Section \ref{sec2_2_2}, the value of prior $p\left(L\middle|\omega,\mathcal{T}\right)$ on $L$ is invariant under the transformation $L\rightarrow CLC^\top$ for a $K\times K$ invertible matrix $C$. Also, we consider $B$ to follow an improper prior $p\left(B\right)\propto 1$ so any transformation of $B$ does not change its value. In this paper, we identify $\left(L,B\right)\rightarrow\left(\hat{L},\hat{B}\right)$ by the following steps, once they are estimated:
\begin{enumerate}
  \item Rewrite $B^\top B = VDV^\top$ as its eigenvalue decomposition so that $D$ is diagonal and $V$ is unitary.
  \item Rearrange $V$ and $D$ so that the diagonal entries of $D$ are in order from smallest to largest.
  \item Transform $B\rightarrow\widetilde{B}=BVD^{-\frac{1}{2}}$ and all $\lambda_m\rightarrow{\widetilde{\lambda}}_m \triangleq D^\frac{1}{2}V^\top\lambda_m VD^\frac{1}{2}$ in $L$, so columns of $\widetilde{B}$ are orthonormal.
  \item Find a $K\times K$ invertible matrix $A = \hat{A}$ under the constraint ${\Vert A \Vert}_{\rm{F}}=\sqrt{K}$, which maximizes the following objective function for a positive scalar hyperparameter $\tau$:
\begin{equation}
\begin{aligned}
\mathcal{L}\left(A\middle|\tau\right) & \triangleq-\sum_{q=1}^{Q}\sum_{l=1}^{K}\left|{\widetilde{B}}_{q,:}A_{:,k}\right|-\frac{\tau}{K}\cdot \KLD{\mathcal{N}\left(0,A^\top{\widetilde{B}}^\top\widetilde{B}A\right)}{\mathcal{N}\left(0,\mathbb{I}_K\right)} \\ & =-\sum_{q=1}^{Q}\sum_{l=1}^{K}\left|{\widetilde{B}}_{q,:}A_{:,k}\right|+\frac{\tau}{K}\log{\left|A\right|}.
\end{aligned}
\label{equation_E_0_1}
\end{equation}
  \item Transform $\widetilde{B}\rightarrow\hat{B}\triangleq\widetilde{B}\hat{A}$ and all ${\widetilde{\lambda}}_m\rightarrow{\hat{\lambda}}_m\triangleq{\hat{A}}^{-1}{\widetilde{\lambda}}_m\left({\hat{A}}^{-1}\right)^\top$ in $\widetilde{L}$.
\end{enumerate}

Note that the objective function \eqref{equation_E_0_1} seeks for a sparser matrix $\hat{B}=\widetilde{B}\hat{A}$ such that ${\hat{B}}^\top\hat{B}={\hat{A}}^\top{\widetilde{B}}^\top\widetilde{B}\hat{A}\approx\mathbb{I}_K$, in other words, ${\hat{B}}^\top\hat{B}$ has a bigger nonzero smallest eigenvalue so that each row of $\hat{B}$ is to have at least one “active” entry: because ${\widetilde{B}}_{q,:}A_{:,k}\equiv{\hat{B}}_{qk}$ by definition and, for the eigenvalues $\left\{\varsigma_k\right\}_{k=1}^K$ of ${\hat{B}}^\top\hat{B}$, by the Jensen’s inequality,
\begin{equation}
\begin{aligned}
\frac{2}{K}\log{\left|A\right|} & =\frac{1}{K}\log{\left|A^\top A\right|}=\frac{1}{K}\log{\left|A^\top {\widetilde{B}}^\top \widetilde{B}A\right|}=\frac{1}{K}\log{\left|{\hat{B}}^\top \hat{B}\right|}=\frac{1}{K}\sum_{k=1}^{K}\log{\varsigma_k} \\ & \le\log{\sum_{k=1}^{K}{\frac{1}{K}\varsigma_k}}=\log{\sum_{k=1}^{K}\varsigma_k}-\log{K} = \log{{\Vert\hat{B}\Vert}_{\rm{F}}^{2}}-\log{K} \\ & =\log{trace\left({\hat{B}}^\top\hat{B}\right)}-\log{K} =\log{trace\left(A^\top{\widetilde{B}}^\top\widetilde{B}A\right)}-\log{K} \\ & = \log{trace\left(A^\top A\right)}-\log{K} = \log{{\Vert A \Vert}_{\rm{F}}^{2}} -\log{K}=0,
\end{aligned}
\label{equation_E_0_2}
\end{equation}
and the equality condition of the above inequality $\log{\left|A\right|}\le 0$ is that $\varsigma_k$’s are identical. Moreover, ${\Vert B \Vert}_{\rm{F}}^{2} = K$ in the above derivation means that $\sum_{k=1}^{K}\varsigma_k=K$, so maximizing $\frac{\tau}{K}\log{\left|A\right|}$ in \eqref{equation_E_0_1} implies maximizing the nonzero smallest eigenvalue of ${\hat{B}}^\top\hat{B}$.

The choice of $\tau$ in Equation \eqref{equation_E_0_1} significantly affects the estimation of $A$. We estimate $A$ by gradient ascent, $\tau$ is initialized by $K$, and, per every a few iterations, we increase $\tau\rightarrow \tau+K$ if the smallest nonzero eigenvalue of the resulting ${\hat{B}}^\top\hat{B}$ does not exceed the half of its largest eigenvalue.

\section{Benchmark: EWMA covariance model} \label{secF}

For evaluating our proposed model, here we define the exponentially-weighted moving average (EWMA) covariance model on the factored data as a benchmark. 

Let $Y$ be the $T \times Q$ matrix where each row is $y_t^\top$ and $Y = U D W^\top$ be its singular value decomposition (SVD). For each number of factors $K$, let $W_K$ be the $T \times K$ matrix that gathers K columns of $W$ corresponding to the K biggest singular values of $Y$. Then, for each $t$, $z_t = W_K^\top y_t$ is the $K$-dimensional linear transformation of $y_t$ obtained by the principal component analysis (PCA) dimensionality reduction and $w_t = W_K z_t$ is the $Q$-dimensional reconstruction.

Let $K$ be fixed. We restate $y_t = w_t + \epsilon_t = W_K z_t + \epsilon_t$ for each $t$ to gather $Z = \left\{z_t\right\}_{t=1}^{T}$ and $E = \left\{\epsilon_t\right\}_{t=1}^{T}$. For a given hyperparameter $\alpha \in \left[0,1\right]$, we define $\Lambda_t$ and $\Sigma$ as follows:

\begin{equation}
\begin{aligned}
\Lambda_t \triangleq \frac{\sum_{s=1}^{T}{{\alpha}^{\left|t-s\right|} \cdot z_s z_s^\top}}{\sum_{s=1}^{T}{\alpha}^{\left|t-s\right|}} \ \ \mbox{and} \ \ \Sigma \triangleq \frac{1}{T} \sum_{t=1}^{T}{diag\left(\epsilon_t \epsilon_t^\top\right)}.
\end{aligned}
\label{equation_F_0_1}
\end{equation}
That is, $z_t \sim \mathcal{N} \left(0,\Lambda_t\right)$ and $\epsilon_t \sim \mathcal{N} \left(0,\Sigma\right)$ with the EWMA covariance models $\Lambda_t$ and $\Sigma$. Clearly, we have a resulting model $y_t \sim \mathcal{N} \left(0,W_K \Lambda_t W_K^\top + \Sigma \right)$ by definition. To decide $K$, we apply the same approach discussed in Sections \ref{sec2_4} and \ref{sec3}, i.e., the leave-one-out cross-validation.

\section{S\&P 100 Stock Return with Outlier Trimming} \label{secH}

In Section \ref{sec6_1}, we applied three factor analysis models to the raw log-returns of the S\&P 100 stocks. In practice, trimming down extreme values is an usual refining technique to deal with outliers. Here, we did the same work as in Section \ref{sec6_1} to the refined data to compare performances of different methods. To be specific, we first estimated standard deviations of the raw log-returns of each stock in the training dataset. Then, we use  three times the standard deviation as the  threshold, and trim all outliers lying beyond the threshold to the threshold value. 

\begin{figure}
\renewcommand{\baselinestretch}{1}
\centering
\includegraphics[width=0.32\textwidth]{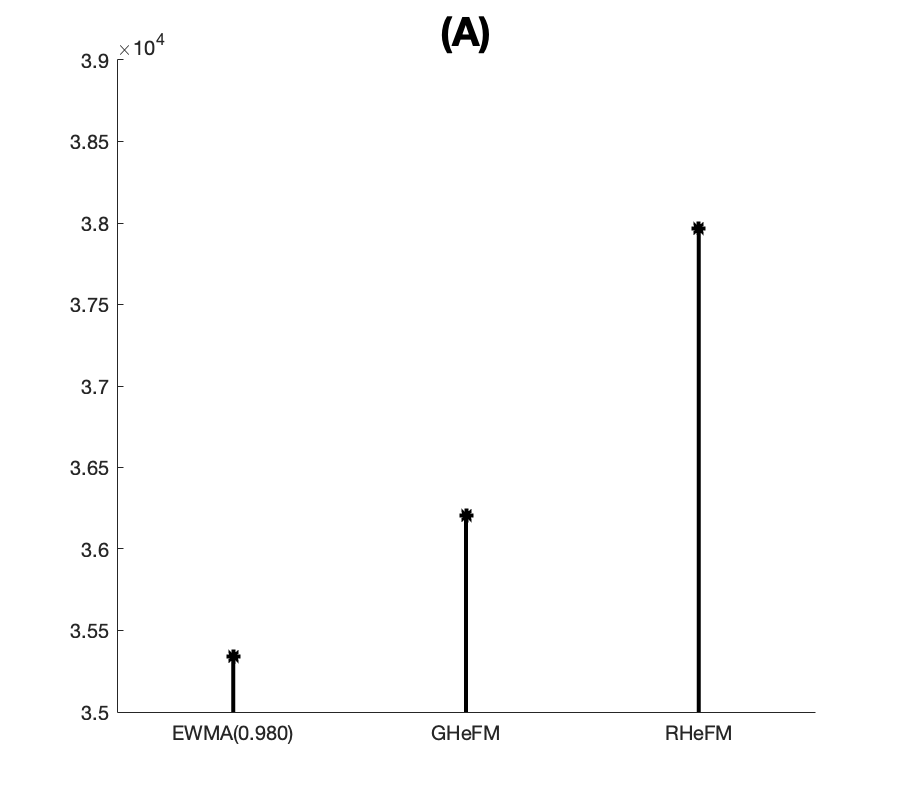}
\includegraphics[width=0.32\textwidth]{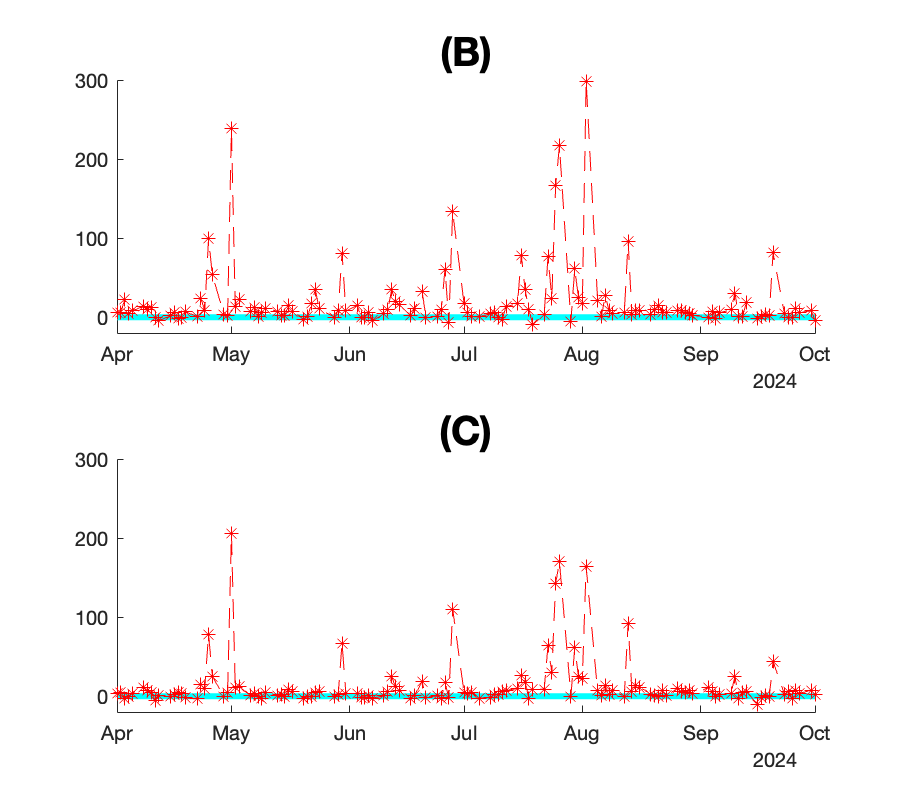}
\includegraphics[width=0.32\textwidth]{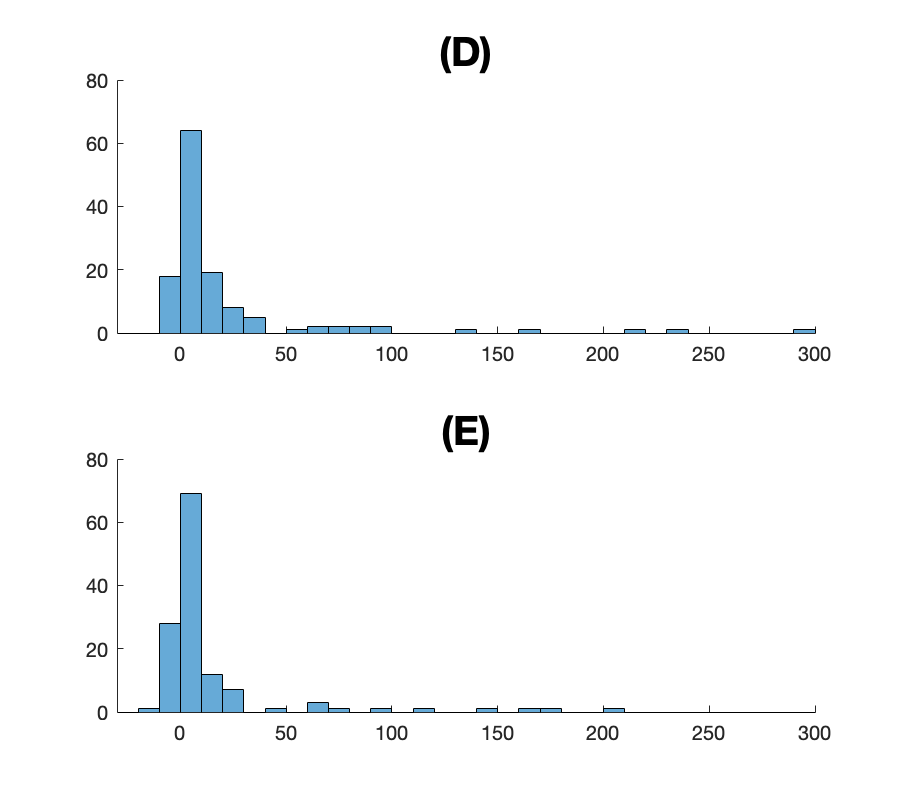}
\caption{(A) Sums of the predictive log-likelihoods (i.e., that evaluated by the true log-daily returns) of various models. (B): log-likelihoods of the true log-daily returns of the RHeFM minus those of the EWMA covariance model with $\alpha = 0.980$ over dates. (D): log-likelihoods of the true log-daily returns of the RHeFM minus those of the GHeFM over dates. (C) and (E): Histograms of the respective log-likelihood differences in the middle panels. Among 128 test business days, the RHeFM beats the GHeFM and EWMA with $\alpha=0.980$ for 93 and 110 days, respectively.}
\label{fig_H_1}
\end{figure}

The estimated number of factors is $\hat{K}=31$. Figure \ref{fig_H_1} shows the comparison between the predictive performance of RHeFM and those of the GHeFM and EWMA covariance models, the same way  as Figure \ref{fig_6_4_1} in the main manuscript. We estimated $\hat{\alpha} = 0.980 \in \left\{1.000, 0.999, 0.998, \cdots , 0.950 \right\}$ and $\hat{K} = 24$ by the leave-one-out cross-validation. For  GHeFM, the estimated number of factors was $\hat{K}=25$.

Panel (A) of Figure \ref{fig_H_1} supports that the performance of RHeFM is significantly better than the common EWMA approach in terms of their predictive likelihoods of the raw log-returns in the test dataset. RHeFM outperformed the EWMA model consistently and significantly, as shown in panels (B) and (D). Also,  RHeFM performed remarkably better than GHeFM, as shown in (C) and (E). All of these factor analysis models underperform the corresponding models trained on the raw log-returns in Section \ref{sec6_1}, though two RHeFMs show similar performances.

\begin{table}
\small
\begin{center}
\begin{tabular}{ |c||c|c||c|c||c|c| } 
  \hline
   & EWMA (1.000) & EWMA (0.980) & GHoFM & GHeFM & RHoFM & RHeFM \\ \hline 
  $\hat{K}$ & 19 & 24 & 24 & 25 & 33 & 31 \\
  \hline Scores & 35590 & 35339 & 36336 & 36203 & 37836 & \textbf{37964} \\
  \hline
  
\end{tabular}
\caption{Performance comparison between homoscedastic and heteroscedastic factor analysis models trained with the trimmed-down log-returns. Scores are the sums of the predictive log-likelihoods (the higher is better).}
\label{table_H_1}
\end{center}
\end{table}

We also compare the performance between each model and its time-invariant (homoscedastic) alternative, as shown in Table \ref{table_H_1}, just as Section \ref{sec6_1}. While EWMA (0.980) and GHeFM show worse performance than their time-invariant versions EWMA (1.000) and GHoFM, RHeFM still outperforms RHoFM in this case.

\newpage

\section{Auxiliary Figures} \label{secG}

\begin{figure}[!htb]
\renewcommand{\baselinestretch}{1}
\centering
\includegraphics[width=0.95\textwidth]{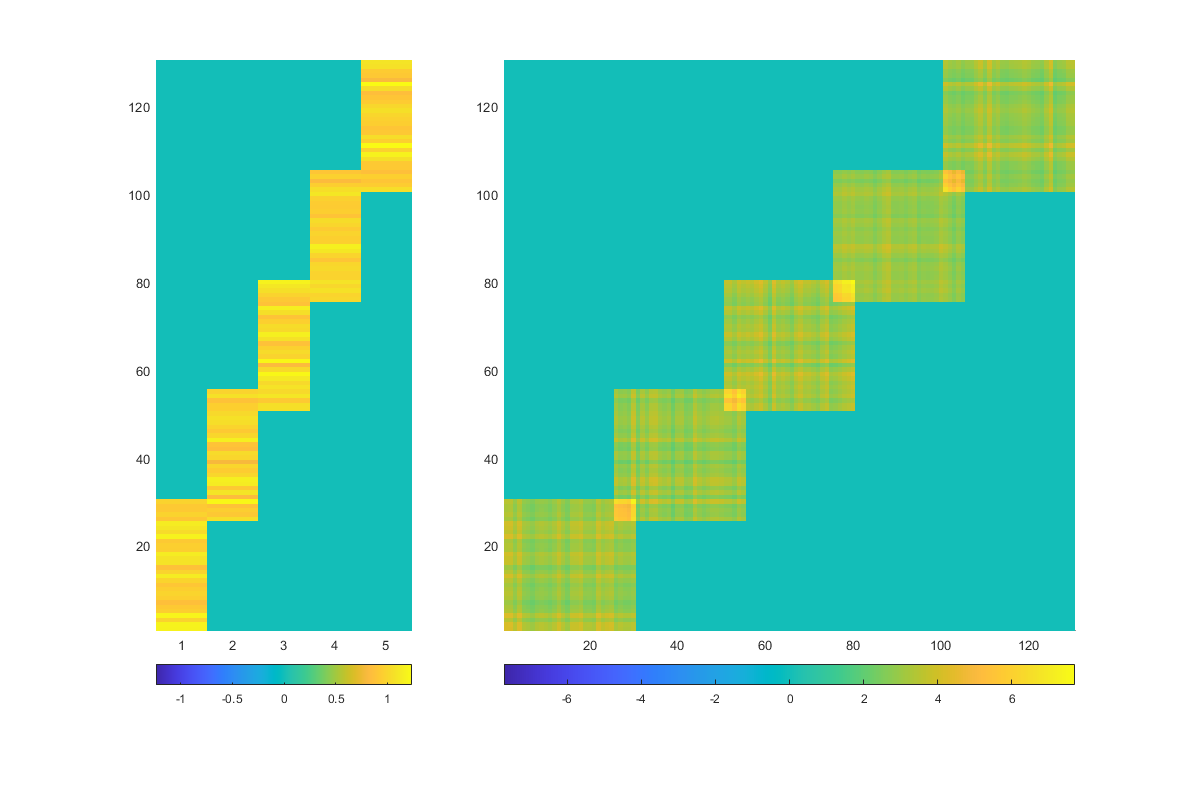}
\caption{(Left) Factor loading matrix $B$. (Right) Covariance in a segment $B\Lambda_t B^\top$ where $\Lambda_t=3\cdot\mathbb{I}_5$.}
\label{fig_5_1_1}
\end{figure}

\begin{figure}[!htb]
\renewcommand{\baselinestretch}{1}
\centering
\includegraphics[width=0.32\textwidth]{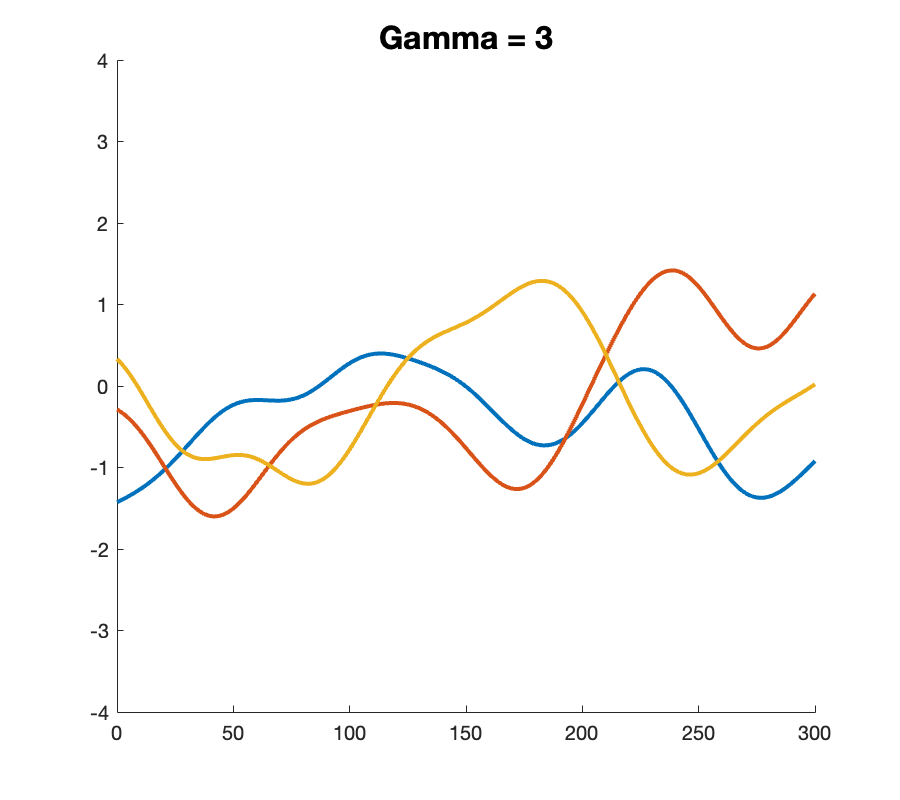}
\includegraphics[width=0.32\textwidth]{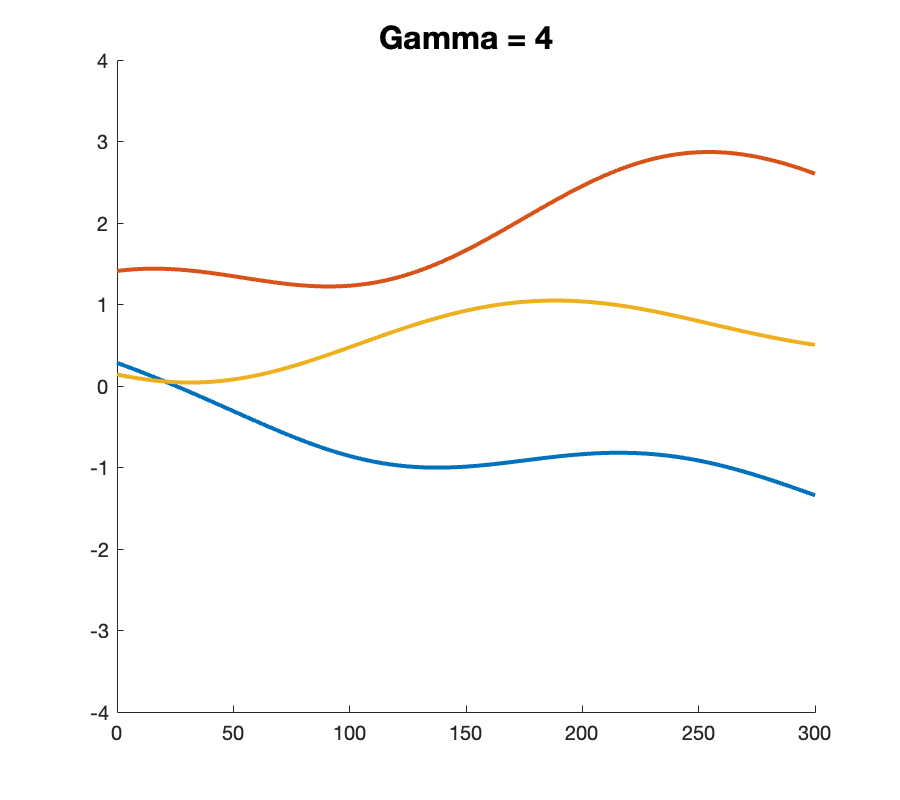}
\includegraphics[width=0.32\textwidth]{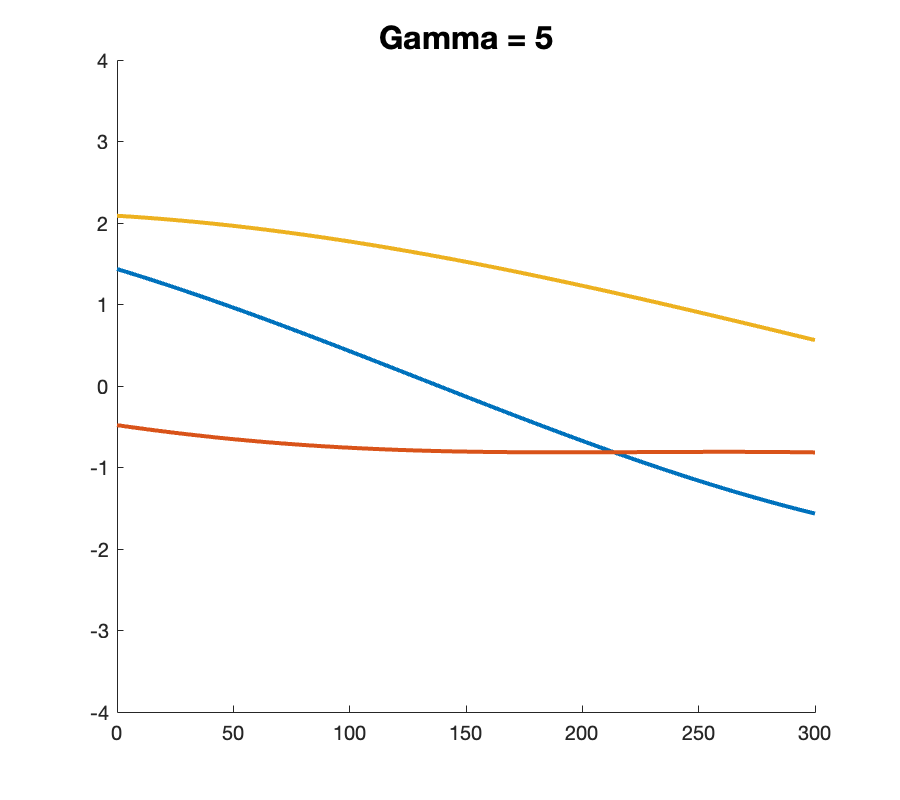}
\caption{Examples of random functions drawn from Gaussian process priors with different choices of $\gamma \in \left\{ 3, 4, 5 \right\} $ of the squared-expoential kernel in Section \ref{sec5}.}
\label{fig_5_1_0}
\end{figure}

\begin{figure}[!htb]
\renewcommand{\baselinestretch}{1}
\centering
\includegraphics[width=0.24\textwidth]{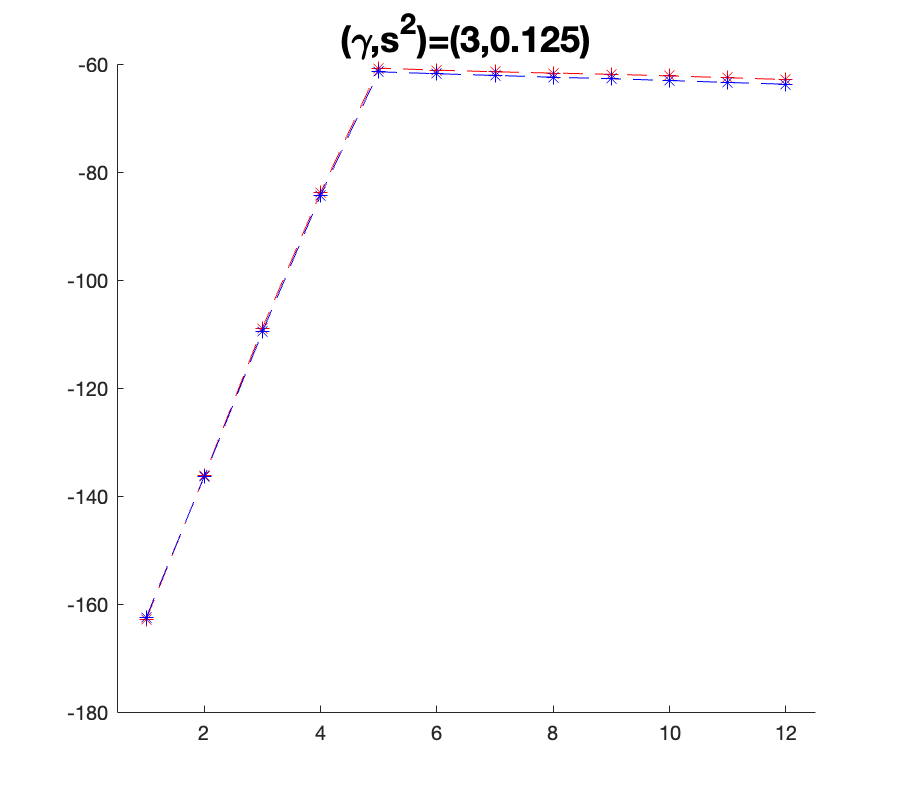}
\includegraphics[width=0.24\textwidth]{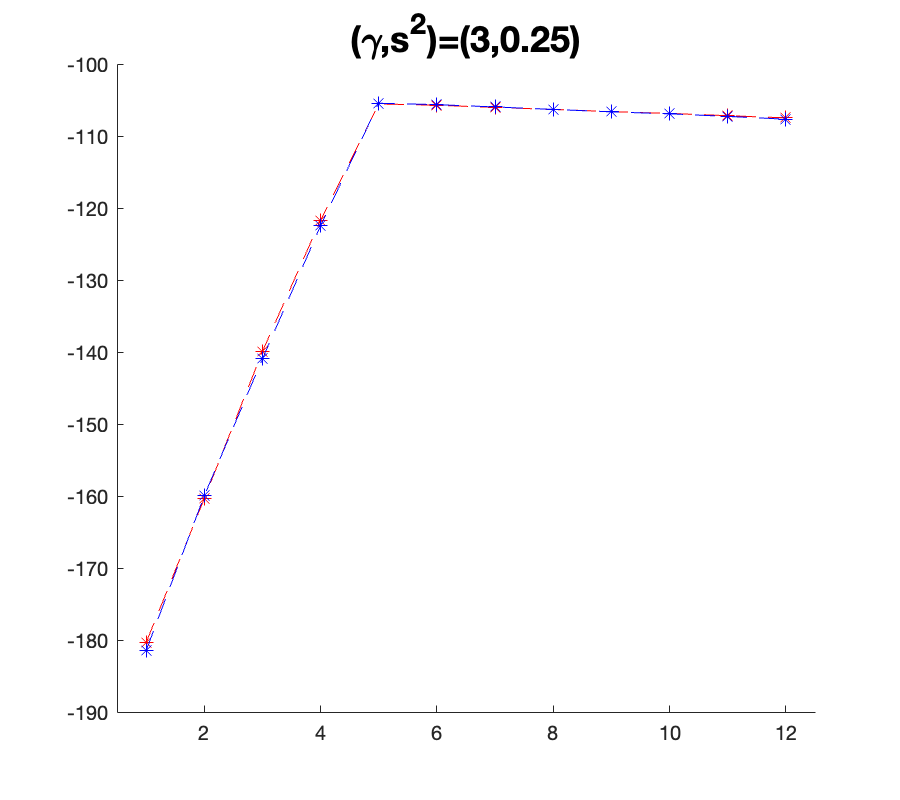}
\includegraphics[width=0.24\textwidth]{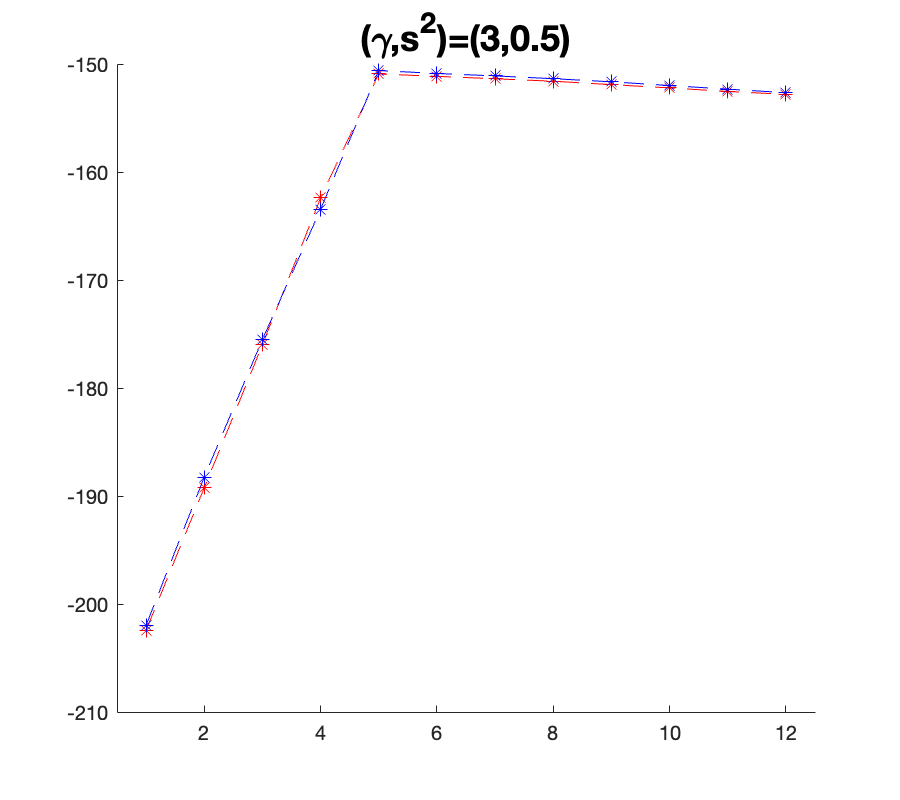}
\includegraphics[width=0.24\textwidth]{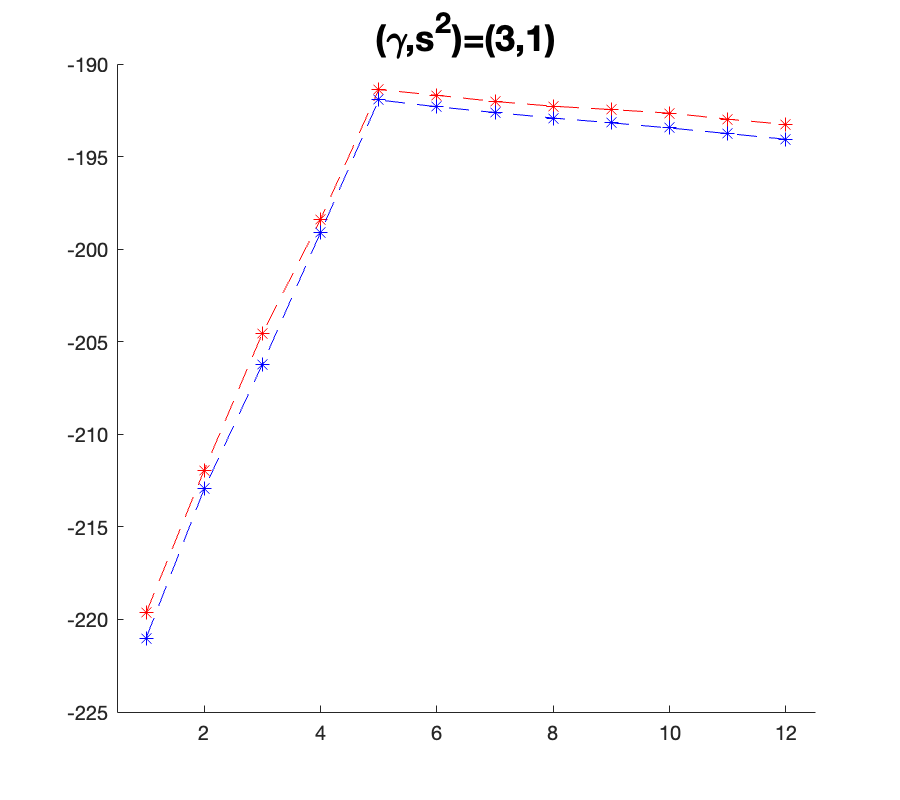}
\includegraphics[width=0.24\textwidth]{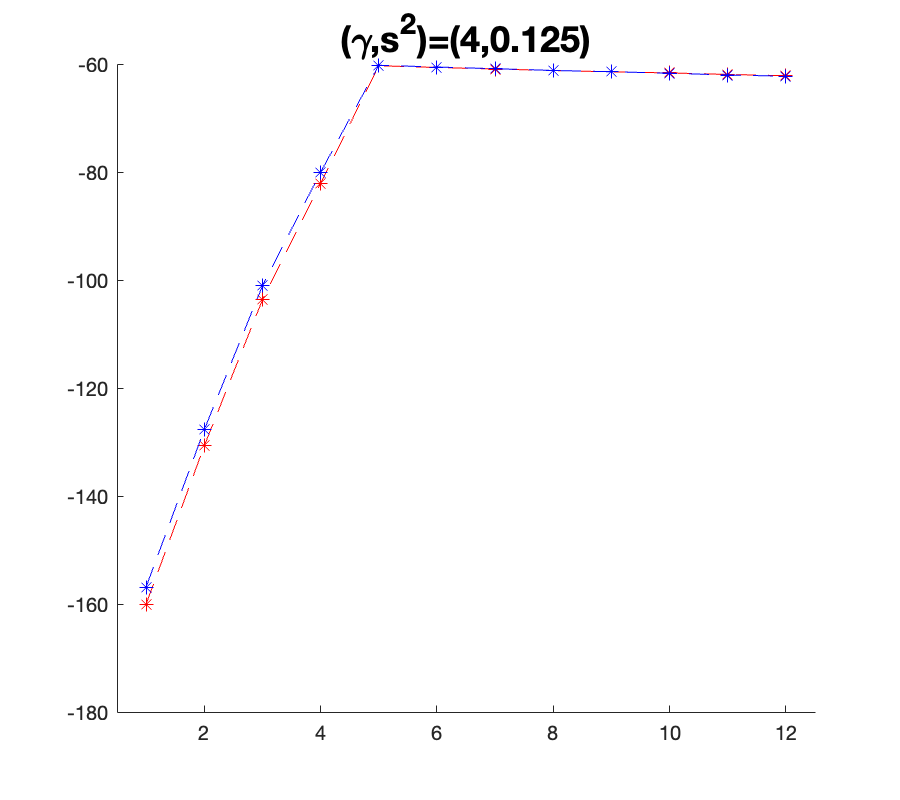}
\includegraphics[width=0.24\textwidth]{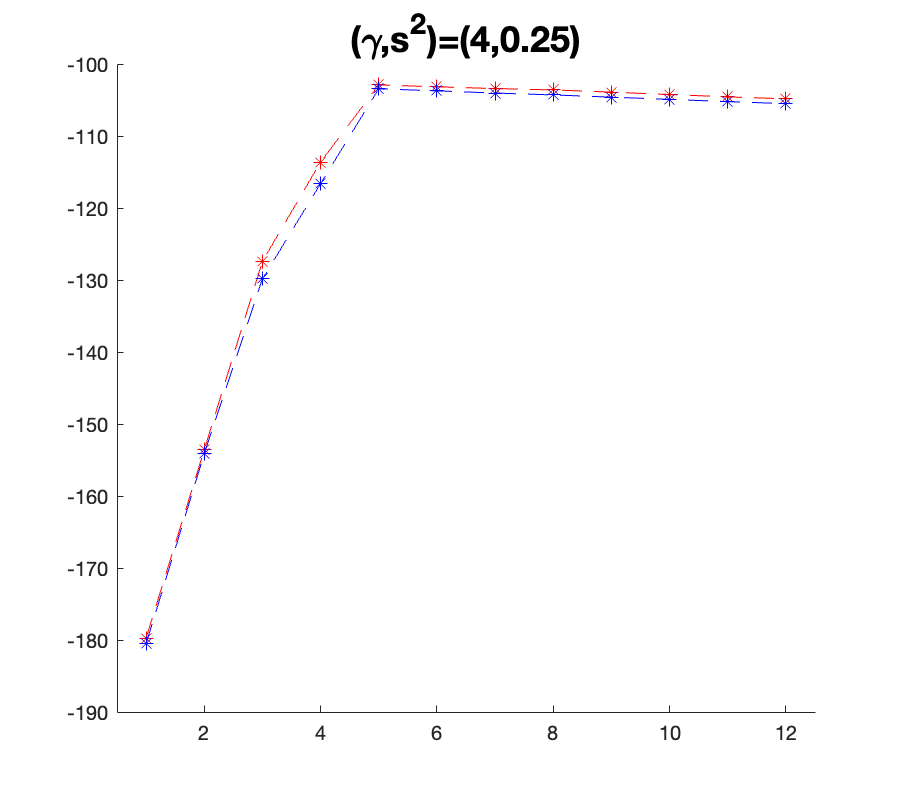}
\includegraphics[width=0.24\textwidth]{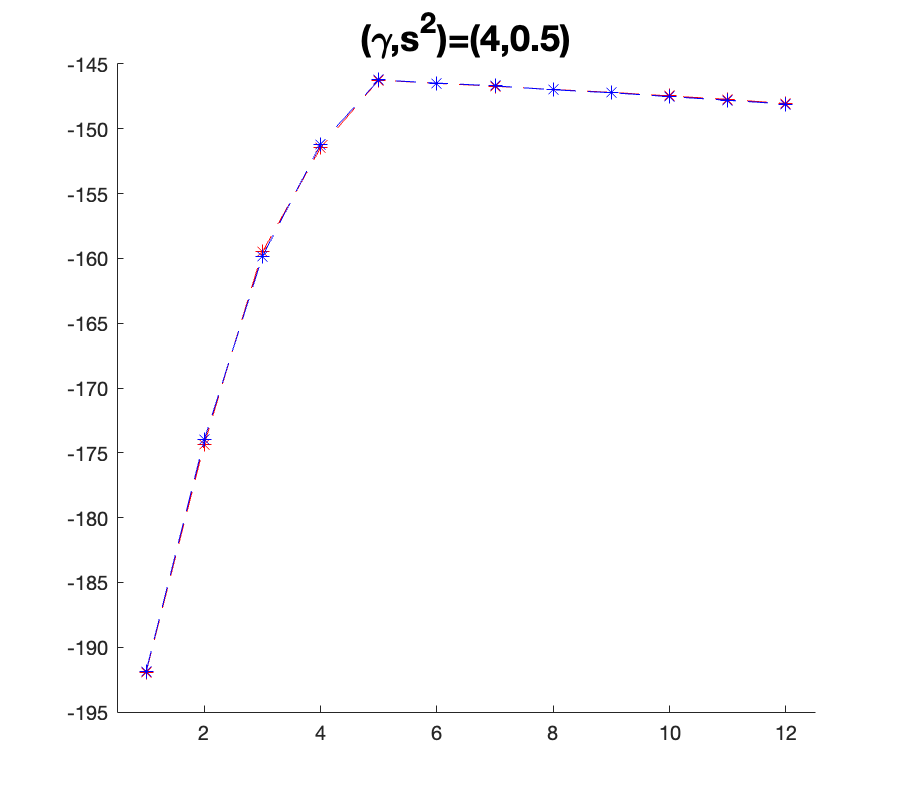}
\includegraphics[width=0.24\textwidth]{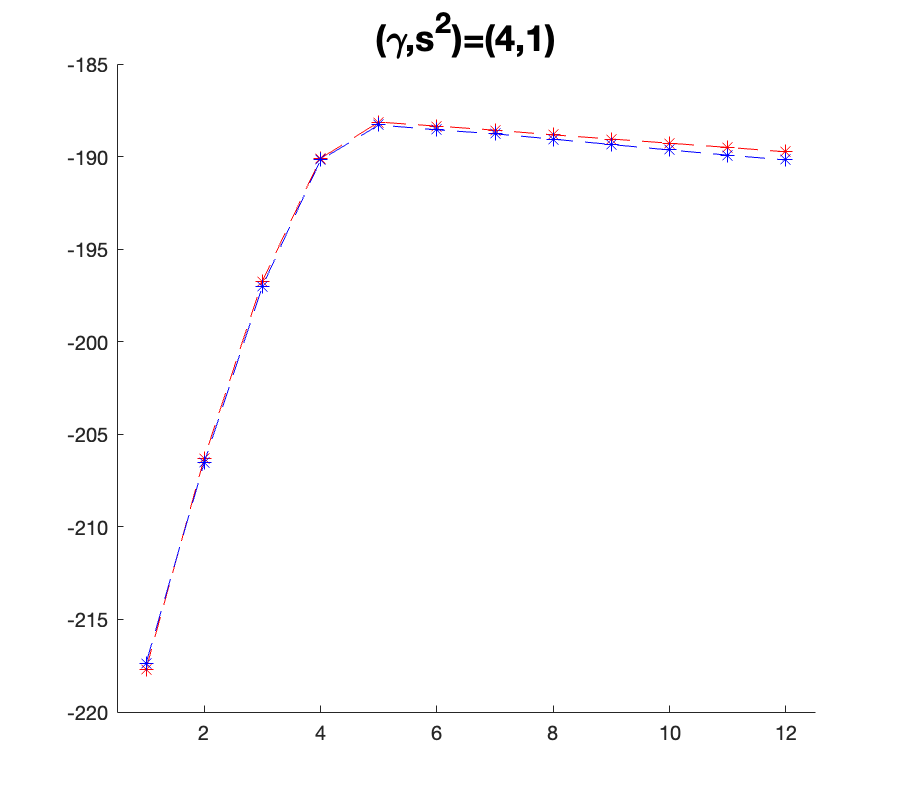}
\includegraphics[width=0.24\textwidth]{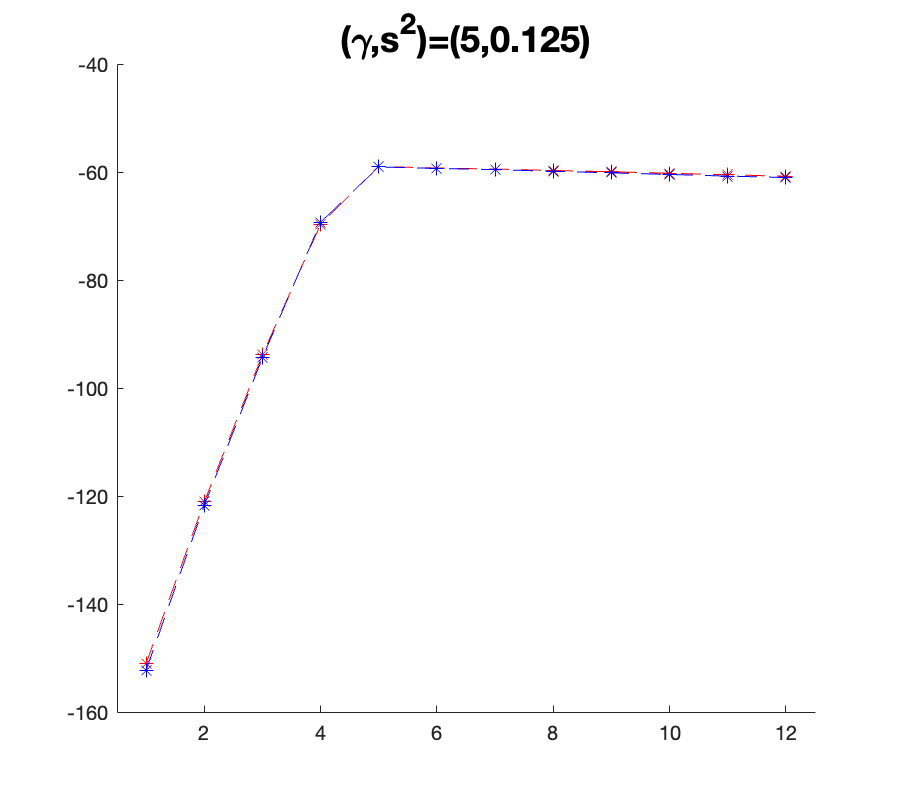}
\includegraphics[width=0.24\textwidth]{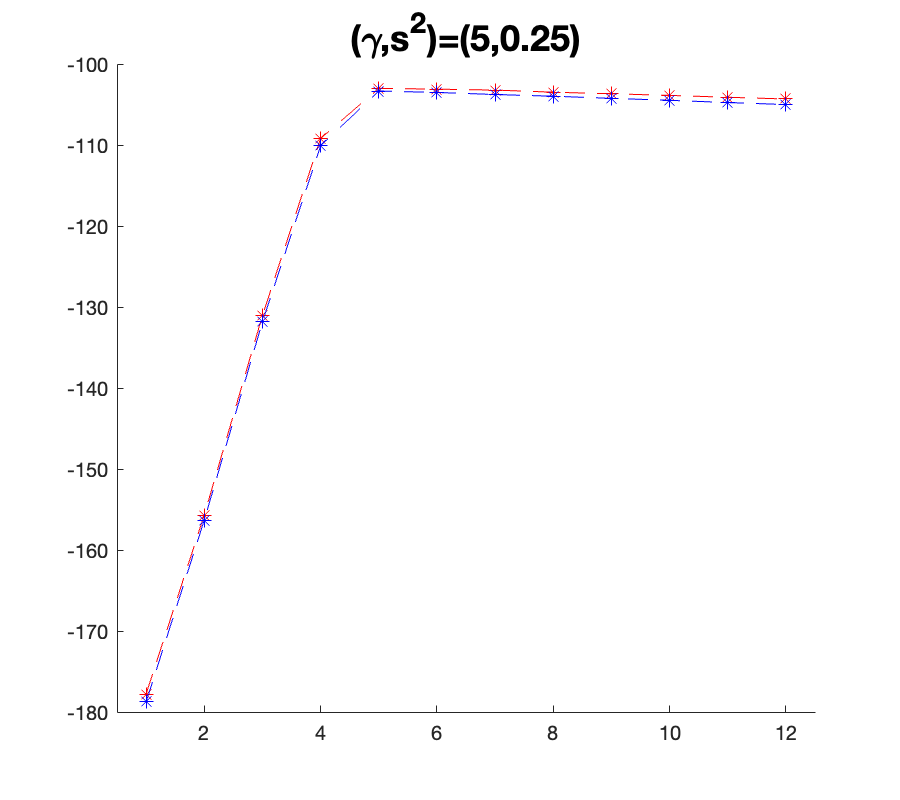}
\includegraphics[width=0.24\textwidth]{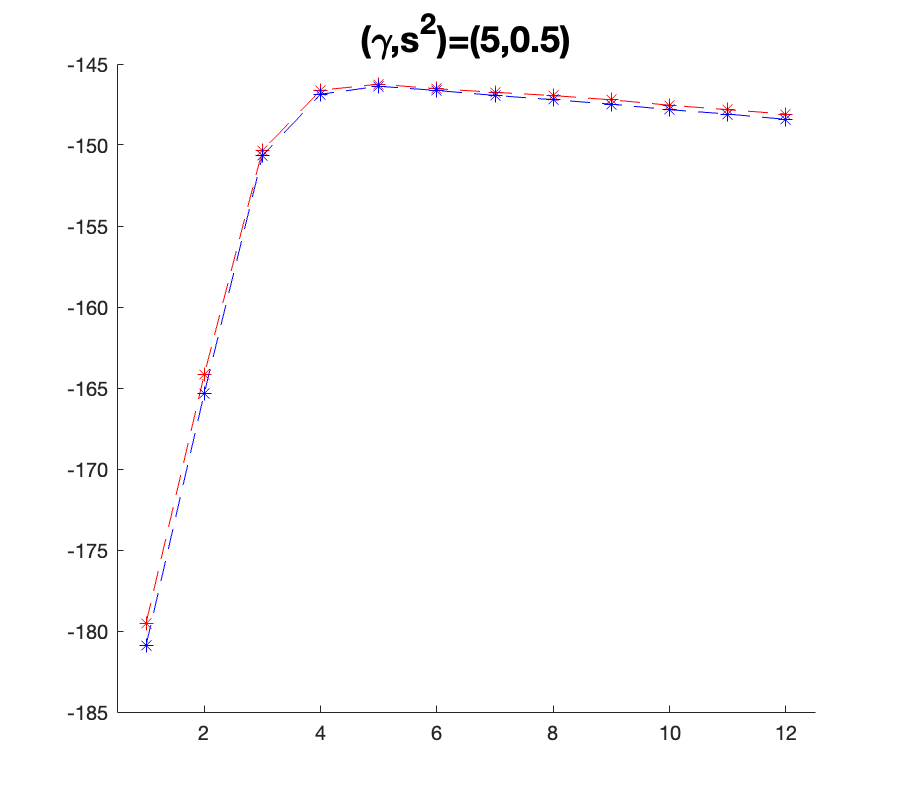}
\includegraphics[width=0.24\textwidth]{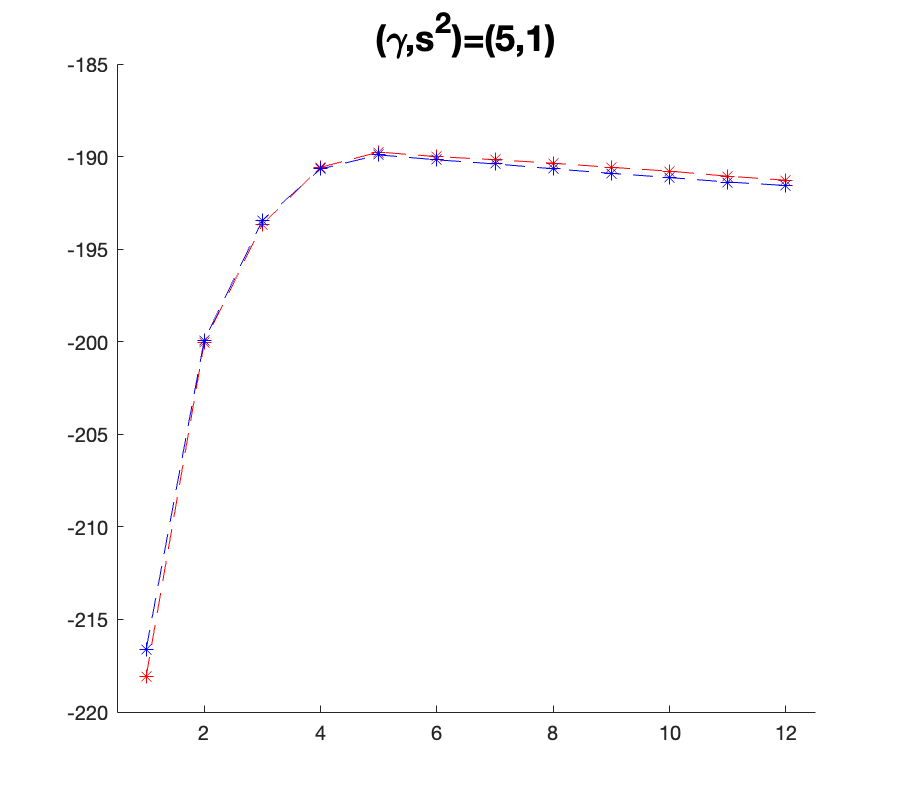}
\caption{The standardized $\mathcal{V}\left(K\right)$ by HeF over the combinations of $\left(\gamma,s^2\right)$. To compare the values of $\mathcal{V}\left(K\right)$ from two different validation ratios of 10\% and 20\%, we divided the original $\mathcal{V}\left(K\right)$ defined in \eqref{equation_3_4_1} by the size of the validation set. Red and blue curves indicate those of 10\% and 20\% validation ratios, respectively.}
\label{fig_5_1_2}
\end{figure}

\begin{figure}
\renewcommand{\baselinestretch}{1}
\centering
\includegraphics[width=0.24\textwidth]{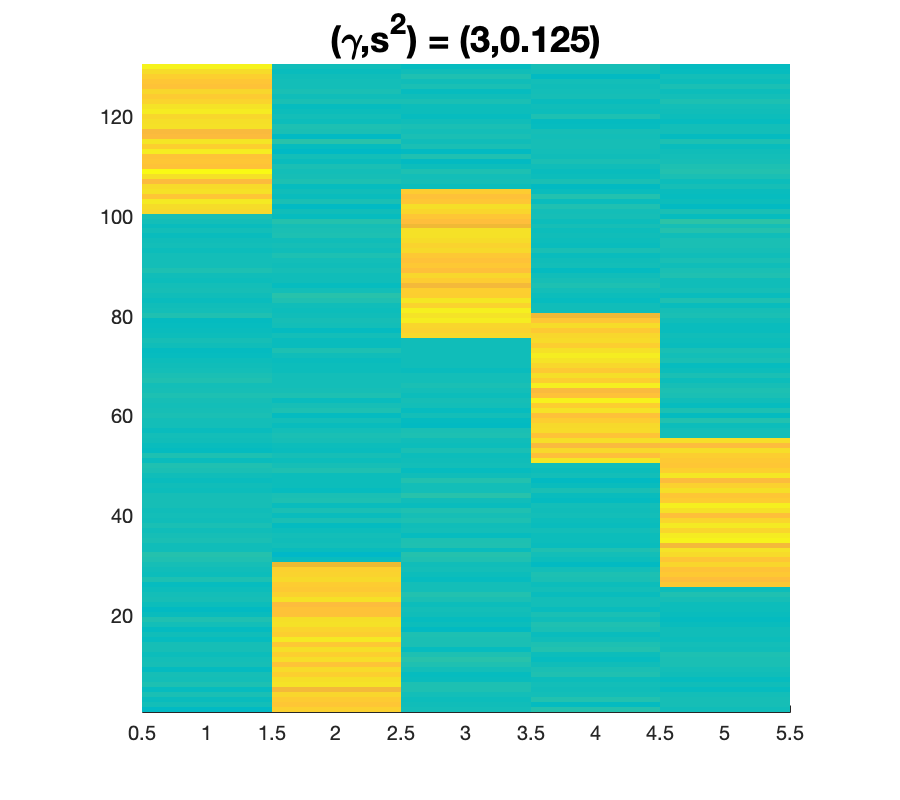}
\includegraphics[width=0.24\textwidth]{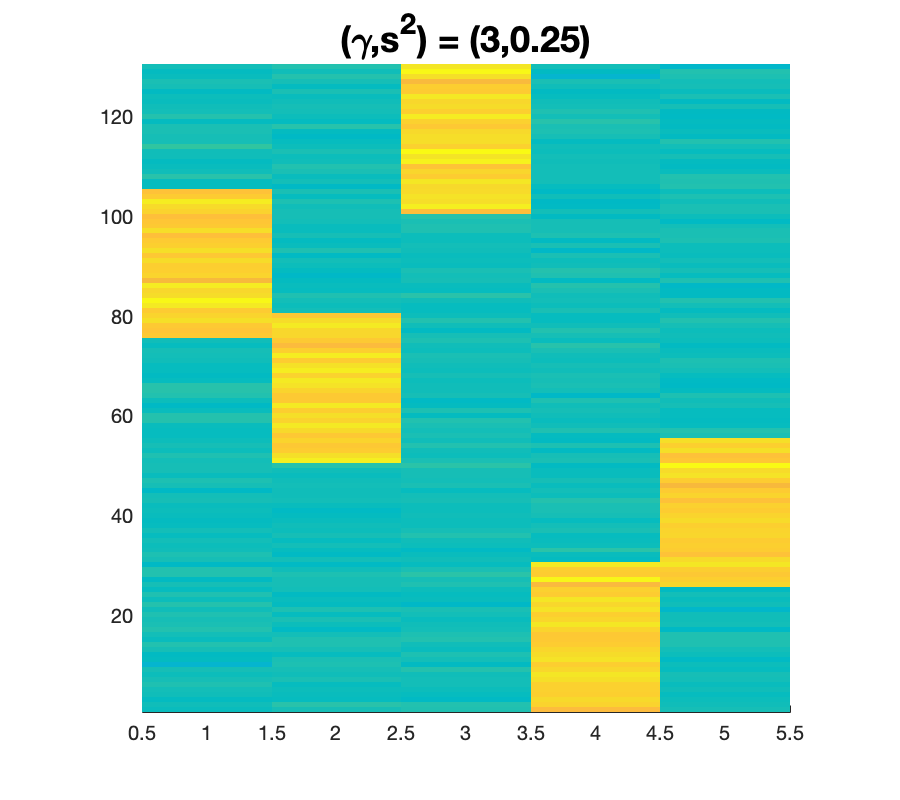}
\includegraphics[width=0.24\textwidth]{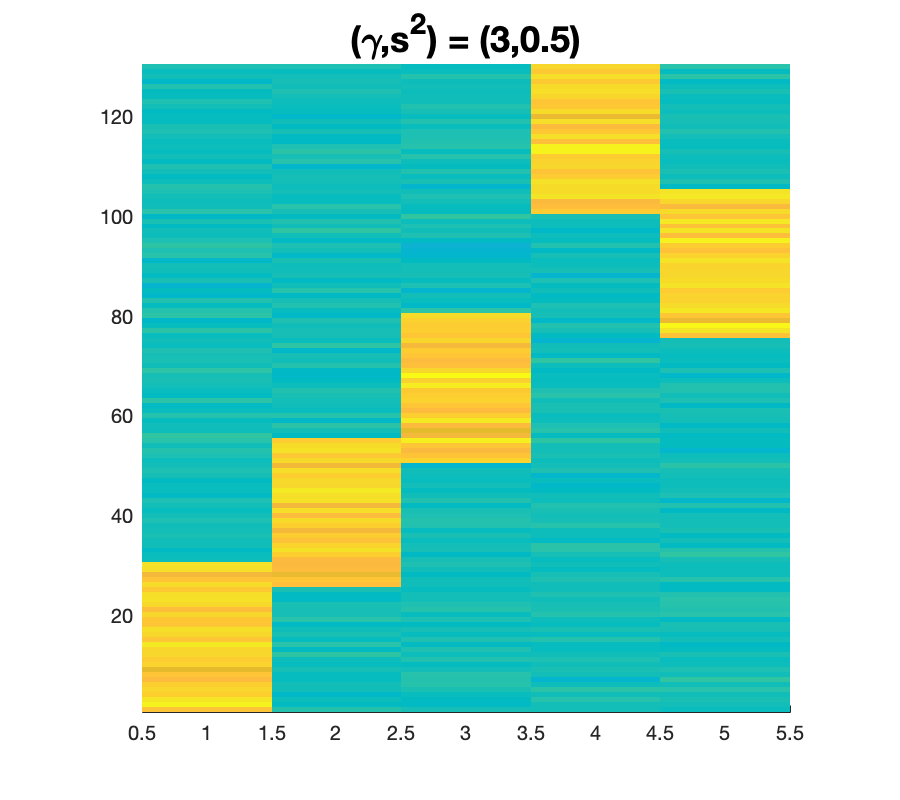}
\includegraphics[width=0.24\textwidth]{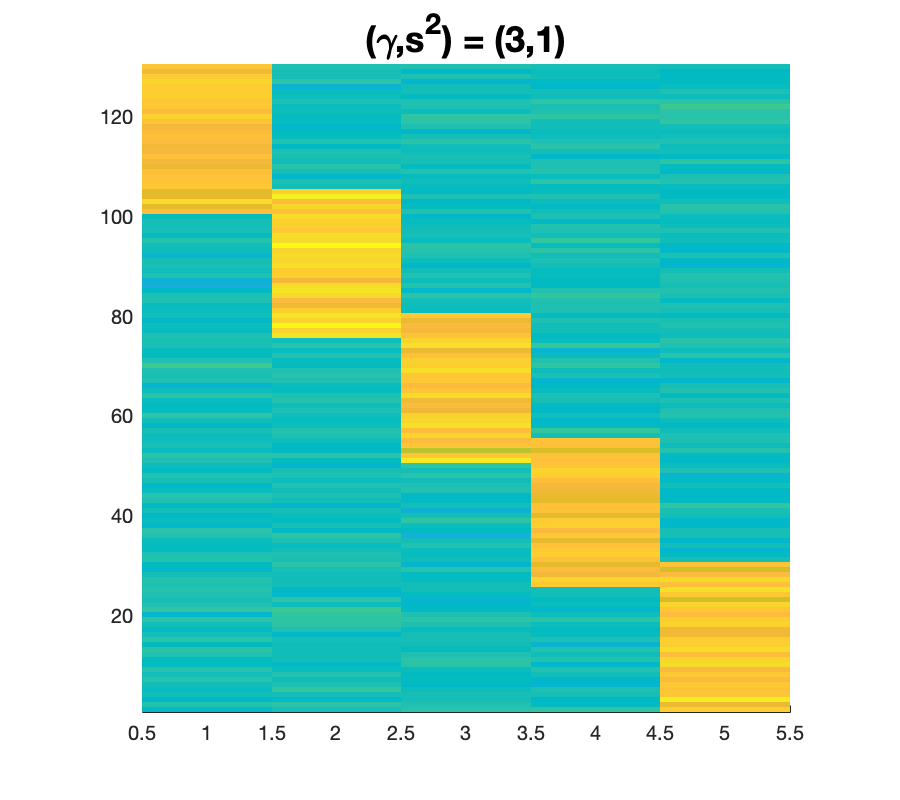}
\includegraphics[width=0.24\textwidth]{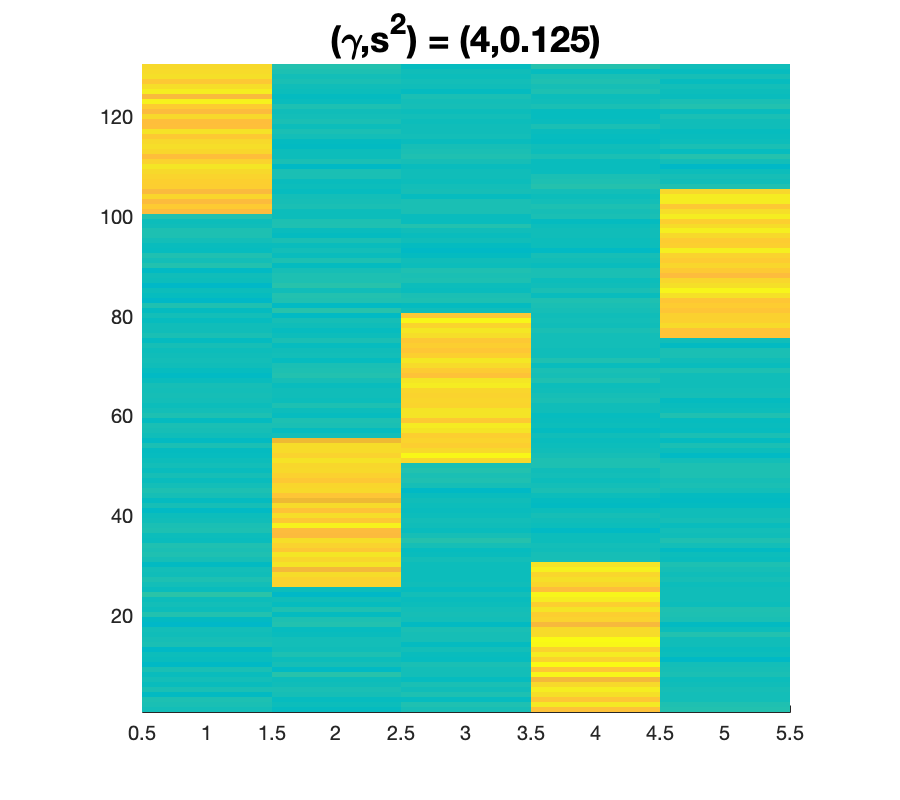}
\includegraphics[width=0.24\textwidth]{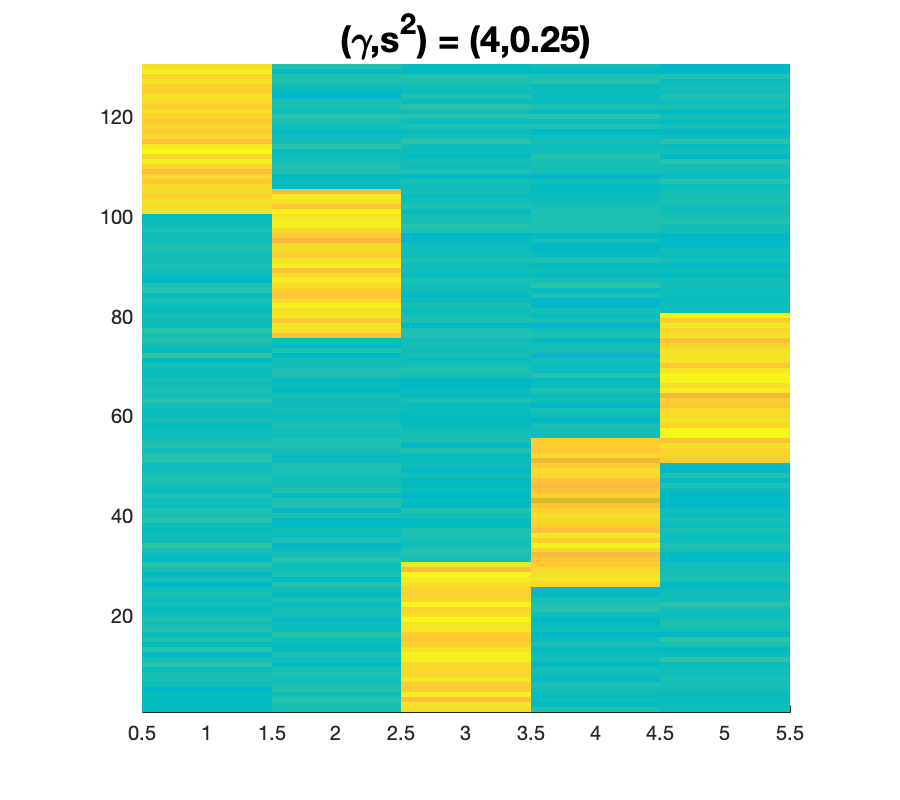}
\includegraphics[width=0.24\textwidth]{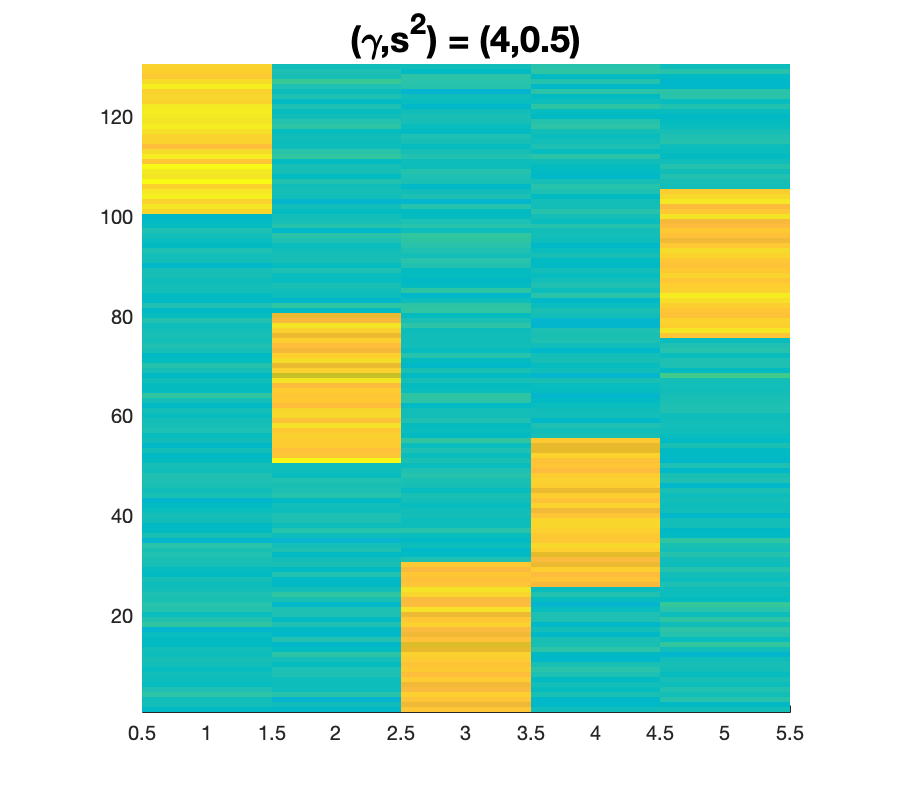}
\includegraphics[width=0.24\textwidth]{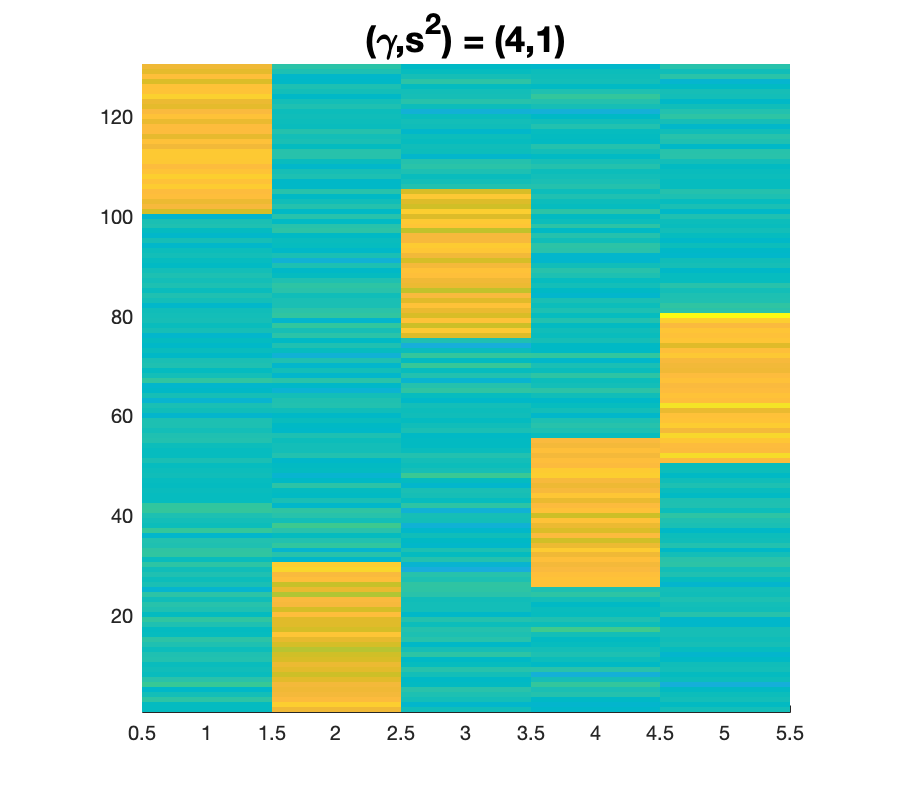}
\includegraphics[width=0.24\textwidth]{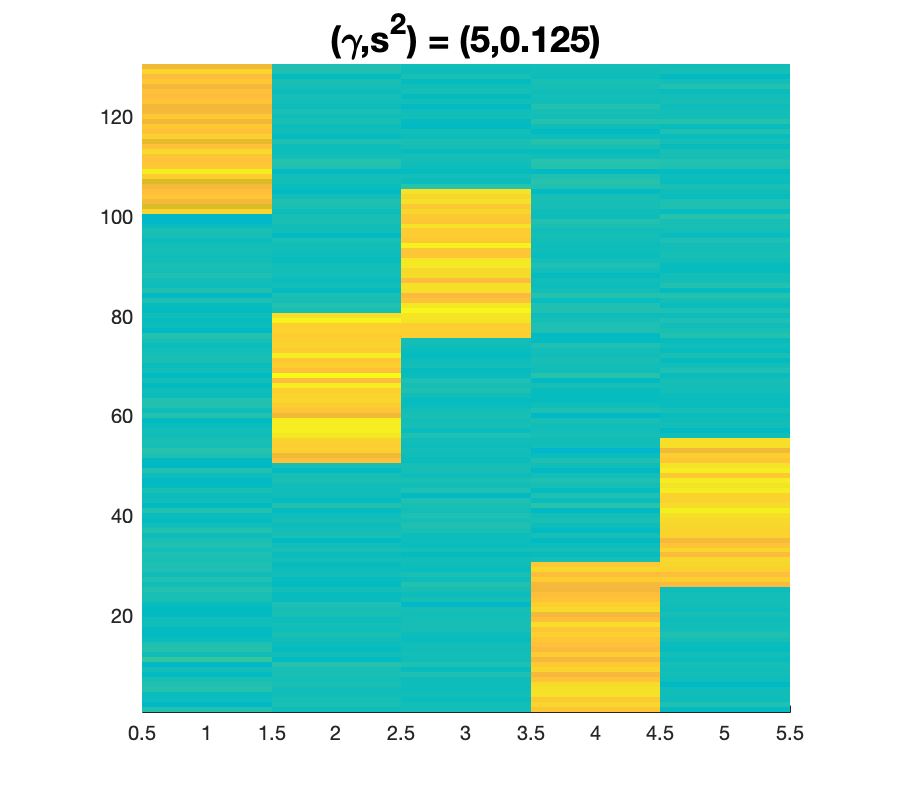}
\includegraphics[width=0.24\textwidth]{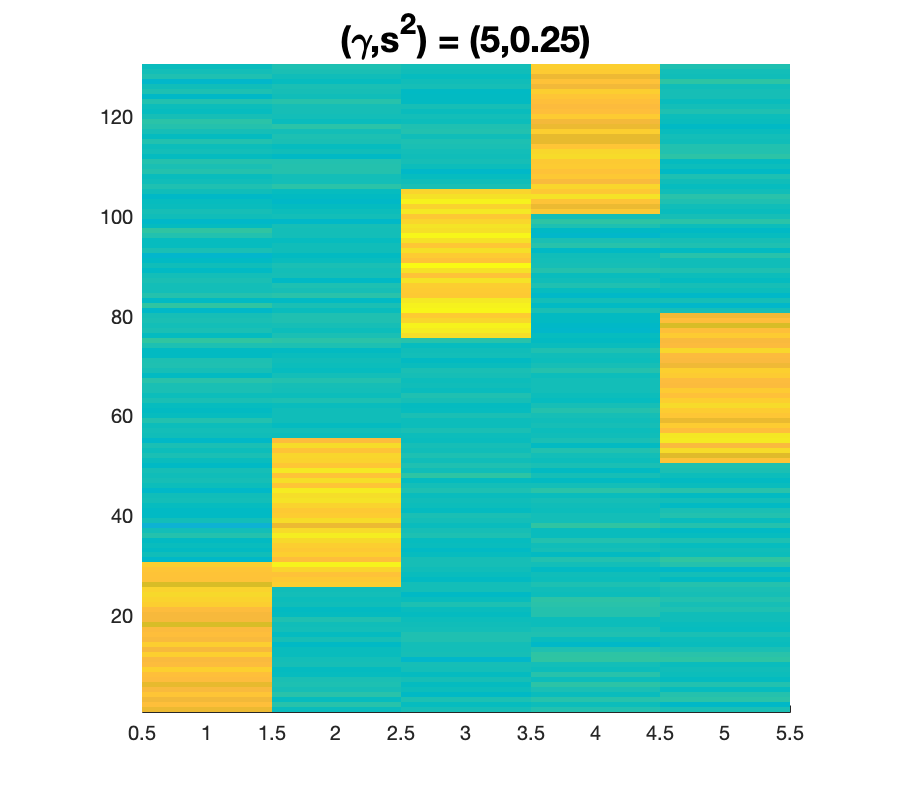}
\includegraphics[width=0.24\textwidth]{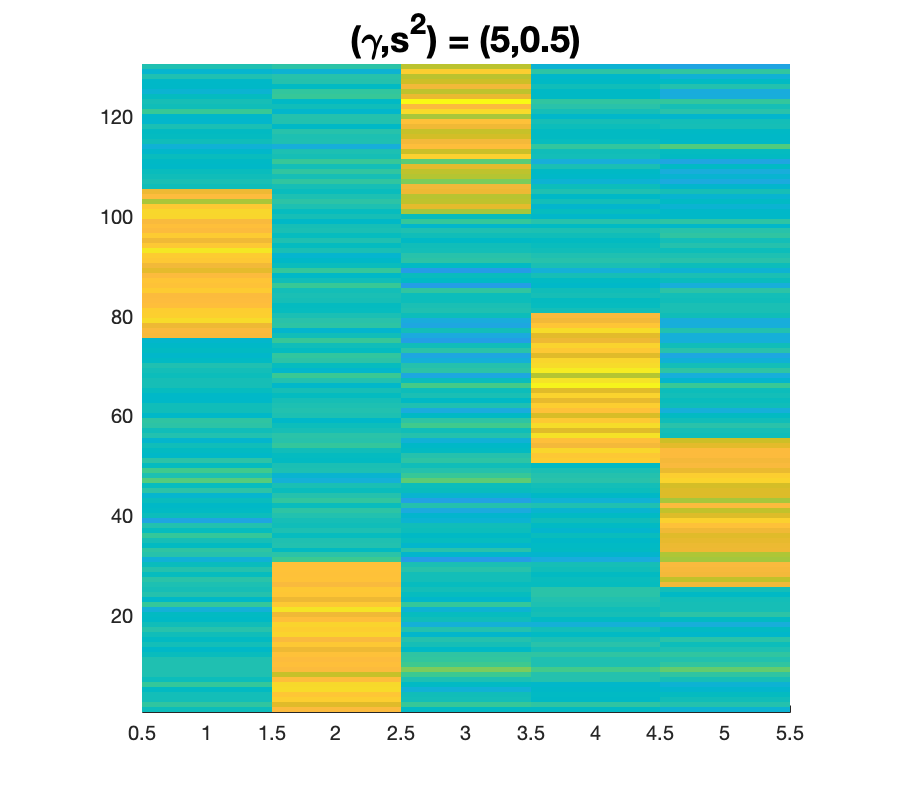}
\includegraphics[width=0.24\textwidth]{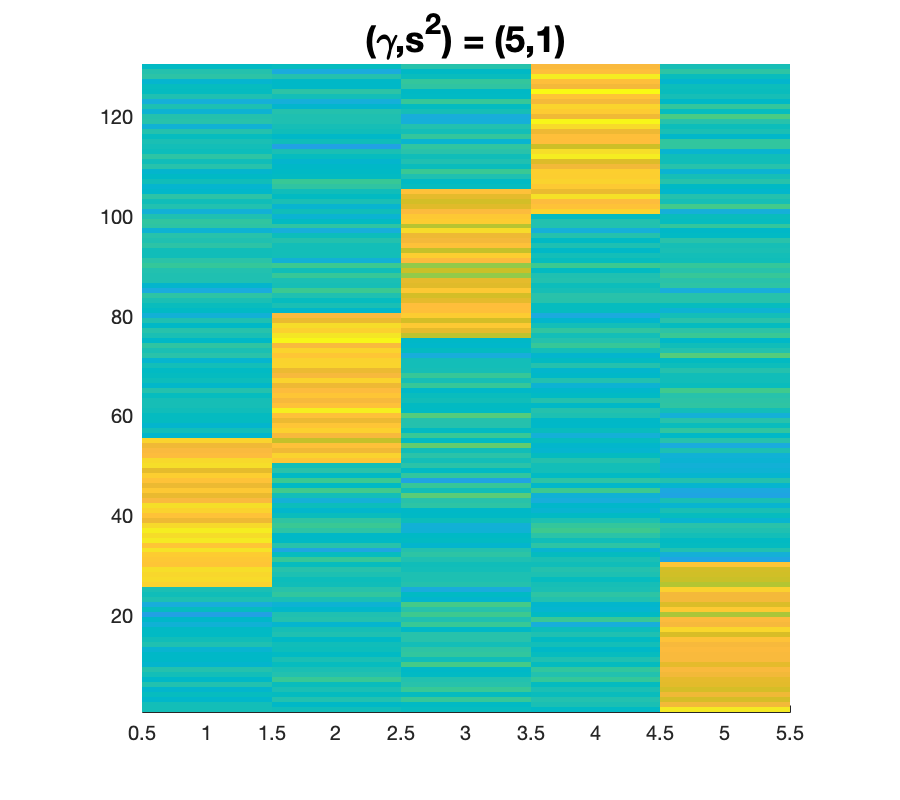}
\caption{Estimated factor loading matrices over the combinations of $\left(\gamma,s^2\right)$.}
\label{fig_5_1_3}
\end{figure}

\begin{table}
\small
\begin{center}
\begin{tabular}{ |c||c|c||c|c|c|c|c|c|c|c|c|c|c| } 
  \hline
  $\gamma,s^{2}$ & HeF & HoF & 0.98 & 0.96 & 0.94 & 0.92 & 0.90 & 0.88 & 0.86 & 0.84 & 0.82 & 0.80 \\ \hline \hline 
  $\left(3,0.125\right)$ & \textbf{2.70} & 3.76 & 3.40 & 3.11 & 2.95 & 2.85 & 2.79 & 2.76 & 2.75 & 2.76 & 2.79 & 2.83  \\ 
  \hline
  $\left(3,0.25\right)$ & \textbf{2.29} & 3.07 & 2.85 & 2.62 & 2.49 & 2.41 & 2.37 & 2.35 & 2.35 & 2.37 & 2.41 & 2.47 \\ 
  \hline
  $\left(3,0.5\right)$ & \textbf{2.24} & 3.35 & 3.09 & 2.82 & 2.64 & 2.51 & 2.42 & 2.36 & 2.33 & 2.33 & 2.34 & 2.36 \\ 
  \hline
  $\left(3,1\right)$ & \textbf{2.19} & 3.39 & 2.95 & 2.65 & 2.49 & 2.39 & 2.32 & 2.29 & 2.28 & 2.28 & 2.29 & 2.32 \\ 
  \hline
  $\left(4,0.125\right)$ & \textbf{2.12} & 3.76 & 2.99 & 2.61 & 2.42 & 2.32 & 2.28 & 2.27 & 2.30 & 2.34 & 2.40 & 2.48  \\ 
  \hline
  $\left(4,0.25\right)$ & \textbf{2.23} & 3.39 & 2.80 & 2.49 & 2.36 & 2.32 & 2.32 & 2.34 & 2.39 & 2.45 & 2.52 & 2.61 \\ 
  \hline
  $\left(4,0.5\right)$ & \textbf{1.71} & 2.95 & 2.26 & 1.94 & 1.85 & 1.83 & 1.84 & 1.88 & 1.92 & 1.98 & 2.04 & 2.11  \\ 
  \hline
  $\left(4,1\right)$ & \textbf{1.67} & 2.58 & 2.10 & 1.94 & 1.91 & 1.93 & 1.95 & 1.99 & 2.03 & 2.08 & 2.14 & 2.20  \\ 
  \hline
  $\left(5,0.125\right)$ & \textbf{1.94} & 2.74 & 2.25 & 2.10 & 2.09 & 2.11 & 2.14 & 2.19 & 2.25 & 2.32 & 2.40 & 2.48  \\ 
  \hline
  $\left(5,0.25\right)$ & \textbf{1.55} & 2.13 & 1.80 & 1.72 & 1.73 & 1.77 & 1.82 & 1.87 & 1.94 & 2.01 & 2.08 & 2.17  \\ 
  \hline
  $\left(5,0.5\right)$ & \textbf{1.76} & 2.26 & 2.14 & 2.11 & 2.15 & 2.19 & 2.23 & 2.28 & 2.34 & 2.40 & 2.46 & 2.54  \\ 
  \hline
  $\left(5,1\right)$ & \textbf{1.47} & 1.80 & 1.73 & 1.71 & 1.73 & 1.76 & 1.79 & 1.83 & 1.87 & 1.92 & 1.97 & 2.02  \\ 
  \hline
  
\end{tabular}
\caption{(Ratio 9:1) The average KL divergences from the simulations.}
\label{table_5_1_1}
\end{center}
\end{table}

\begin{table}
\small
\begin{center}
\begin{tabular}{ |c||c|c||c|c|c|c|c|c|c|c|c|c|c| } 
  \hline
  $\gamma,s^{2}$ & HeF & HoF & 0.98 & 0.96 & 0.94 & 0.92 & 0.90 & 0.88 & 0.86 & 0.84 & 0.82 & 0.80 \\ \hline \hline 
  $\left(3,0.125\right)$ & \textbf{3.01} & 3.95 & 3.62 & 3.35 & 3.21 & 3.12 & 3.08 & 3.06 & 3.07 & 3.09 & 3.14 & 3.21  \\ 
  \hline
  $\left(3,0.25\right)$ & \textbf{2.63} & 3.29 & 3.09 & 2.86 & 2.74 & 2.67 & 2.64 & 2.64 & 2.66 & 2.70 & 2.77 & 2.85 \\ 
  \hline
  $\left(3,0.5\right)$ & \textbf{2.47} & 3.52 & 3.27 & 3.01 & 2.84 & 2.72 & 2.64 & 2.59 & 2.56 & 2.56 & 2.58 & 2.62 \\ 
  \hline
  $\left(3,1\right)$ & \textbf{2.50} & 3.60 & 3.17 & 2.87 & 2.71 & 2.62 & 2.58 & 2.56 & 2.56 & 2.58 & 2.61 & 2.65  \\ 
  \hline
  $\left(4,0.125\right)$ & \textbf{2.29} & 3.85 & 3.12 & 2.75 & 2.57 & 2.48 & 2.45 & 2.45 & 2.48 & 2.53 & 2.60 & 2.69  \\ 
  \hline
  $\left(4,0.25\right)$ & \textbf{2.48} & 3.60 & 3.02 & 2.70 & 2.59 & 2.55 & 2.56 & 2.60 & 2.66 & 2.73 & 2.82 & 2.93  \\ 
  \hline
  $\left(4,0.5\right)$ & \textbf{1.91} & 3.13 & 2.47 & 2.16 & 2.06 & 2.05 & 2.07 & 2.11 & 2.16 & 2.23 & 2.30 & 2.39  \\ 
  \hline
  $\left(4,1\right)$ & \textbf{1.87} & 2.74 & 2.31 & 2.16 & 2.14 & 2.16 & 2.19 & 2.23 & 2.28 & 2.34 & 2.40 & 2.47  \\ 
  \hline
  $\left(5,0.125\right)$ & \textbf{2.16} & 2.95 & 2.48 & 2.33 & 2.32 & 2.34 & 2.39 & 2.45 & 2.51 & 2.59 & 2.68 & 2.77  \\ 
  \hline
  $\left(5,0.25\right)$ & \textbf{1.77} & 2.36 & 2.03 & 1.94 & 1.96 & 2.01 & 2.07 & 2.13 & 2.21 & 2.29 & 2.39 & 2.49  \\ 
  \hline
  $\left(5,0.5\right)$ & \textbf{1.98} & 2.49 & 2.38 & 2.36 & 2.40 & 2.45 & 2.51 & 2.57 & 2.64 & 2.72 & 2.80 & 2.89  \\ 
  \hline
  $\left(5,1\right)$ & \textbf{1.65} & 1.97 & 1.92 & 1.90 & 1.93 & 1.96 & 2.00 & 2.05 & 2.10 & 2.15 & 2.21 & 2.28  \\ 
  \hline
  
\end{tabular}
\caption{(Ratio 8:2) The average KL divergences from the simulations.}
\label{table_5_1_2}
\end{center}
\end{table}

\begin{figure}
\renewcommand{\baselinestretch}{1}
\centering
\includegraphics[width=0.24\textwidth]{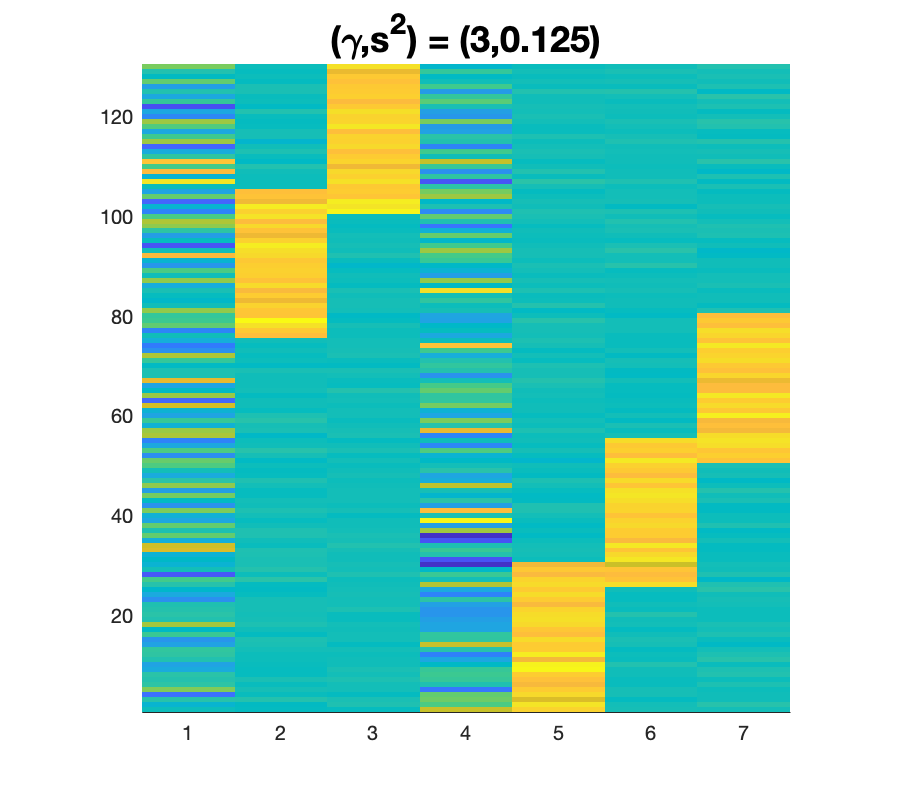}
\includegraphics[width=0.24\textwidth]{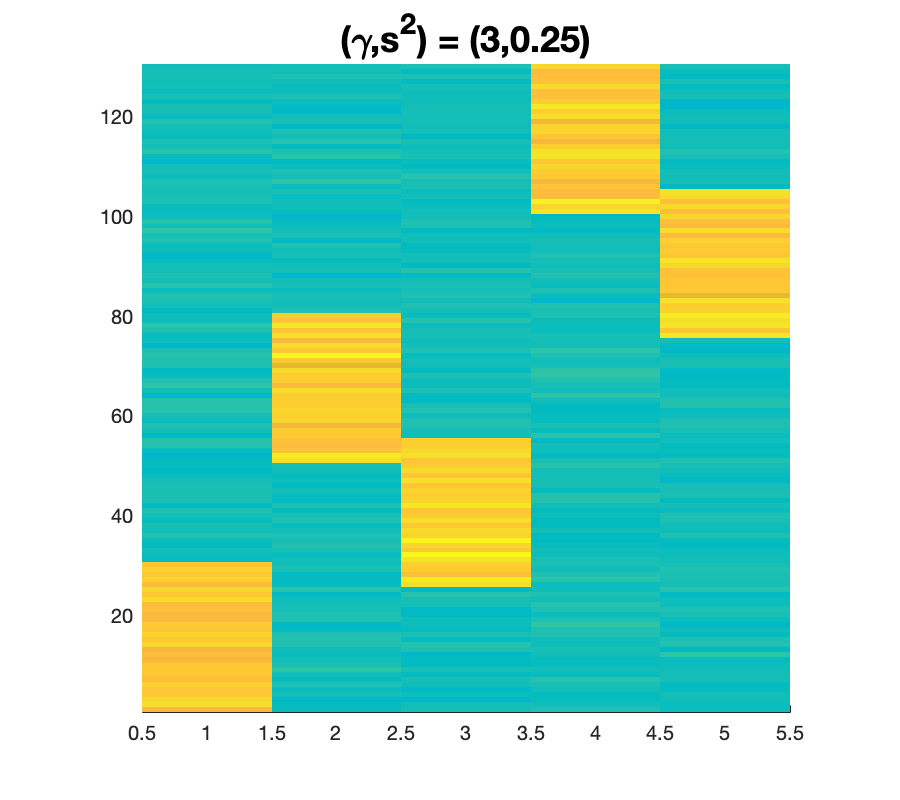}
\includegraphics[width=0.24\textwidth]{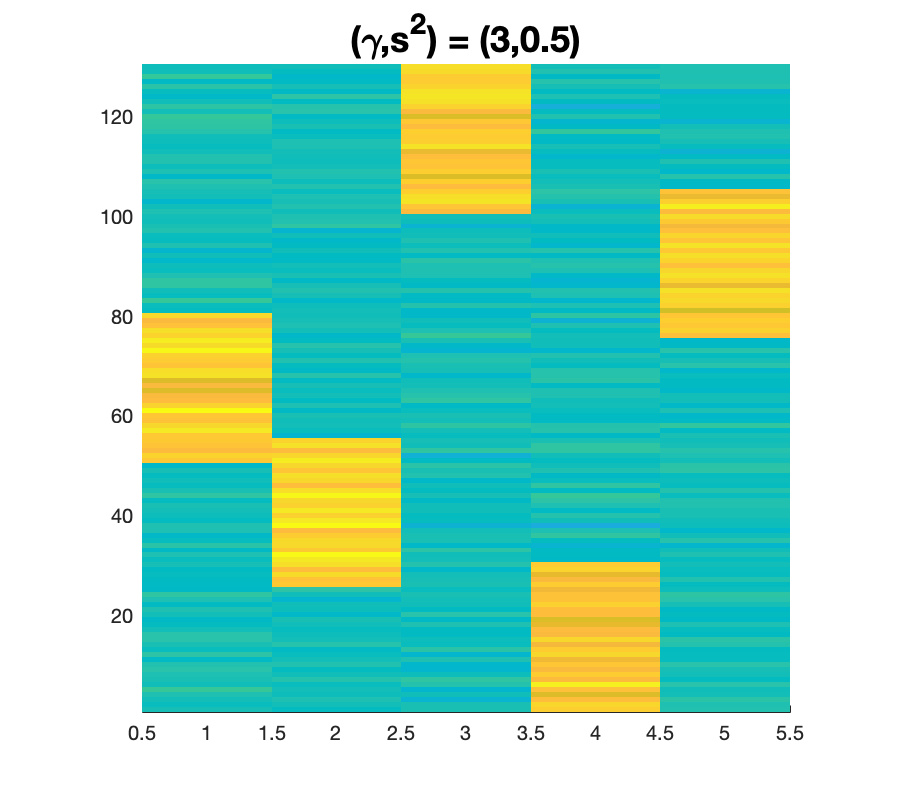}
\includegraphics[width=0.24\textwidth]{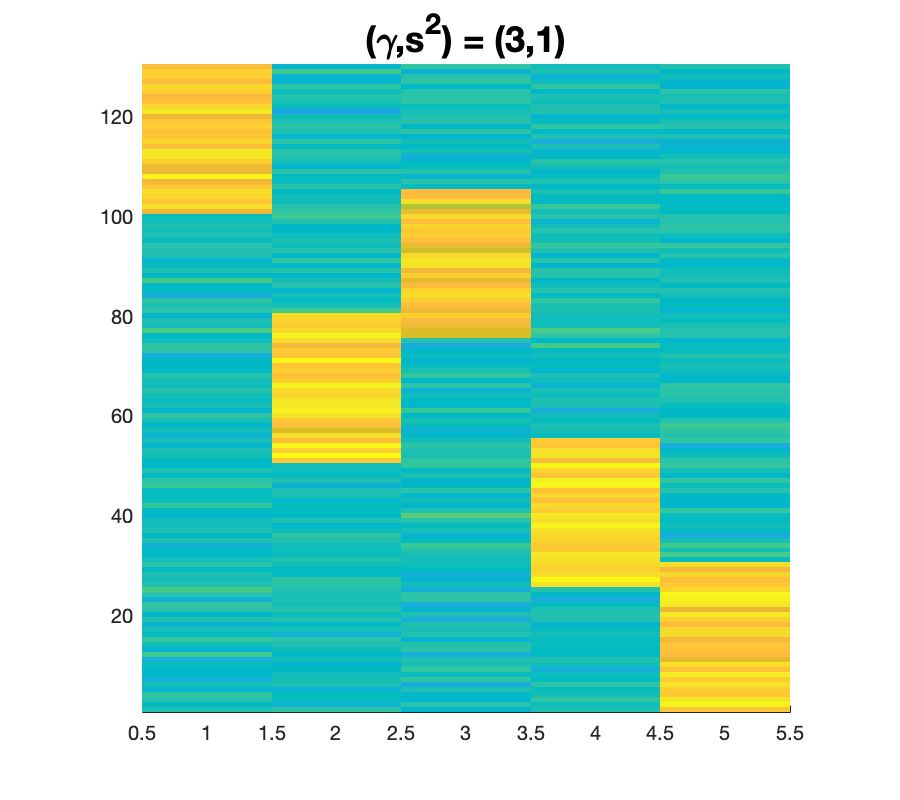}
\includegraphics[width=0.24\textwidth]{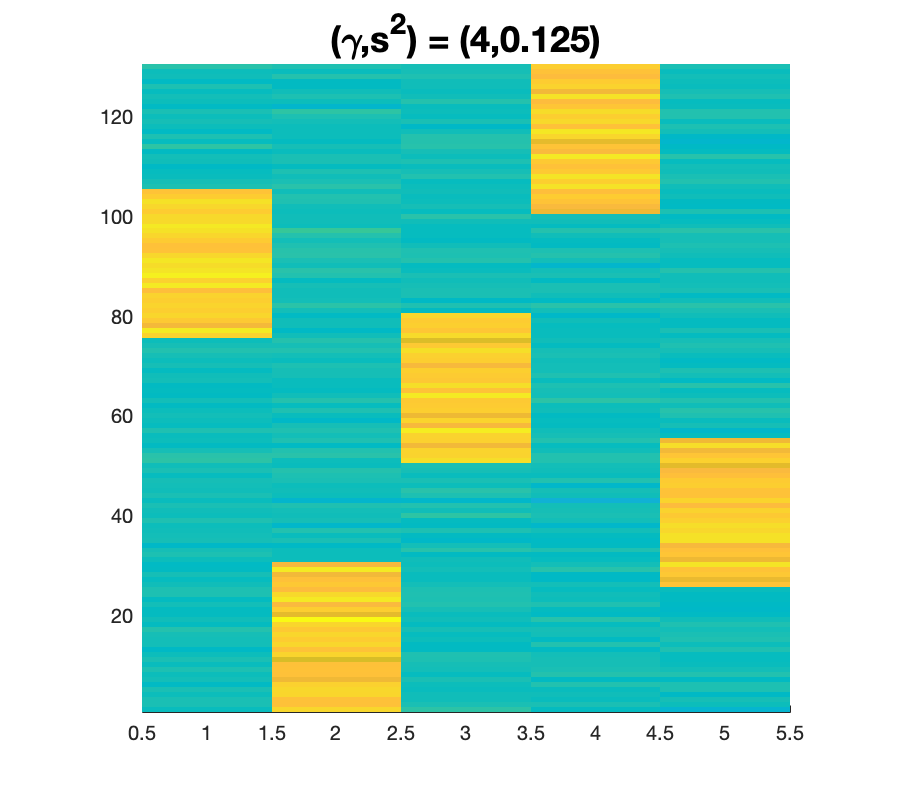}
\includegraphics[width=0.24\textwidth]{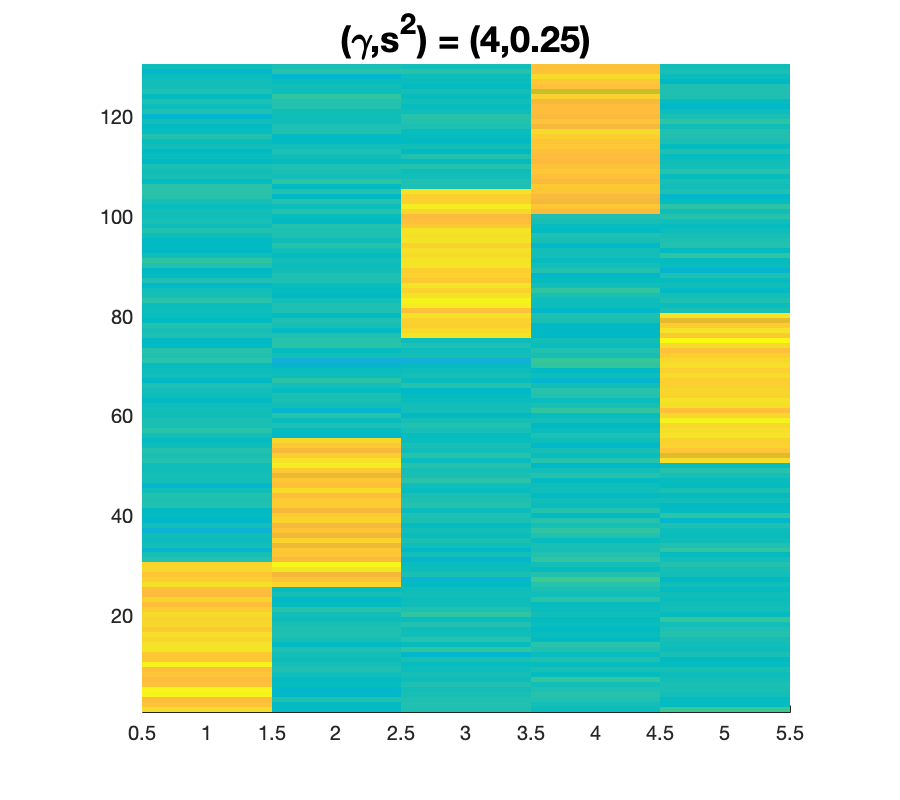}
\includegraphics[width=0.24\textwidth]{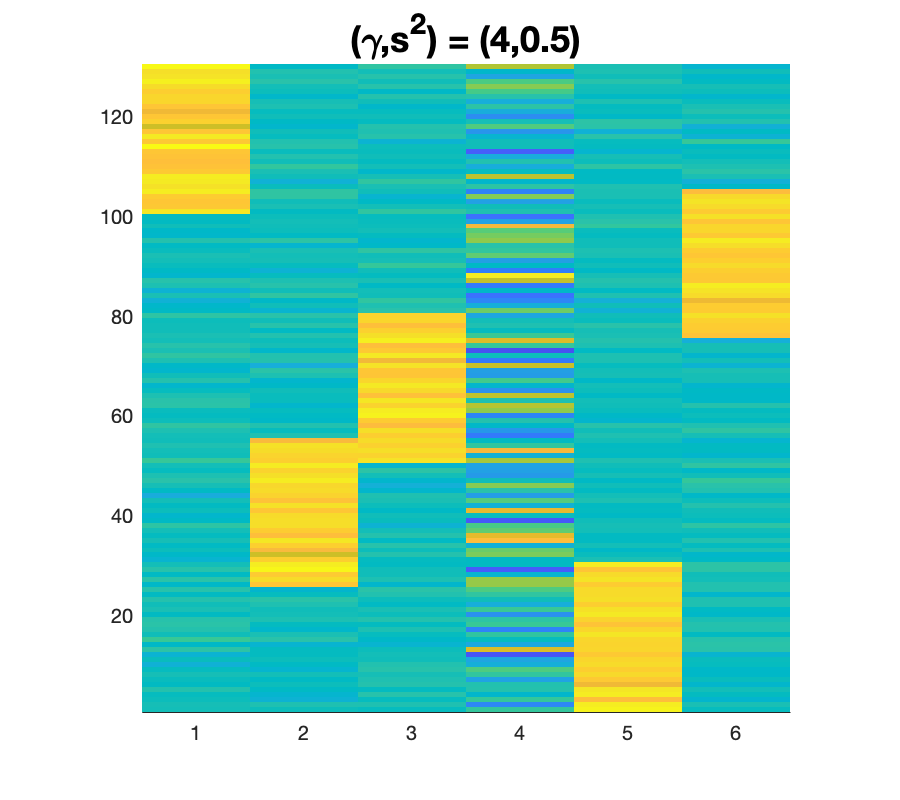}
\includegraphics[width=0.24\textwidth]{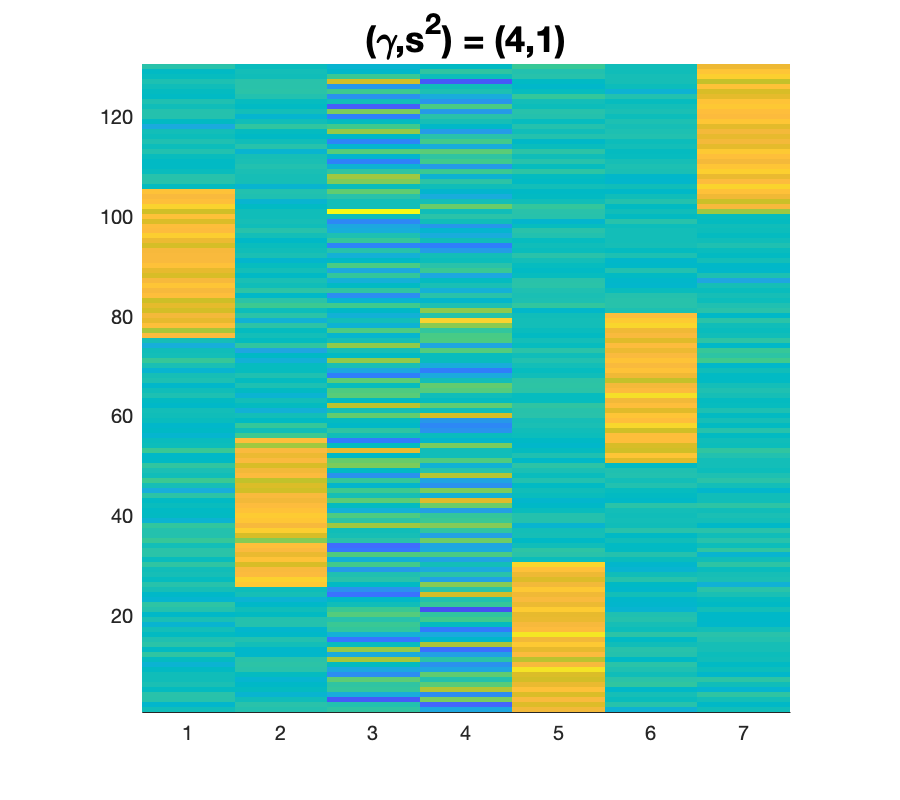}
\includegraphics[width=0.24\textwidth]{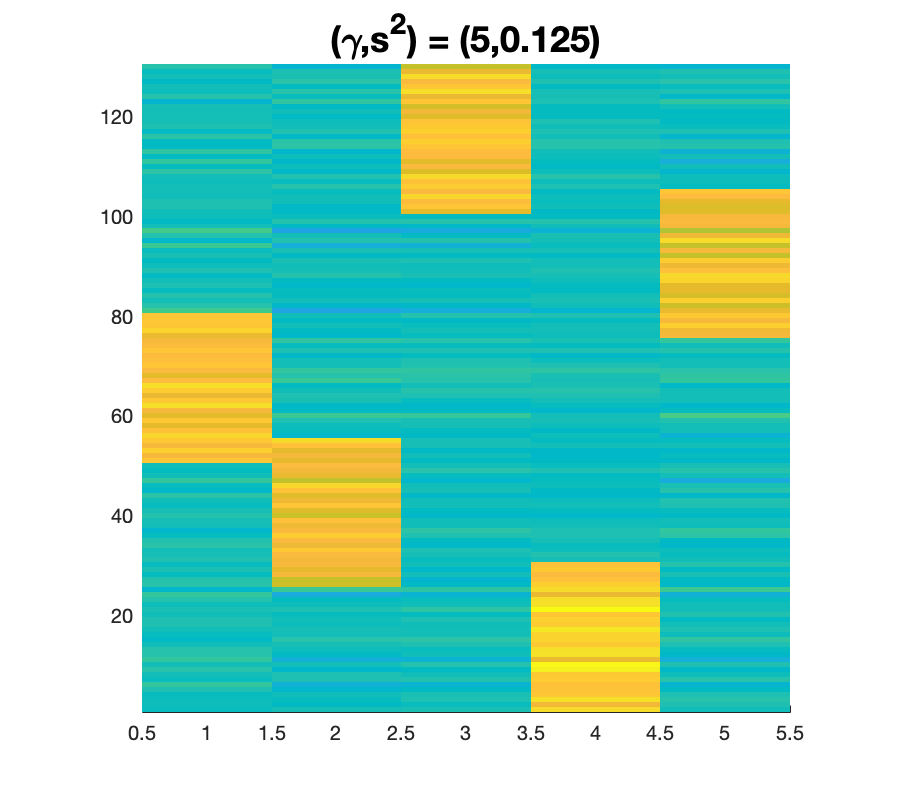}
\includegraphics[width=0.24\textwidth]{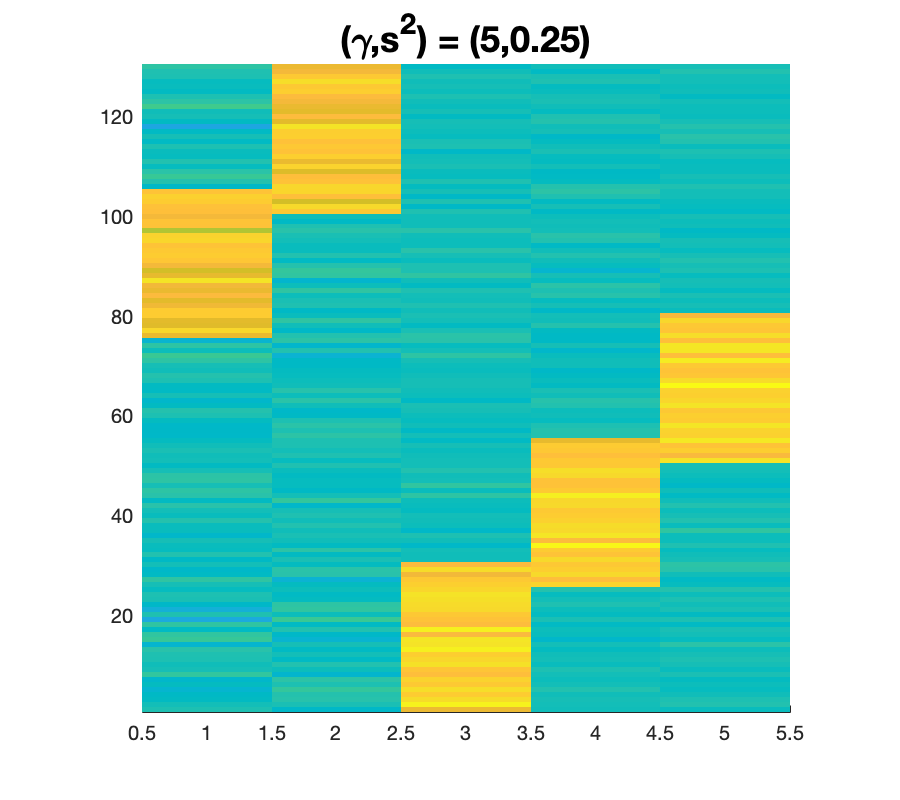}
\includegraphics[width=0.24\textwidth]{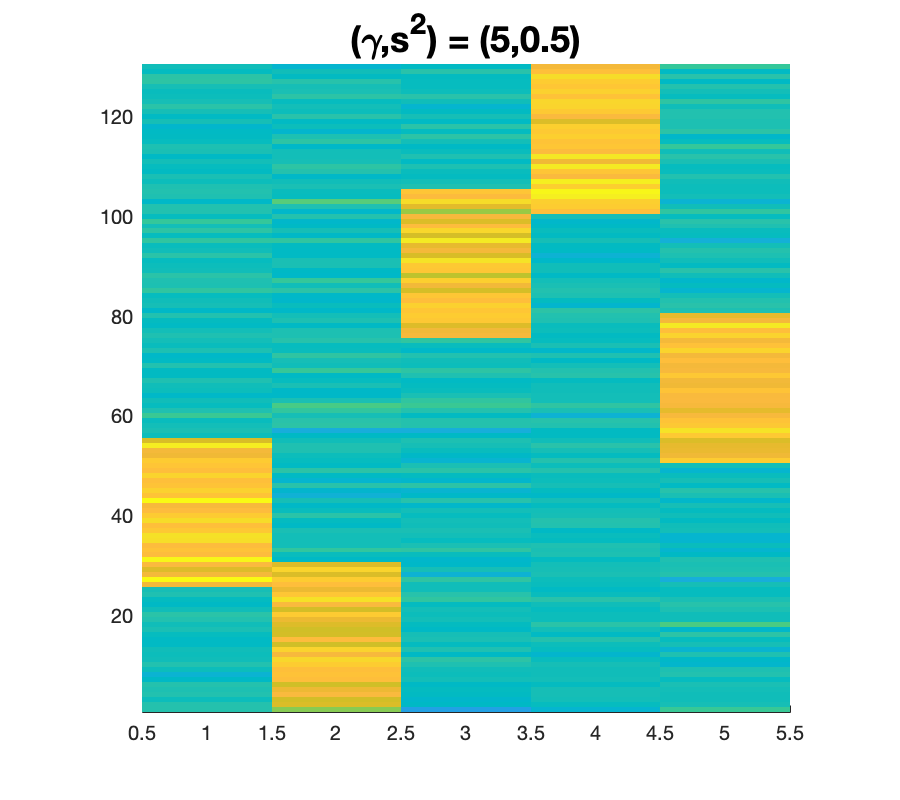}
\includegraphics[width=0.24\textwidth]{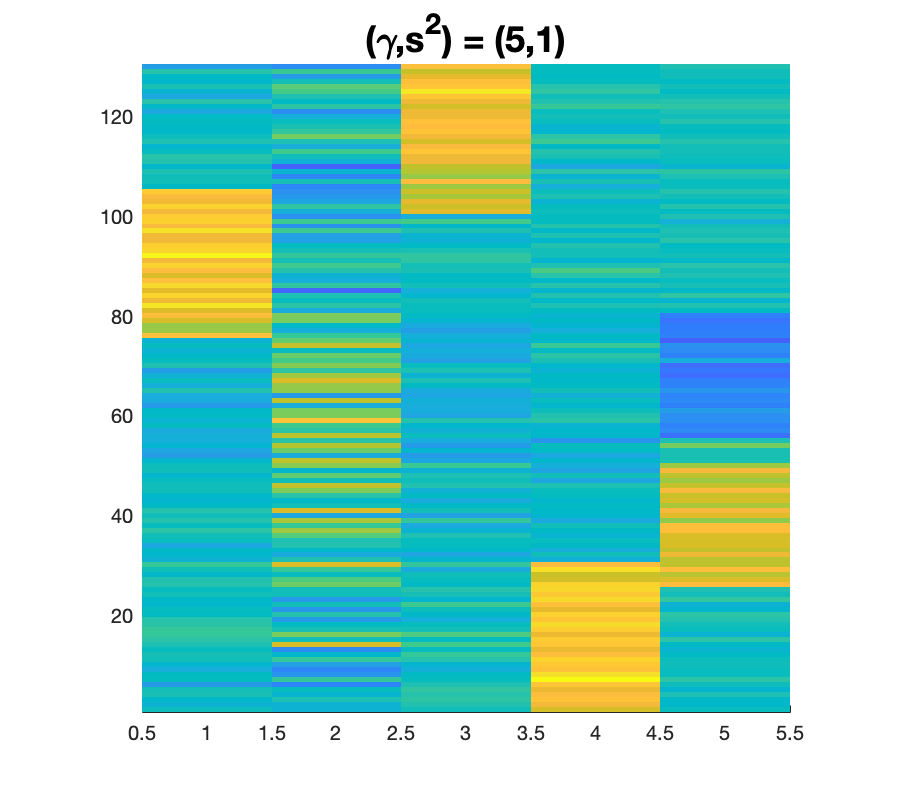}
\caption{Estimated factor loading matrices by Algorithm \ref{algorithm1} (Gaussian factor model) for the simulation with the Student's t-distribution in Section \ref{sec5}.}
\label{fig_5_2_1}
\end{figure}

\begin{figure}
\renewcommand{\baselinestretch}{1}
\centering
\includegraphics[width=0.24\textwidth]{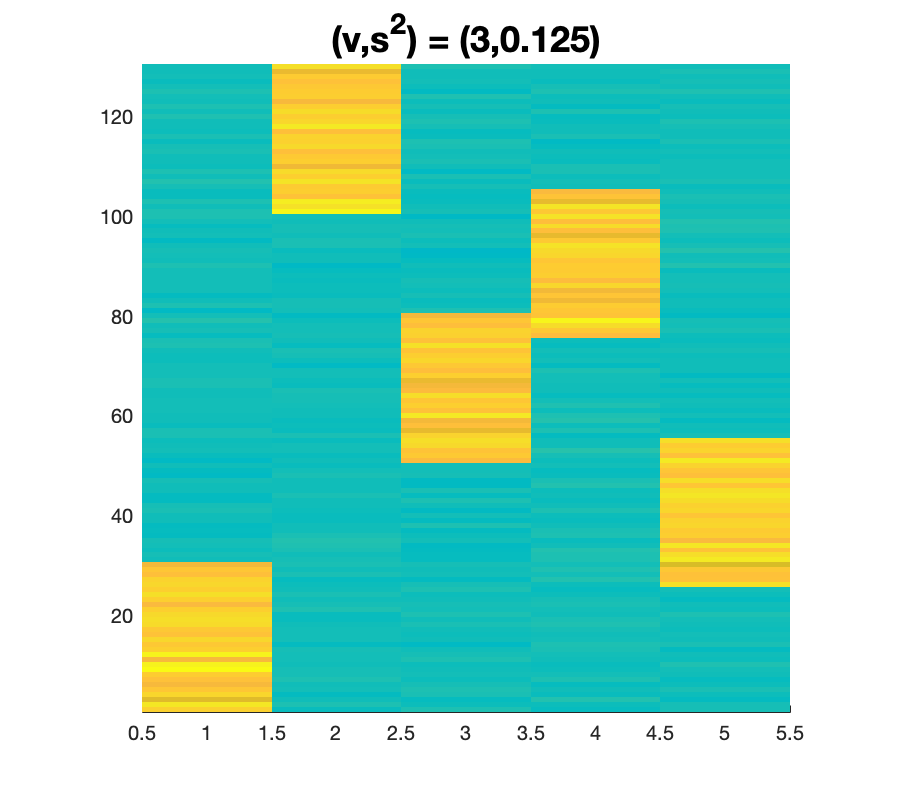}
\includegraphics[width=0.24\textwidth]{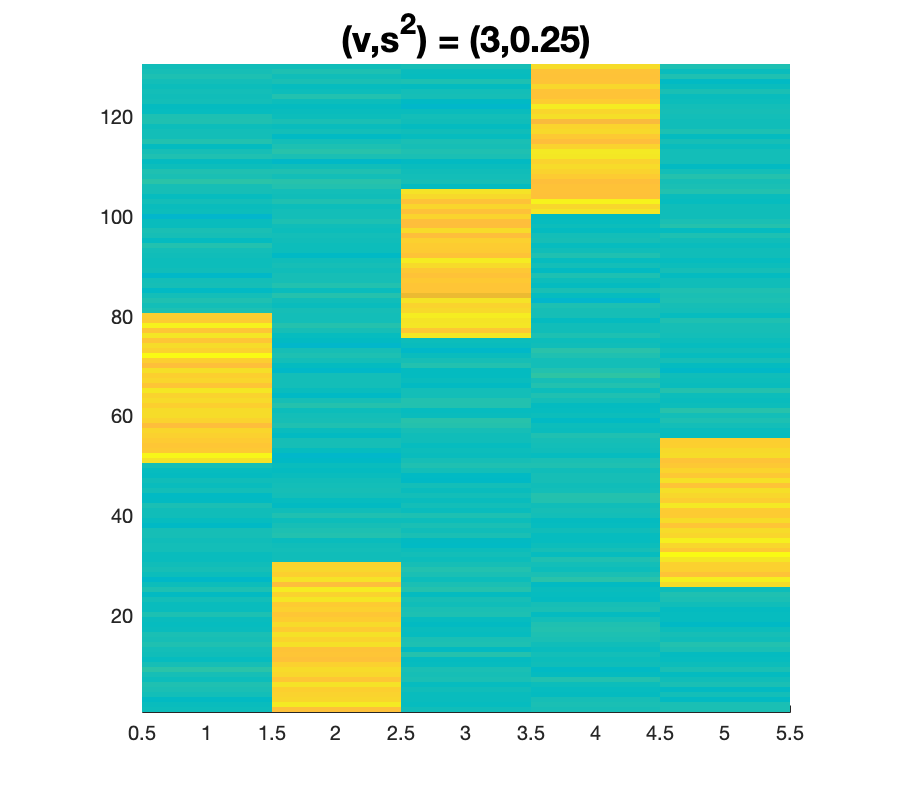}
\includegraphics[width=0.24\textwidth]{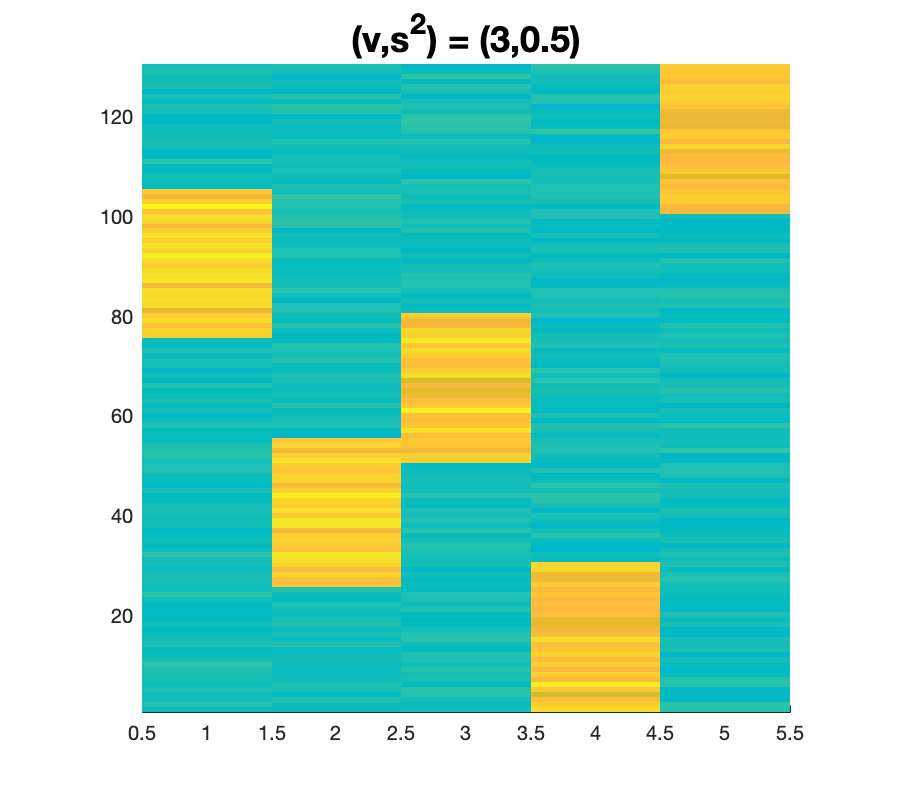}
\includegraphics[width=0.24\textwidth]{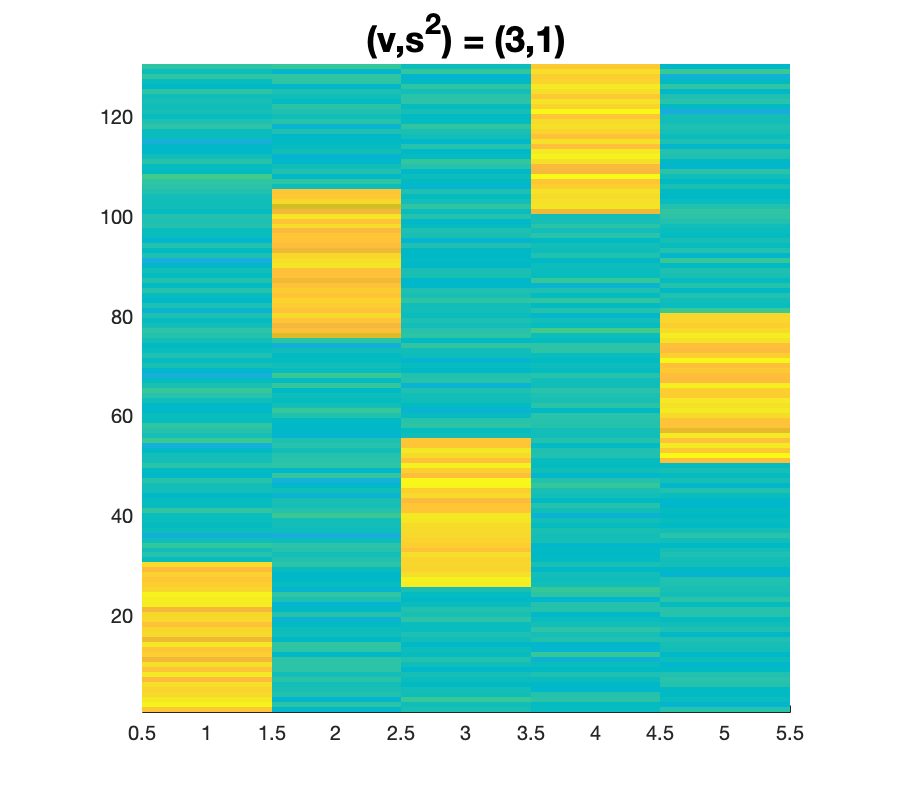}
\includegraphics[width=0.24\textwidth]{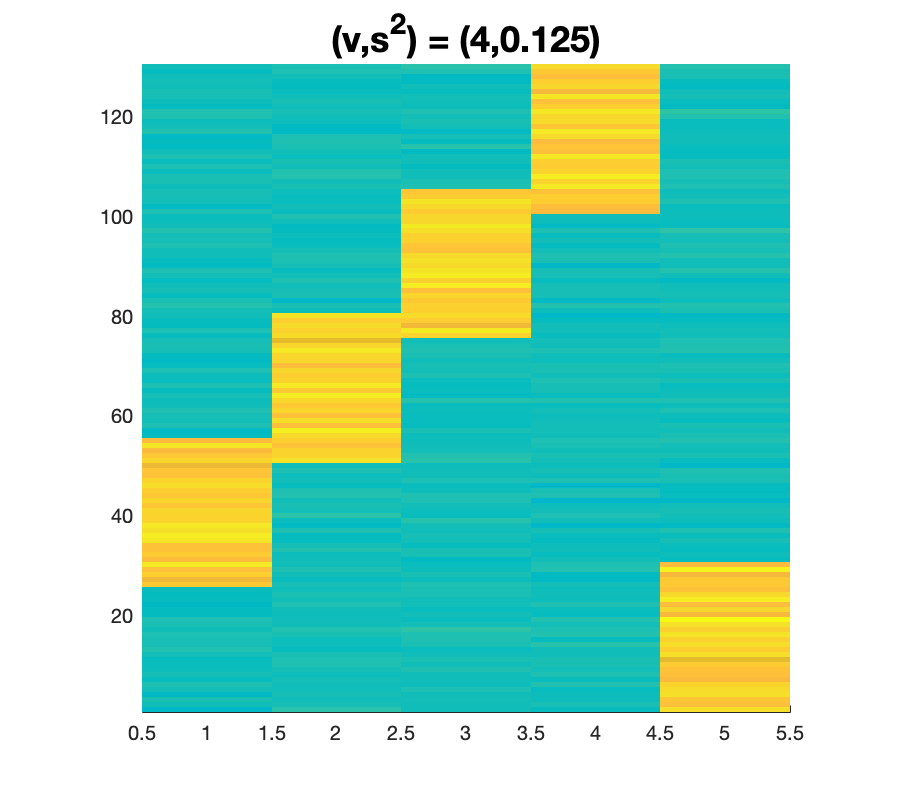}
\includegraphics[width=0.24\textwidth]{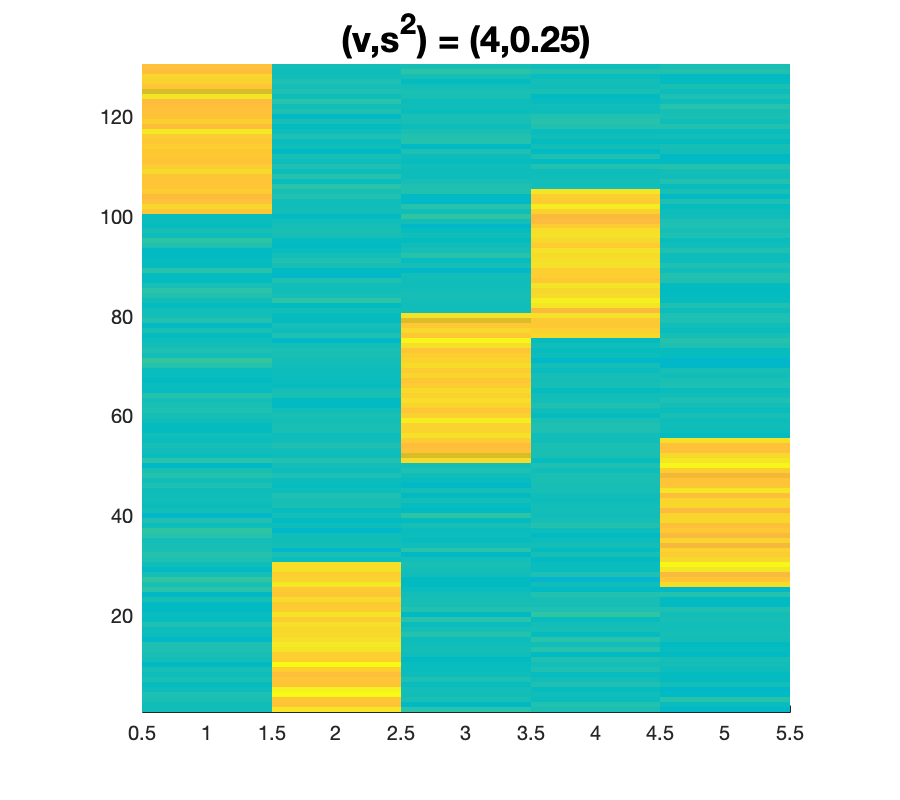}
\includegraphics[width=0.24\textwidth]{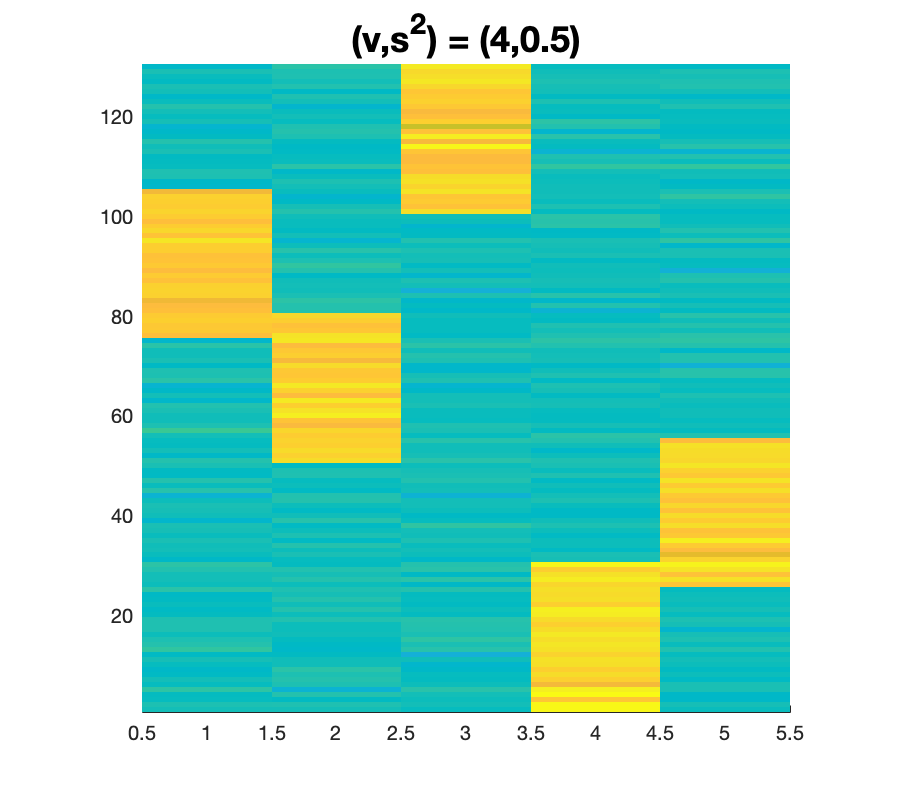}
\includegraphics[width=0.24\textwidth]{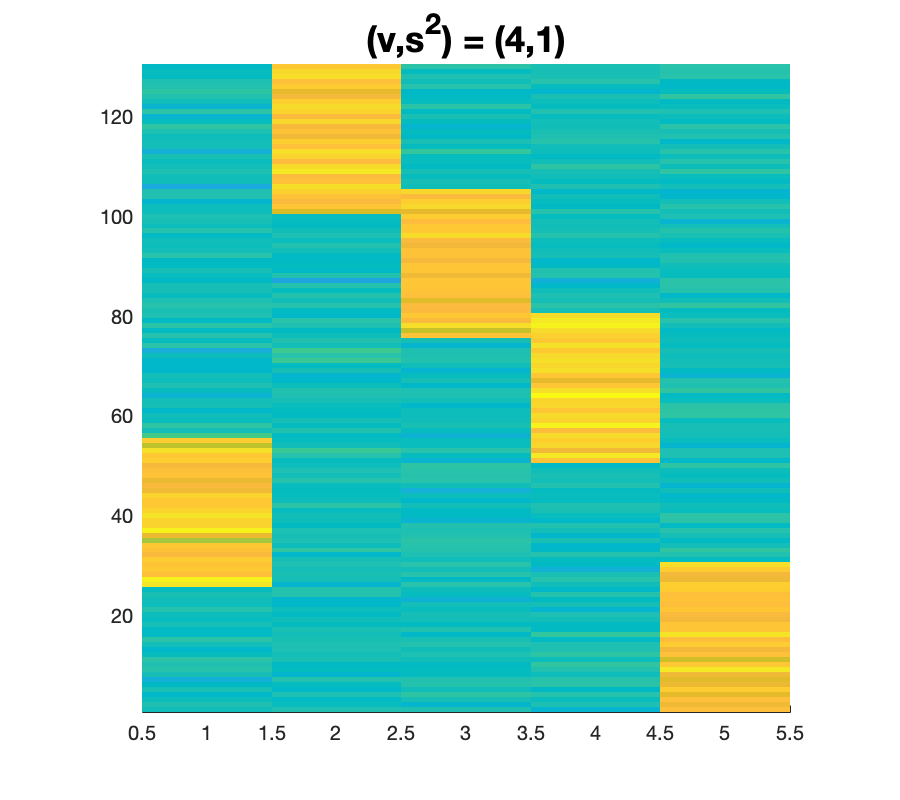}
\includegraphics[width=0.24\textwidth]{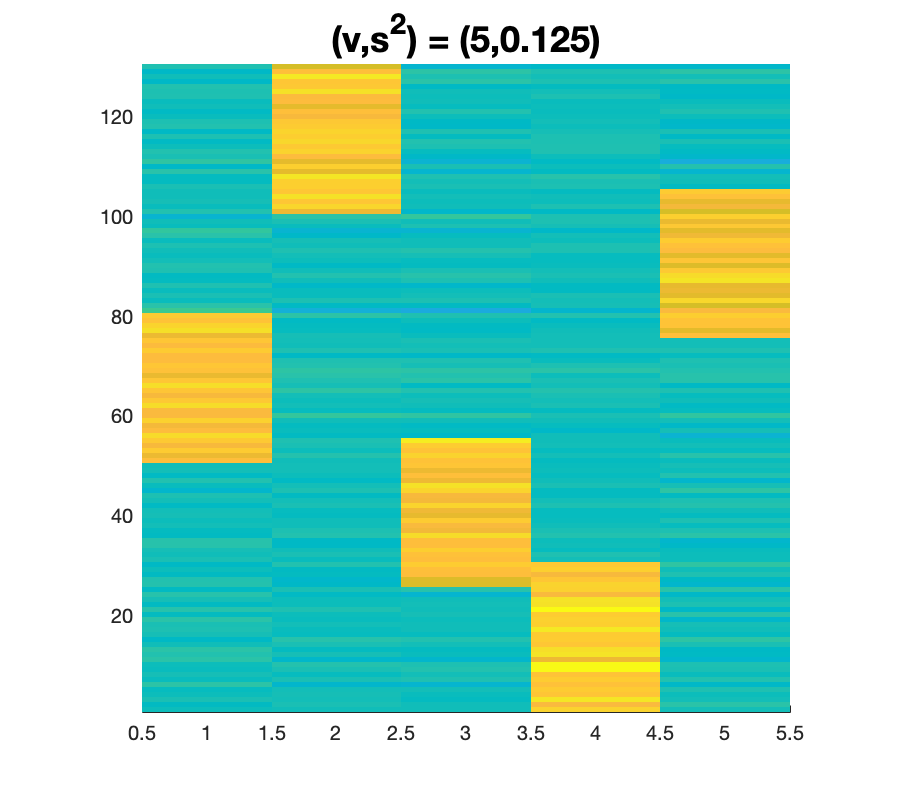}
\includegraphics[width=0.24\textwidth]{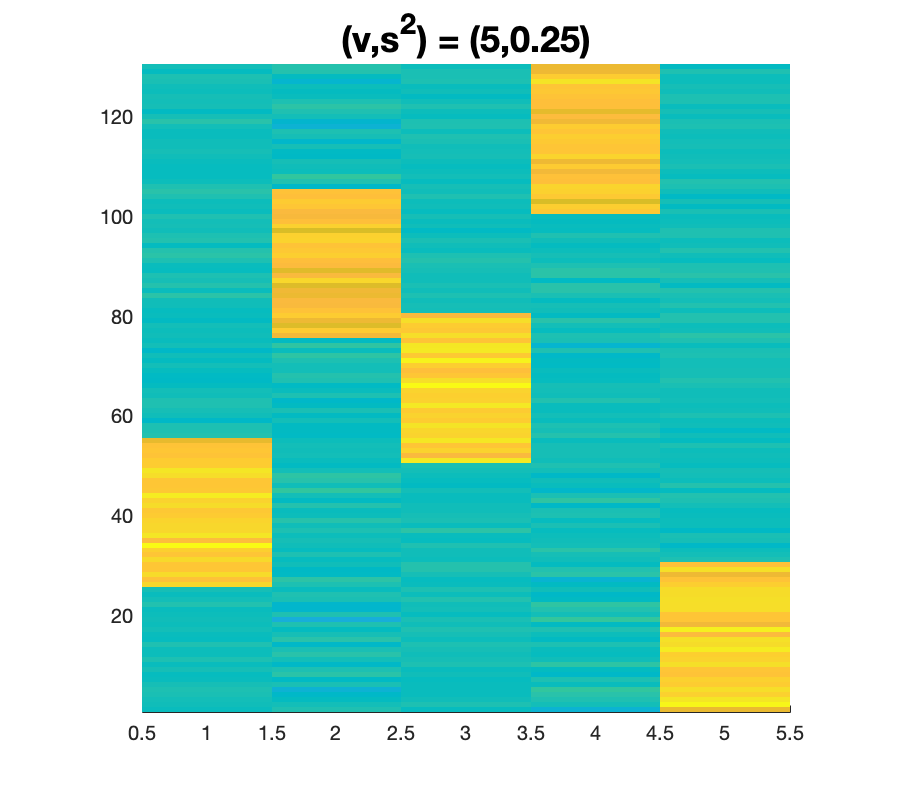}
\includegraphics[width=0.24\textwidth]{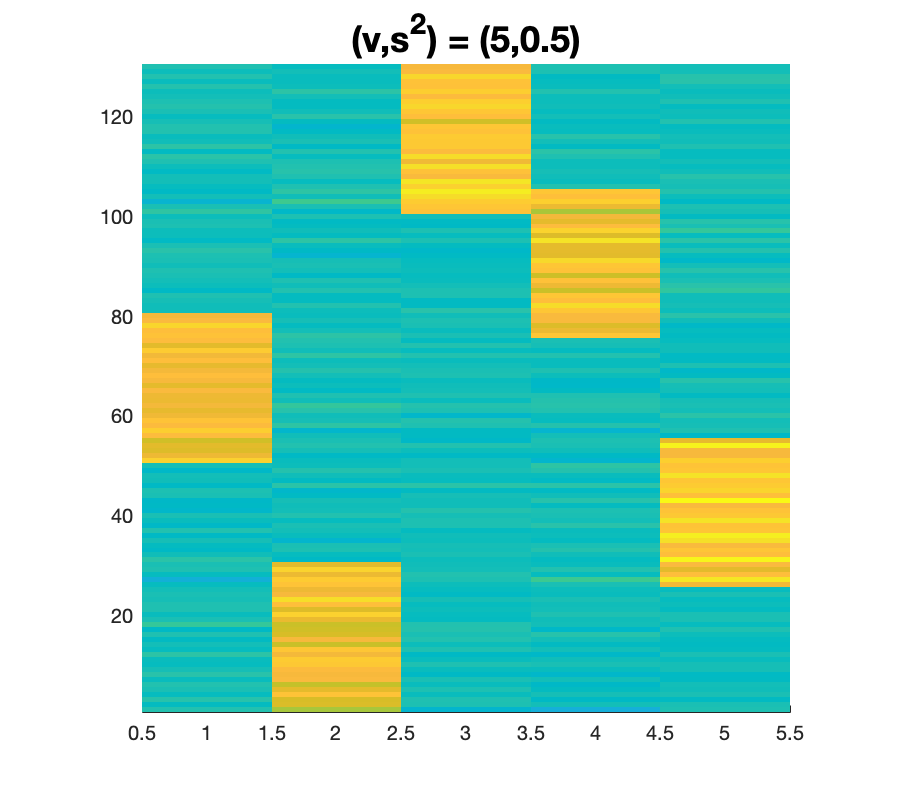}
\includegraphics[width=0.24\textwidth]{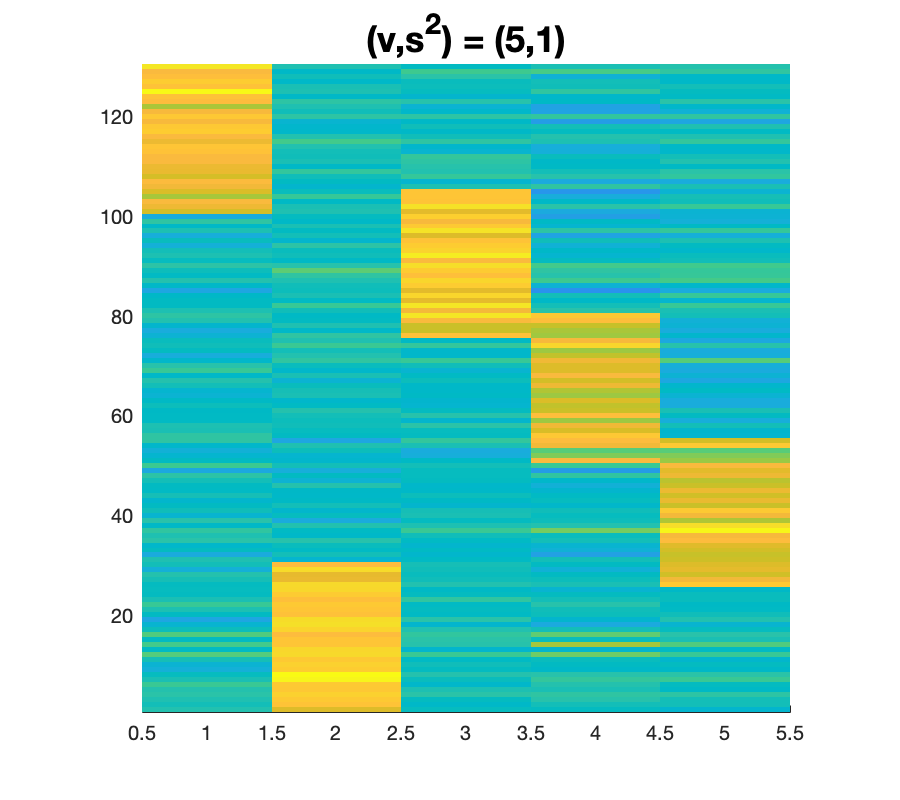}
\caption{Estimated factor loading matrices by the Robust Factor Model for the simulation with the Student's t-distribution in Section \ref{sec5}.}
\label{fig_5_2_2}
\end{figure}

\begin{figure}
\renewcommand{\baselinestretch}{1}
\centering
\includegraphics[width=0.24\textwidth]{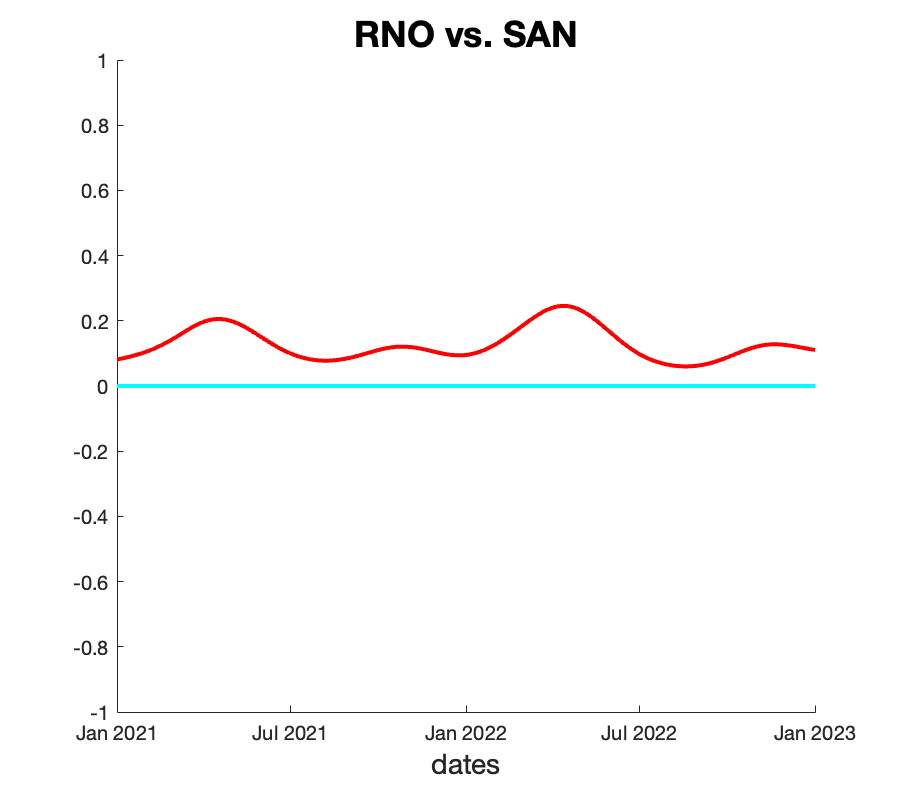}
\includegraphics[width=0.24\textwidth]{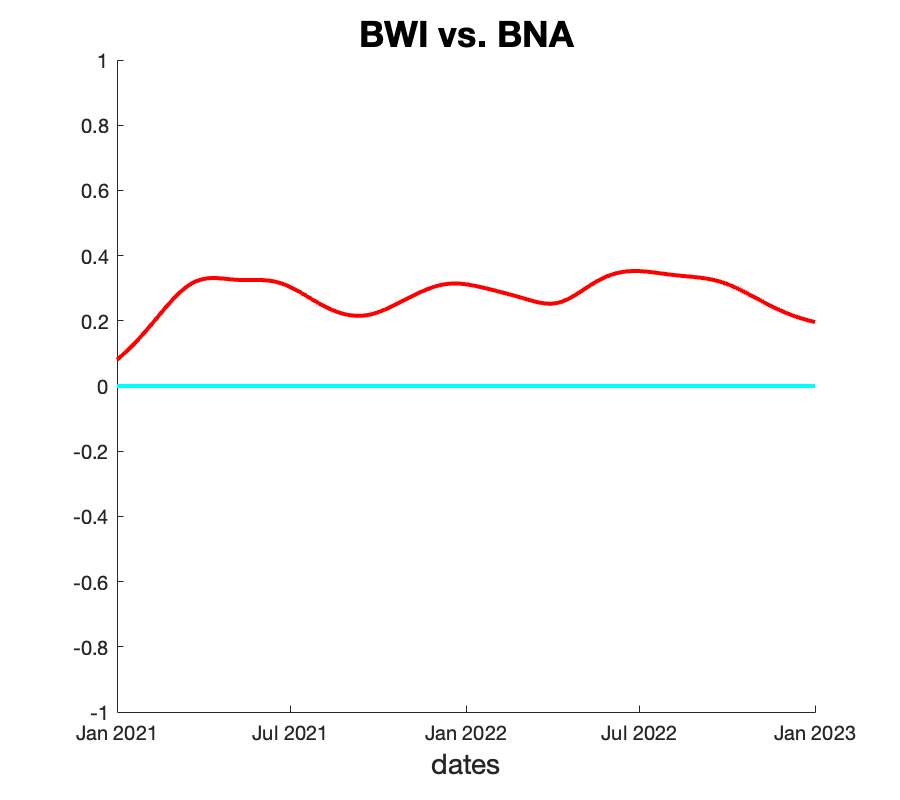}
\includegraphics[width=0.24\textwidth]{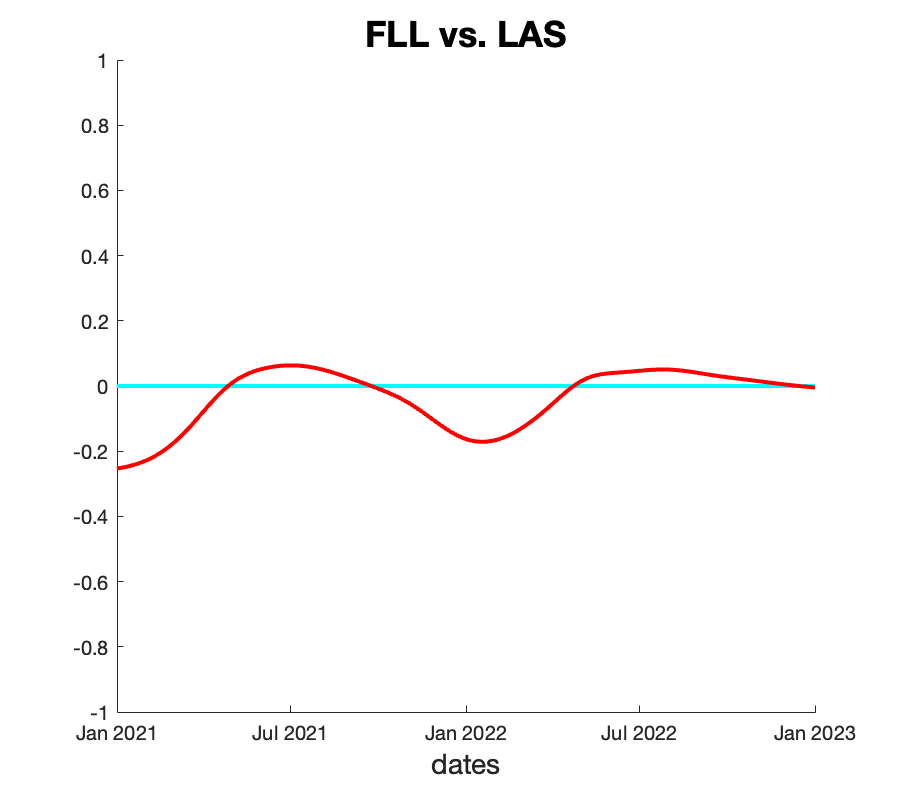}
\includegraphics[width=0.24\textwidth]{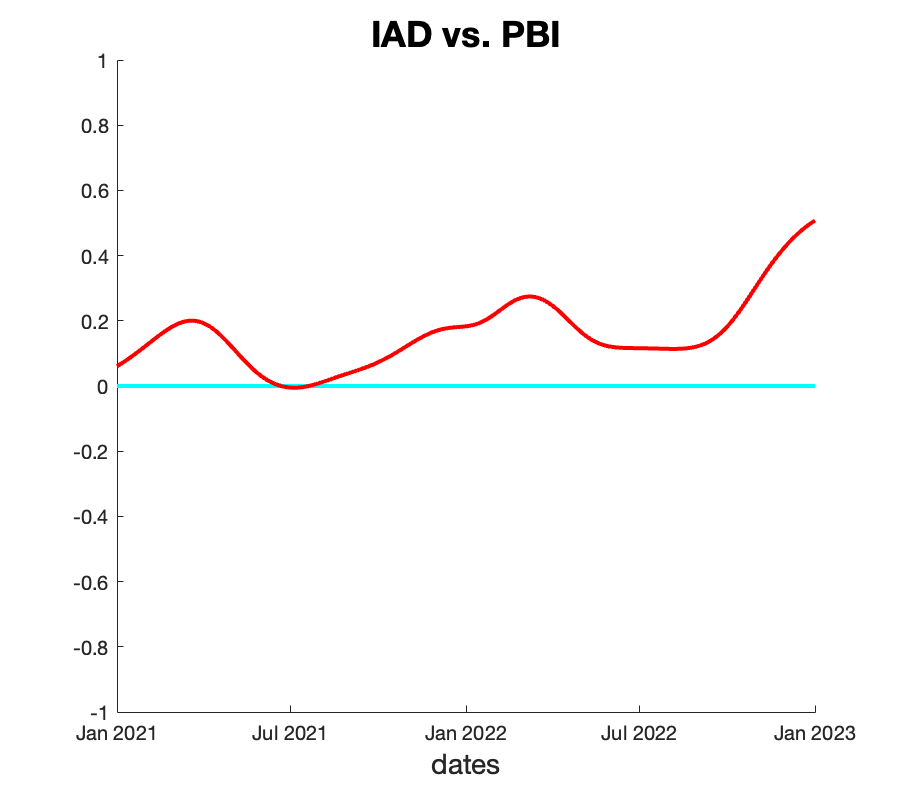}
\includegraphics[width=0.24\textwidth]{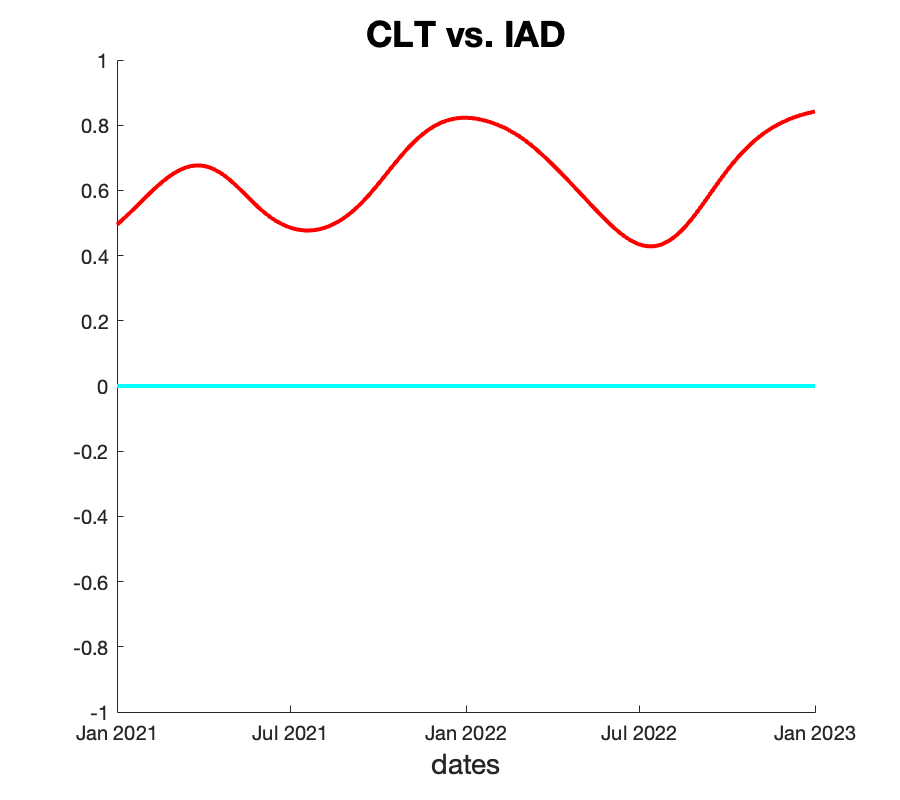}
\includegraphics[width=0.24\textwidth]{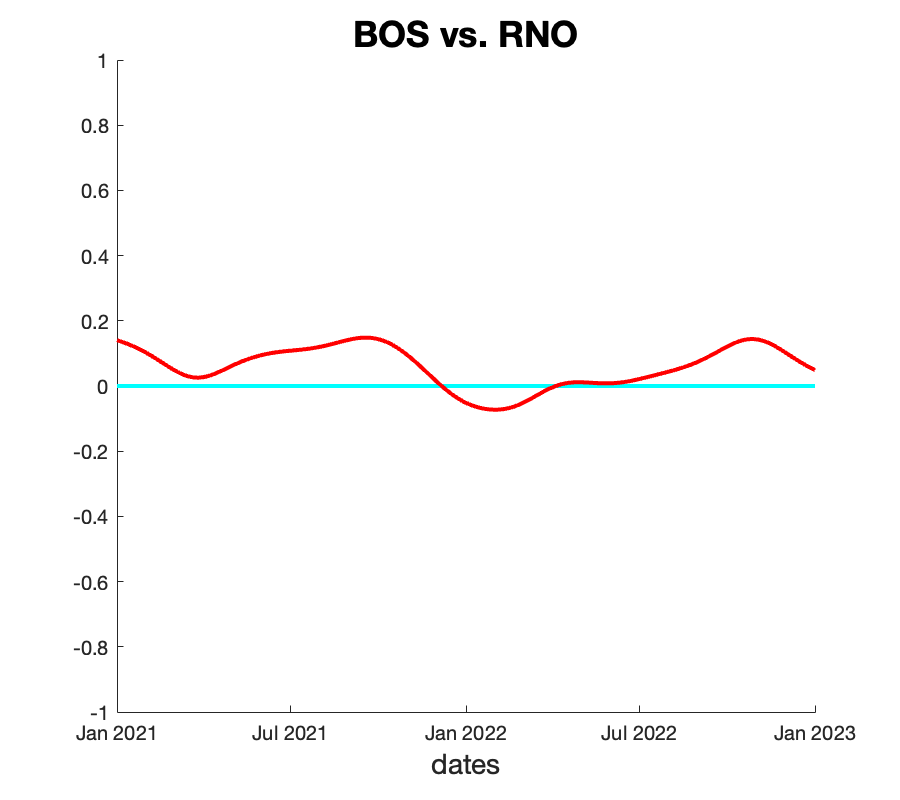}
\includegraphics[width=0.24\textwidth]{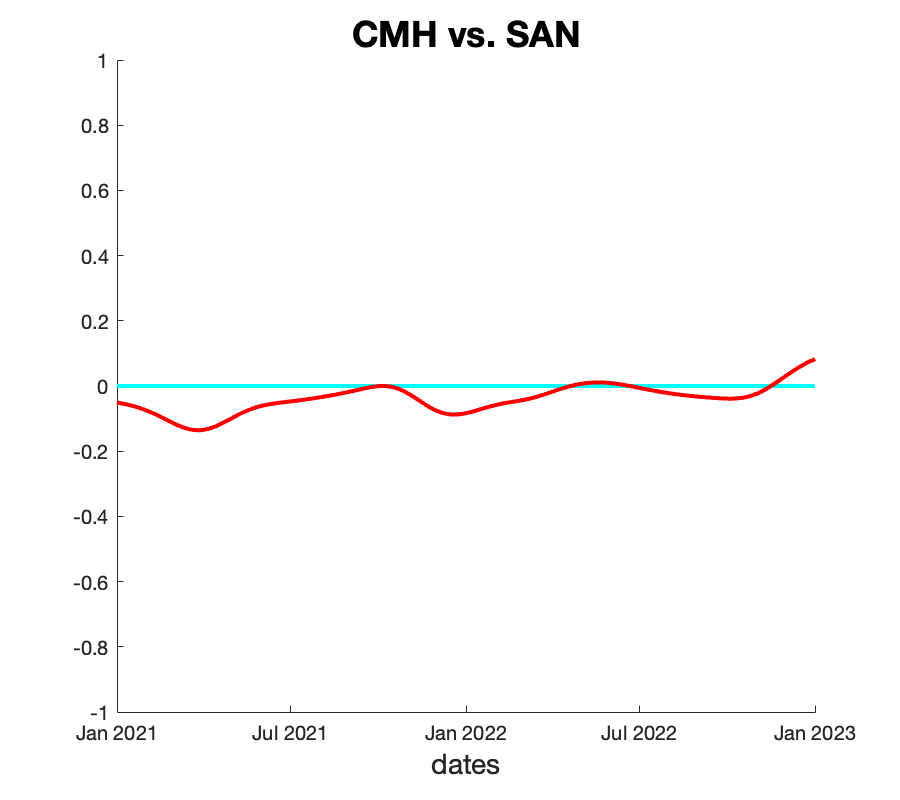}
\includegraphics[width=0.24\textwidth]{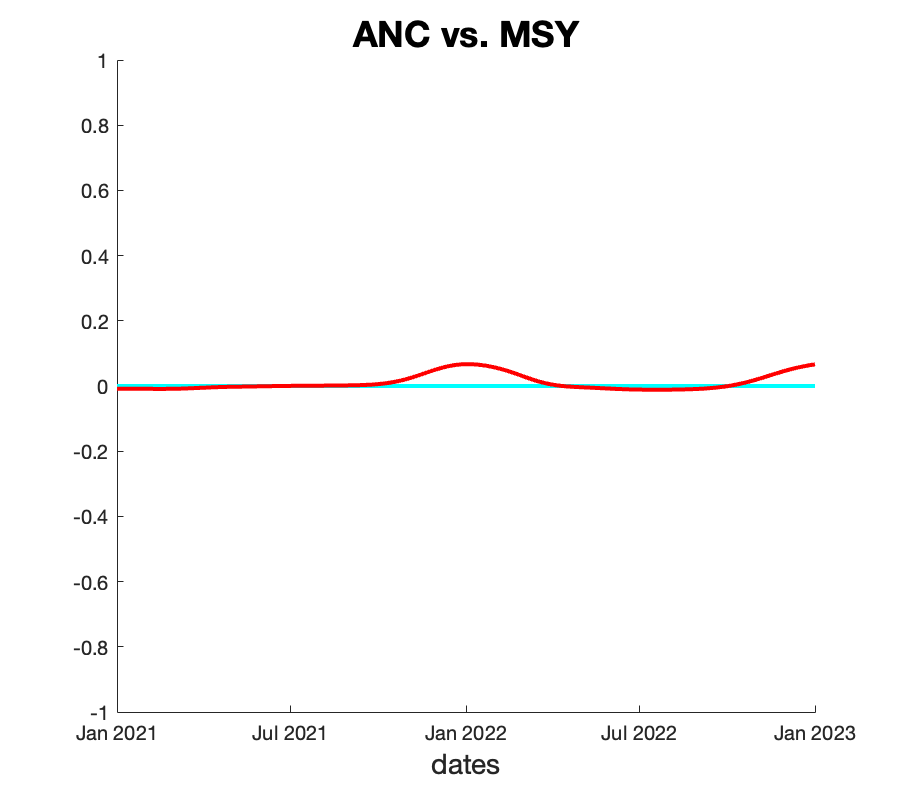}
\includegraphics[width=0.24\textwidth]{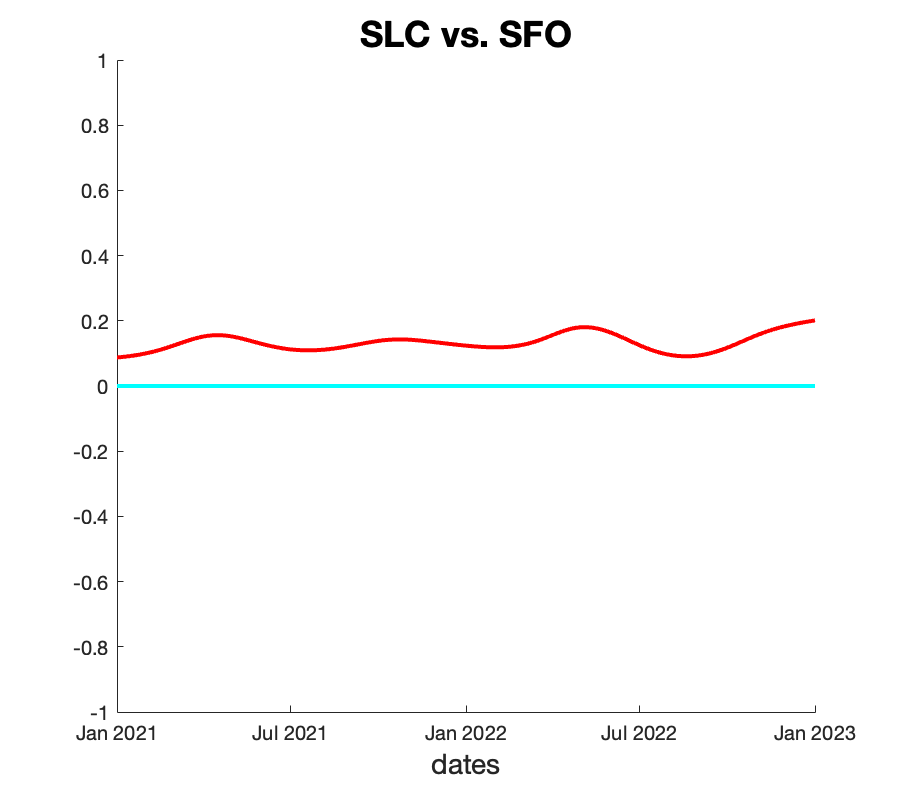}
\includegraphics[width=0.24\textwidth]{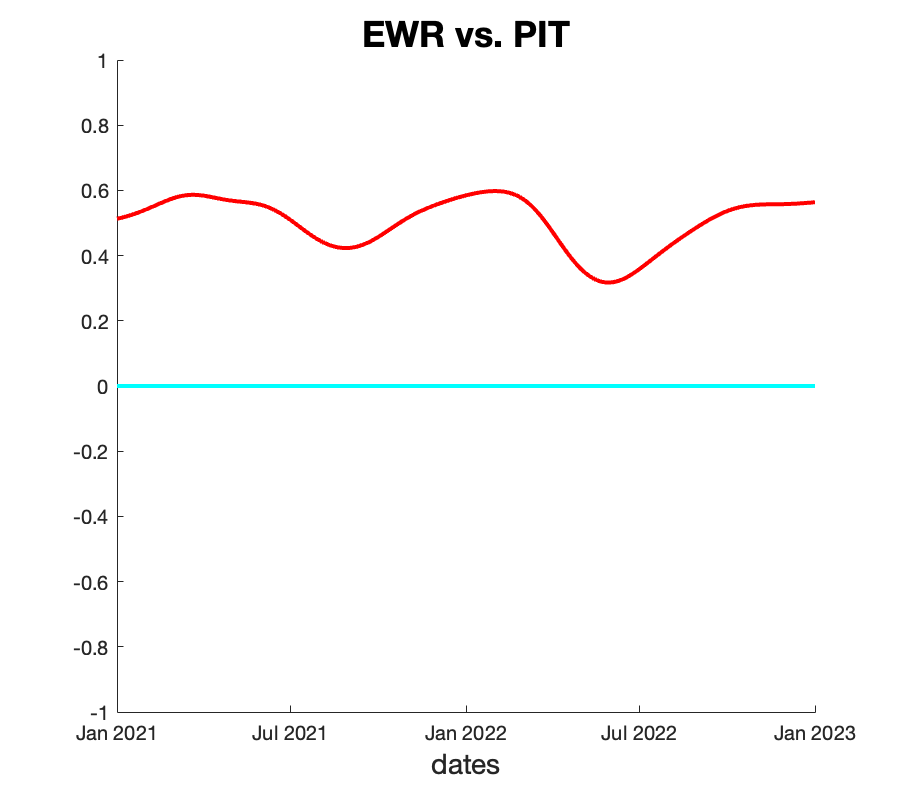}
\includegraphics[width=0.24\textwidth]{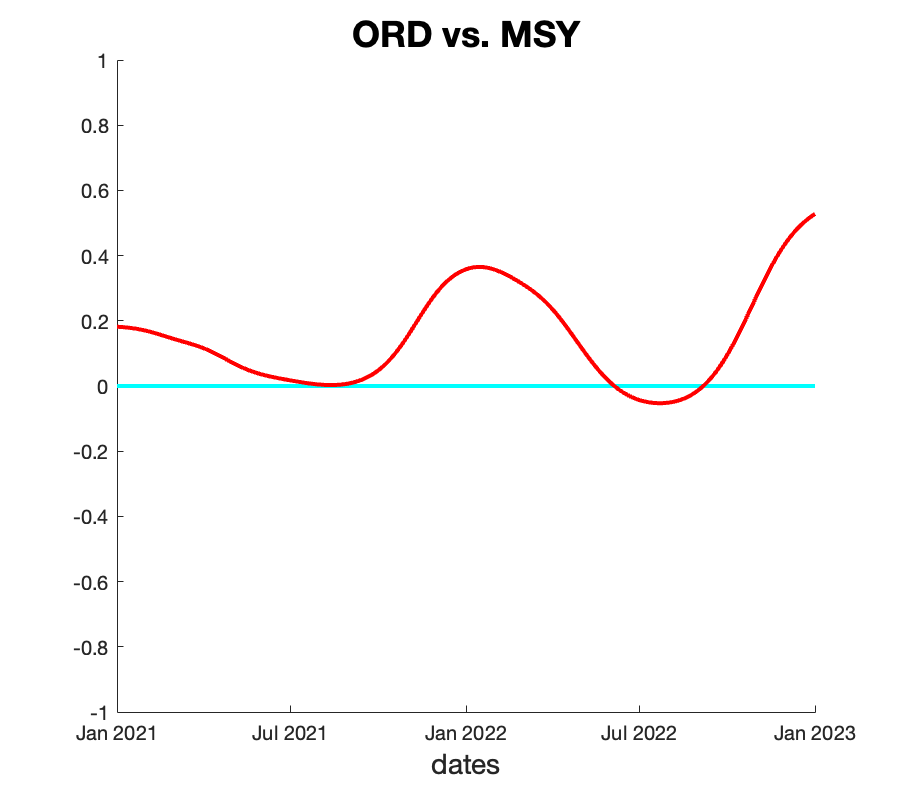}
\includegraphics[width=0.24\textwidth]{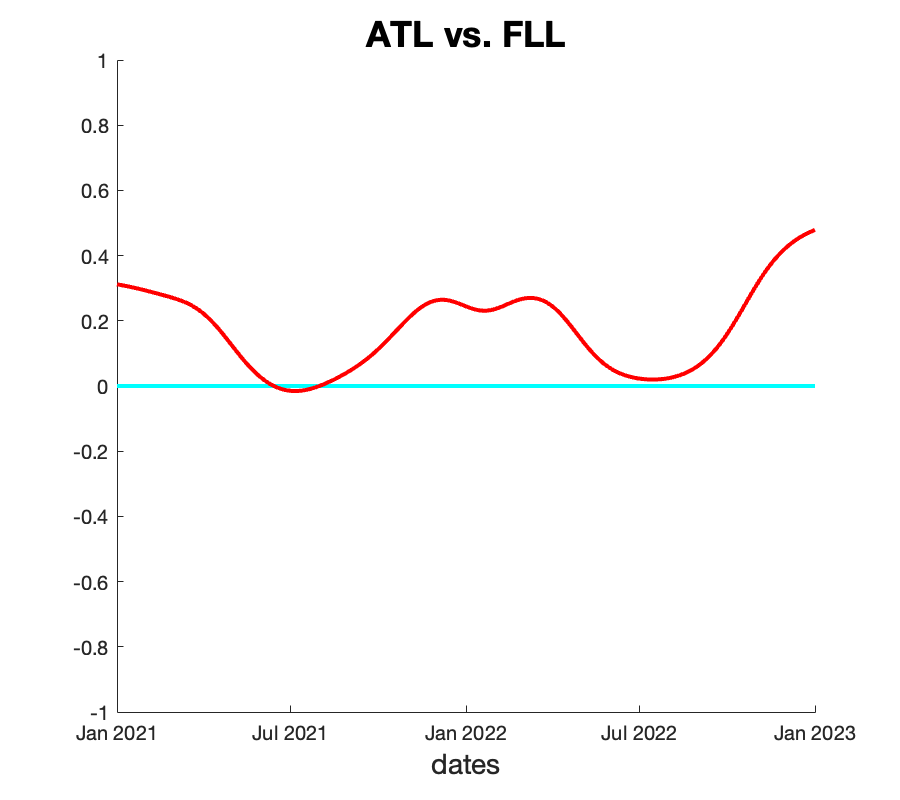}
\includegraphics[width=0.24\textwidth]{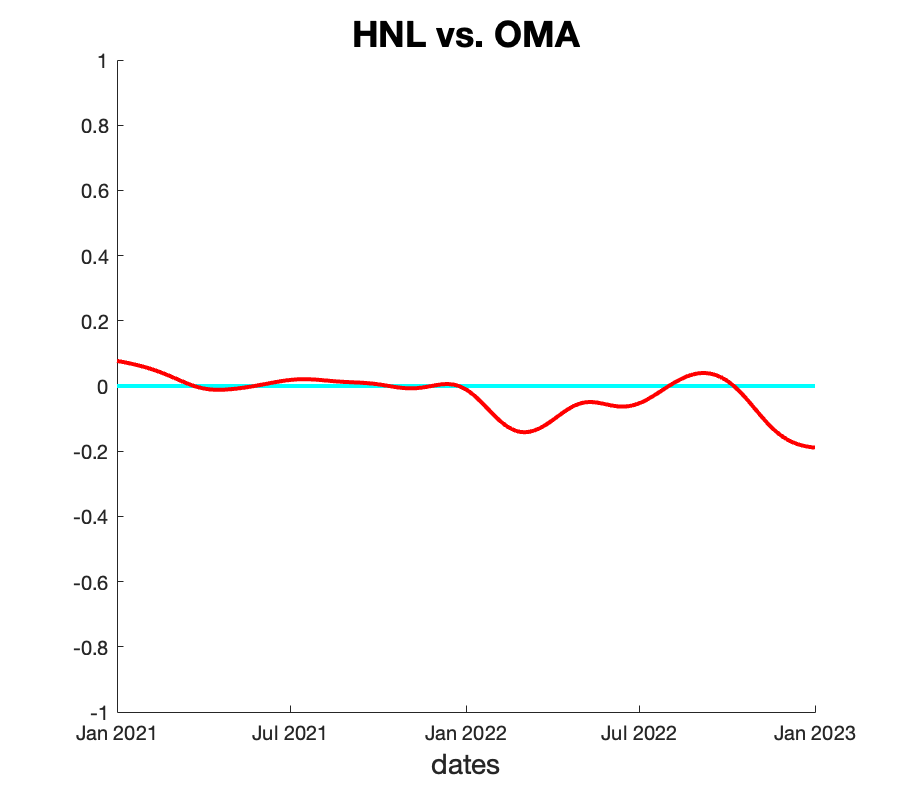}
\includegraphics[width=0.24\textwidth]{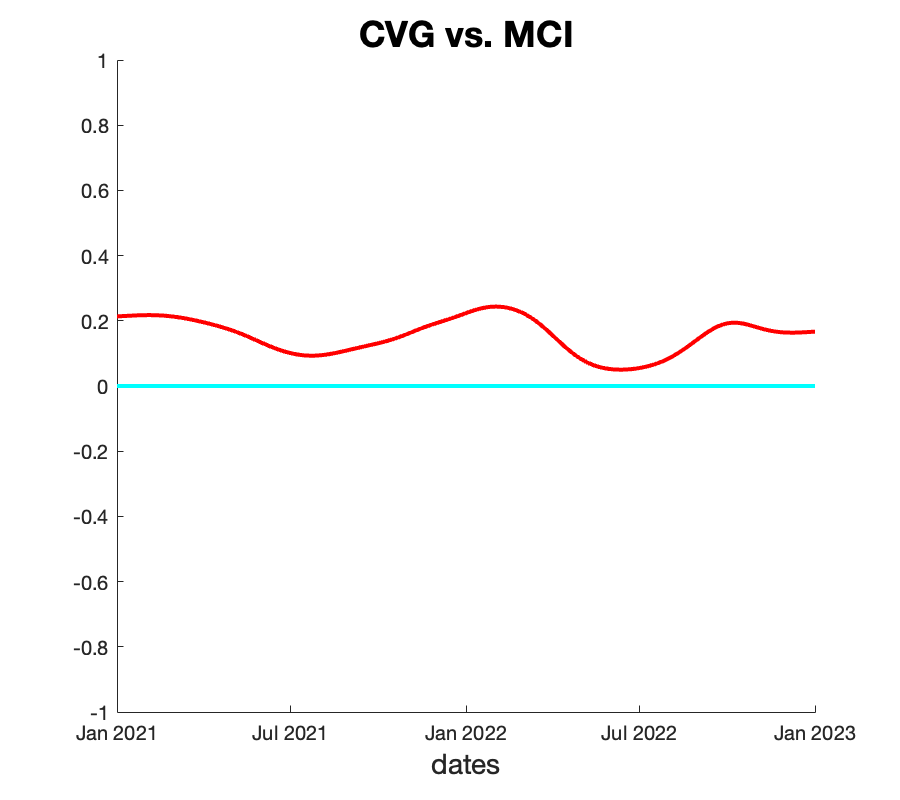}
\includegraphics[width=0.24\textwidth]{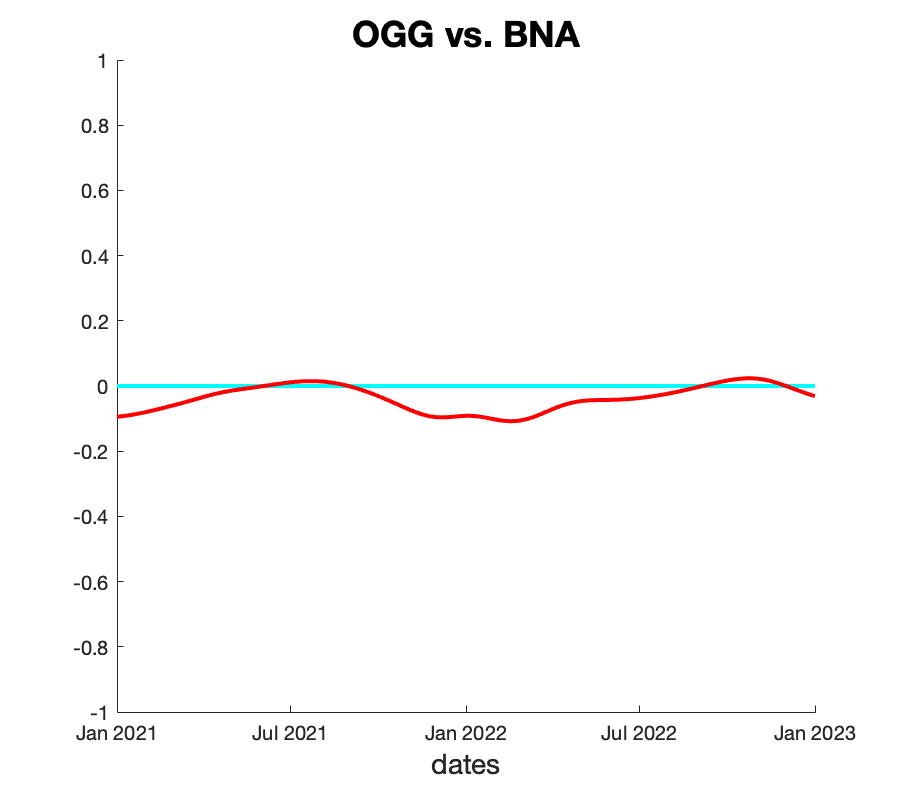}
\includegraphics[width=0.24\textwidth]{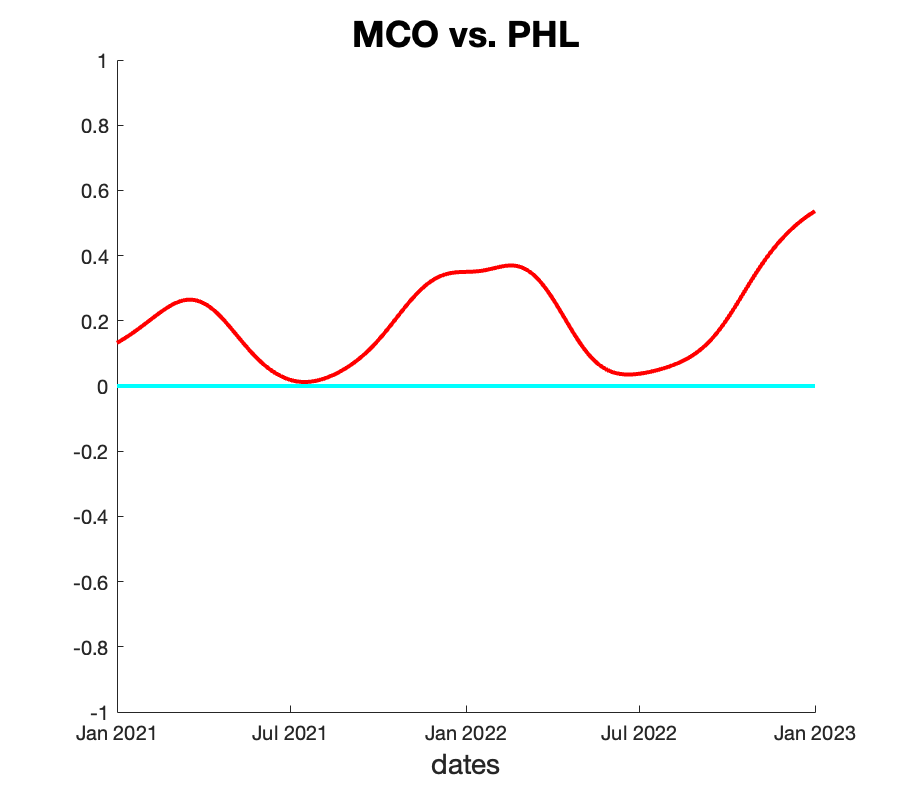}
\caption{Some examples of the estimate correlation between two randomly chosen airports over dates.}
\label{fig_6_2_1}
\end{figure}

\begin{figure}
\renewcommand{\baselinestretch}{1}
\centering
\includegraphics[width=0.24\textwidth]{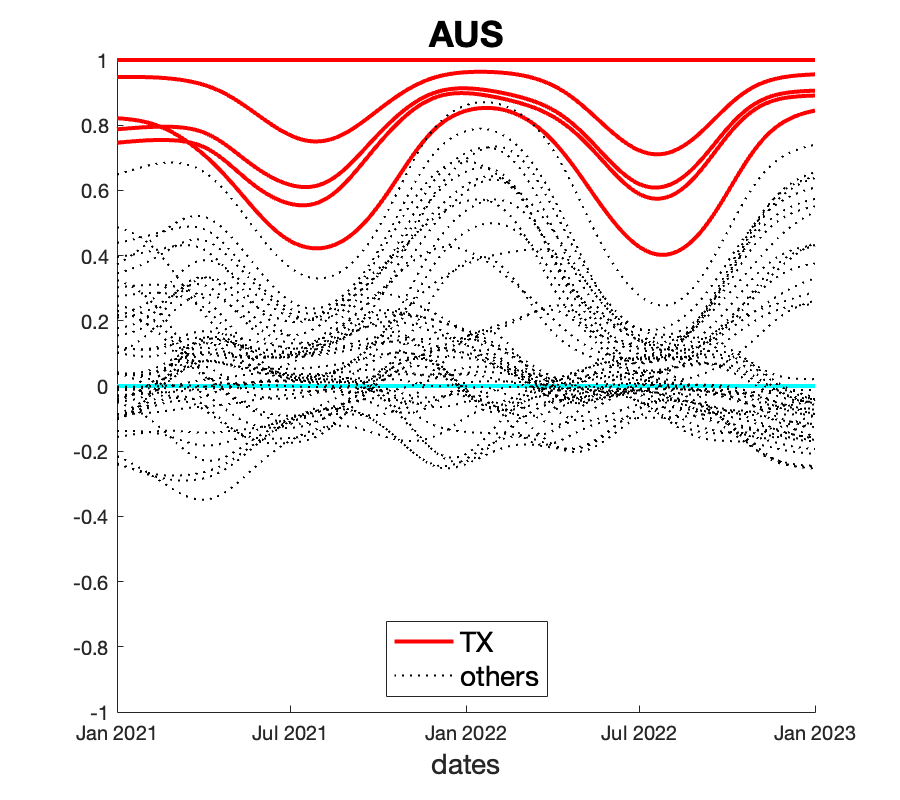}
\includegraphics[width=0.24\textwidth]{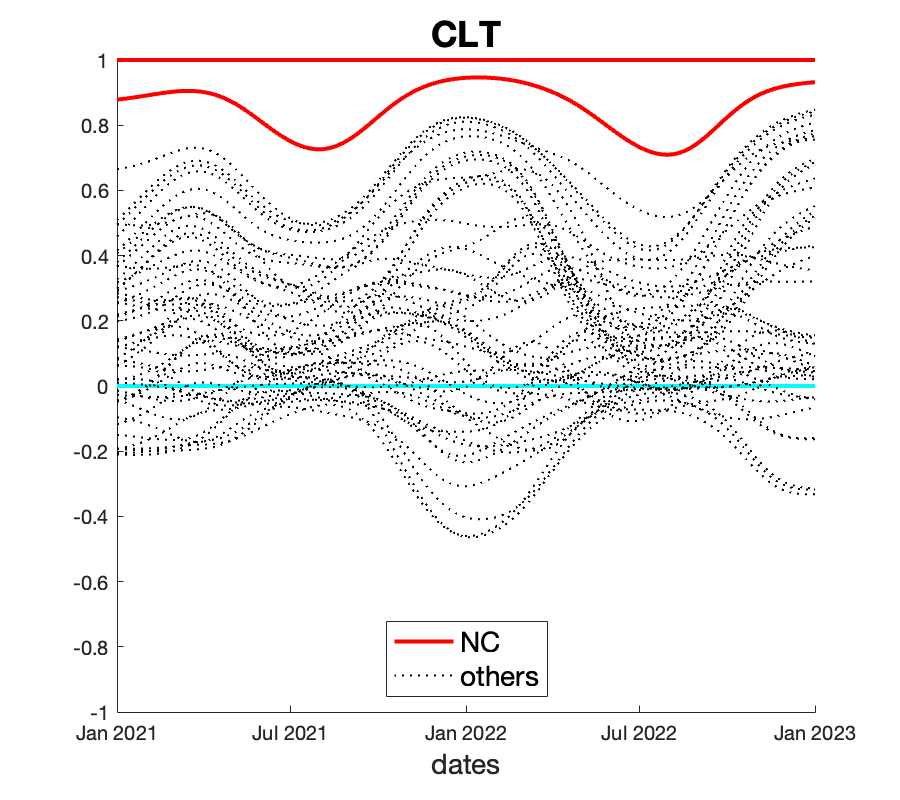}
\includegraphics[width=0.24\textwidth]{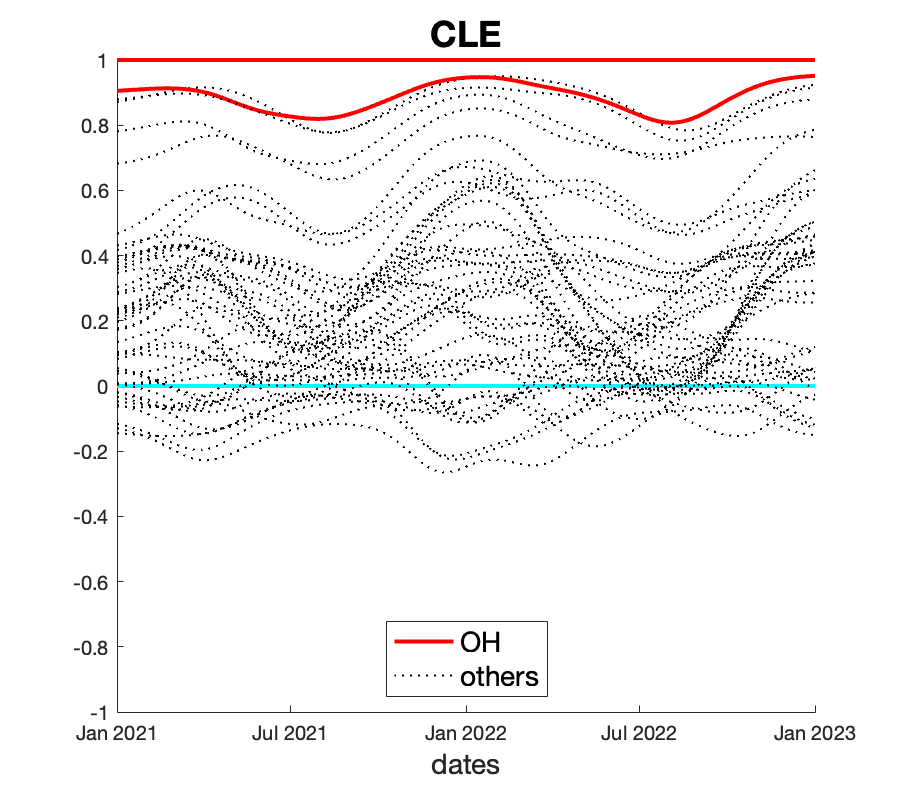}
\includegraphics[width=0.24\textwidth]{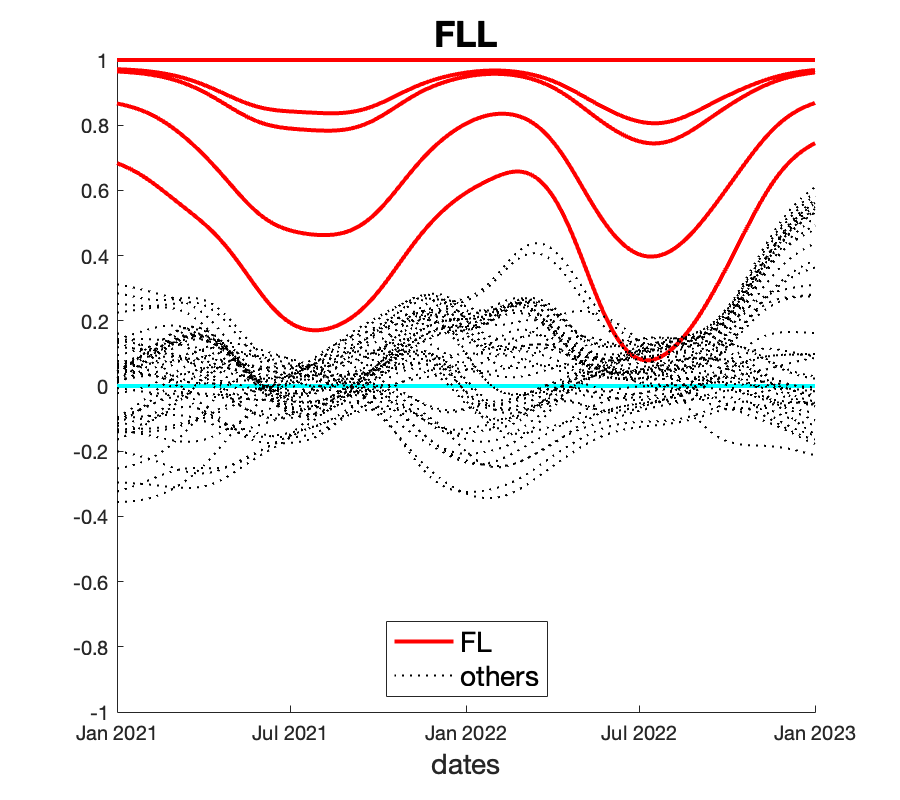}
\includegraphics[width=0.24\textwidth]{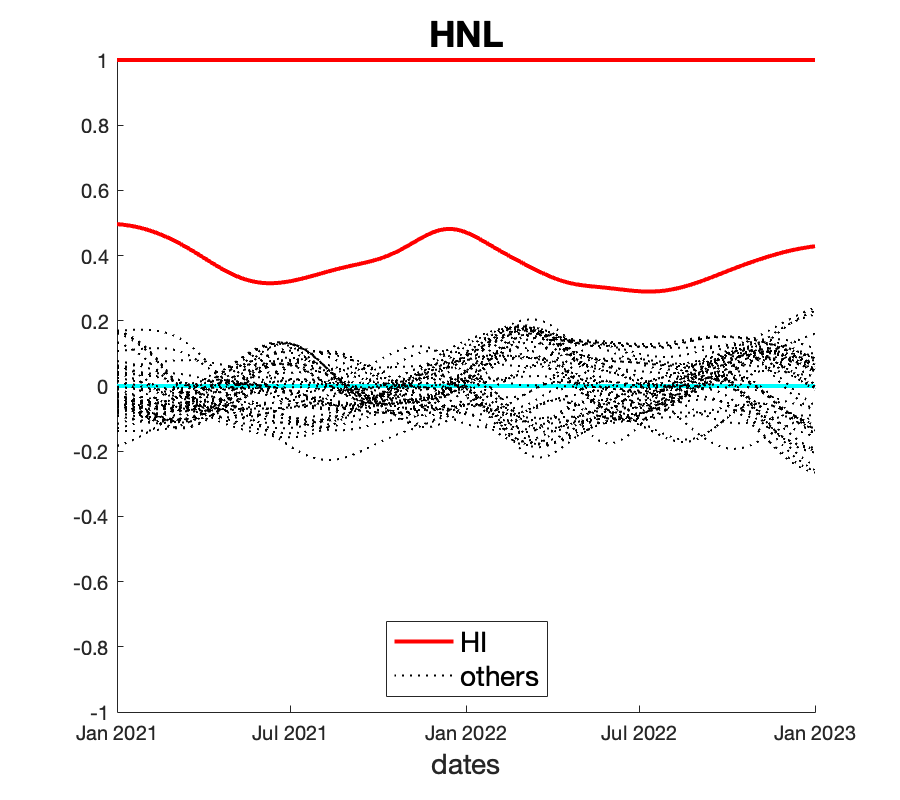}
\includegraphics[width=0.24\textwidth]{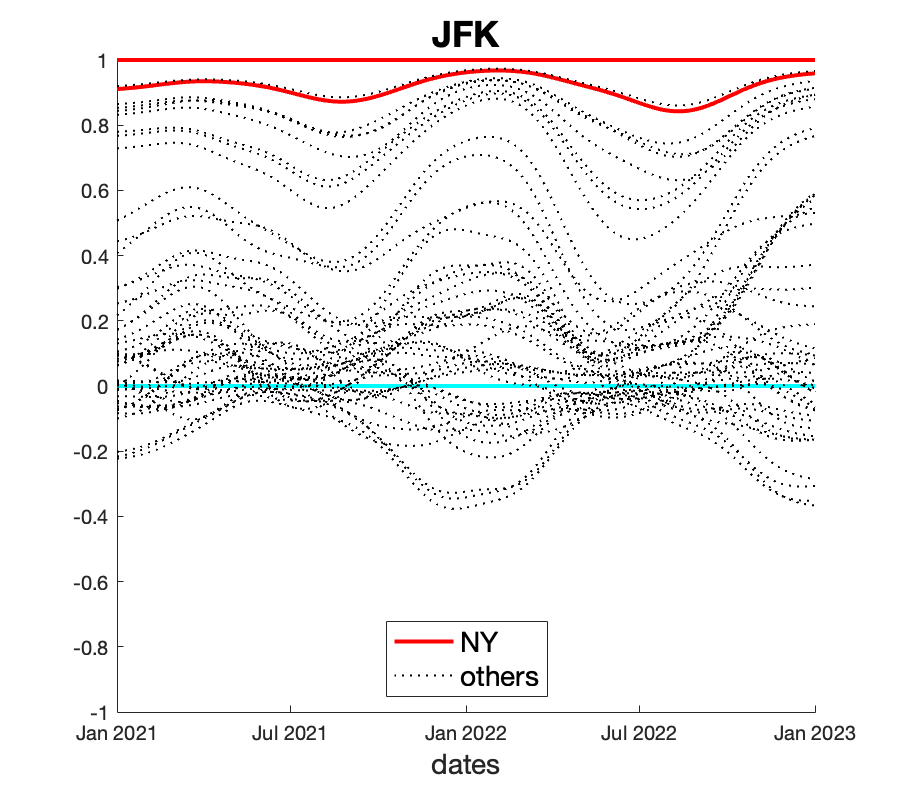}
\includegraphics[width=0.24\textwidth]{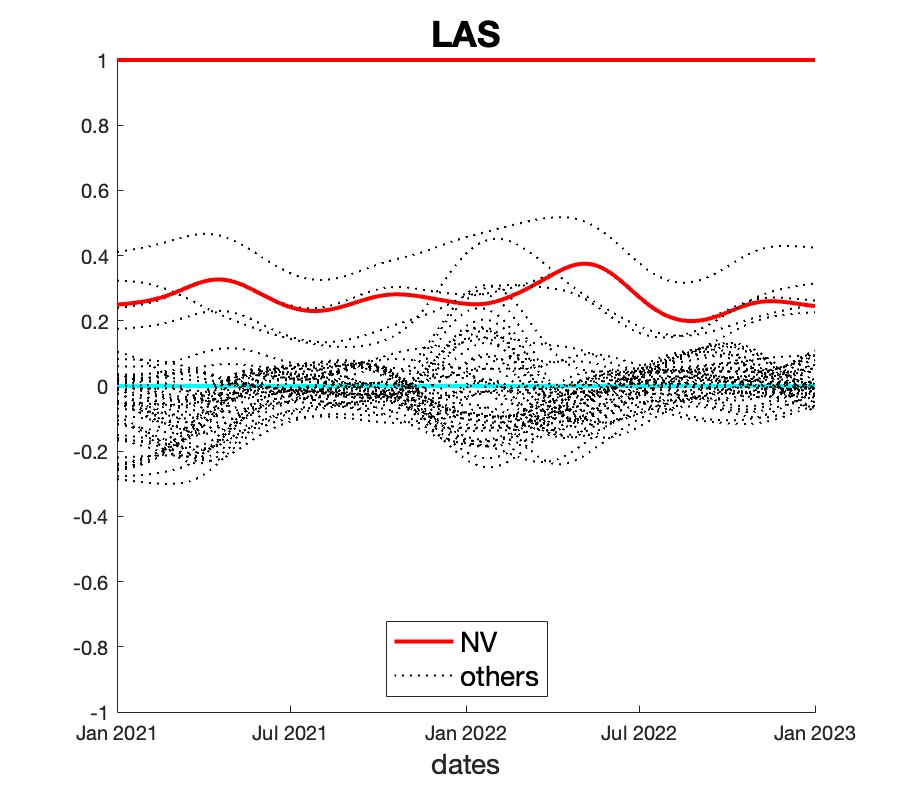}
\includegraphics[width=0.24\textwidth]{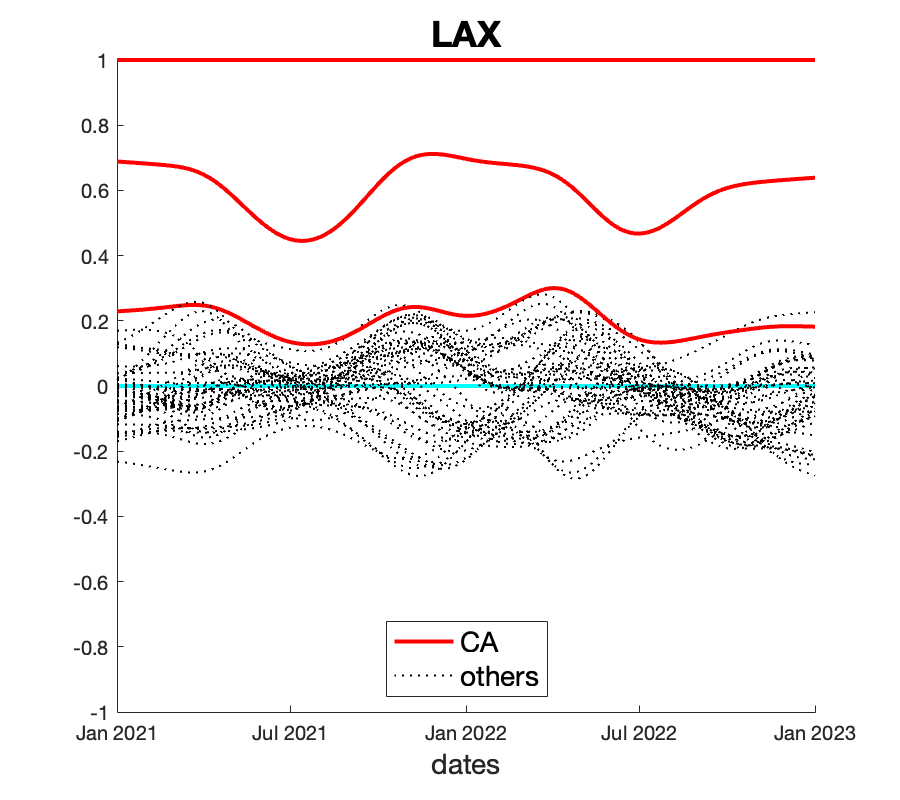}
\includegraphics[width=0.24\textwidth]{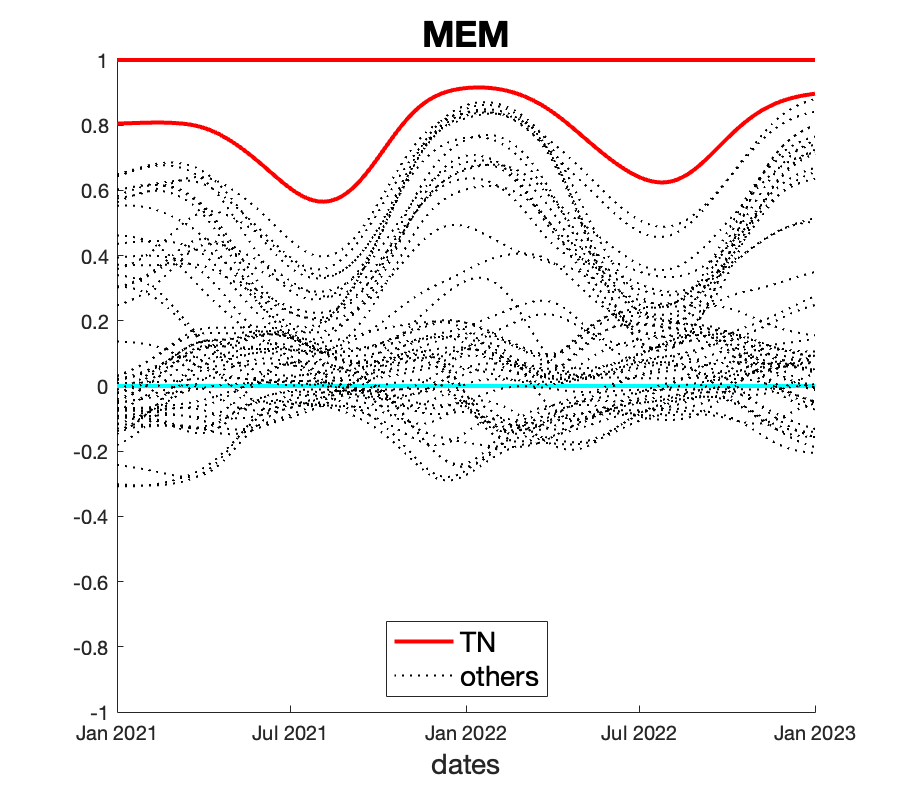}
\includegraphics[width=0.24\textwidth]{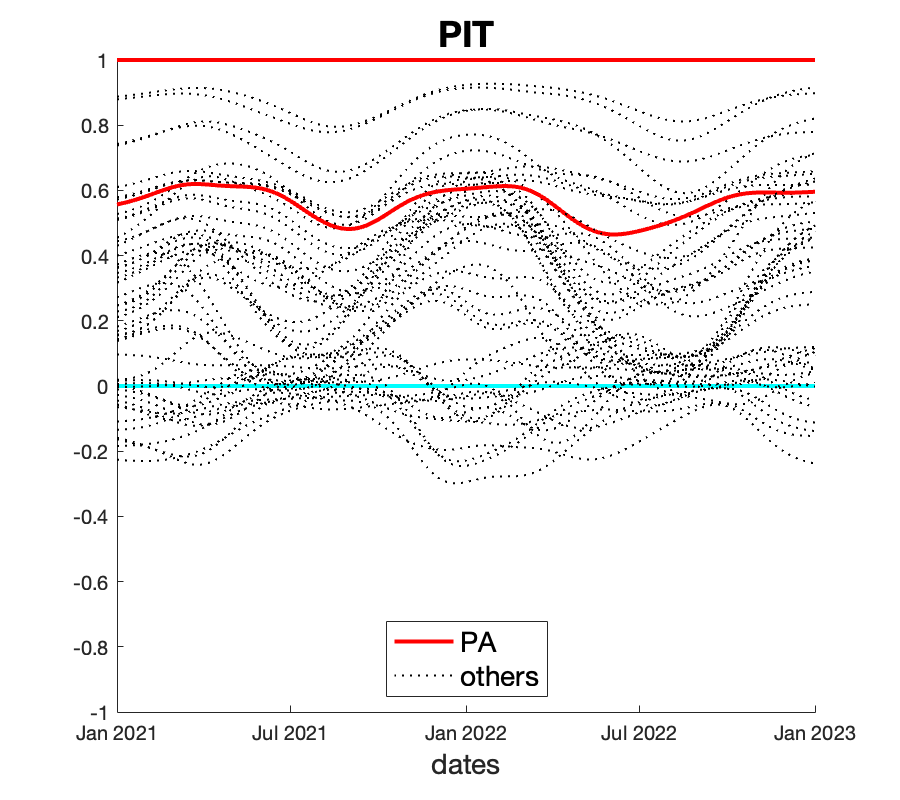}
\includegraphics[width=0.24\textwidth]{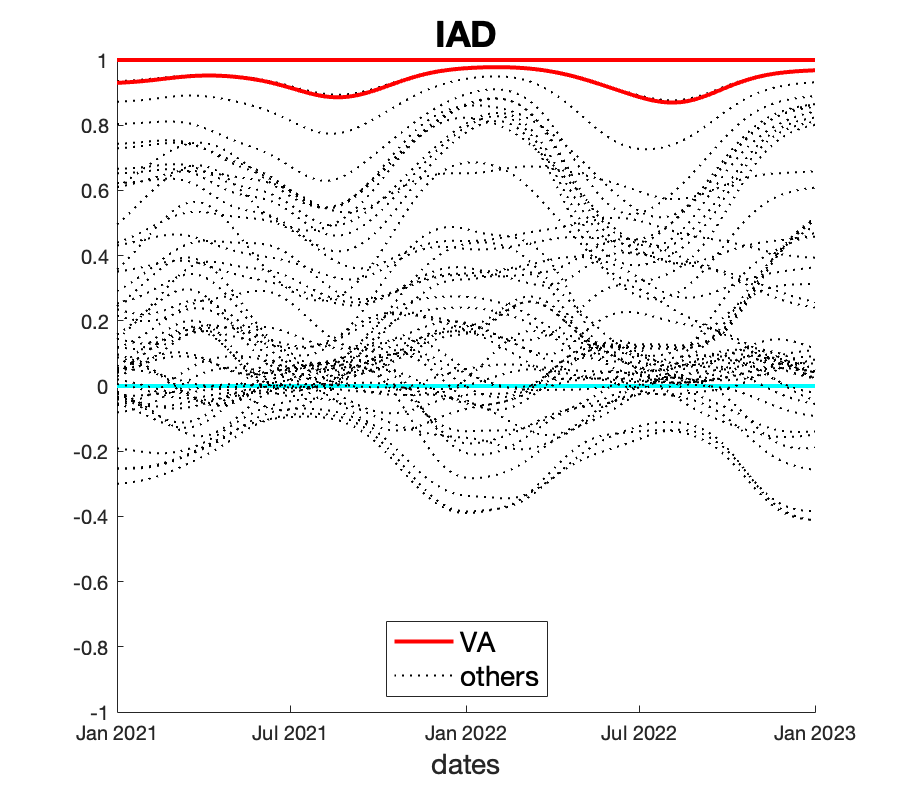}
\includegraphics[width=0.24\textwidth]{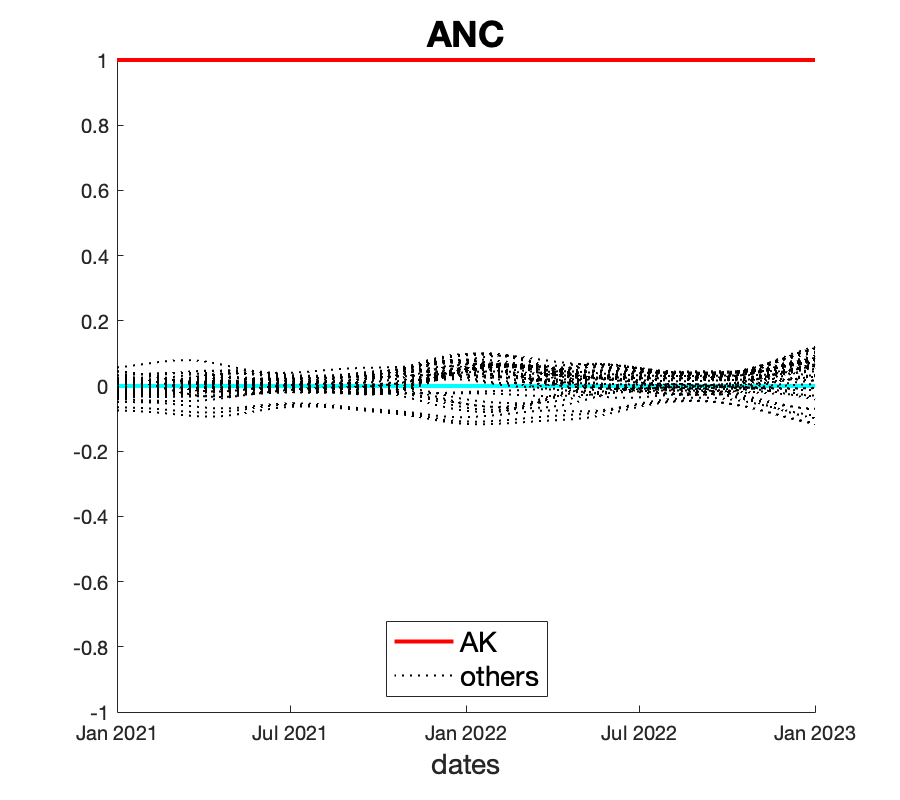}
\caption{Some examples of the estimate correlation between an airport and all the others over dates. Red curves indicate the correlation curves for those in the same state.}
\label{fig_6_2_2}
\end{figure}

\begin{figure}
\renewcommand{\baselinestretch}{1}
\centering
\includegraphics[width=1.0\textwidth]{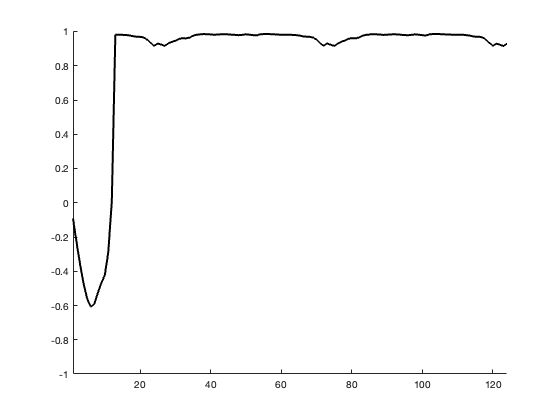}
\caption{The running correlation (with the window size of 15) between ETR and ETRN for Section \ref{sec6_3} in the main manuscript.}
\label{fig_6_2_3}
\end{figure}

\begin{figure}
\renewcommand{\baselinestretch}{1}
\centering
\includegraphics[width=0.24\textwidth]{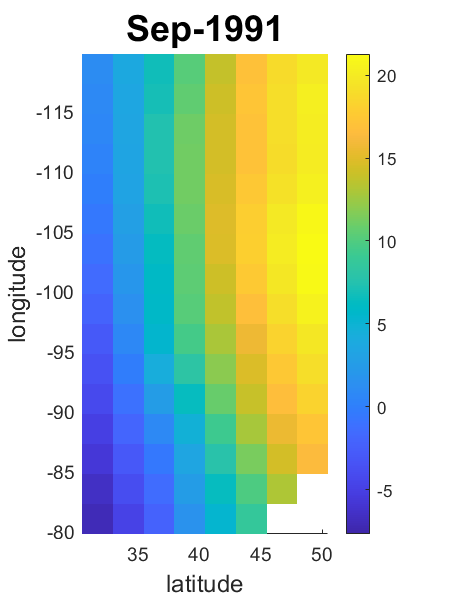}
\includegraphics[width=0.24\textwidth]{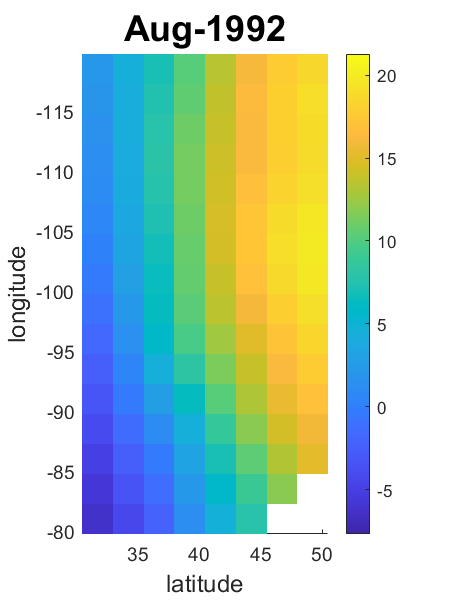}
\includegraphics[width=0.24\textwidth]{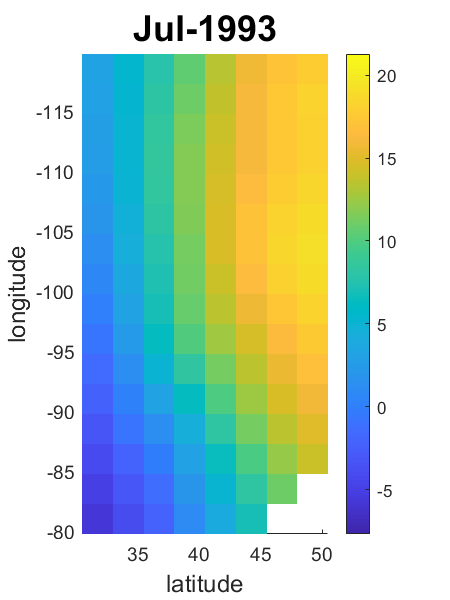}
\includegraphics[width=0.24\textwidth]{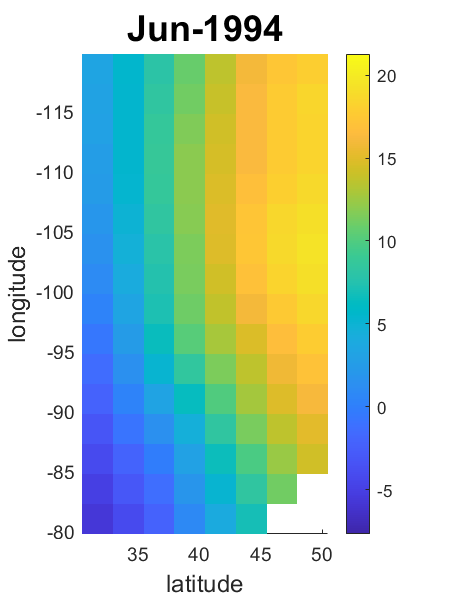}
\includegraphics[width=0.24\textwidth]{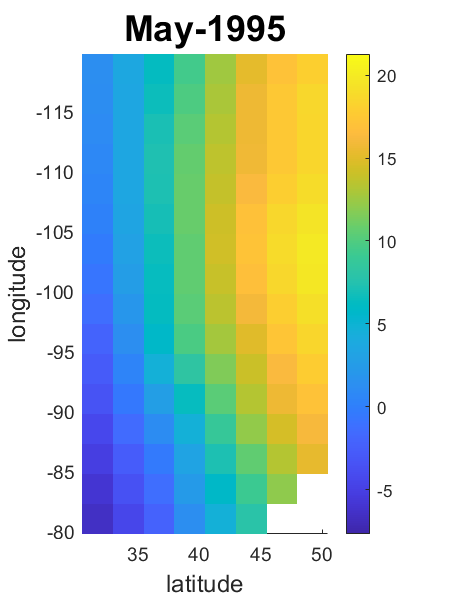}
\includegraphics[width=0.24\textwidth]{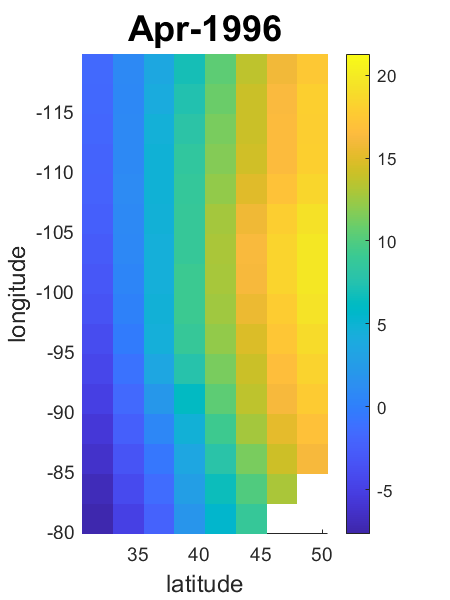}
\includegraphics[width=0.24\textwidth]{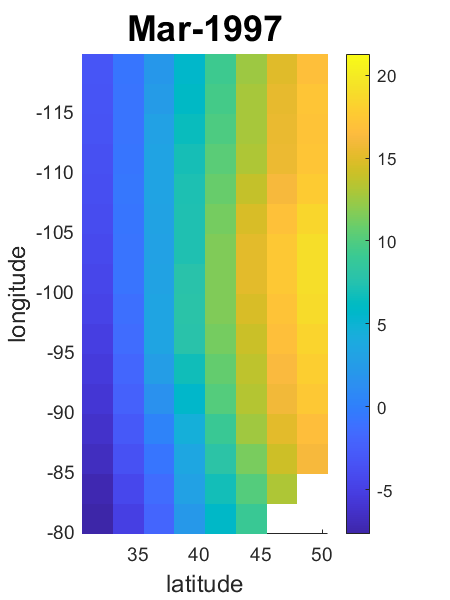}
\includegraphics[width=0.24\textwidth]{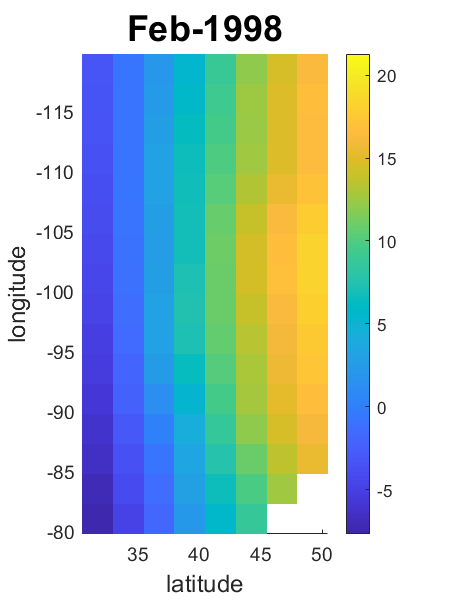}
\includegraphics[width=0.24\textwidth]{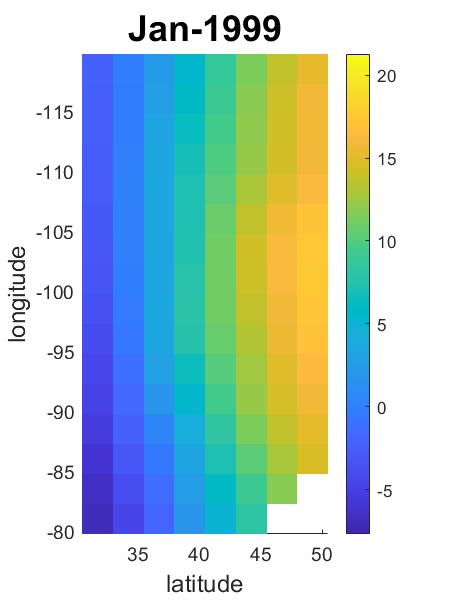}
\includegraphics[width=0.24\textwidth]{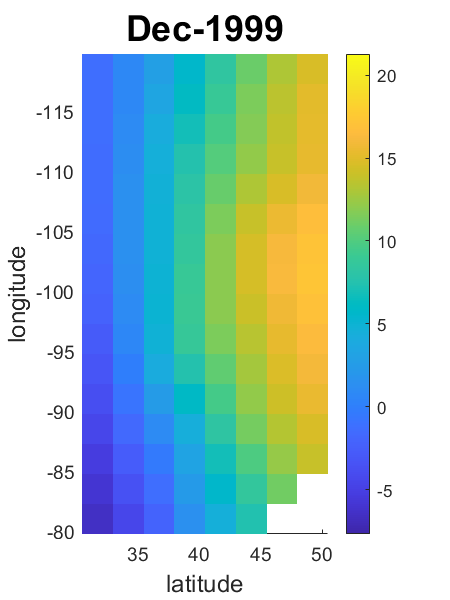}
\includegraphics[width=0.24\textwidth]{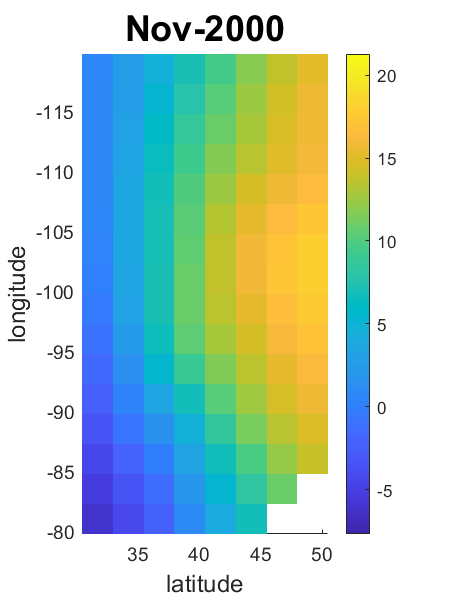}
\includegraphics[width=0.24\textwidth]{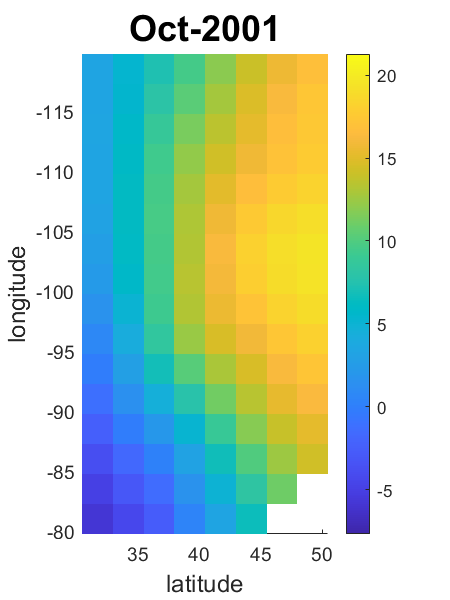}
\caption{The second spatial factor loadings over locations and time in Section \ref{sec6_3}.}
\label{fig_6_3_1_2}
\end{figure}

\begin{figure}
\renewcommand{\baselinestretch}{1}
\centering
\includegraphics[width=0.24\textwidth]{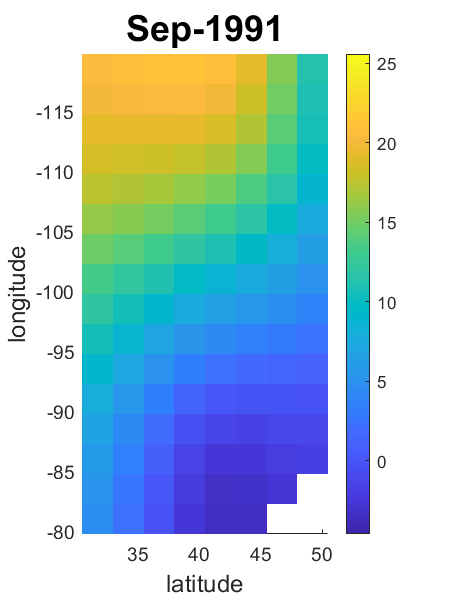}
\includegraphics[width=0.24\textwidth]{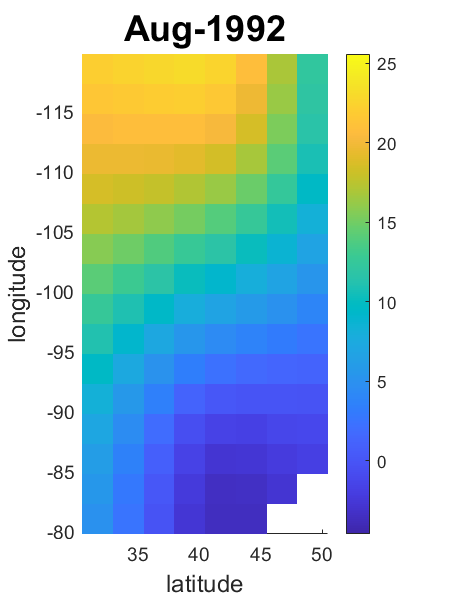}
\includegraphics[width=0.24\textwidth]{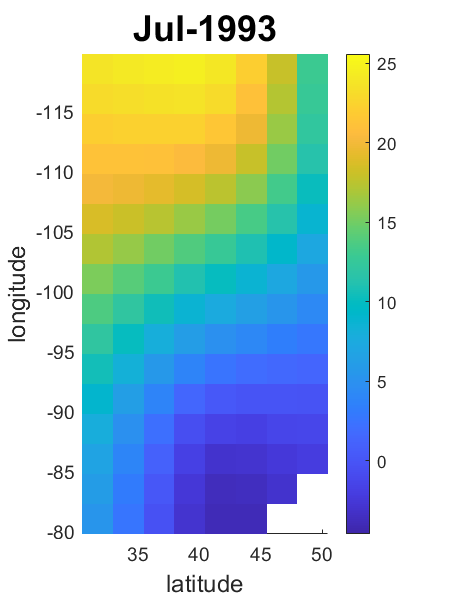}
\includegraphics[width=0.24\textwidth]{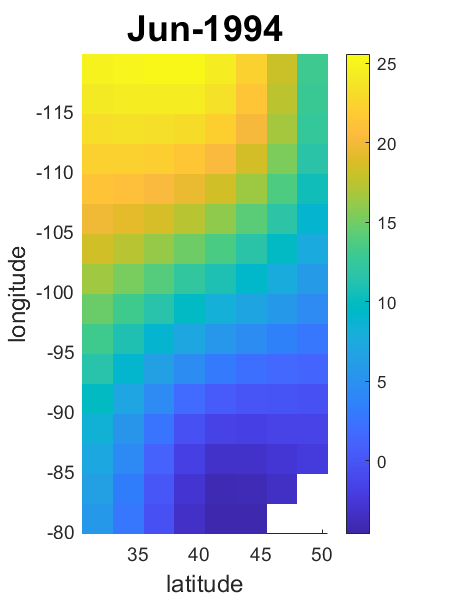}
\includegraphics[width=0.24\textwidth]{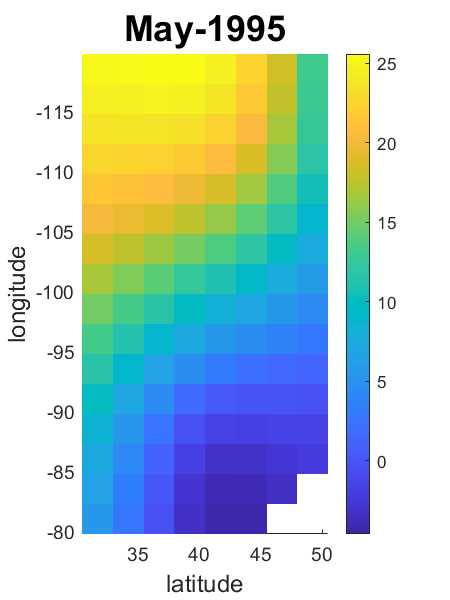}
\includegraphics[width=0.24\textwidth]{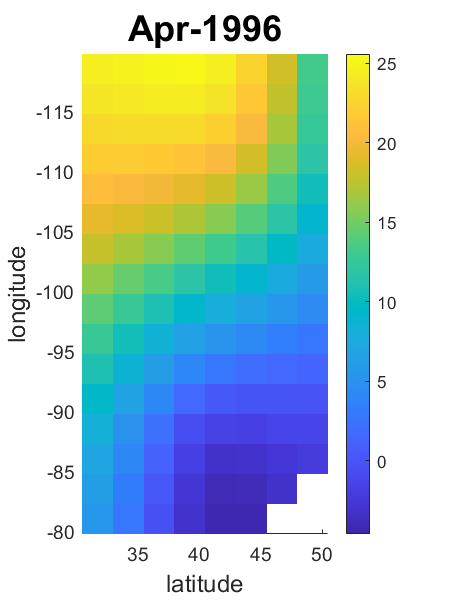}
\includegraphics[width=0.24\textwidth]{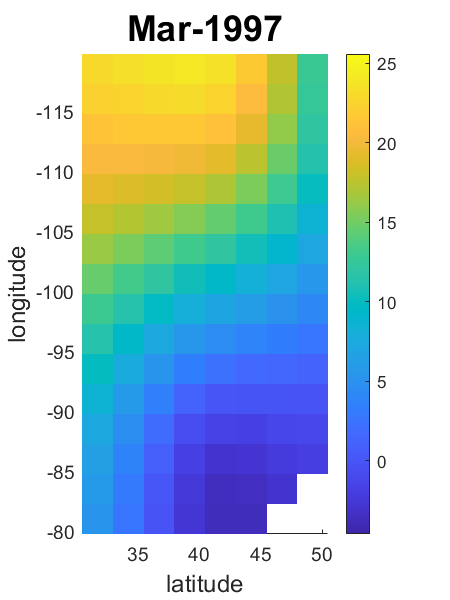}
\includegraphics[width=0.24\textwidth]{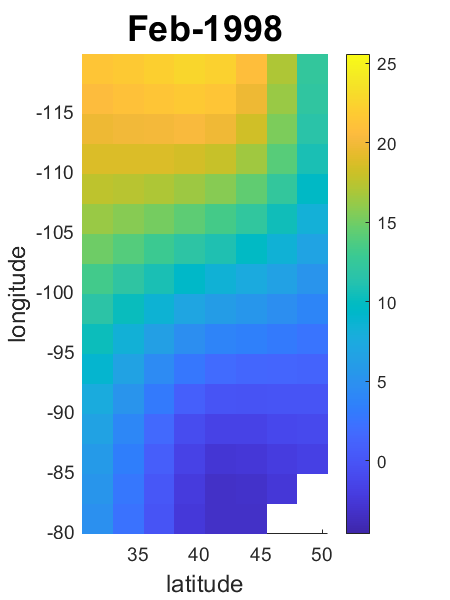}
\includegraphics[width=0.24\textwidth]{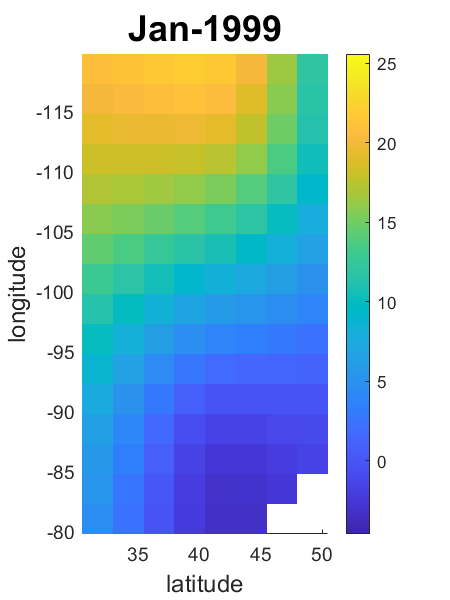}
\includegraphics[width=0.24\textwidth]{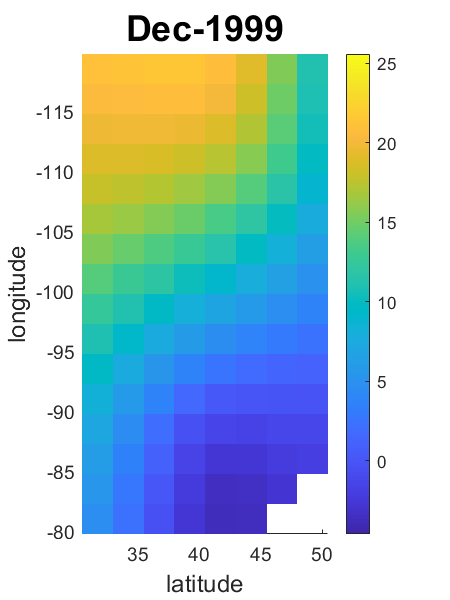}
\includegraphics[width=0.24\textwidth]{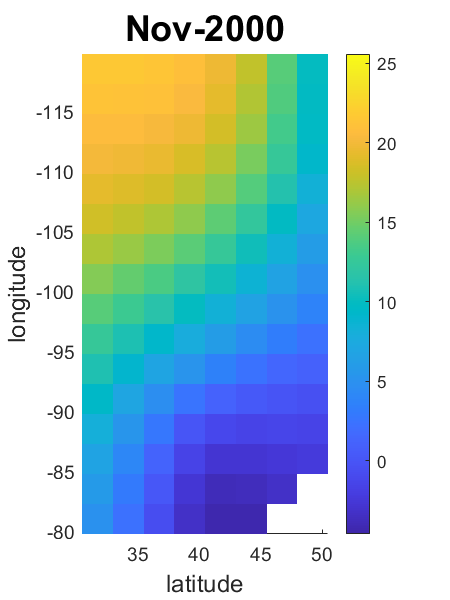}
\includegraphics[width=0.24\textwidth]{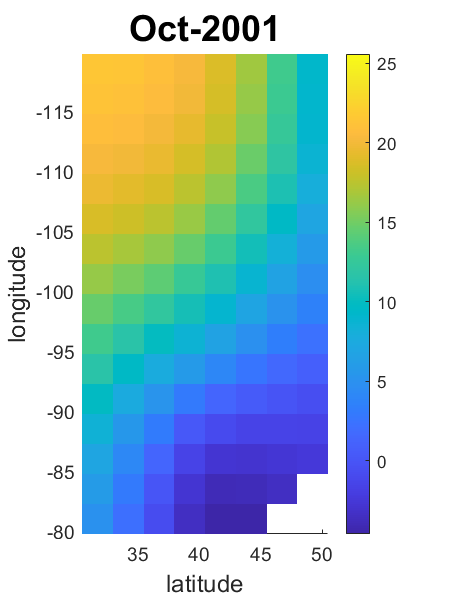}
\caption{The third spatial factor loadings over locations and time in Section \ref{sec6_3}.}
\label{fig_6_3_1_3}
\end{figure}

\begin{figure}
\renewcommand{\baselinestretch}{1}
\centering
\includegraphics[width=0.24\textwidth]{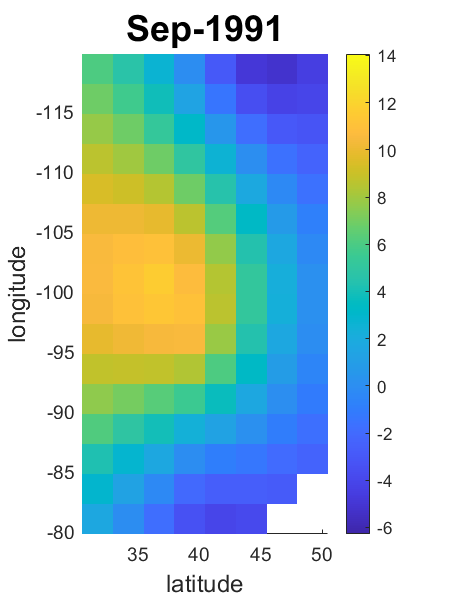}
\includegraphics[width=0.24\textwidth]{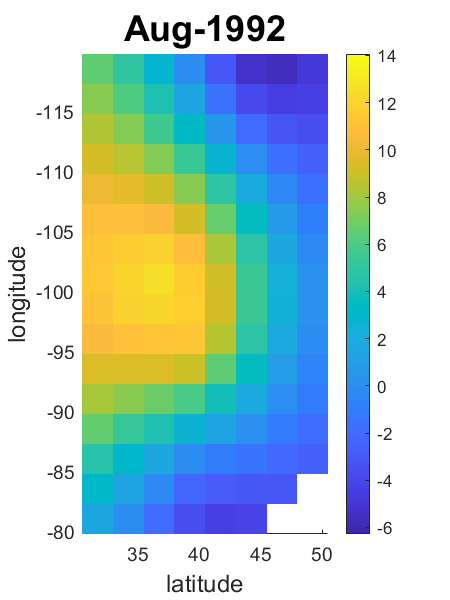}
\includegraphics[width=0.24\textwidth]{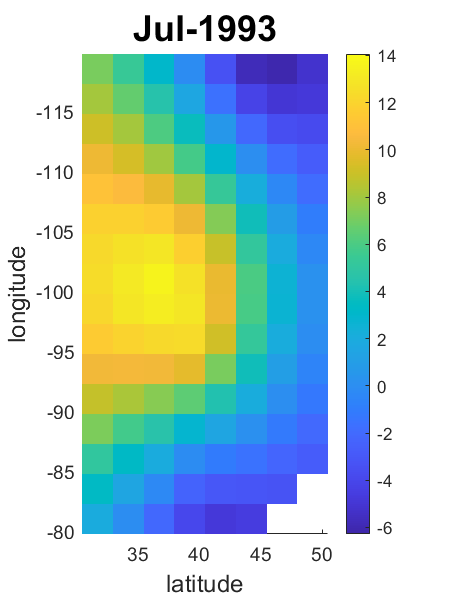}
\includegraphics[width=0.24\textwidth]{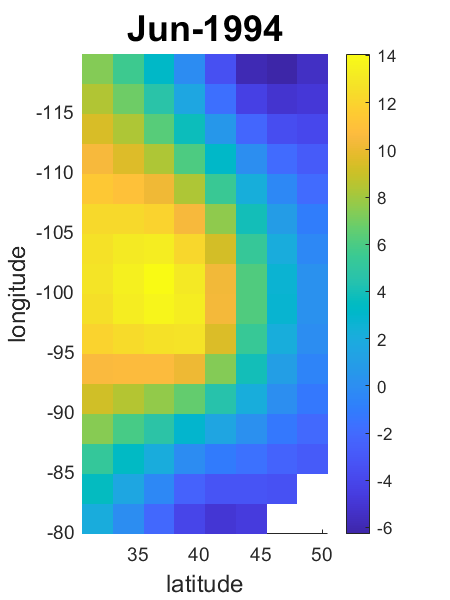}
\includegraphics[width=0.24\textwidth]{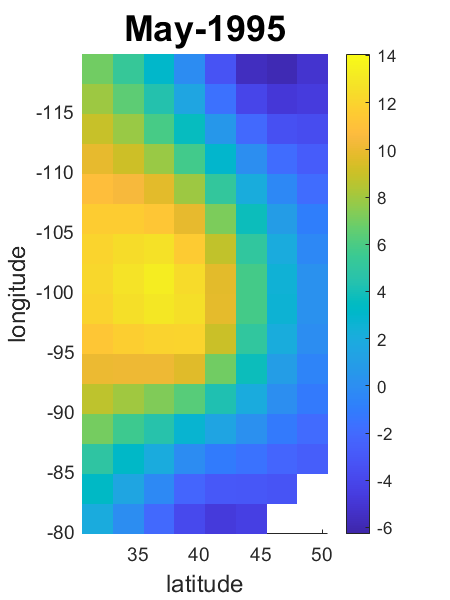}
\includegraphics[width=0.24\textwidth]{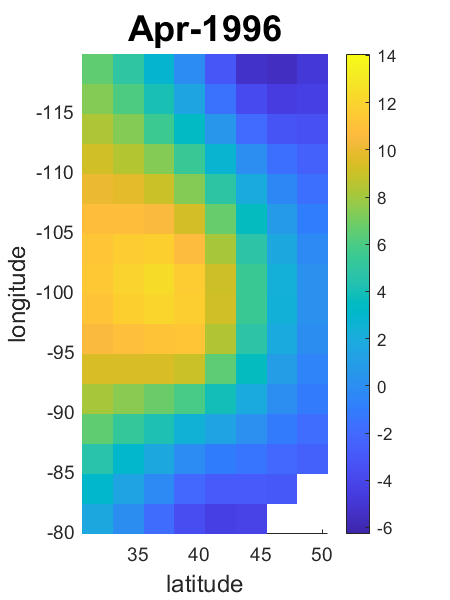}
\includegraphics[width=0.24\textwidth]{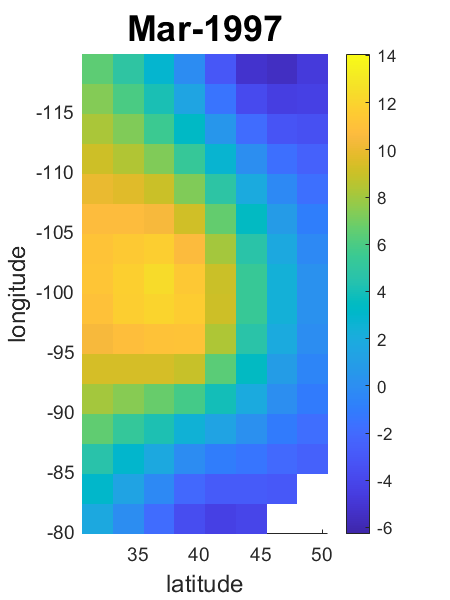}
\includegraphics[width=0.24\textwidth]{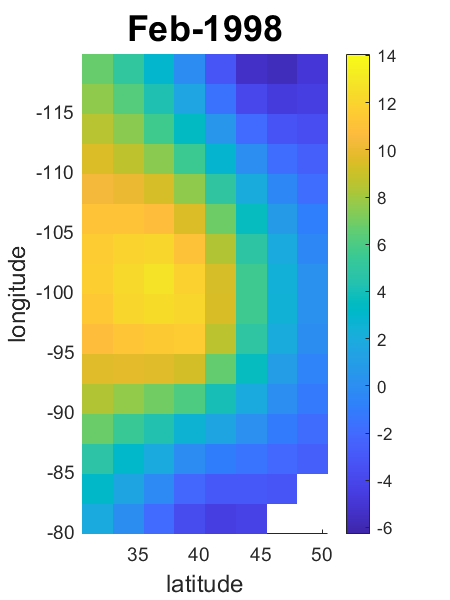}
\includegraphics[width=0.24\textwidth]{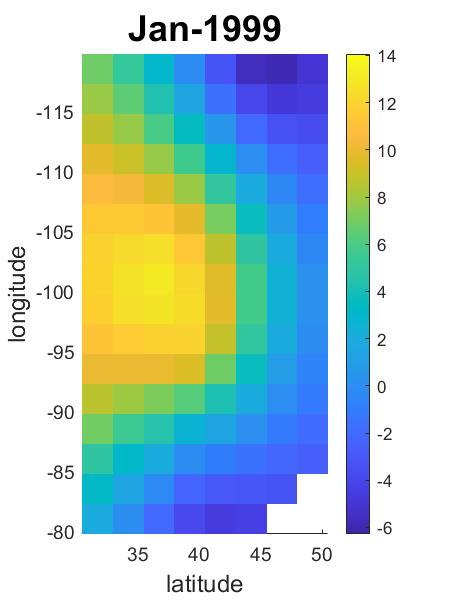}
\includegraphics[width=0.24\textwidth]{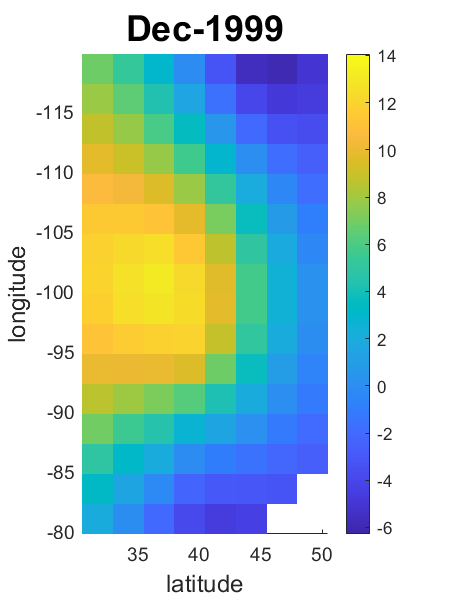}
\includegraphics[width=0.24\textwidth]{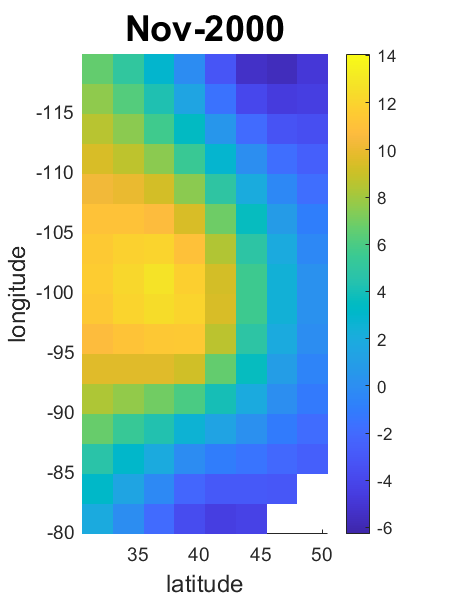}
\includegraphics[width=0.24\textwidth]{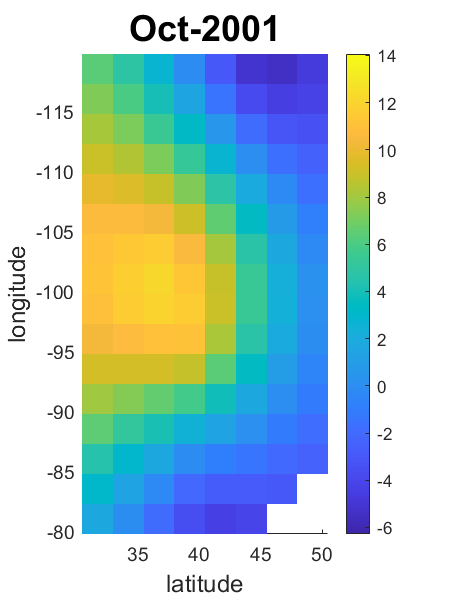}
\caption{The fourth spatial factor loadings over locations and time in Section \ref{sec6_3}.}
\label{fig_6_3_1_4}
\end{figure}

\begin{figure}
\renewcommand{\baselinestretch}{1}
\centering
\includegraphics[width=0.24\textwidth]{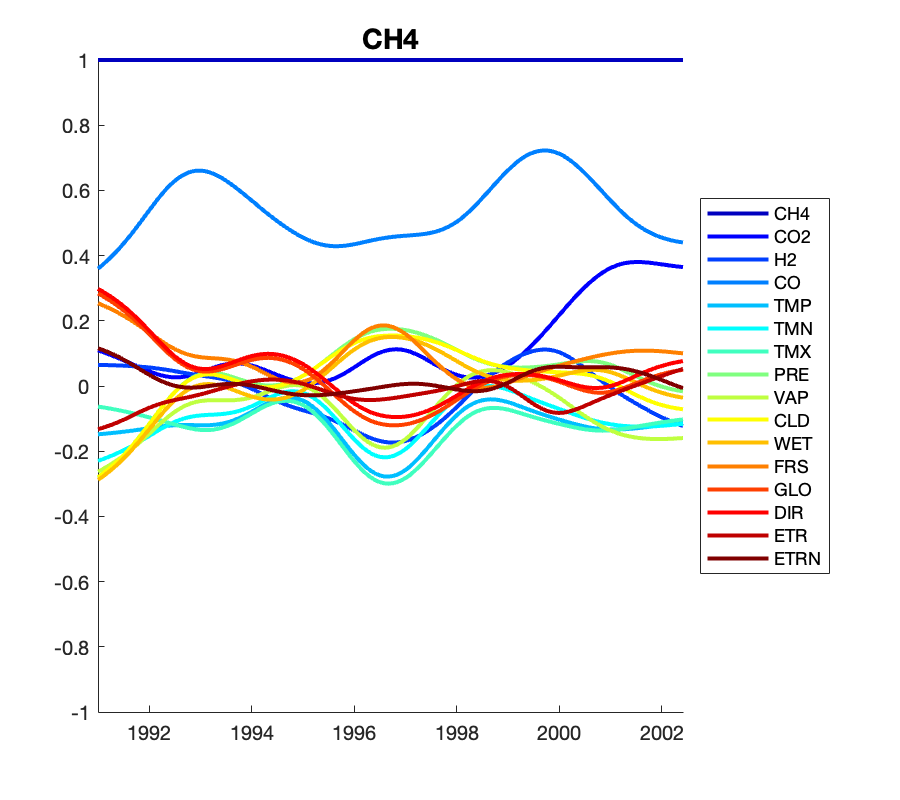}
\includegraphics[width=0.24\textwidth]{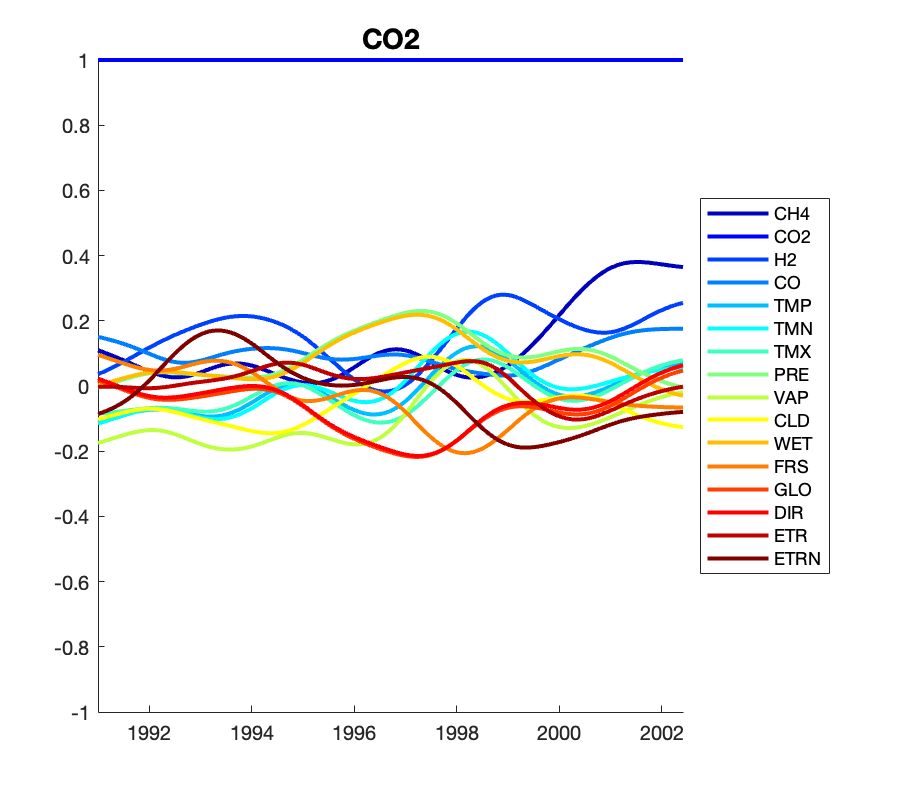}
\includegraphics[width=0.24\textwidth]{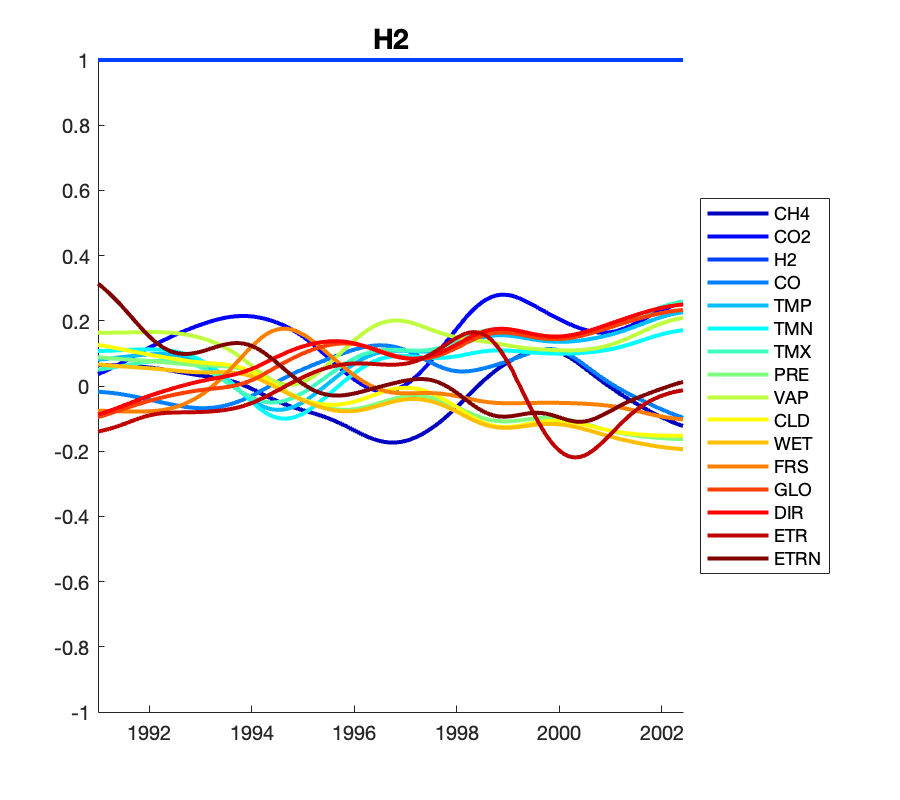}
\includegraphics[width=0.24\textwidth]{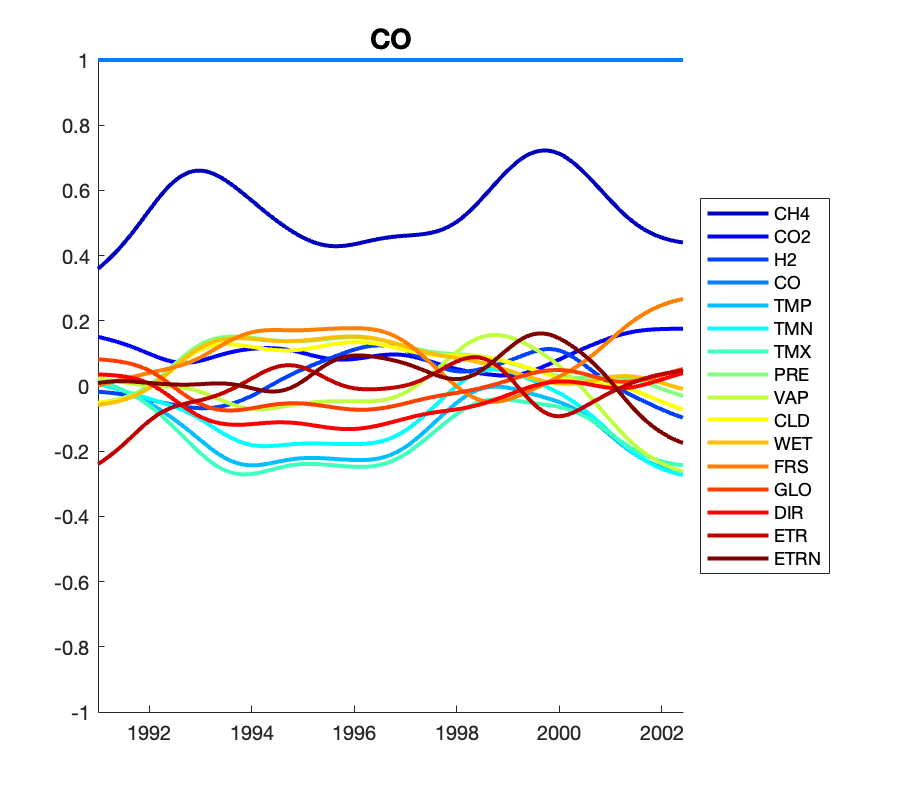}
\includegraphics[width=0.24\textwidth]{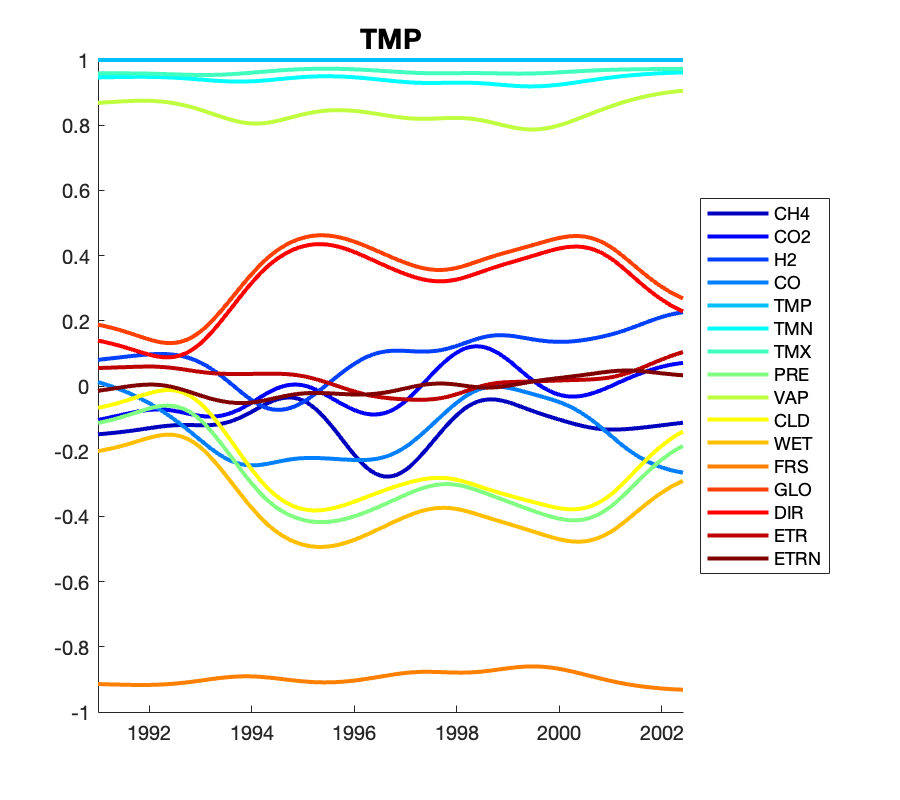}
\includegraphics[width=0.24\textwidth]{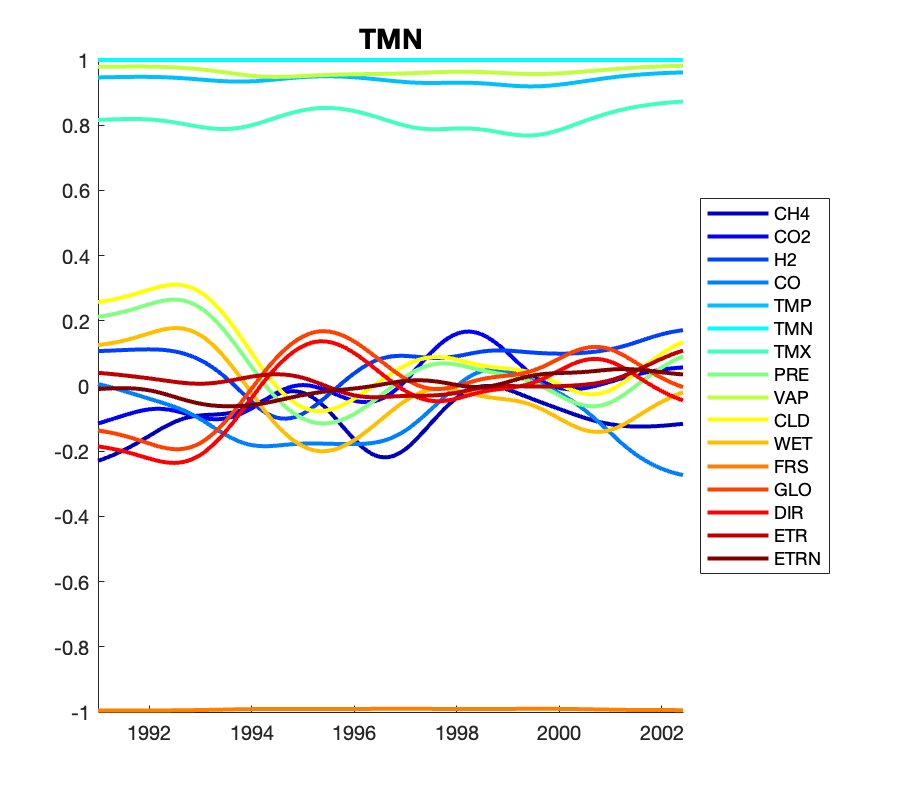}
\includegraphics[width=0.24\textwidth]{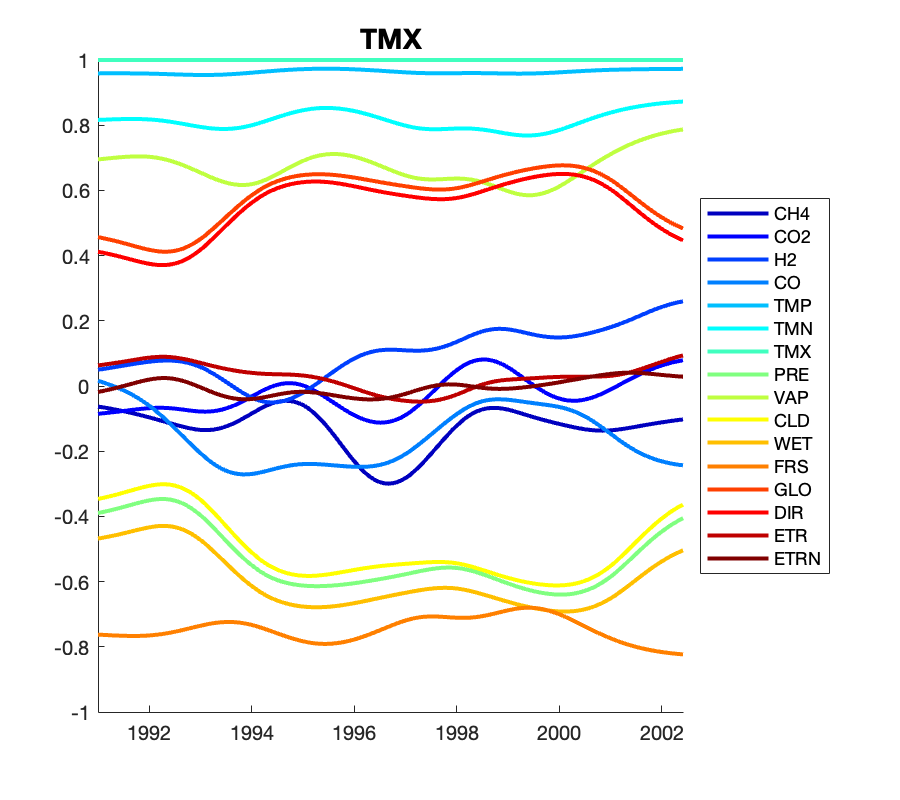}
\includegraphics[width=0.24\textwidth]{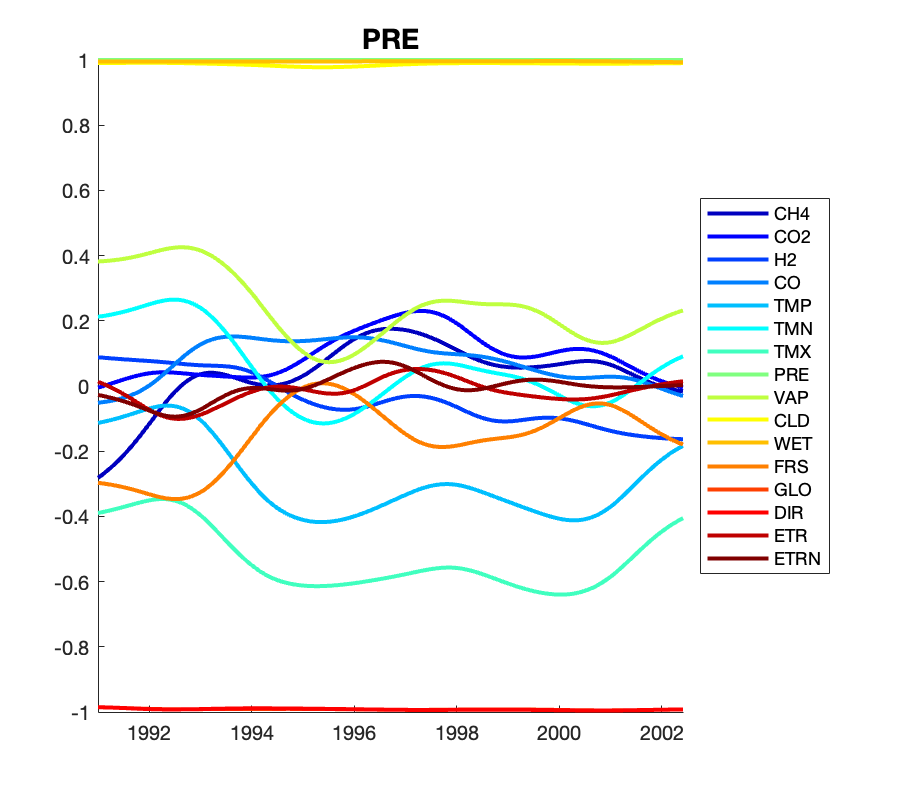}
\includegraphics[width=0.24\textwidth]{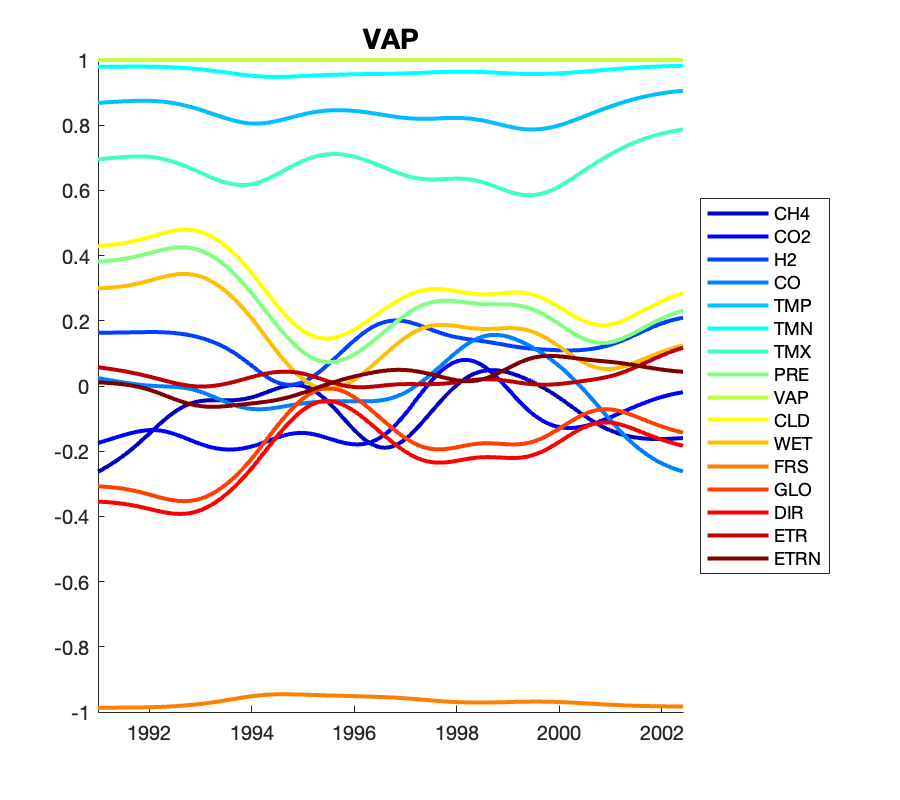}
\includegraphics[width=0.24\textwidth]{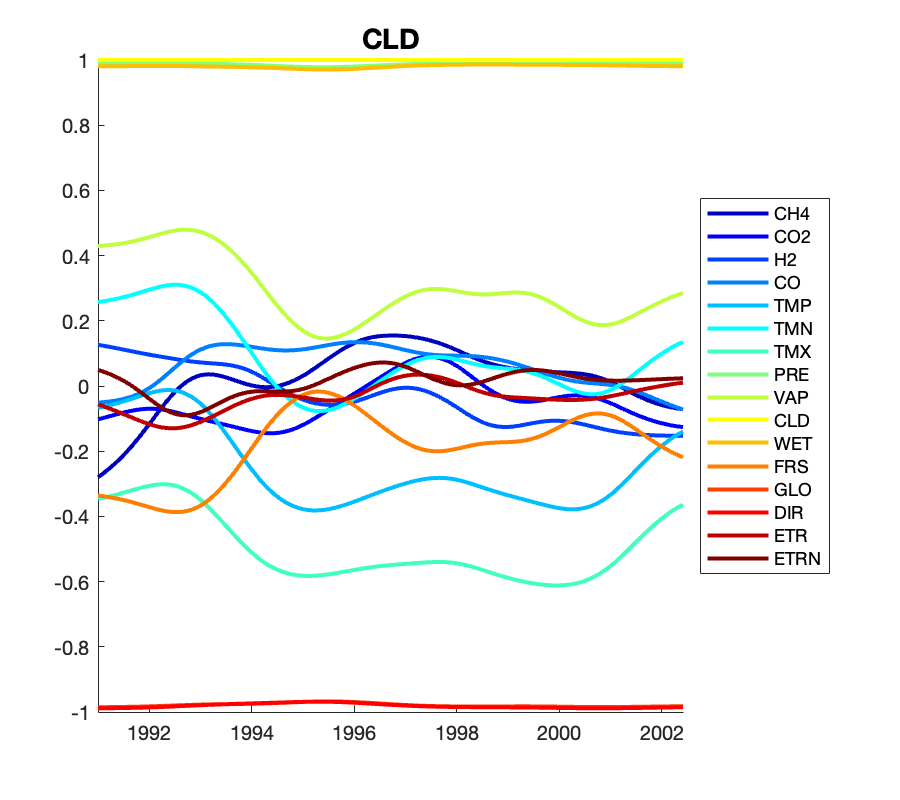}
\includegraphics[width=0.24\textwidth]{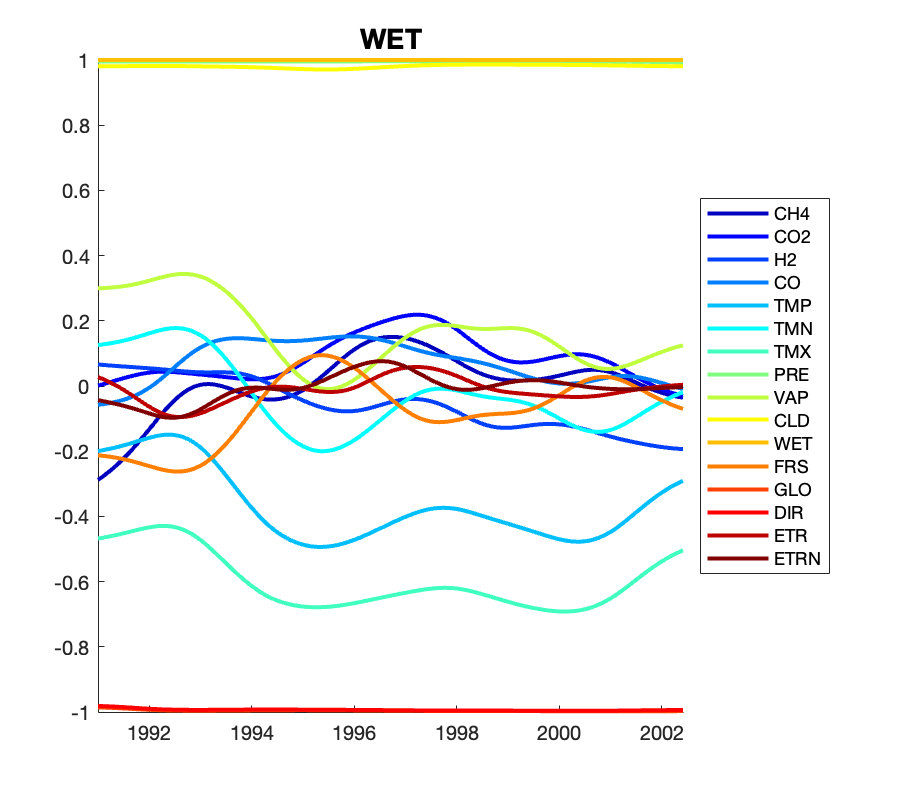}
\includegraphics[width=0.24\textwidth]{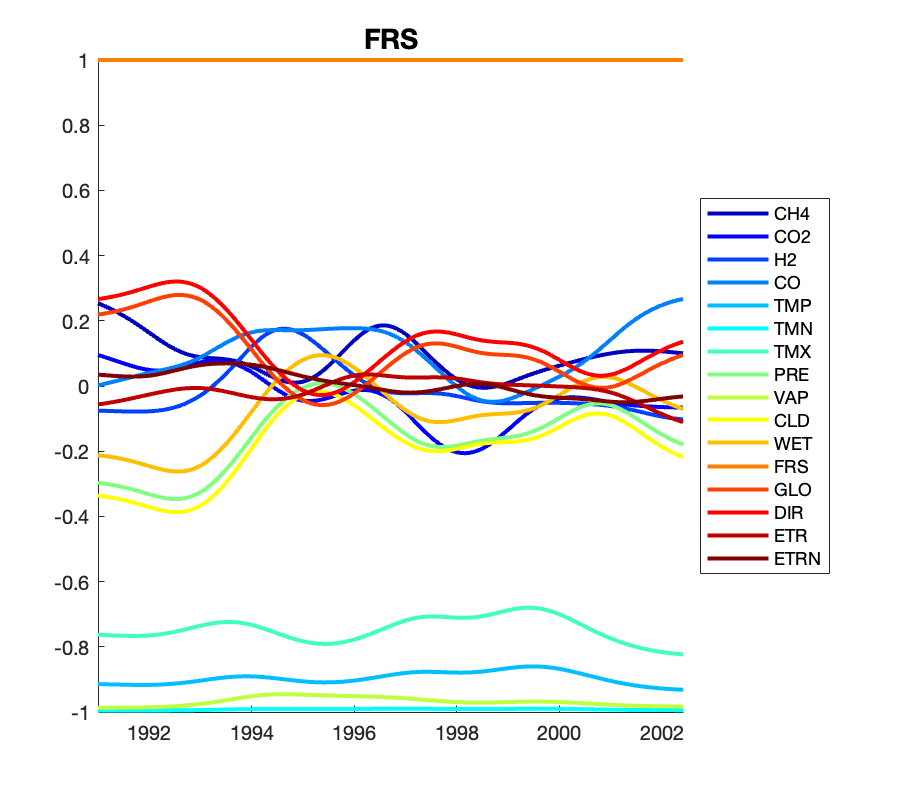}
\includegraphics[width=0.24\textwidth]{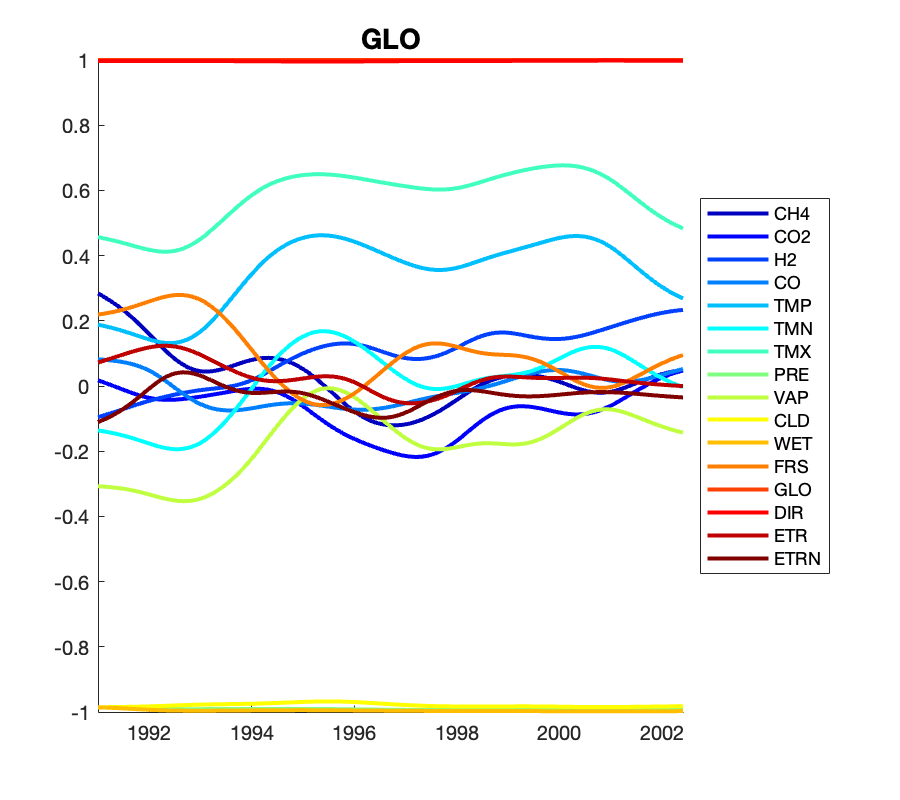}
\includegraphics[width=0.24\textwidth]{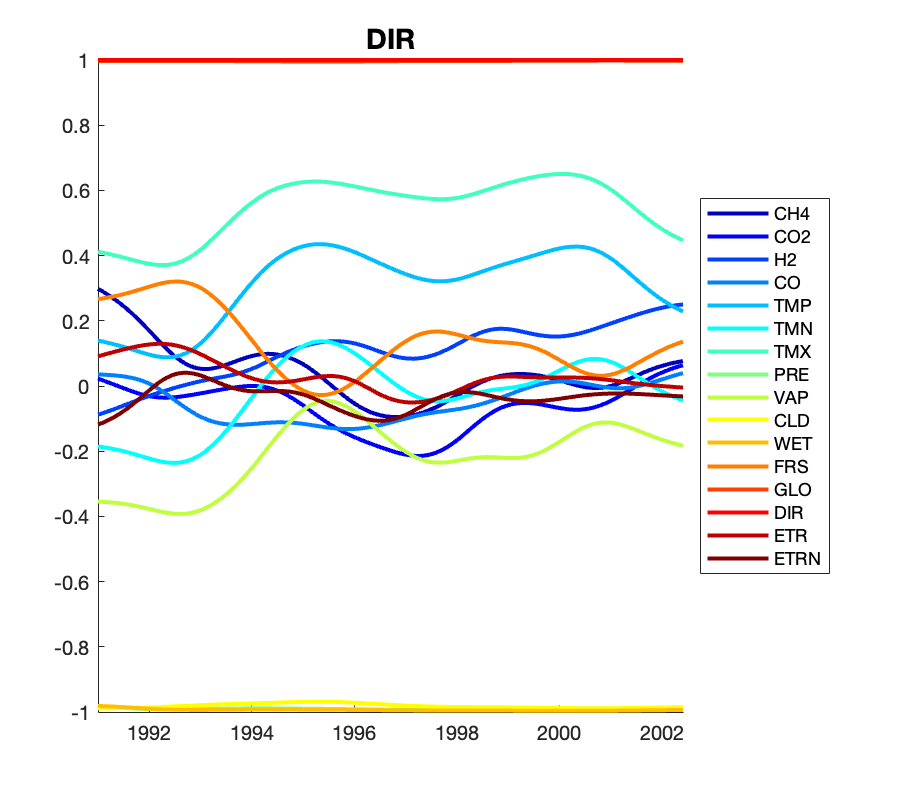}
\includegraphics[width=0.24\textwidth]{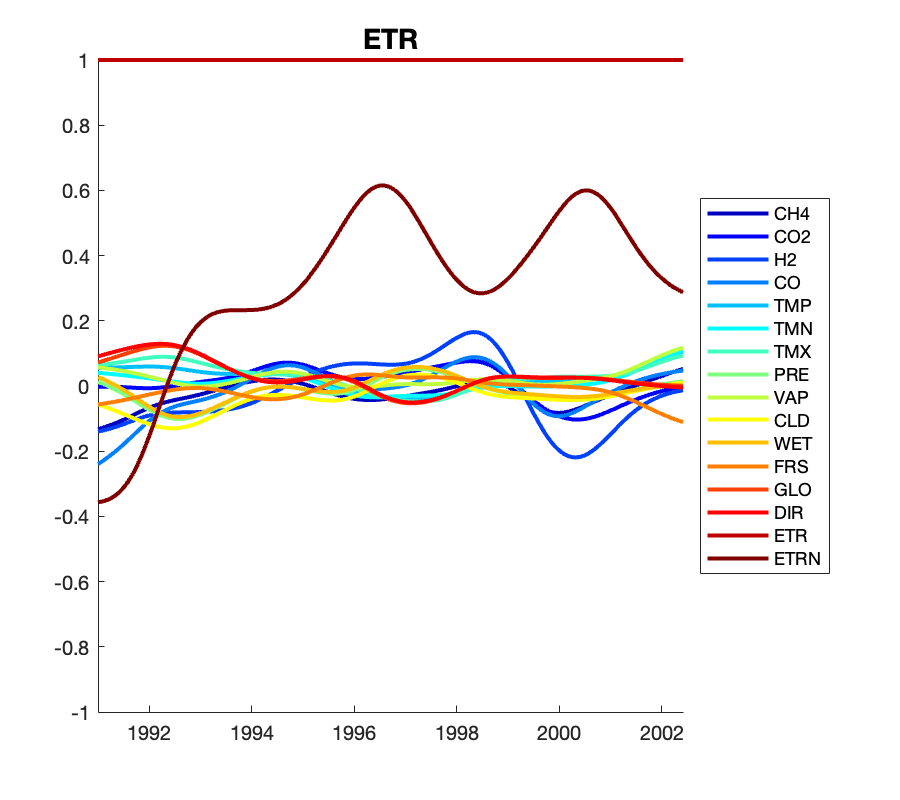}
\includegraphics[width=0.24\textwidth]{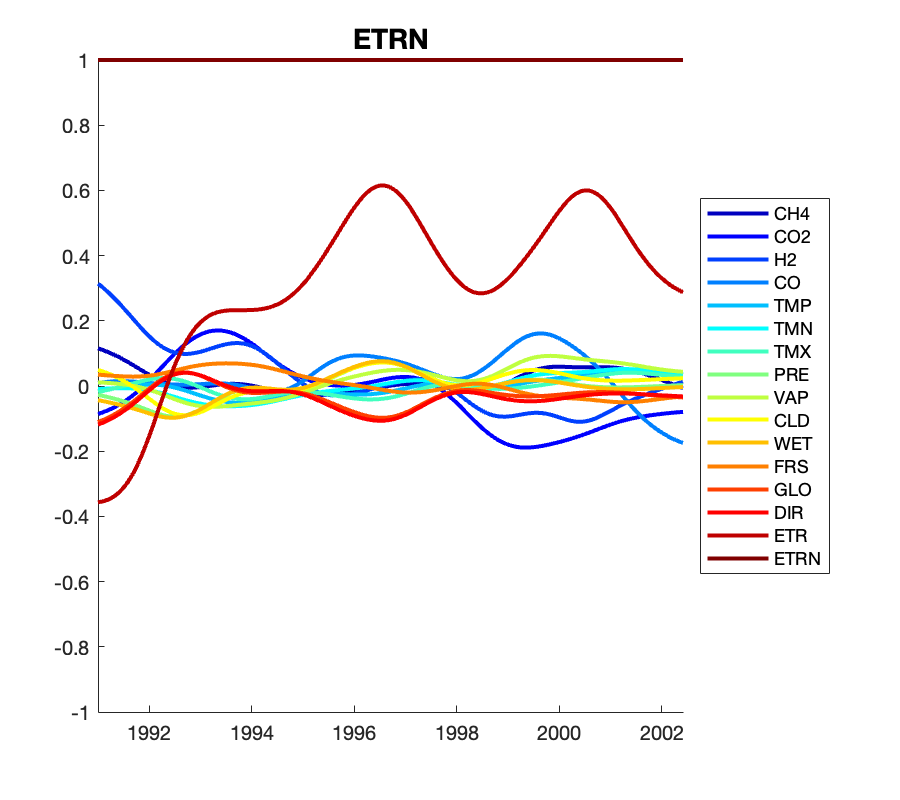}
\caption{Estimated correlations between a variable and the rest over time.}
\label{fig_6_3_4}
\end{figure}

\begin{figure}
\renewcommand{\baselinestretch}{1}
\centering
\includegraphics[width=0.24\textwidth]{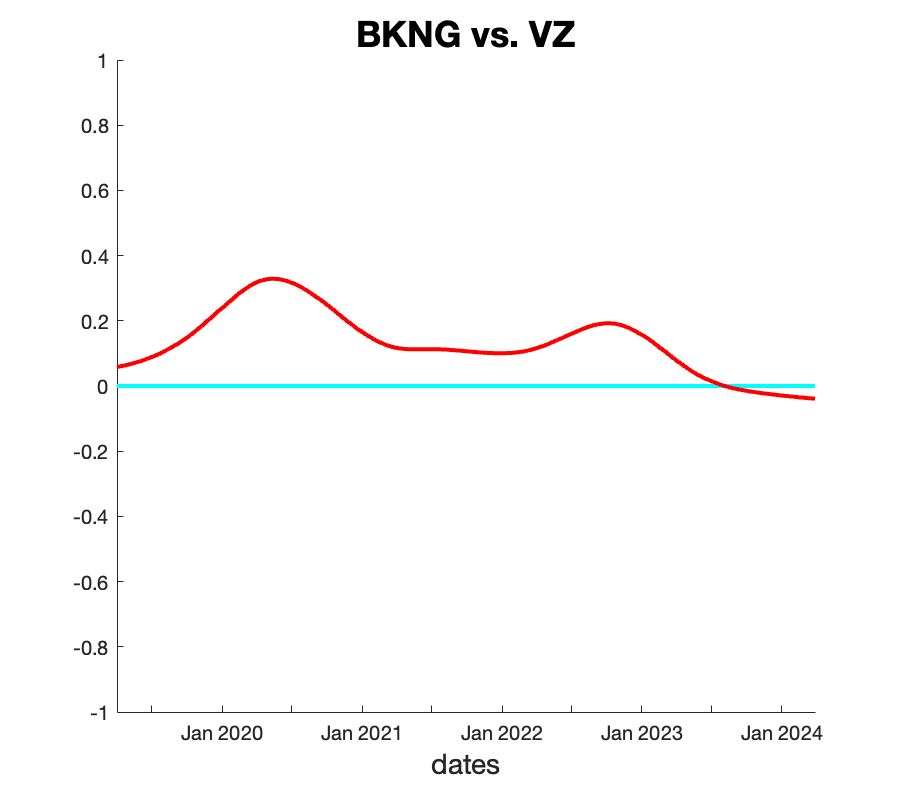}
\includegraphics[width=0.24\textwidth]{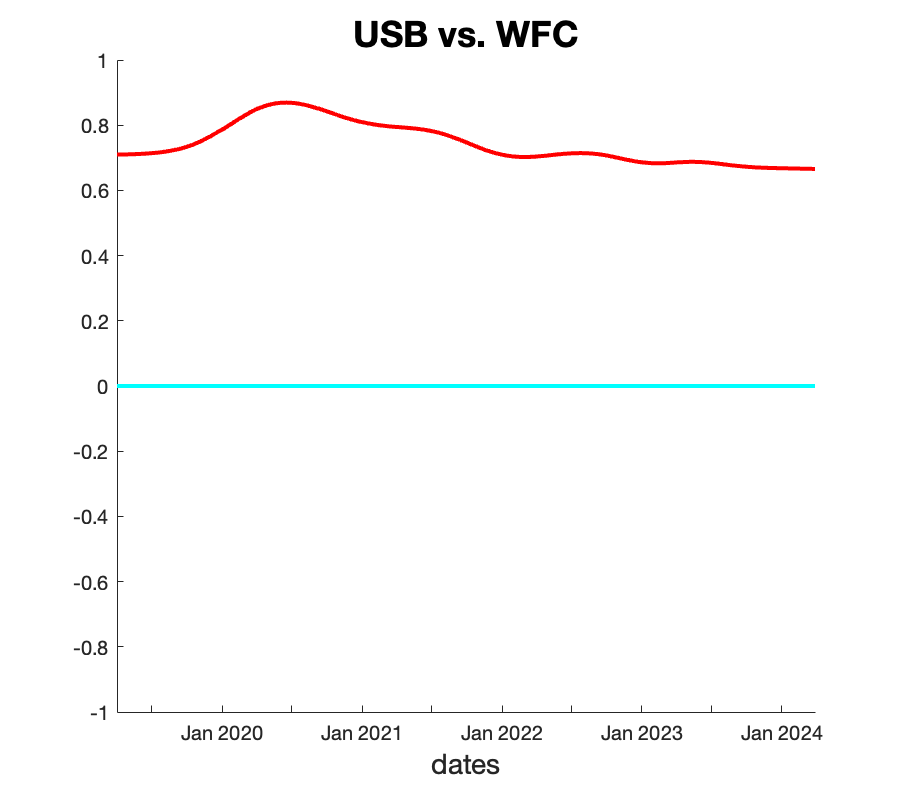}
\includegraphics[width=0.24\textwidth]{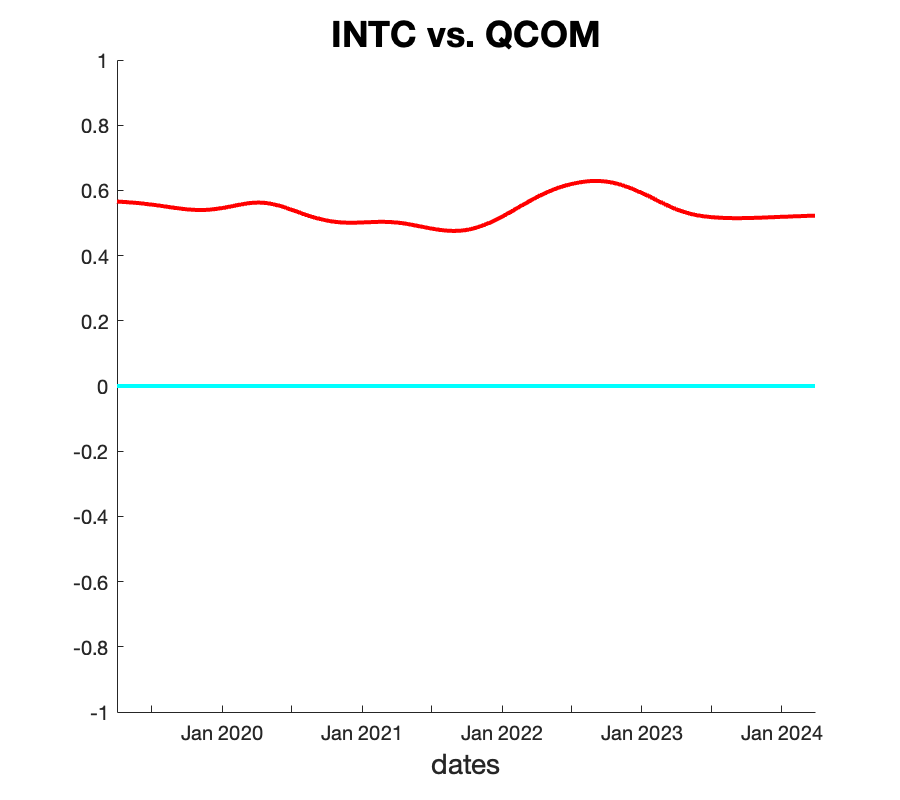}
\includegraphics[width=0.24\textwidth]{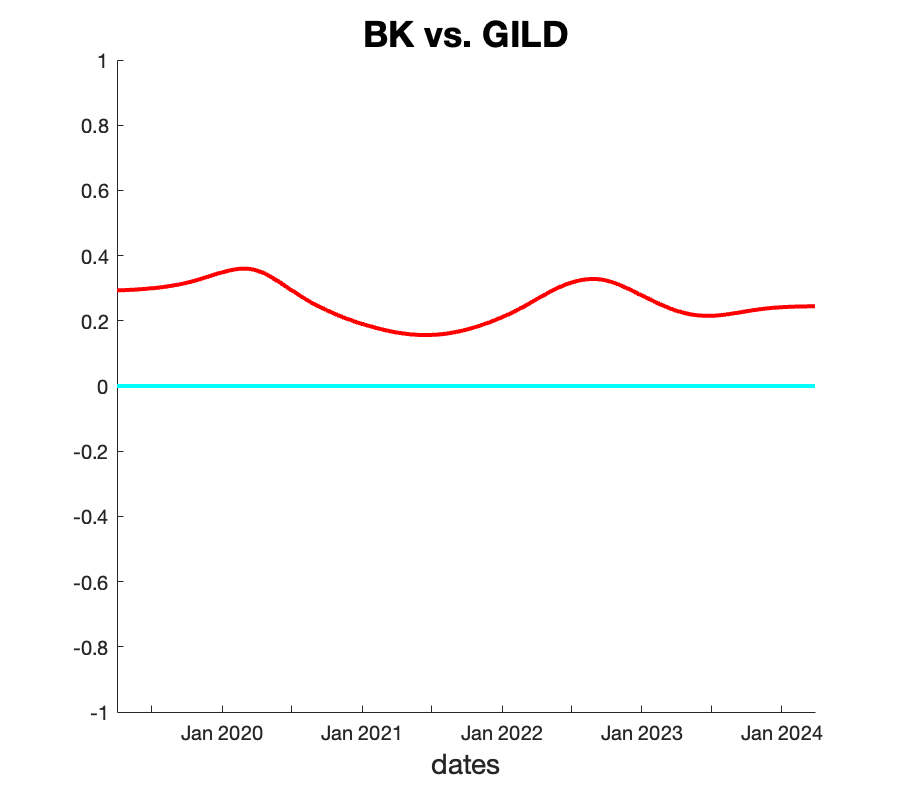}
\includegraphics[width=0.24\textwidth]{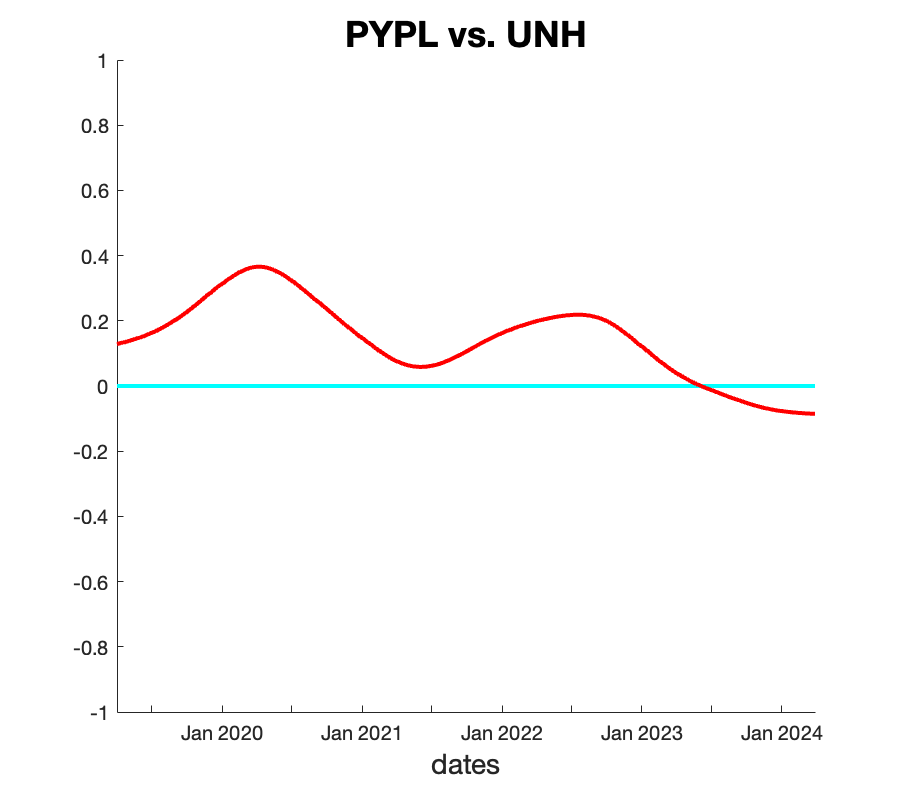}
\includegraphics[width=0.24\textwidth]{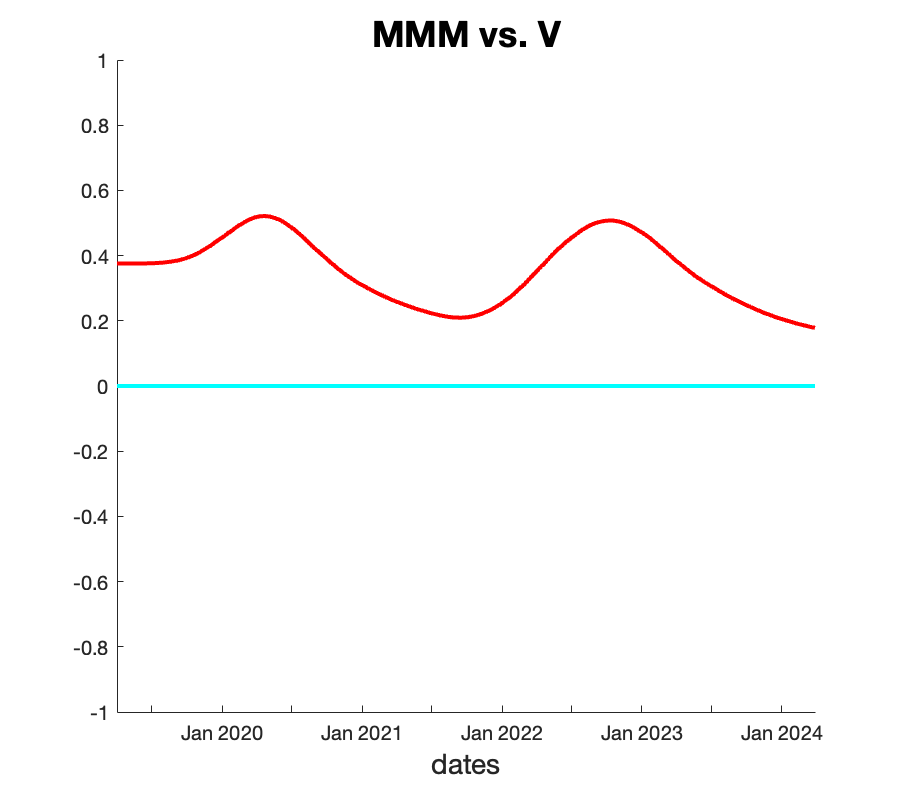}
\includegraphics[width=0.24\textwidth]{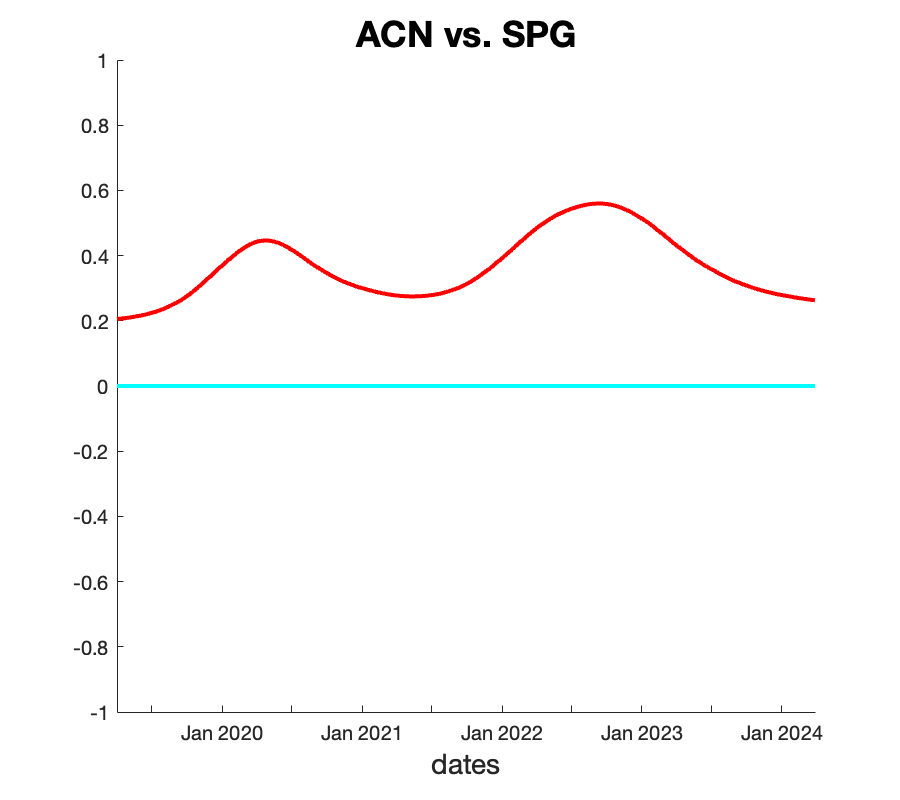}
\includegraphics[width=0.24\textwidth]{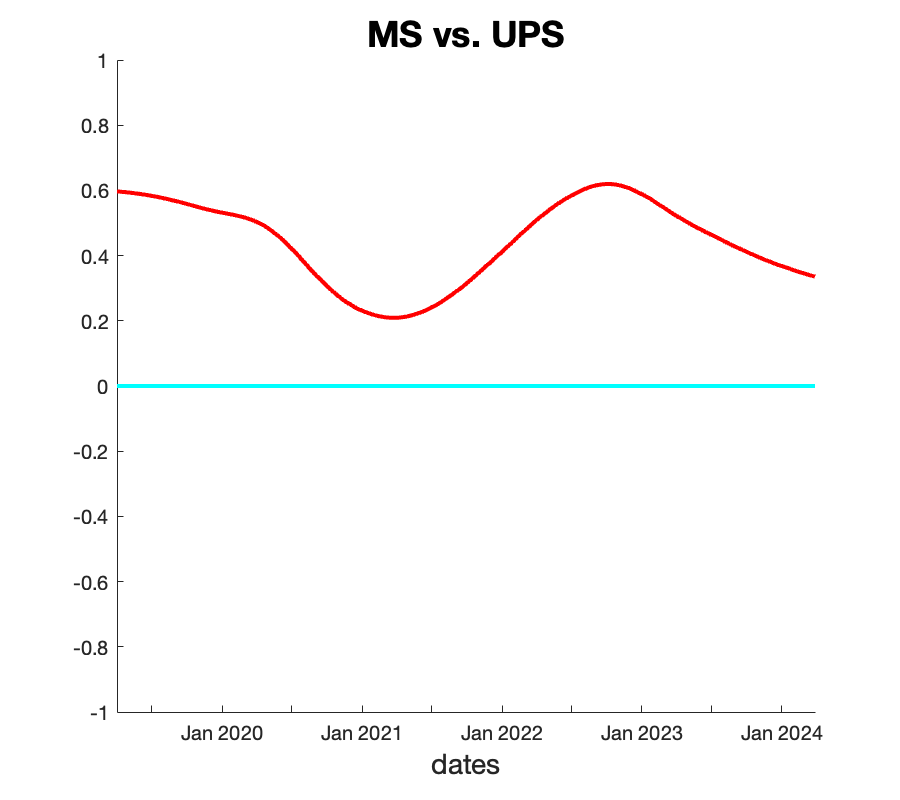}
\includegraphics[width=0.24\textwidth]{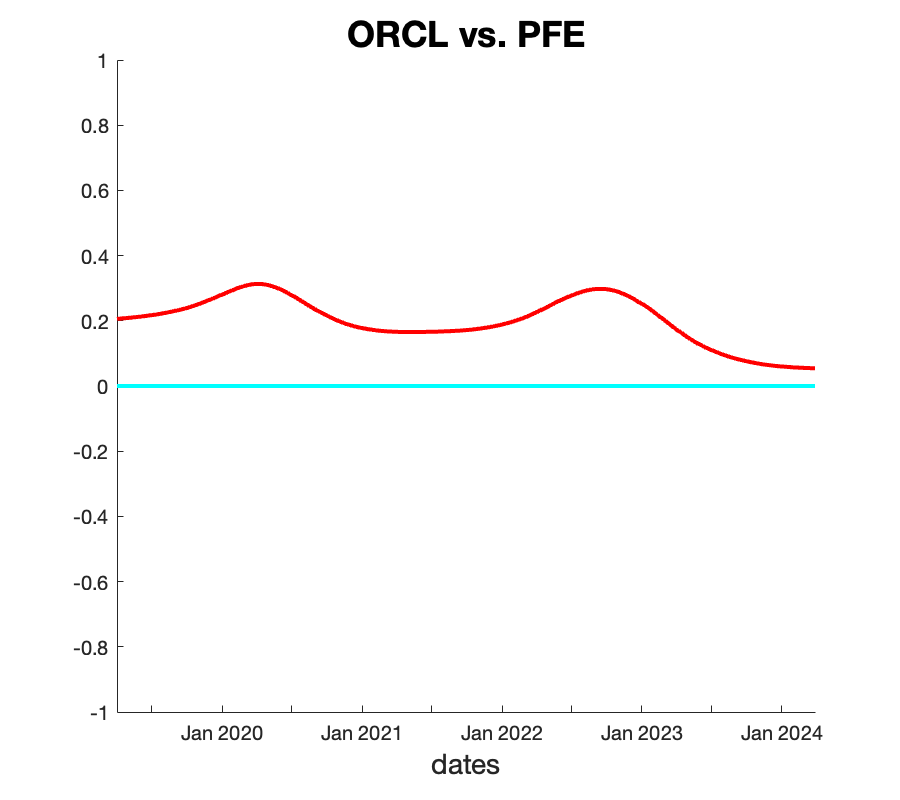}
\includegraphics[width=0.24\textwidth]{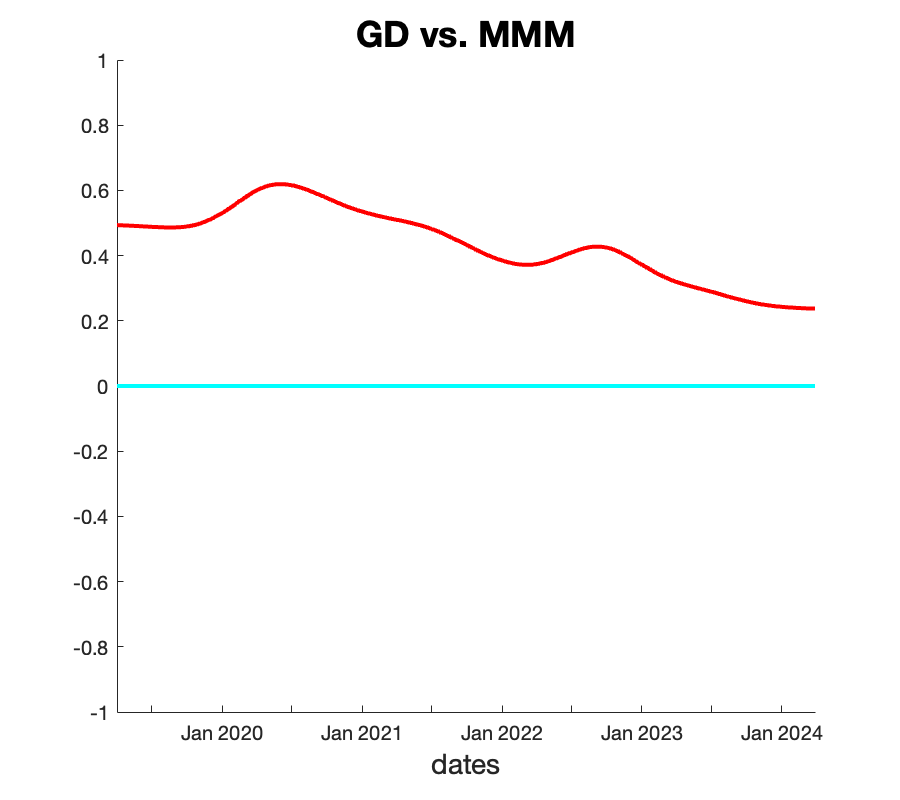}
\includegraphics[width=0.24\textwidth]{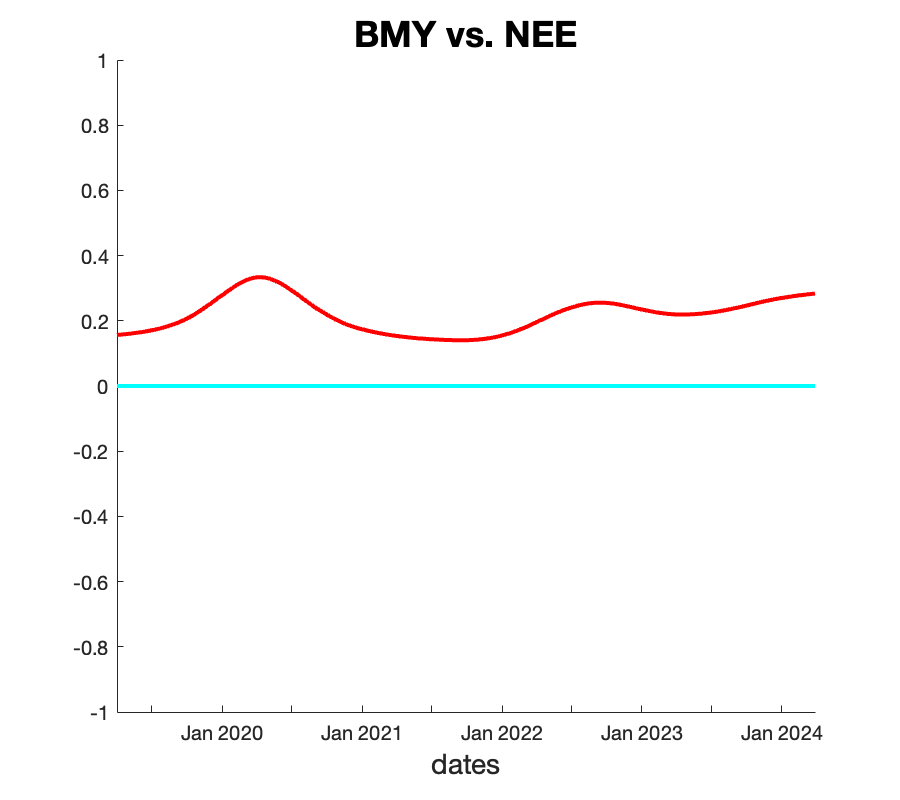}
\includegraphics[width=0.24\textwidth]{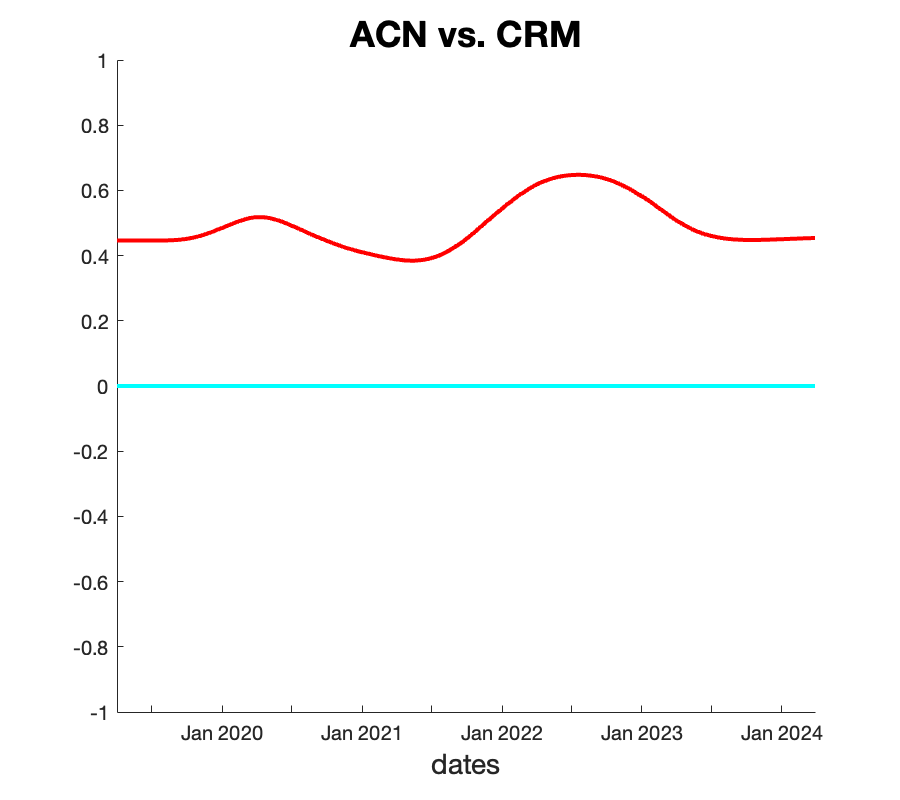}
\includegraphics[width=0.24\textwidth]{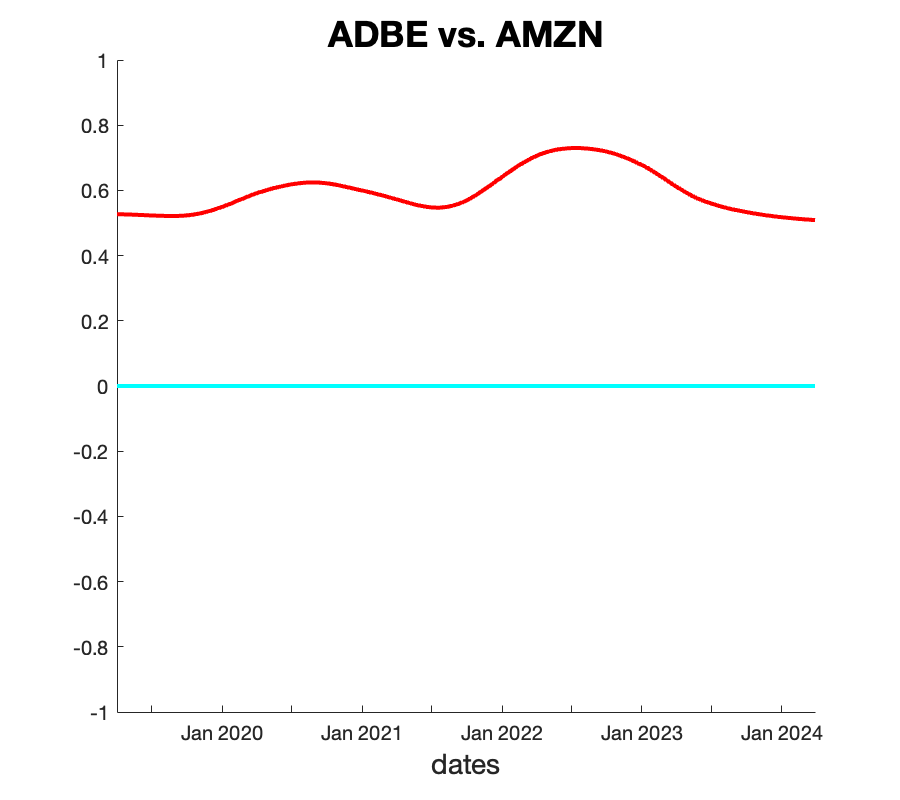}
\includegraphics[width=0.24\textwidth]{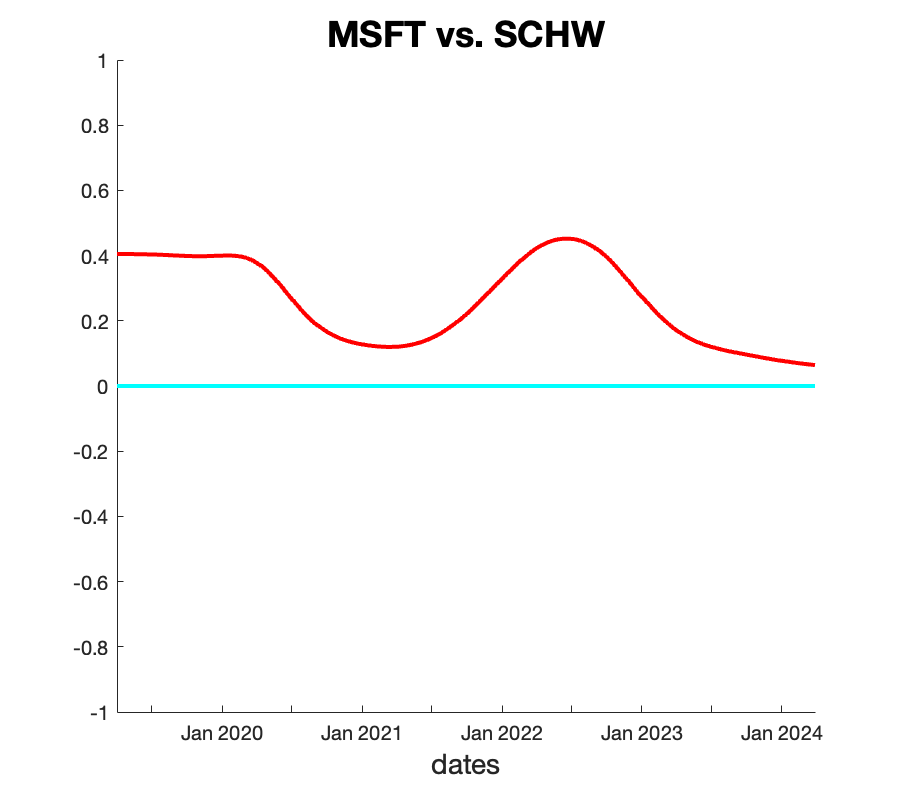}
\includegraphics[width=0.24\textwidth]{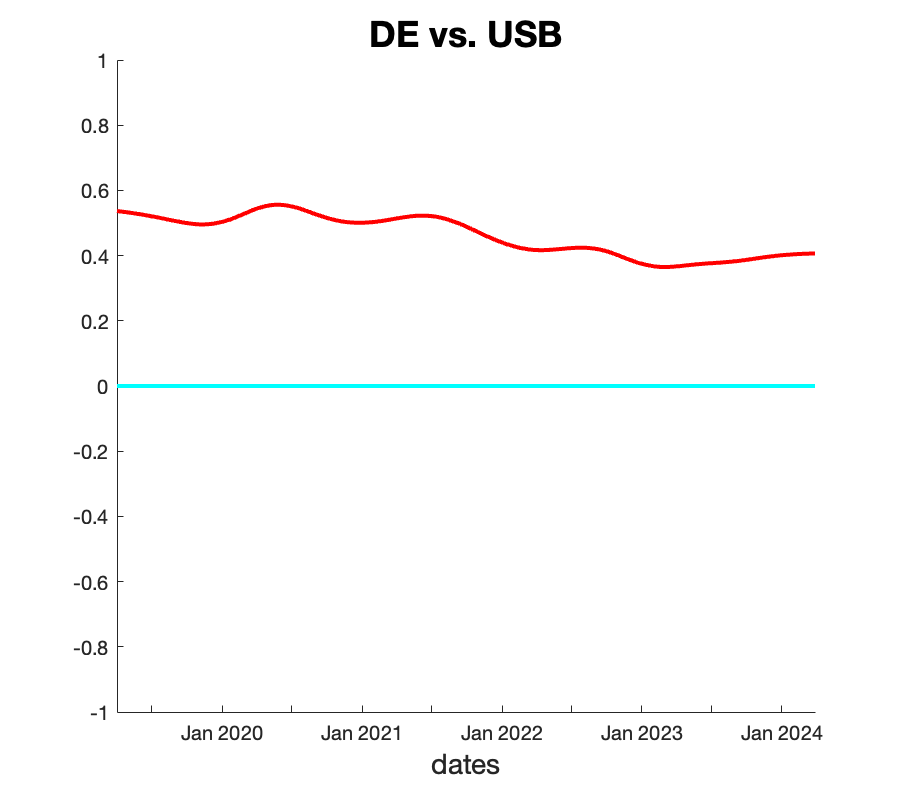}
\includegraphics[width=0.24\textwidth]{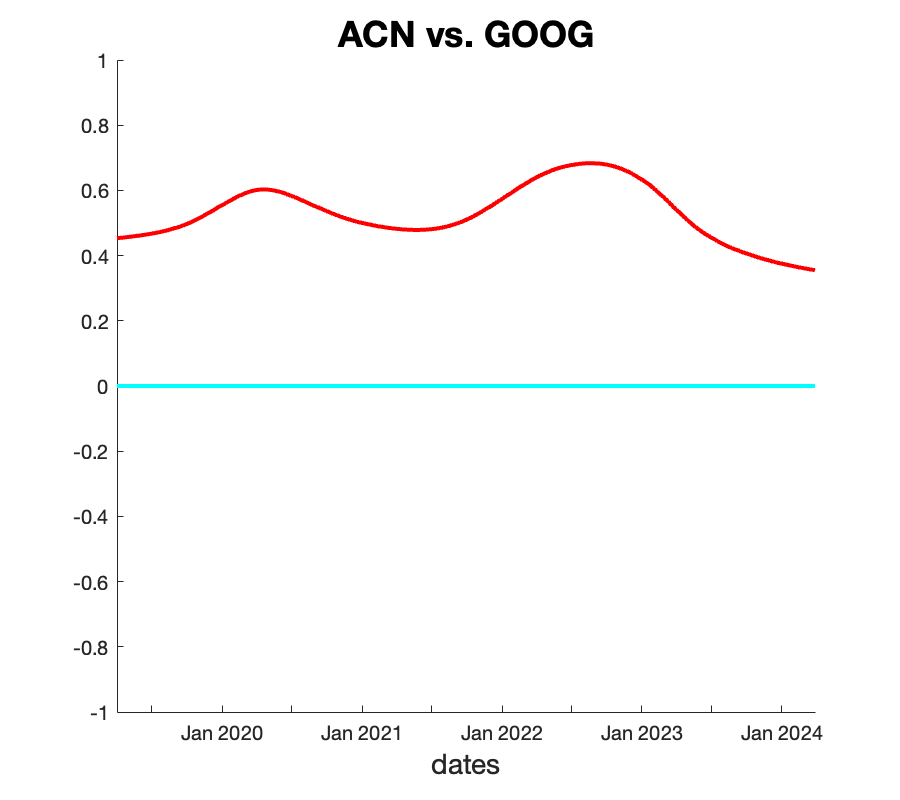}
\caption{Some examples of the estimate correlation between two randomly chosen companies over dates.}
\label{fig_6_4_3}
\end{figure}

\begin{figure}
\renewcommand{\baselinestretch}{1}
\centering
\includegraphics[width=0.24\textwidth]{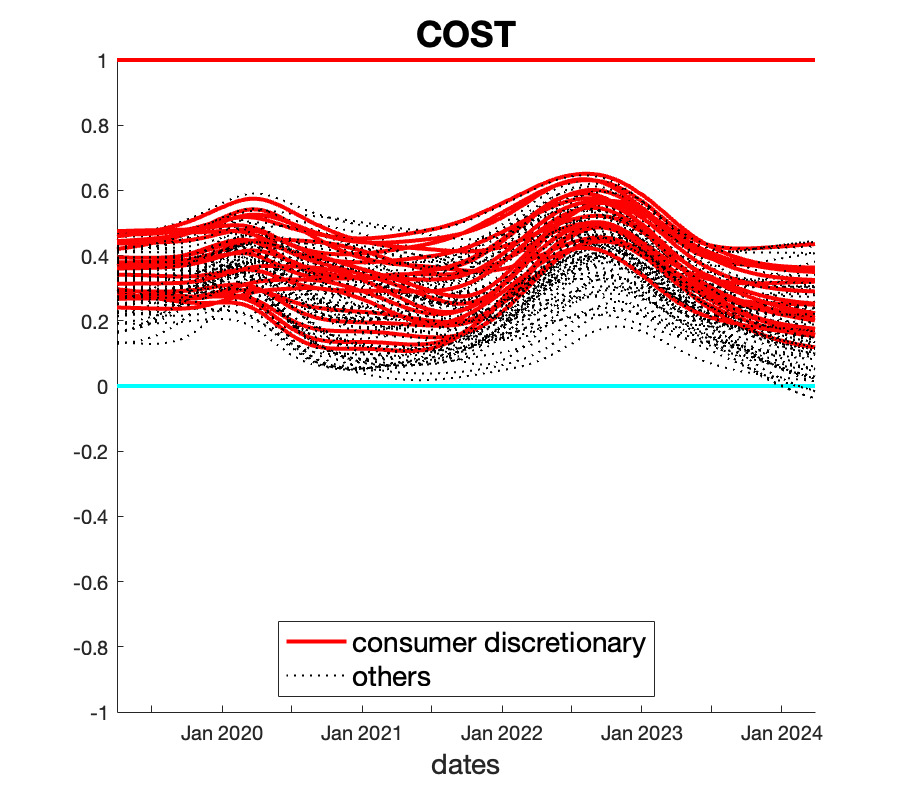}
\includegraphics[width=0.24\textwidth]{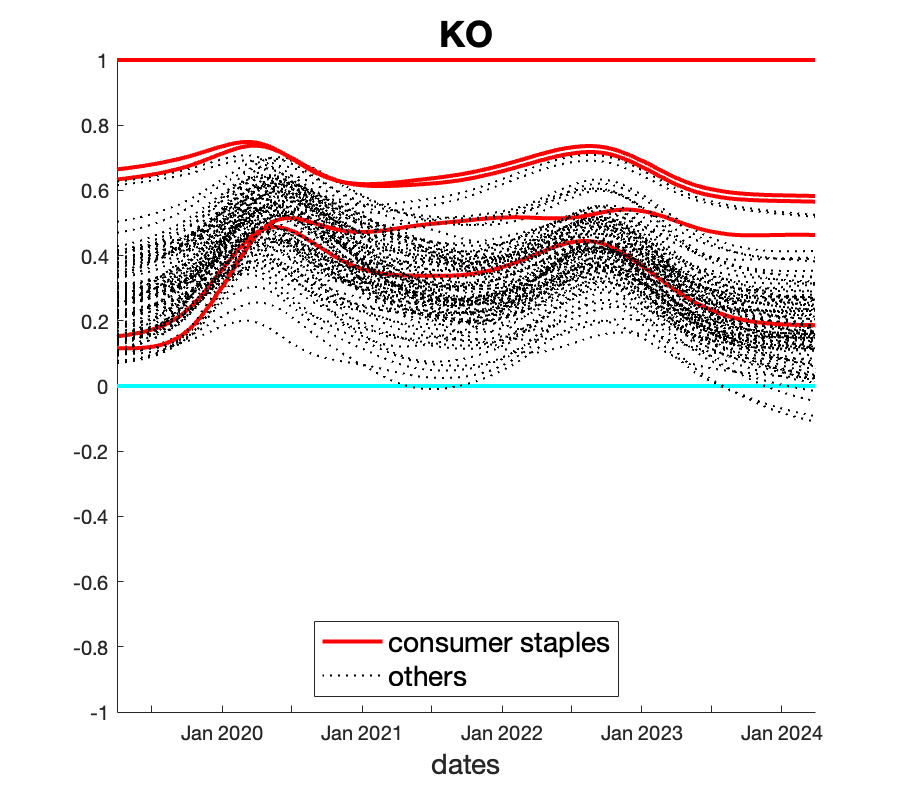}
\includegraphics[width=0.24\textwidth]{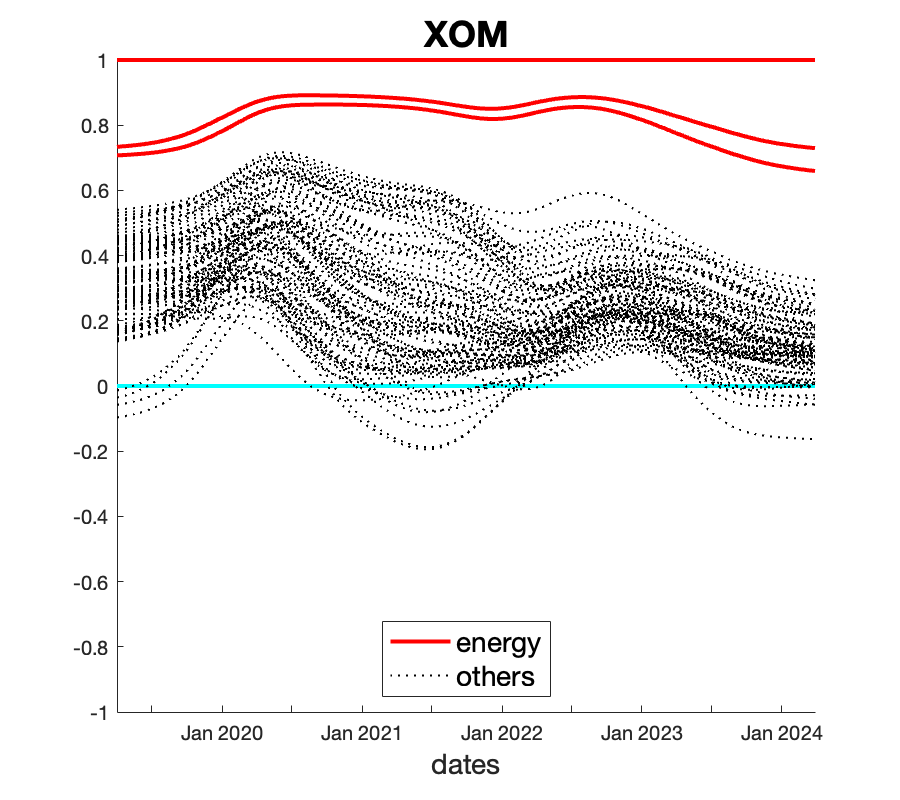}
\includegraphics[width=0.24\textwidth]{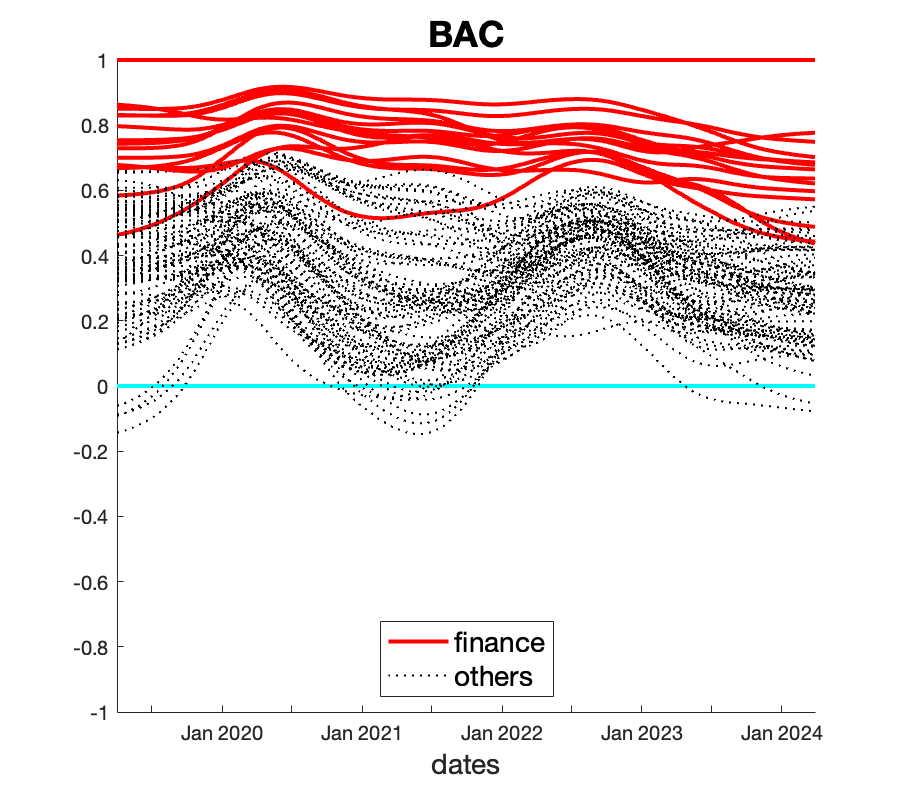}
\includegraphics[width=0.24\textwidth]{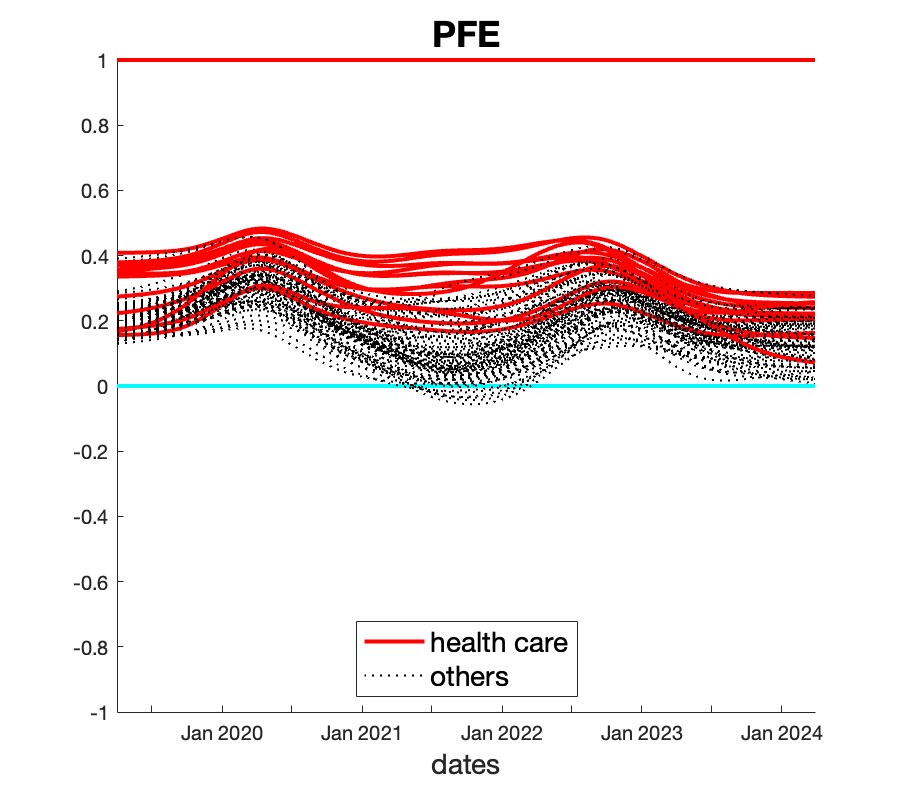}
\includegraphics[width=0.24\textwidth]{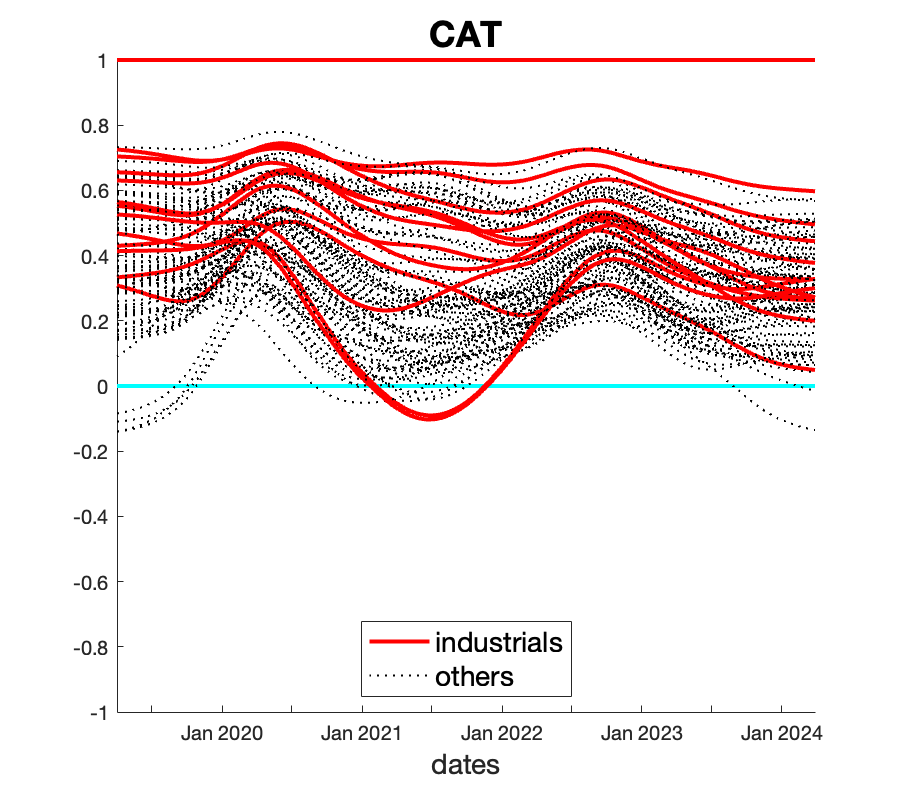}
\includegraphics[width=0.24\textwidth]{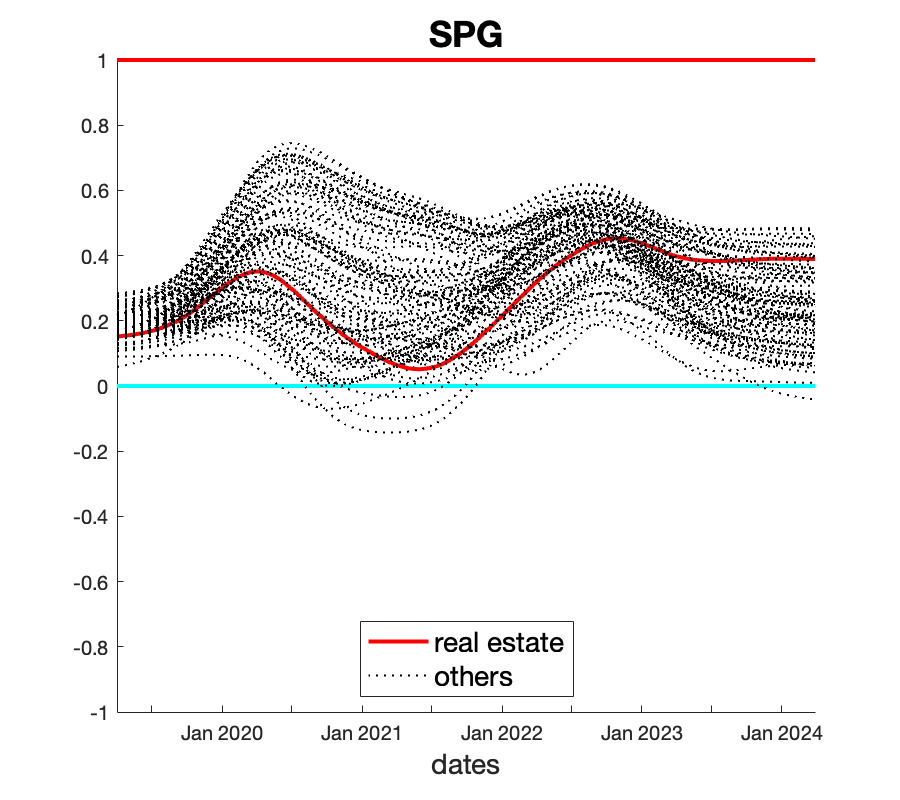}
\includegraphics[width=0.24\textwidth]{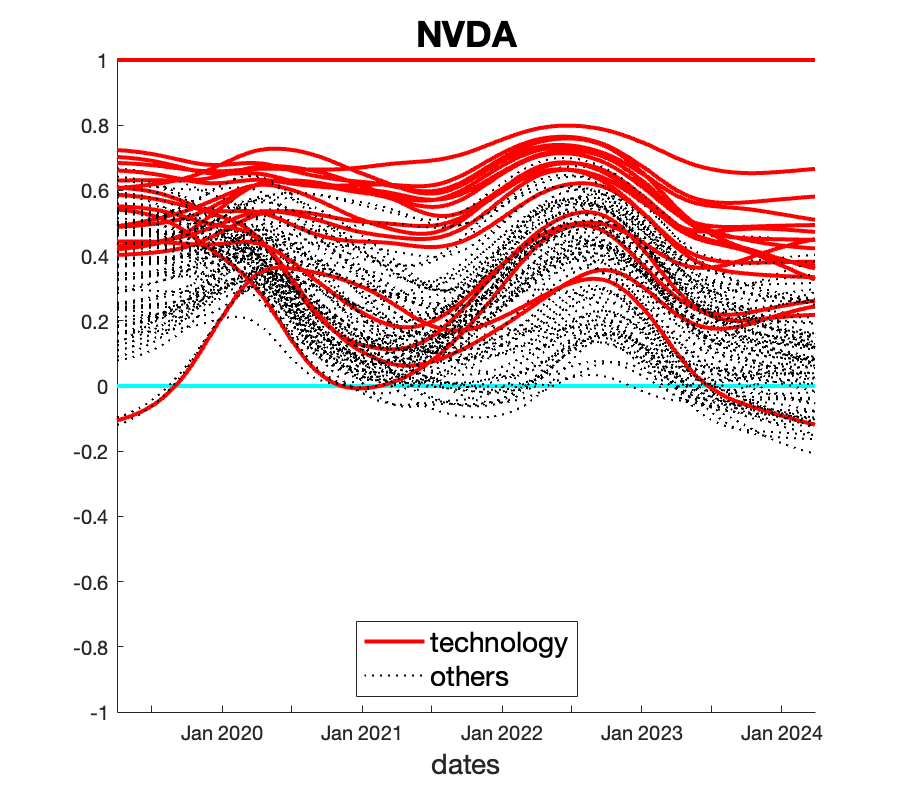}
\includegraphics[width=0.24\textwidth]{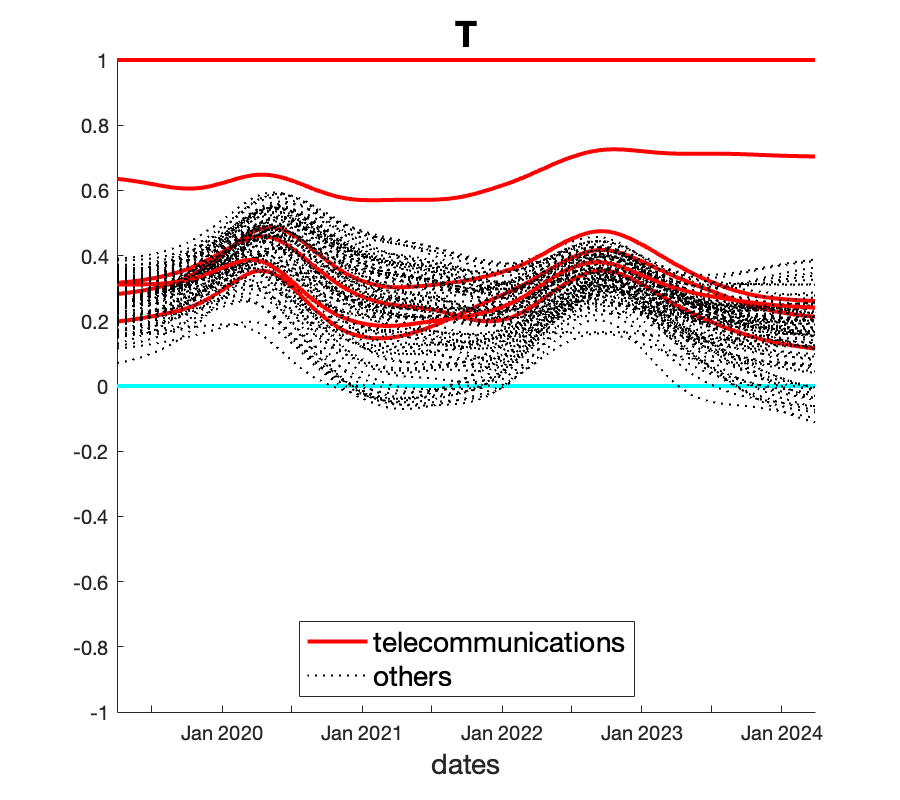}
\includegraphics[width=0.24\textwidth]{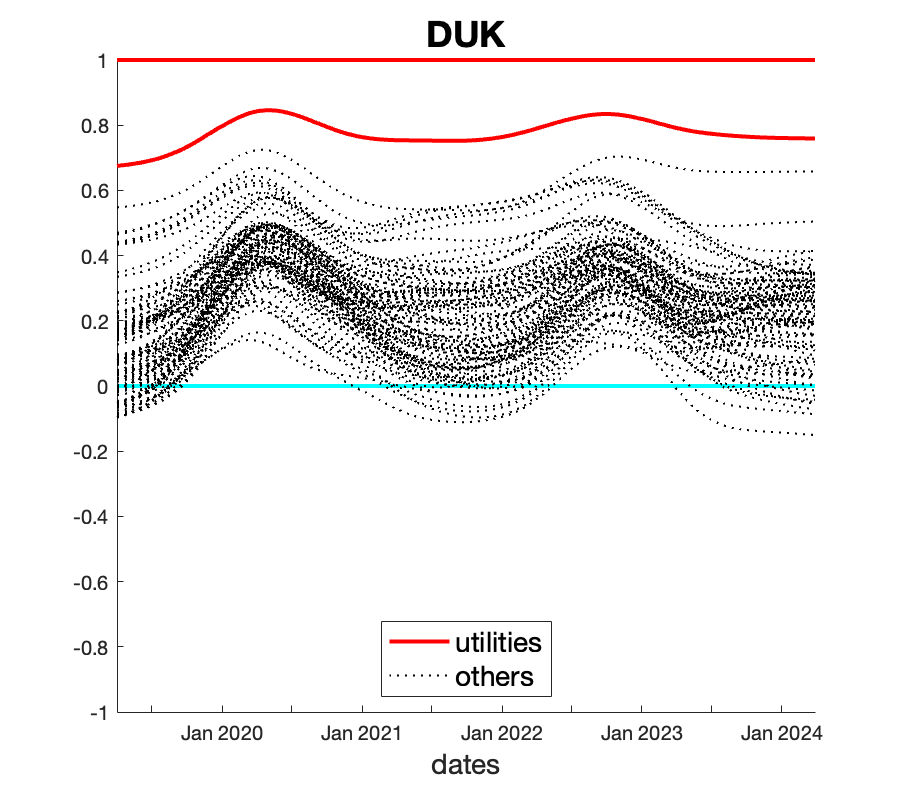}
\includegraphics[width=0.24\textwidth]{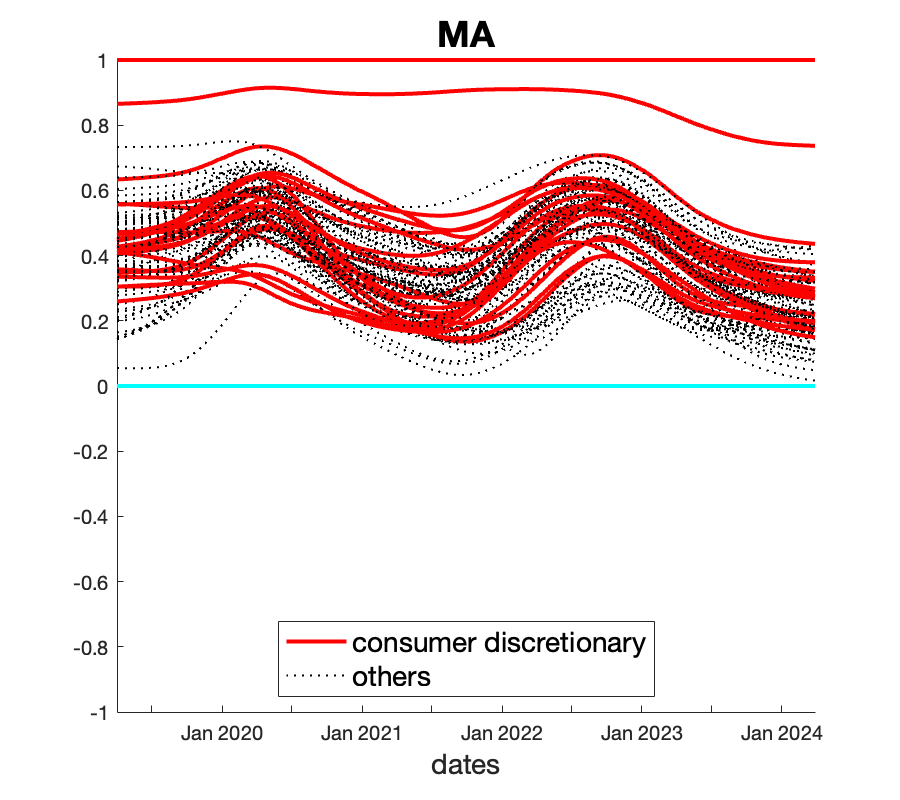}
\includegraphics[width=0.24\textwidth]{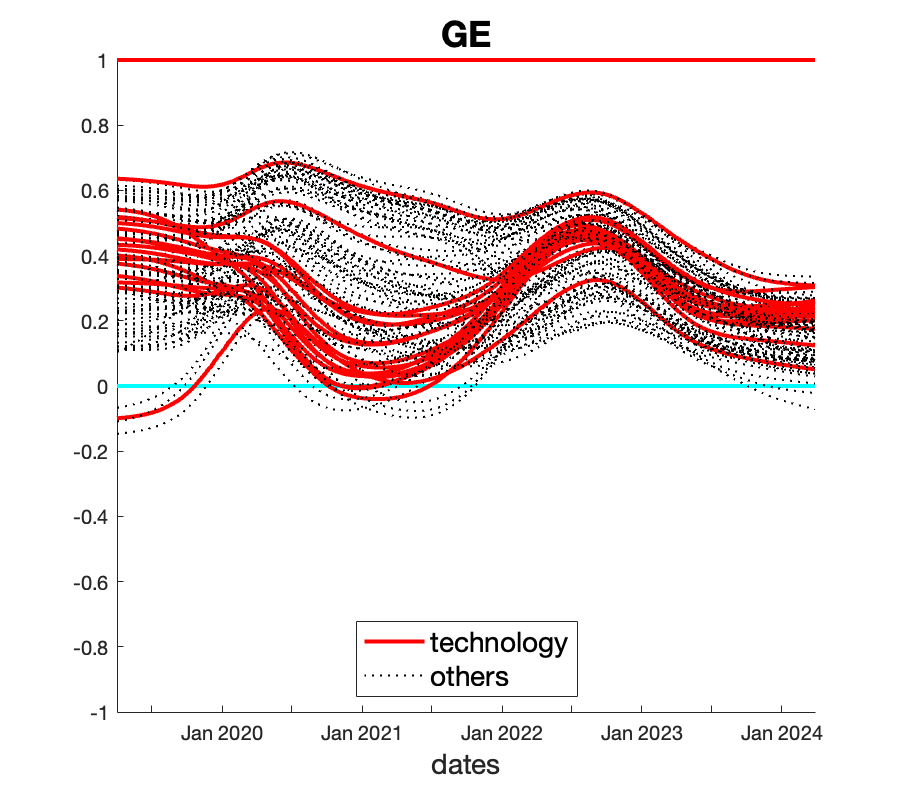}
\includegraphics[width=0.24\textwidth]{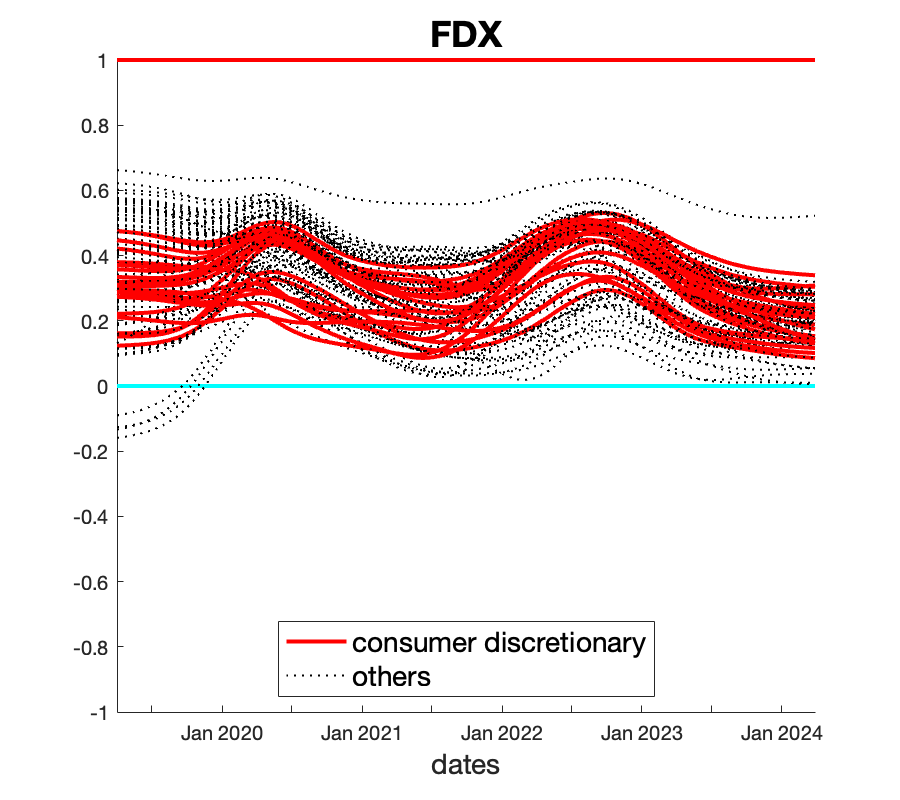}
\includegraphics[width=0.24\textwidth]{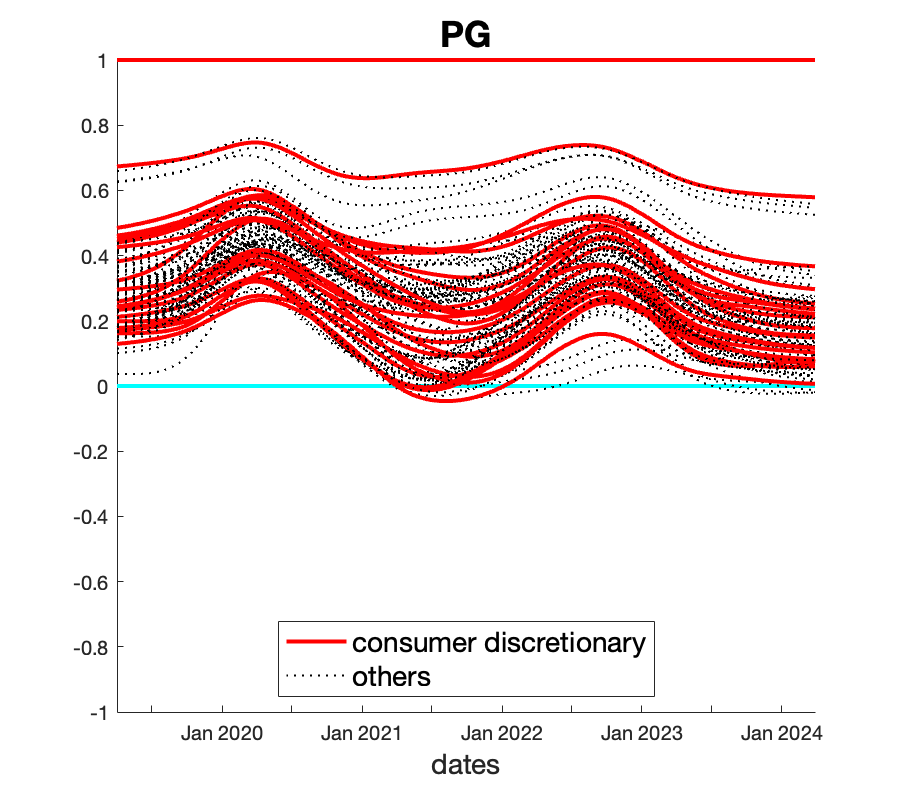}
\includegraphics[width=0.24\textwidth]{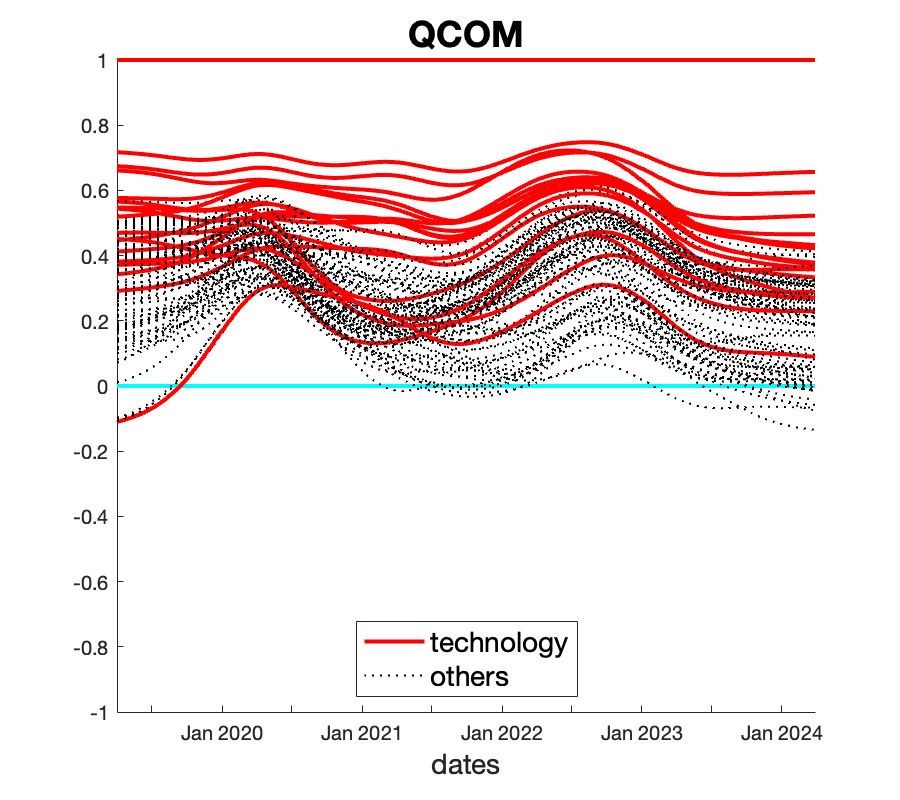}
\includegraphics[width=0.24\textwidth]{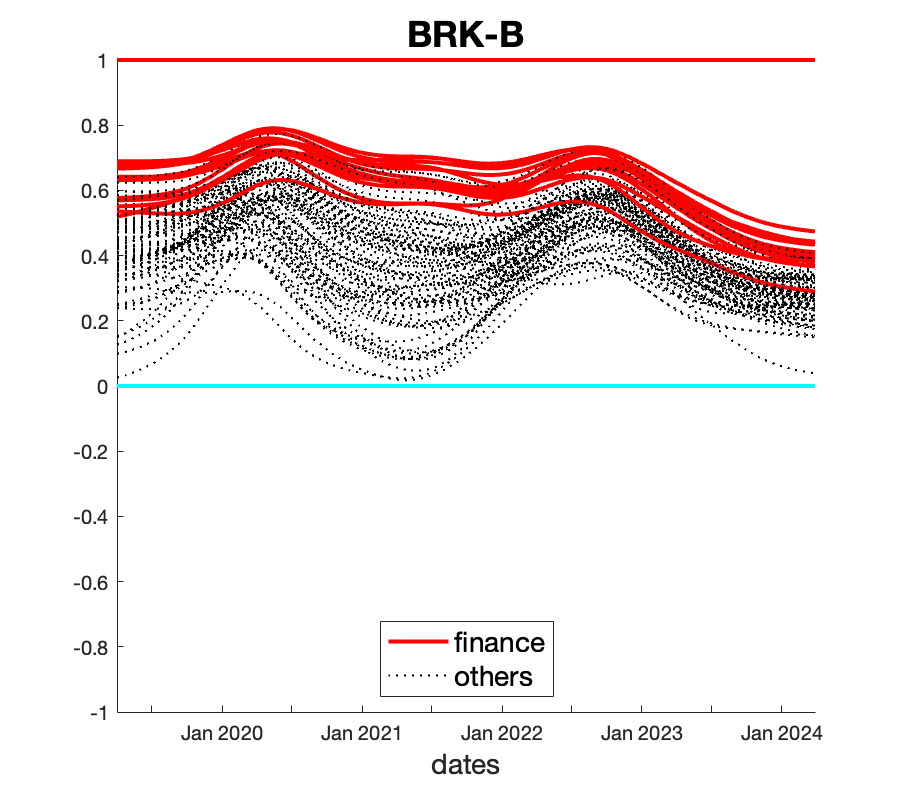}
\caption{Some examples of the estimated coefficients of correlation between a company and all the others over dates. Red curves indicate the correlation curves for those in the same sector.}
\label{fig_6_4_4}
\end{figure}

\begin{table}[!htb]
\small
\begin{center}
\begin{tabular}{ |c|c| } 
  \hline
  Sectors & Companies \\ \hline \hline 
  Consumer Discretionary & \makecell{ ACN, AMZN, BKNG, CL, COST, DIS, F, \\ FDX, GM, HD, LOW, MA, MCD, NFLX, NKE, \\ PG, PYPL, SBUX, TGT, TSLA, V, WMT }  \\ 
  \hline
  Consumer Staples & CVS, KHC, KO, MDLZ, PEP \\ 
  \hline
  Energy & COP, CVX, XOM \\
  \hline
  Finance & \makecell{ AIG, AXP, BAC, BK, BLK, BRK-B, C, \\ COF, GS, JPM, MET, MS, SCHW, USB, WFC } \\ \hline
  Health Care & \makecell{ ABBV, ABT, AMGN, BMY, GILD, JNJ, \\ LLY, MDT, MMM, MO, MRK, PFE, PM, UNH } \\ \hline
  Industrials & \makecell{ BA, CAT, DE, DHR, DOW, GD, HON, \\ LIN, LMT, RTX, TMO, UNP, UPS } \\ \hline
  Real Estate & AMT, SPG \\ \hline
  Technology & \makecell{ AAPL, ADBE, AMD, AVGO, CRM, EMR, \\ GE, GOOG, IBM, INTC, INTU, META, \\ MSFT, NEE, NVDA, ORCL, QCOM, TXN } \\ \hline
  Telecommunications & CHTR, CMCSA, CSCO, T, TMUS, VZ \\ \hline
  Utilities & DUK, SO \\ \hline
\end{tabular}
\caption{List of S\&P 100 companies in https://www.barchart.com/stocks/indices/sp/sp100, as of December 1st, 2024.}
\label{table_6}
\end{center}
\end{table}

\end{document}